\documentclass{template/uiucthesis2021}
\usepackage[utf8]{inputenc}
\usepackage[english]{babel}
\usepackage{csquotes}
\usepackage{microtype}
\usepackage{amsmath,amsthm,amssymb}
\usepackage[bookmarksdepth=3,linktoc=all,colorlinks=true,urlcolor=blue,linkcolor=blue,citecolor=blue]{hyperref}
\usepackage[capitalize]{cleveref}
\usepackage[bibstyle=ieee,citestyle=numeric-comp,minnames=3,maxnames=3]{biblatex}
\usepackage{graphicx}
\usepackage{longtable}
\usepackage{multirow}
\usepackage{subcaption}
\usepackage[dvipsnames]{xcolor}
\usepackage{lineno}
\usepackage{comment}
\usepackage{lipsum}  

\newcounter{counterforappendices}

\begin{document}

\title{Measurement of Transverse Single-Spin Asymmetries in $\pi^0$ and $\eta$ Meson Production in $\sqrt{s}$ = 200 GeV $p^\uparrow+p$ Collisions with sPHENIX}
\author{Gregory Mattson}
\department{Physics}
\phdthesis
\degreeyear{2025}
\committee{
    Professor Anne Sickles, Chair\\
    Research Professor Caroline Riedl, Director of Research\\
    Professor Matthias Gro\ss{}e Perdekamp\\
    Professor Jorge Noronha\\
    Research Professor James Eckstein}
\maketitle

\frontmatter

\begin{abstract}
The sPHENIX experiment is a next-generation collider detector at the Relativistic Heavy Ion Collider (RHIC) designed for rare jet and heavy-flavor probes of Au + Au, $p$ + Au, and polarized $p+p$ collisions. The experiment includes a large acceptance, granular electromagnetic calorimeter and very high-rate data acquisition plus trigger system. In RHIC Run-24, sPHENIX sampled 107 $\mathrm{pb}^{-1}$ of collision data with transversely polarized protons at $\sqrt{s}=200$~GeV using an efficient high-$p_T$ photon trigger. This dissertation describes the extraction of transverse single-spin asymmetries in inclusive production of $\pi^0$ and $\eta$ mesons decaying into two photons. Such observables are sensitive to multi-parton correlations in the proton, which are related to transverse-momentum dependent (TMD) effects. The new sPHENIX data set allows for significant extension of the kinematic range covered by previous RHIC mid-rapidity measurements. The results are corrected for background contributions and three different sources of systematic uncertainties are considered: the calculation method, the method of background subtraction, and contributions from possible false asymmetries due to instrumental effects. The results are presented and compared to existing measurements from the PHENIX experiment. 
\end{abstract}

\begin{dedication}
To my dad: we miss you.
\end{dedication}

\begin{acknowledgments}
My experience in research at Illinois would not have been a success without help and guidance from countless members of the faculty; research, support, and technical staff; colleagues in my experimental collaborations; and friends and family. An exhaustive list of acknowledgments would run for many pages, but I would like to give special thanks to my most important sources of support here.

First, I want to thank my advisor Caroline Riedl for her years of patient guidance and encouragement. You have taught me so much about our field and the more general ebb and flow of research work; supported me through challenges both in research and in life with understanding and compassion; and have given me an outstanding model for what an effective mentor looks like, which I will carry with me throughout my career.

Next, thank you to the UIUC Spin group: faculty members Matthias Gro\ss{}e Perdekamp and Jen-Chieh Peng, post-docs and staff scientists Vincent Andrieux and Riccardo Longo, and fellow graduate students Robert Heitz, April Townsend and Athira Vijayakumar. All of you have helped me learn about nuclear physics, and shaped the direction of my work, by sharing your abundant knowledge and keen insights in our many fruitful discussions.

Similarly, thanks to the UIUC sPHENIX group: faculty member Anne Sickles, post-doc Anthony Hodges, and fellow grad students Apurva Narde, Justin Bennett and Grace Garmire. Your regular feedback provided helpful guidance throughout my work, and valuable lessons on sharing it with a wider audience outside of spin physics.

More broadly, thanks are in order for a large number of my sPHENIX collaborators. Thanks to Ralf Seidl, Sasha Bazilevsky, Itaru Nakagawa, Devon Loomis, and Jaein Hwang, for your invaluable help as colleagues in the sPHENIX Cold QCD group; and to Blair Seidlitz, JaeBeom Park, Tim Rinn, Joe Osborn, and Chris Pinkenburg, for sharing a wealth of sPHENIX software knowledge. Special thanks to the Saclay sPHENIX group -- Nicole D'Hose, Virgile Mahaut, Dylan Neff, and Audrey Francisco -- for all your hard work making our shared analysis a success. Lastly, I would like to acknowledge the welcoming and congenial atmosphere created by the sPHENIX collaboration as a whole; it has been a delight to work with such outstanding individuals.

Finally and most importantly, I would like to thank my primary source of support: my wife, Jaspreet Kaur. Every step of the way, you've known exactly how to encourage me. You've pushed me to achieve my loftiest goals, even when they seemed most out of reach. I cannot imagine my life without you, much less completing this thesis without you. I am more grateful than I can express in words.
\end{acknowledgments}

{
    \hypersetup{linkcolor=black}  
    \tableofcontents
}

\mainmatter

\chapter{Introduction}\label{ch:intro}
The internal structure of protons and neutrons has been a key focus of nuclear physics since the first deep inelastic scattering (DIS) experiments in the late 1960s \cite{SLAC1969}. These experiments indicated that protons and neutrons, collectively termed nucleons, are not elementary particles but are instead composed of point-like constituent particles. The constituent particles, called partons, are now understood as quarks and gluons in Quantum Chromodynamics (QCD), the quantum field theory of the strong nuclear interaction. Since the details of nucleon structure are not readily calculable within QCD, however, our current understanding of this structure remains incomplete.

The proton-proton ($p+p$) collision process is an important contemporary experimental probe of nucleon structure. The measurement of several transverse single-spin asymmetries (TSSAs) in the cross sections of this process can contribute to our understanding of the spin structure of the nucleon. In particular, TSSAs shed light on how parton spin and transverse momentum correlate with nucleon spin, and on the role of multi-parton interactions in both nucleon structure and the process by which partons form QCD bound states.

The following chapters describe a measurement of TSSAs in inclusive $\pi^0$- and $\eta$-meson production performed using the sPHENIX detector at the Relativistic Heavy Ion Collider (RHIC). Chapter~\ref{ch:phys} provides an overview of the experimental and theoretical context of this work. Chapter~\ref{ch:sphenix} describes RHIC and the sPHENIX experiment. Chapter~\ref{ch:data} discusses the selection of high-quality data collected by sPHENIX and the process used to reconstruct $\pi^0$- and $\eta$-mesons from this data. Chapter~\ref{ch:asym} details the extraction of TSSAs from the data. Chapter~\ref{ch:syst} describes the determination of systematic uncertainties associated with the asymmetry measurement. Chapter~\ref{ch:results} shows the key results of the measurement, while Chapter~\ref{ch:summary} provides both a summary of the results and outlook for future improvements to the analysis.

Note that an earlier iteration of this analysis formed the basis for the first public sPHENIX result using the 2024 data set~\cite{ConfNote}. A future iteration will constitute the final sPHENIX physics publication.

\chapter{Physics Motivation}\label{ch:phys}
\section{Quantum Chromodynamics}\label{sec:QCD}

\subsection{General Features}\label{sec:QCD_general}
Quantum Chromodynamics (QCD) is the quantum field theory of the strong nuclear force, one of the four fundamental forces of nature \cite{Thompson, HalzenMartin}. The fundamental particles of QCD are \textit{quarks} and \textit{gluons}; the former are massive spin-1/2 fermions which come in six flavors (called up, down, strange, charm, bottom and top), while the latter are massless spin-1 bosons. The strong interaction is responsible for binding quarks and gluons into bound states called \textit{hadrons}, which include protons and neutrons. This interaction is also what binds protons and neutrons together into atomic nuclei. Thus the strong force is responsible for producing around 98\% of the mass in the visible universe\footnote{This accounts for only ``ordinary'' matter, which is believed to make up 5\% of the observable universe. The remaining 95\% is composed of dark matter (27\%) and dark energy (68\%)~\cite{DarkMatter}.}. Despite its ubiquitous influence, we will see that several aspects of QCD remain poorly understood as of today.

In QCD, quarks and gluons carry a fundamental charge called \textit{color}, of which there are three types. These are called ``red,'' ``blue'' and ``green'' (hence ``chromo'' in the name QCD). Interactions between objects with color charge are governed by an SU(3) gauge symmetry. Quarks carry a red, blue or green charge while antiquarks carry an antired, antiblue or antigreen charge. They can change color via interaction with gluons, which carry one color and one anticolor. Quarks and antiquarks may be combined to form color-neutral QCD bound states, collectively called hadrons. In analogy with light, where combining red, blue and green together yields neutral white light, combining a red quark, blue quark and green quark together yields a color-neutral bound state; such three-quark states are called baryons. Similarly, a quark of one color and an antiquark of the corresponding anticolor may form a color-neutral bound state; these quark-antiquark states are called mesons.

One key feature of QCD is the strength of the strong coupling constant, $\alpha_s$. The value of $\alpha_s$ depends on the interaction energy. This is a common feature among quantum field theories, including Quantum Electrodynamics (QED), the quantum field theory of the electromagnetic force. However unlike in QED where the coupling constant $\alpha$ increases with increasing interaction energy, in QCD, $\alpha_s$ \textit{decreases}. As the interaction energy increases (or equivalently as the distance scale decreases), quarks and gluons become closer and closer to behaving as free, non-interacting particles. This property is known as \textit{asymptotic freedom}. A crucial consequence of asymptotic freedom is that at low energies (or large distance scales), $\alpha_s$ is large and successive powers of $\alpha_s$ do not converge to zero quickly. Therefore perturbative methods are limited to energies significantly larger than the characteristic QCD scale $\Lambda_\mathrm{QCD} \approx 200$ MeV. As the scale of interactions within the nucleon is given by its mass of $\sim 1$ GeV, perturbative QCD (pQCD) cannot be applied to the study of nucleon structure. As a matter of terminology, we say that such low-energy, long-distance interactions are \textit{soft} interactions, while high-energy interactions are described as \textit{hard} interactions.

Another distinguishing feature of QCD is that isolated objects with non-neutral color are prohibited. This is closely related to asymptotic freedom and may be understood intuitively by the following thought experiment. Consider attempting to pull apart the quark and antiquark in a meson. As the distance between them increases, so does the coupling constant and therefore the energy in the gluon field holding them together. Eventually, the gluon field will have enough energy to produce a new quark-antiquark pair from vacuum. The new quark will then pair together with the original antiquark; the new antiquark will pair with the original quark. The result is a pair of mesons, rather than an isolated quark and antiquark. This property is called \textit{color confinement}; the process by which individual quarks and gluons form hadrons is known variably as \textit{hadronization} or \textit{fragmentation}. The preceding thought experiment is not a rigorous proof of color confinement, but does serve to illustrate the concept intuitively. As a result of confinement, despite more than five decades of experiments studying the strong force, no isolated single quarks or gluons have ever been observed. In experiment, we may only access information on QCD interactions through their hadronic initial and final states.

\subsection{Non-Perturbative QCD and Cross Sections}\label{sec:QCD_xsections}

As noted above, high-energy scattering cross sections involving nucleon beams or targets are not directly calculable using pQCD due to soft initial- and final-state interactions. However, these cross sections may be described using the framework of QCD factorization theorems \cite{QCD_factorization}. In this framework, cross sections are separated into a perturbatively-calculable hard scattering portion, and non-perturbative portions of two types. The first type encodes information on the partonic structure of the nucleon in objects called parton distribution functions (PDFs) \cite{PDFs_whitepaper}. The second type, relevant to scattering processes with a hadronic final state, encodes information on the fragmentation process in objects known as fragmentation functions (FFs) \cite{FFs}. An example of a cross section decomposed in this way can be found in Equation~\ref{eq:CT3_xsec}.

To understand the nature of PDFs and FFs, we begin by considering the nucleon as a bound state of partons. This includes three ``valence'' quarks, as well as a ``sea'' of virtual gluons and quark-antiquark pairs. The valence quarks determine the electric charge and quantum numbers of the nucleon according to the quark model of Gell-Mann and Zweig. The sea quarks and gluons do not contribute to the nucleon quantum numbers, but they do carry momentum, spin, and orbital angular momentum.

In the simplest model for nucleon structure, called the (naive) quark parton model (QPM), the partons are viewed as free, non-interacting point particles \cite{Thompson, Longo}. Note this approximation has some basis for validity in the limit of highly relativistic nucleons, due to the asymptotic freedom of QCD. In the center-of-mass frame, the nucleon travels in the longitudinal direction and its partons are assumed to move collinearly. Thus the interacting parton in a hard scattering process carries only longitudinal momentum and is described completely by the fraction of nucleon momentum it carries, denoted by $x$ and called the Bjorken-$x$ variable. The nucleon structure can then be encoded in parton distribution functions $f^q(x)$ with the following interpretation: $f^q(x)dx$ is the number of partons of flavor $q$ carrying longitudinal momentum fraction in $[x, x + dx]$. The flavors $q$ refer to the six known quark flavors as well as the gluon. Thus PDFs describe the momentum distribution of partons within the nucleon. Analogously, fragmentation functions describe the momentum distribution of final-state hadrons produced by the fragmentation of a parton participating in the hard scattering. The FF $D^{h/q}(z)$ can be interpreted as the probability density for a parton of flavor $q$ fragmenting into a hadron $h$ carrying a fraction $z$ of the parton's momentum.

Note that this description of PDFs and FFs is spin-independent. For a more complete description, we must account for the spin states of the nucleon, the interacting parton, and the final-state hadron. The nucleon, hadron, and an interacting quark may be unpolarized, longitudinally polarized, or transversely polarized. An interacting gluon may be unpolarized, circularly polarized, or linearly polarized. The basic PDFs and FFs given above are for the unpolarized case. A set of directly analogous PDFs and FFs are defined for the remaining combinations of nucleon, parton and hadron polarizations; see Figures~\ref{fig:TMDs} and \ref{fig:CT3_table} for examples.

The QPM is adequate to explain some features of the cross sections for high-energy nucleon scattering processes, but fails to predict other features. For a more complete picture, we must consider the soft-scale interactions between partons. The simplest such extension to the QPM is known as the QCD-improved parton model. In this model, partons are still assumed to move collinearly within the nucleon, but their soft interactions are considered at leading twist in the QCD Operator Product Expansion. PDFs and FFs may be given formal field-theoretic definitions and, unlike in the QPM, depend on the scale $Q^2$ of the hard scattering interaction. Several theoretical schemes exist for translating a PDF from one $Q^2$ to another, known as $Q^2$-evolution; the most prominent of these are the Dokshitzer–Gribov–Lipatov–Altarelli–Parisi (DGLAP) equations \cite{Thompson}. However, at leading twist, these functions retain their straightforward probabilistic interpretations.

The QCD-improved parton model is successful in describing key features of the spin-independent, or unpolarized, cross sections for high-energy scattering processes; particularly the scale-dependence of PDFs is well-described. However, the leading-twist, collinear QCD analysis disagrees with experimental results for the case of polarized nucleon scattering processes. These experimental results are described in the following section. We subsequently return to the theoretical interpretation of the experimental data in Section~\ref{sec:QCD_extensions}, which describes two contemporary extensions to the collinear leading-twist QCD factorization framework.

\section{Transverse Spin-Dependent Asymmetries in Experiments}
For the last four decades, experimental and theoretical interest in the transverse spin structure of the nucleon have been driven in large part by the experimental observation of transverse spin-dependent asymmetries (TSAs)\footnote{Here ``transverse'' is defined as perpendicular to the longitudinal direction, i.e. perpendicular to the beamline.}. TSAs are asymmetries in the azimuthal distributions of final-state particles. In the case of scattering processes involving one polarized probe, such asymmetries are called transverse single-spin asymmetries (TSSAs) and denoted by $A_N$. These asymmetries are the focus of this thesis. Although beyond the scope of this work, note that in scattering processes between two polarized nucleons or one polarized nucleon and one polarized lepton, one may also measure double-spin asymmetries. We further note that while this work focuses on the case of transverse nucleon polarization, analogous longitudinal single- and double-spin asymmetries are an important and complementary area of contemporary study.

\begin{figure}
    \centering
    \includegraphics[width=0.8\textwidth]{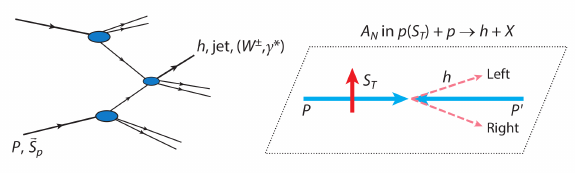}
    \caption{Definition of the left-right asymmetry $A_N$ in $p^\uparrow+p$ collisions. A transversely polarized proton (spin $S_T$) colliding with an unpolarized proton can produce a hadron ($h$), jet, or electroweak boson (left). The distribution of these products is asymmetric to the left and right of the transverse spin direction (right). From \cite{Perdekamp2015TransverseSS}.}
    \label{fig:pp_AN_diagram}
\end{figure}

Transverse single-spin asymmetries were first observed in polarized proton-proton scattering experiments at Argonne National Laboratory in the late 1970s \cite{ArgonneTSA}. These experiments found that the distribution of charged pions produced in the process $p^{\uparrow} + p \rightarrow \pi^{\pm} + X$ is asymmetric to the left and to the right of the plane spanned by the proton's momentum and spin directions (see Fig. \ref{fig:pp_AN_diagram}). The asymmetry may be defined as
\begin{equation}
    A_N = \frac{\mathrm{d}\sigma_L - \mathrm{d}\sigma_R}{\mathrm{d}\sigma_L + \mathrm{d}\sigma_R},
\end{equation}
where $\mathrm{d}\sigma_{L(R)}$ denotes the portion of the cross section to the left (right) of the proton's spin. Therefore $A_N$ is often referred to as a left-right asymmetry. Equivalently, and more formally, $A_N$ may be defined in terms of the two antiparallel spin states of the proton:
\begin{equation}\label{eq:AN_def}
    A_N \equiv \frac{\mathrm{d}\sigma(S_T) - \mathrm{d}\sigma(-S_T)}{\mathrm{d}\sigma(S_T) + \mathrm{d}\sigma(-S_T)},
\end{equation}
where $\mathrm{d}\sigma(\pm S_T)$ denote the cross sections for spin-up and spin-down protons. Later experiments at both fixed-target and collider facilities have confirmed the early Argonne result \cite{Antille_pi0_AN, E704_charged_pion_AN, PHENIX_hadron_AN, BRAHMS_charged_AN, PHENIX_charged_pion_AN}. These experiments have further shown nonzero asymmetries in inclusive neutron, $\pi^0$, $K^\pm$, and $\eta$-meson production \cite{E704_pi0_AN, STAR_pi0_AN, PHENIX_A_N, RHICf_neutron_AN}. Large TSSAs have also been observed in jet\footnote{A jet is a cone of particles with momentum in approximately the same direction, resulting from the fragmentation of an initial quark or gluon.}, photon, $Z$ and $W^{\pm}$ boson production in polarized $p+p$ collisions \cite{STAR_jet_AN, STAR_jet_AN2, STAR_Z0, PHENIX_photon_AN, STAR_dijet_AN}. The measured TSSAs are largely independent of center-of-mass energy over two orders of magnitude ($\sqrt{s}$ from 6 to 500 GeV) \cite{Perdekamp2015TransverseSS, RevModPhys.85.655}.

\begin{figure}
    \centering
    \includegraphics[width=0.8\textwidth]{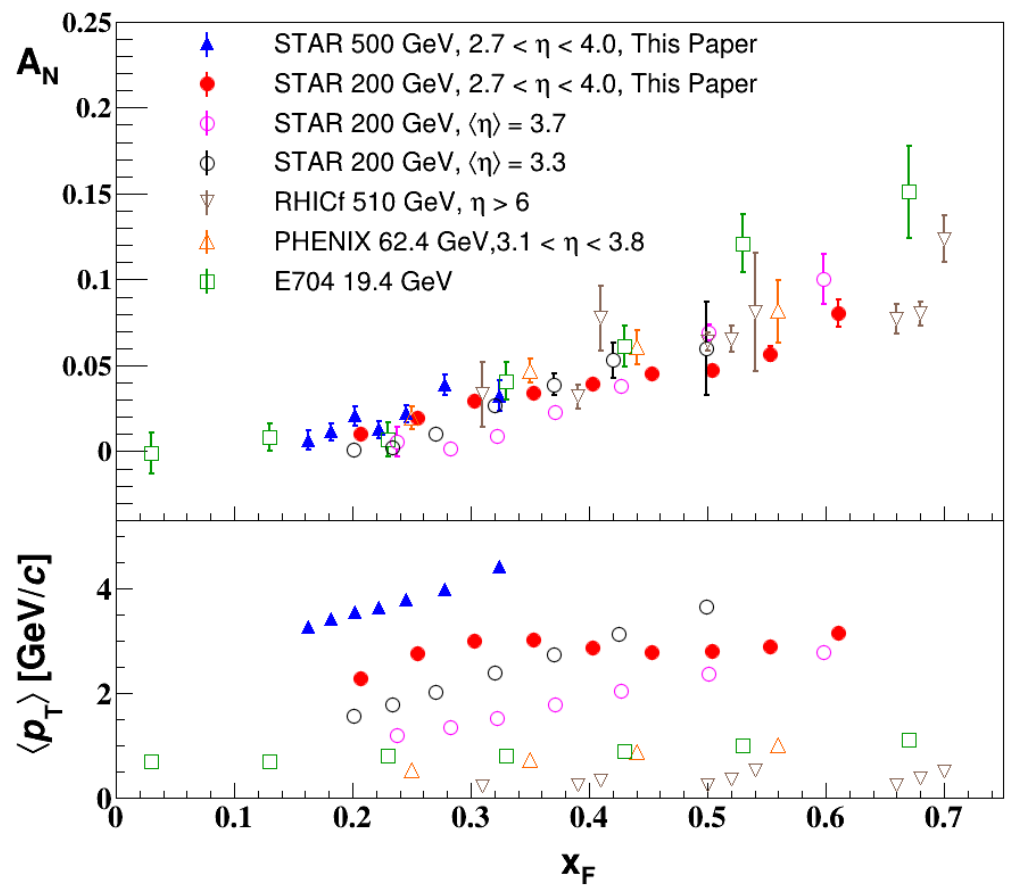}
    \caption{Overview of existing measurements of $A_N$ in the process $p^\uparrow + p \rightarrow \pi^0 + X$. The top panel shows measured $A_N$ as a function of $x_F = 2p_z/\sqrt{s}$. The bottom panel shows the average pion transverse momentum $p_T$ in each $x_F$ bin, indicating the kinematic region of each measurement. From \cite{STAR_jet_AN2}.}
    \label{fig:global_pi0}
\end{figure}

The global data for $A_N$ in inclusive $\pi^0$ production is shown in Figure~\ref{fig:global_pi0}. Here $A_N$ is shown as a function of the Feynman-$x$ variable $x_F = 2p_z/\sqrt{s}$, where $p_z$ is the component of $\pi^0$ momentum along the longitudinal direction. We see that $A_N$ increases with increasing $x_F$, corresponding to pions carrying a large (collinear) fraction of the polarized proton's momentum; at small $x_F$, $A_N$ is consistent with zero. Here we emphasize that most existing measurements of the $\pi^0$ asymmetry cover the moderate- to large-$x_F$ kinematic region. Measurements below $x_F \approx 0.1$ are sparsely populated. As we will see, the new sPHENIX measurement will contribute valuable new data in this kinematic region. The results of this measurement will be further discussed in the context of the global data in Chapter~\ref{ch:results}.

Beyond $p+p$ scattering, analogous TSSAs have also been observed in lepton-nucleon scattering. In the deep inelastic scattering (DIS) process, an incoming lepton $(\ell)$ scatters off a nucleon $(N)$, which fragments into a hadronic final state $(X)$: $\ell + N \rightarrow \ell' + X$. In semi-inclusive DIS (SIDIS), one or more final-state hadrons $(h)$ are also measured; the reaction then becomes $\ell + N \rightarrow \ell' + h + X$. The COMPASS, HERMES and JLab collaborations have measured large asymmetries in the hadron counting rates for the two antiparallel nucleon spin states, which depend on the transverse momentum of the produced hadron \cite{Perdekamp2015TransverseSS}.

The Drell-Yan (DY) process provides another example of TSSAs in hadron-hadron scattering. In DY, a quark from one initial-state hadron annihilates with an antiquark from the other to form an intermediate virtual photon $\gamma^*$ or $Z$ boson. The boson subsequently decays to a lepton-antilepton pair detected in the final state. The hadronic part of the final state, resulting from the fragmentation of the initial-state hadrons, is not measured. The reaction may be written as $h + N \rightarrow \ell + \bar{\ell} + X$, where $\ell$ and $\bar{\ell}$ are the lepton and antilepton, respectively. In DY with a transversely polarized nucleon target and unpolarized pion beam, the COMPASS collaboration has measured a non-zero TSSA in the distribution of the transverse momentum of the final-state di-lepton \cite{COMPASS_DY}.

While beyond the scope of this work, we note that azimuthal asymmetries have also been studied in di-hadron production in high-energy $e^+ e^-$ annihilation, by the Belle and BaBar experiments \cite{Perdekamp2015TransverseSS}.

The observed TSSAs represent a unique challenge to the framework of QCD factorization. In particular, the collinear, leading-twist model predicts that these asymmetries should be vanishingly small \cite{Sivers_TMD}:
\begin{equation}
    A_N \propto \alpha_s \frac{m_q}{\sqrt{s}},
\end{equation}
where $m_q$ is the quark mass. As the proton is primarily composed of up and down quarks, with masses on the order of several MeV, this model predicts TSSAs no larger than $10^{-3}$ -- in stark contrast with experimental results. Understanding the physical content of these TSSAs requires that we extend the collinear leading-twist approach. We now describe the two such extensions commonly applied to contemporary QCD factorization.

\section{Extensions to QCD Factorization}\label{sec:QCD_extensions}
The collinear, leading-twist factorization framework may be extended in two ways \cite{TSAorigins}. On one hand, we may forego the assumption of collinearity, and consider transverse degrees of freedom. Taking into account the transverse momentum of partons leads to the transverse-momentum-dependent (TMD) factorization framework. One the other hand, we may assume collinearity, but consider higher-twist effects: this leads to the collinear twist-3 (CT3) factorization framework.

The TMD scheme is applicable to processes with two scales: one hard scale $Q_1$, and one soft scale $Q_2$, with $Q_1 \gg Q_2 \sim \Lambda_\mathrm{QCD}$. These processes include SIDIS and DY, where the hard scale is determined by the momentum of the exchanged electroweak boson and the soft scale by the transverse momentum of the final-state hadron or di-lepton. The TMD scheme can also be used to study the azimuthal distribution of hadrons within jets produced in $p+p$ collisions; in this case the hard scale is provided by the jet transverse momentum, and the soft scale by the component of hadron momentum transverse to the jet axis. In contrast, the CT3 scheme is applicable to processes involving only a single hard scale $Q_1 \gg \Lambda_\mathrm{QCD}$. This includes inclusive hadron and jet production in high-energy $p+p$ collisions, where the hard scale is set by the hadron or jet transverse momentum. As such a process is the main focus of this thesis, the CT3 scheme will be described in detail. However, we begin with a brief overview of the TMD scheme, as this scheme lends itself to a more intuitive understanding of nucleon structure and fragmentation effects, and shares important connections with the CT3 scheme.

\subsection{Transverse-Momentum-Dependent Factorization}
\begin{figure}
    \centering
    \includegraphics[totalheight=0.8\textwidth]{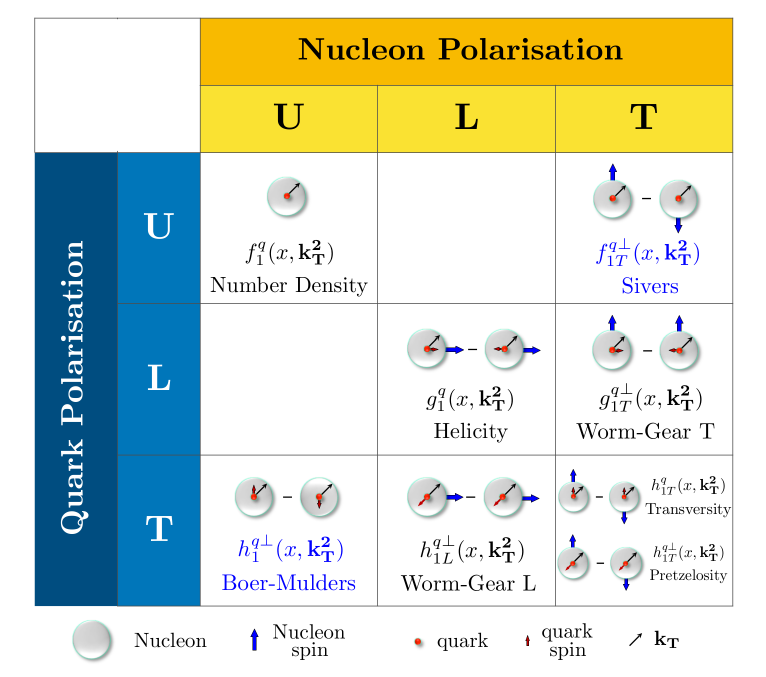}
    \caption{TMD PDFs organized according to quark and nucleon polarization. The labels U, L and T refer to unpolarized (i.e. integrated over spin), longitudinally polarized, and transversely polarized, respectively. A $\perp$ superscript in the PDF name indicates explicit dependence on quark transverse momentum $k_T$; the PDFs $f_1$, $g_1$ and $h_{1T}$ along the main diagonal are integrated over $k_T$. From \cite{Longo}.}
    \label{fig:TMDs}
\end{figure}

\begin{figure}
    \centering
    \includegraphics[width=0.8\textwidth]{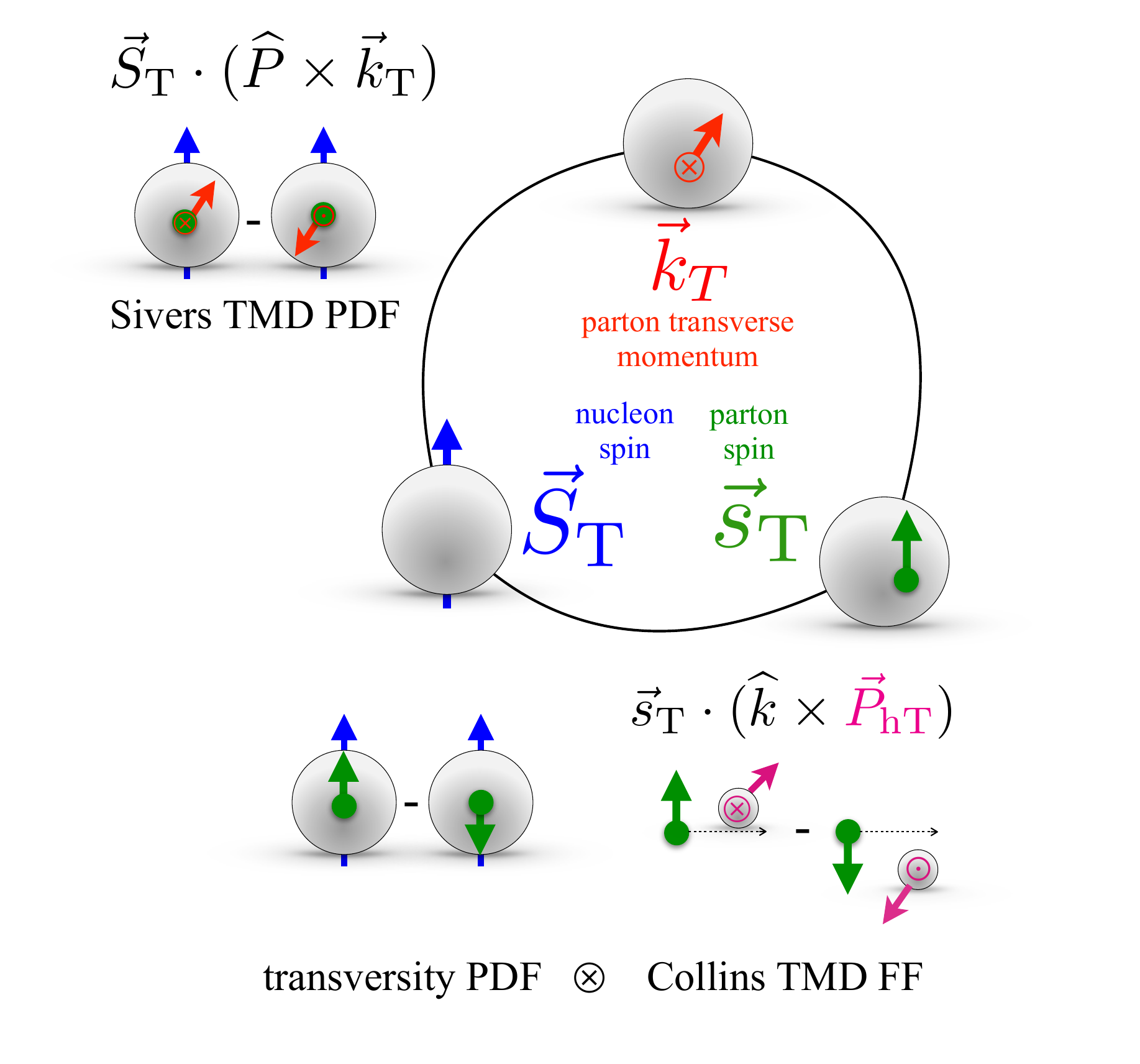}
    \caption{Diagrammatic depiction of the nucleon spin, parton spin, and parton transverse momentum correlations encoded by TMD PDFs and FFs.}
    \label{fig:TMD_correlations}
\end{figure}

In the TMD factorization scheme, PDFs depend on the longitudinal momentum fraction $x$ and hard scale $Q^2$ as before, but also the transverse momentum $k_T$ carried by the parton: $f^q(x, Q^2) \rightarrow f^q(x, Q^2, k_T^2)$. At leading twist in the QCD operator product expansion, there are eight quark TMD PDFs, summarized in Fig.~\ref{fig:TMDs}. Each of the TMD PDFs encodes a correlation between nucleon spin, parton spin, and/or parton intrinsic transverse momentum, as depicted schematically in Figure~\ref{fig:TMD_correlations}. It is these correlations that generate TSAs.

For example, the distribution $f_{1T}^{q\perp}(x, k_T^2)$ is known as the Sivers function and describes the correlation between transverse nucleon spin and parton transverse momentum. The number density PDF $f^q_1(x, k_T^2)$ may be viewed as the sum of two components: $f^q_1(x, k_T^2) = f^q_{\uparrow}(x, k_T^2) + f^q_{\downarrow}(x, k_T^2)$, where $f^q_{\uparrow, \downarrow}$ are the number densities for the two transverse nucleon spin states. The Sivers function is then the difference between these components: $f_{1T}^{q\perp}(x, k_T^2) = f^q_{\uparrow}(x, k_T^2) - f^q_{\downarrow}(x, k_T^2)$. Thus a nonzero Sivers function indicates a nonzero correlation between nucleon spin and parton transverse momentum, which may in turn be related to the orbital motion of partons within the nucleon. Note that in this discussion the $Q^2$ dependence of the PDFs has been made implicit.

The other TMD PDFs have similar interpretations. Of particular relevance to SIDIS and DY are the Boer-Mulders function, describing parton spin and transverse momentum correlations in an unpolarized nucleon; transversity, describing transverse nucleon spin and collinear transverse parton spin correlations; and pretzelosity, describing transverse nucleon spin and orthogonal transverse parton spin correlations. Finally, an analogous set of TMD FFs describe the probabilities for a quark with given flavor and polarization to fragment into a hadron of specific type, polarization, energy fraction $z$, and transverse momentum. The Collins FF $H_{1}^{h/q\perp}(z, p_{\perp}^2)$, describing the fragmentation of a transversely polarized quark $q$ into an unpolarized hadron $h$ with transverse momentum $p_{\perp}$, is relevant to SIDIS, $e^+ e^-$ annihilation, and identified hadron-in-jet production in $p+p$ collisions.

\subsection{Collinear Twist-3 Factorization}
\begin{figure}
    \centering
    \includegraphics[width=0.8\textwidth]{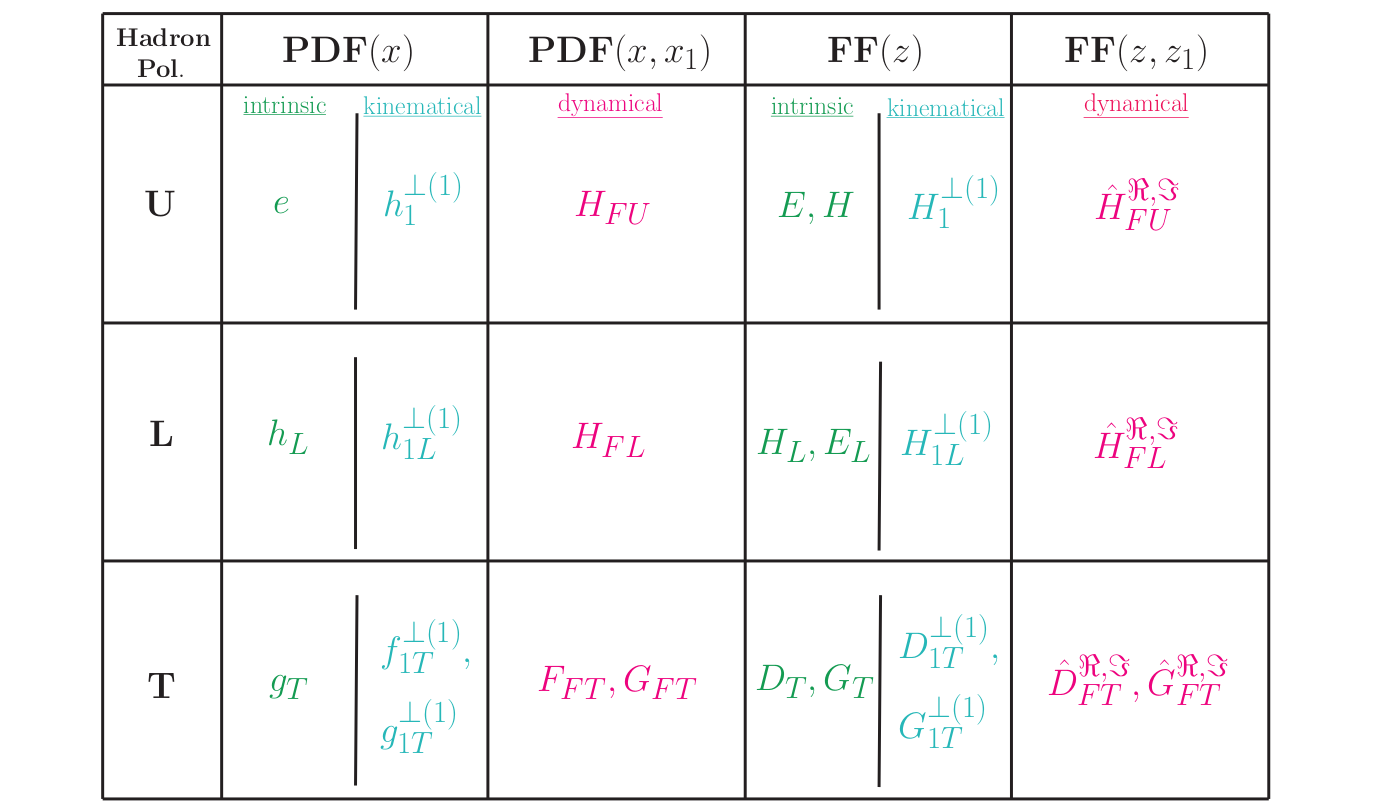}
    \caption{Summary of CT3 correlators for unpolarized (U), longitudinally polarized (L) and transversely polarized (T) hadrons. From \cite{Pitonyak_collinear_TSAs}.}
    \label{fig:CT3_table}
\end{figure}

\begin{figure}
    \centering
    \includegraphics[width=0.8\linewidth]{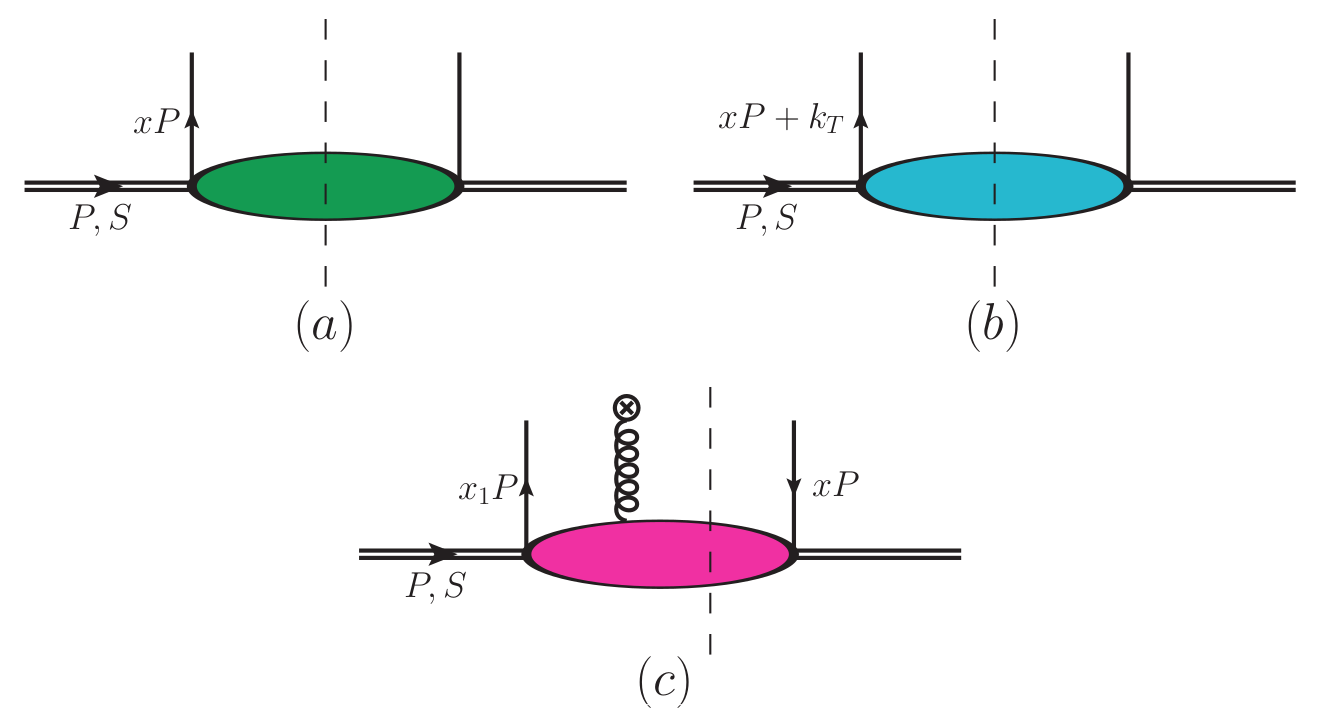}
    \includegraphics[width=0.8\linewidth]{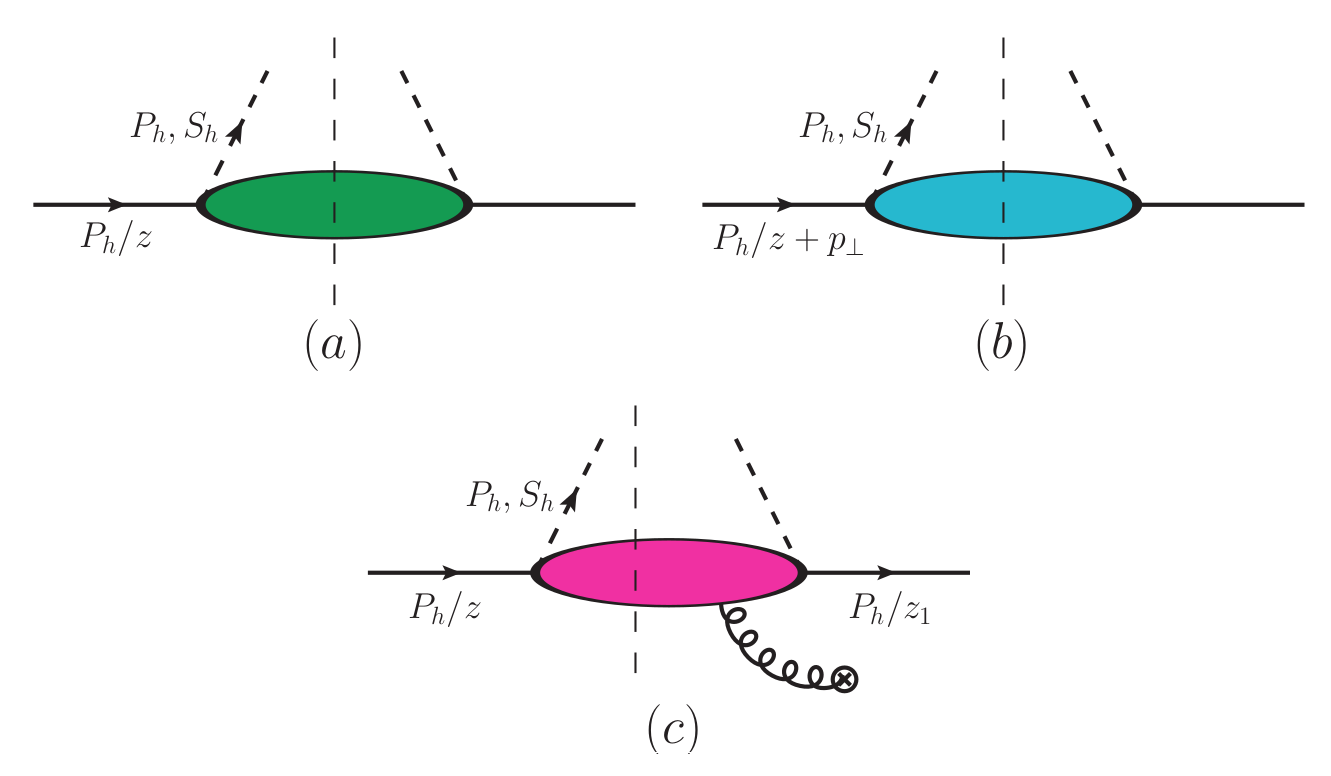}
    \caption{Feynman diagrams for CT3 PDFs (top) and FFs (bottom). Type (a) refers to intrisic correlators; type (b), to kinematical correlators; and type (c), to dynamical correlators. From \cite{Pitonyak_collinear_TSAs}.}
    \label{fig:CT3_feynman}
\end{figure}

In the CT3 factorization scheme, TSAs originate from quantum mechanical interference of multi-parton states. Thus the fundamental nonperturbative objects are multi-parton correlation functions. These may be quark-gluon-quark or tri-gluon correlators. In contrast to the TMD scheme, CT3 correlators depend only on the momentum fractions $x$ or $z$ and interaction scale $Q^2$, but now encode such multi-parton correlations. Figure~\ref{fig:CT3_table} shows a summary of these correlators. Figure~\ref{fig:CT3_feynman} shows the related Feynman diagrams for each type of correlator, using the same color scheme as Figure~\ref{fig:CT3_table}. Note that at orders above leading-twist, these correlators do not have a probabilistic interpretation as leading-twist parton distribution functions and fragmentation functions do; they are not true PDFs and FFs in the sense used up to this point. However, the terms ``PDF'' and ``FF'' are still commonly used in the context of CT3 factorization to distinguish between initial-state interactions within the nucleon and final-state effects related to fragmentation. 

CT3 correlators may be separated into three distinct types. First, so-called ``intrinsic'' correlators are twist-3 Dirac projections of collinear quark-quark correlators (Figure~\ref{fig:CT3_feynman} (a)). Next, ``kinematical'' correlators (Figure~\ref{fig:CT3_feynman} (b)) may be defined as first moments of TMD functions:
\begin{align}
    f^{\perp(1)}_1 &= \int \mathrm{d}^2 k_T \frac{\vec{k}_T^2}{2M^2} f^\perp_1(x, k_T^2) ,\\
    D^{\perp(1)}_1 &= z^2 \int \mathrm{d}^2 p_\perp \frac{\vec{p}_\perp^{\;2}}{2M_h^2} D^\perp_1(z, z^2p_\perp^2) ,
\end{align}
where $M$ is the nucleon mass and $M_h$ is the final-state hadron mass. Lastly, ``dynamical'' functions are quark-gluon-quark correlators (Figure~\ref{fig:CT3_feynman} (c)). Note that these functions depend on \textit{two} momentum fractions $x, x_1$ or $z, z_1$; these correspond to the cases of scattering off a single-parton state or a two-parton state, respectively. Further, while the dynamical PDFs are purely real, the corresponding FFs have both real and imaginary components. Finally, we comment CT3 PDFs and FFs are not all independent. Subsets of these functions may be related to each other through QCD equation of motion relations and Lorentz invariance relations.

In the CT3 scheme, the spin-dependent differential cross section for a generic inclusive hadron scattering process $A^\uparrow + B \rightarrow C + X$ may be written as \cite{Pitonyak_collinear_TSAs}
\begin{align}\label{eq:CT3_xsec}
    \begin{split}
        \mathrm{d}\sigma(S_T) = & H \otimes f_{a/A(3)} \otimes f_{b/B(2)} \otimes D_{C/c(2)} \\
        + & H' \otimes f_{a/A(2)} \otimes f_{b/B(3)} \otimes D_{C/c(2)} \\
        + & H'' \otimes f_{a/A(2)} \otimes f_{b/B(2)} \otimes D_{C/c(3)}.
    \end{split}
\end{align}
Here $H, H', H''$ are the perturbatively-calculable hard scattering terms; $f_{a/A(t)}$ is the twist-$t$ PDF associated with parton $a$ in hadron $A$ (and similarly for $f_{b/B(t)}$); $D_{C/c(t)}$ is the twist-$t$ FF associated with parton $c$ fragmenting into hadron $C$; and $\otimes$ denotes convolution in the relevant momentum fractions. Note that Equation~\ref{eq:CT3_xsec} contains three terms with one CT3 correlator each: one for the polarized hadron $A$, one for the unpolarized hadron $B$, and one for the (unpolarized) final-state hadron $C$. We now restrict the discussion to the case of (charged or neutral) pion production in $p+p$ collisions. In this case, all three terms in Equation~\ref{eq:CT3_xsec} are relevant. 

The first term, including $f_{a/A(3)}$, contains two types of contributions, called soft-gluon poles (SGPs) and soft-fermion poles (SFPs). This is because of the time-reversal symmetry of TSAs: these effects are naive time-reversal odd, resulting in a pole in the hard scattering. Consequently the momentum fraction of either a gluon or quark in the two-parton state vanishes, corresponding respectively to SGP and SFP terms. Of particular interest is the quark-gluon-quark correlator SGP term $F_{FT}(x, x)$\footnote{The notation $(x, x)$ signifies that, because the gluon momentum fraction vanishes, the single-quark state and the quark-gluon state share the same momentum $x$.}, called the Qiu-Sterman or Efremov-Teryaev-Qui-Sterman (ETQS) function. This function is the CT3 analogue to the TMD Sivers function. The two may be related through the equation
\begin{equation}\label{eq:Sivers_ETQS}
    F^q_{FT}(x, x) = \frac{1}{\pi} \int \mathrm{d}^2 k_T \frac{\vec{k}_T^2}{2M^2} f^{q\perp}_{1T}(x, k_T^2).
\end{equation}

For many years, it was believed that the ETQS function was the dominant contribution to $A_N$ in this process. However, recent evidence suggests this is not true. Equation~\ref{eq:Sivers_ETQS} indicates that there are two  independent ways to access the ETQS function: one may access it directly by measuring $A_N$ in $p^\uparrow + p \rightarrow \pi + X$ (and assuming contributions from other CT3 terms are small); or by measuring the Sivers function in SIDIS and taking the first $k_T$-moment. One would expect these two methods to agree, but calculations have found they in fact differ by a sign change \cite{Sivers_ETQS}. Further analysis of the global SIDIS and $p+p$ data corroborate this sign disagreement and suggest that the ETQS function cannot be the only source of TSSAs in inclusive pion production \cite{Small_Sivers_pp}.

Thus, we must examine the second and third terms in Equation~\ref{eq:CT3_xsec} more closely. The second term, including $f_{b/B(3)}$, was calculated over two decades ago and was found to be negligible. This agrees with experimental results from SIDIS suggesting a negligible contribution from the Boer-Mulders TMD PDF. The third term, in which the twist-3 contributions come from the fragmentation process ($D_{C/c(3)}$), contains factors of the unpolarized FFs $H, H_1^{\perp(1)}$, and $\hat{H}_{FU}^\mathfrak{I}$. These three may be related to each other through a QCD equation of motion. The kinematical correlator $H_1^{\perp(1)}$ may also be related to the Collins TMD FF:
\begin{equation}
    H^{q\perp(1)}_1 = z^2 \int \mathrm{d}^2 p_\perp \frac{\vec{p}_\perp^2}{2M_h^2} H^{q\perp}_1(z, z^2p_\perp^2).
\end{equation}
This kinematical term may be accessed by extracting the Collins FF from SIDIS or $e^+ e^-$ data. Thus the full contribution to the cross section from the fragmentation process may be evaluated given knowledge of either $H$ or $\hat{H}_{FU}^\mathfrak{I}$.

\begin{figure}
    \centering
    \includegraphics[width=0.8\textwidth]{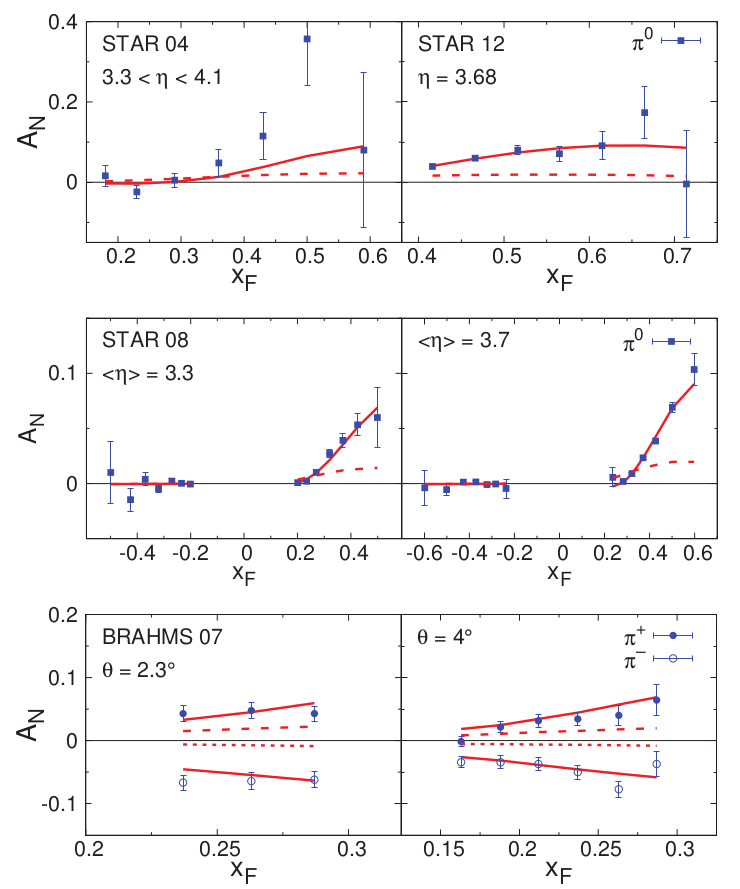}
    \caption{Inclusive pion TSSAs measured by the STAR and BRAHMS experiments, together with CT3 model calculations. The solid lines correspond to the full CT3 model; dotted or dashed lines correspond to the model excluding the $\hat{H}_{FU}^\mathfrak{I}$ contribution. From~\cite{CT3_fragmentation}.}
    \label{fig:pitonyak_ff1}
\end{figure}

\begin{figure}
    \centering
    \includegraphics[width=0.8\textwidth]{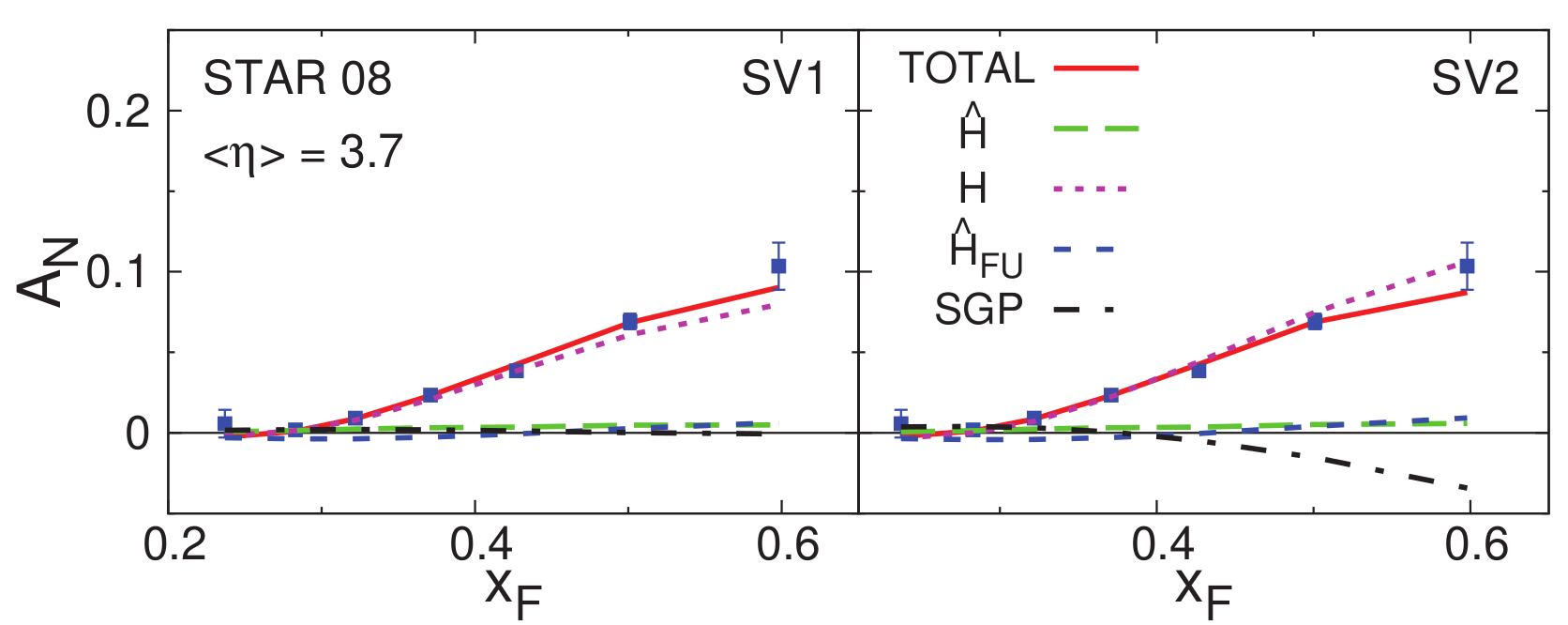}
    \caption{Individual contributions to the inclusive $\pi^0$ TSSAs, compared with data from STAR. Here SV1 (left) and SV2 (right) correspond to two different parametrizations of the Sivers function. From~\cite{CT3_fragmentation}.}
    \label{fig:pitonyak_ff2}
\end{figure}

The latter has been parametrized in terms of the twist-2 unpolarized FF $D_1$ and fit to $p+p$ data, enabling a calculation of the remaining correlator $H$ \cite{CT3_fragmentation}. The authors of Ref. \cite{CT3_fragmentation} further calculated the total contributions to $A_N$ in pion production from the fragmentation and ETQS terms (the latter fixed by extractions of the Sivers function from SIDIS data), and compared their model to $p+p$ data from the STAR and BRAHMS experiments. Their results are shown in Figures~\ref{fig:pitonyak_ff1} and \ref{fig:pitonyak_ff2}. Figure~\ref{fig:pitonyak_ff1} shows good agreement between the model and data, and that the dynamical correlator $\hat{H}_{FU}^\mathfrak{I}$ contributes significantly to the total asymmetry. Figure~\ref{fig:pitonyak_ff2} shows the individual contributions of each term, for two different model-dependent extractions of the Sivers function (labeled SGP in the figure). In one model, the ETQS contribution is negligible; in the other, it has the opposite sign compared to the data. Thus the fragmentation terms are necessary to generate the observed asymmetries. Note in Figure~\ref{fig:pitonyak_ff2} that the direct contribution from $\hat{H}_{FU}^\mathfrak{I}$ is small; however, as this correlator is used to calculate $H$ through the equation of motion relation, it is still crucial to the success of this model.

\section{Summary}
The preceding section describes the current understanding of TSSAs in the process $p^\uparrow + p \rightarrow h + X$. Theoretical considerations, together with experimental evidence, suggest that these TSSAs are the combined product of effects related to nucleon structure and those related to final-state fragmentation. The latter is at this time believed to be the dominating contribution. However, additional experimental data is necessary to better disentangle these effects. More precise measurements of $A_N$ will enable more rigorous tests of QCD predictions, while measurements in new kinematic regions will test the applicability of these predictions. The sPHENIX data analyzed in this thesis offer both of these opportunities.

\chapter{The sPHENIX Experiment}\label{ch:sphenix}
The sPHENIX experiment is the successor to the PHENIX (\textbf{P}ioneering \textbf{H}igh \textbf{E}nergy \textbf{N}uclear \textbf{I}nteraction e\textbf{X}periment) experiment at the Relativistic Heavy Ion Collider (RHIC) at Brookhaven National Laboratory (BNL). sPHENIX represents a major overhaul to the PHENIX detector and is designed for novel precision measurements in both “hot” QCD -- the physics of the quark-gluon plasma formed in heavy ion collisions -- and “cold” QCD, encompassing spin- and transverse momentum-dependent effects in polarized $p+p$ and $p+\mathrm{Au}$ collisions. Construction of sPHENIX concluded in 2022, with initial detector subsystems commissioned during the 2023 $\mathrm{Au}+\mathrm{Au}$ RHIC run. During the 2024 RHIC run, sPHENIX collected transversely-polarized $p+p$ data constituting an integrated luminosity of roughly 107 pb${}^{-1}$. The 2024 data set forms the basis for this thesis. sPHENIX will continue taking $\mathrm{Au}+\mathrm{Au}$ collision data throughout 2025, with the possibility of additional data-taking beyond 2025 before the start of construction on the future Electron-Ion Collider at BNL \cite{BUP}.

The following sections describe RHIC and the sPHENIX detector. Emphasis is placed on those aspects most relevant to this analysis: RHIC's polarized proton beam configuration and the sPHENIX Electromagnetic Calorimeter are discussed in detail. Other detector subsystems not directly used in this analysis are described briefly.

\section{The Relativistic Heavy Ion Collider}
\label{sec:RHIC}
sPHENIX takes advantage of RHIC’s unique capability to collide polarized protons and (unpolarized) nuclei of various species \cite{RHIC}. RHIC is composed of two independent storage rings that circulate protons or nuclei in opposite directions. The clockwise and counterclockwise rings are known as ``Blue'' and ``Yellow,'' respectively. The rings intersect at six interaction points, allowing the counter-rotating beams to collide at variable center-of-mass energy. Protons and nuclei are accelerated in stages: protons originate from a polarized proton source while nuclei originate from the Electron Beam Ion Source. Particles are then accelerated through multiple stages before finally being injected into the RHIC rings. In the case of heavy ions, the electron cloud is progressively stripped away at each successive stage until fully ionized nuclei enter the RHIC rings. The remainder of this section describes the components of RHIC relevant to polarized $p+p$ collisions.

\subsection{Accelerator Chain}
\begin{figure}
    \centering
    \includegraphics[width=0.8\textwidth]{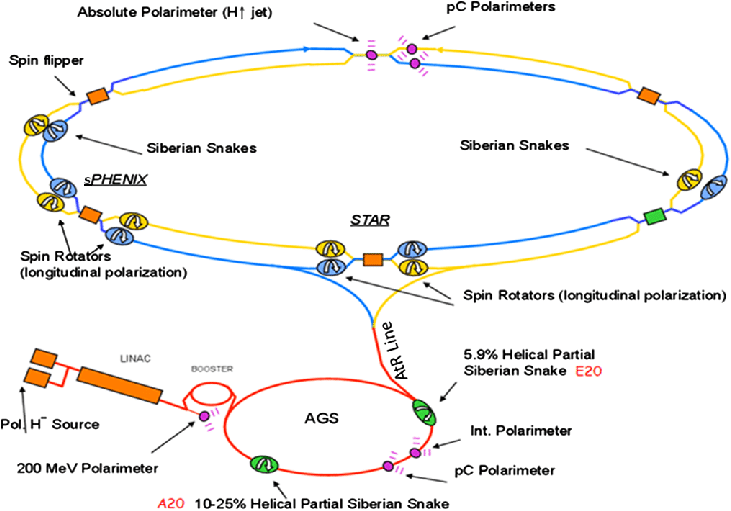}
    \caption{Diagram of the RHIC accelerator complex, including the polarized proton source, linear accelerator, Booster, Alternating Gradient Synchrotron, RHIC rings, and instrumentation for measuring and maintaining beam polarization. Adapted from~\cite{RHIC_complex}.}
    \label{fig:rhic_complex}
\end{figure}

As mentioned above, particle acceleration at RHIC is achieved through a chain of acceleration steps. The main components of this chain are shown in Figure~\ref{fig:rhic_complex}. 

Proton acceleration begins in the Optically-Pumped Polarized Ion Source (OPPIS)~\cite{OPPIS}. OPPIS produces 300 $\mu$s pulses of nuclearly polarized H${}^-$ ions of 750 keV kinetic energy through a multi-step process. First, 3 keV unpolarized protons are passed through rubidium gas in a high magnetic field to produce electronically polarized neutral hydrogen atoms. Residual H${}^+$ ions are removed by electrostatic deflector plates. A Sona transition~\cite{Sona, OPPIS_Sona} is induced in the remaining H${}^0$ atoms, transferring the electron spin to the proton. Finally, the H${}^0$ atoms pass through a sodium-jet vapor cell, acquiring one addition electron to produce H${}^-$ ions, which are further accelerated to 750 keV by a radio frequency quadrupole. The resulting pulse contains 9x10${}^{11}$ ions with polarization of approximately 80\%.

Next, the polarized H${}^-$ ions are accelerated to 200 MeV by a linear accelerator (LINAC). From here the negative ions are strip-injected as a single bunch into the Booster synchrotron. The resulting polarized protons are accelerated to 1.5 GeV in the Booster, then are injected into the Alternating Gradient Synchrotron (AGS). The AGS accelerates proton bunches to 25 GeV. Finally, the bunches are transferred to the RHIC rings via the AGS-to-RHIC (AtR) transfer line. Note that in both AGS and RHIC, the proton polarization is vertical\footnote{``Vertical'' means in the direction normal to the accelerator plane. ``Up'' means towards the sky; ``down'' means towards the ground.}. This orientation is the simplest to maintain in a circular storage ring, as the spin vector and primary bending magnetic fields align, nullifying the effect of spin precession.

\subsection{RHIC Beam Configuration}
\label{sec:rhic_beam}
RHIC accelerates protons to their final energy, which can be variably chosen. Since 2000, RHIC has collided proton beams at center-of-mass energies $\sqrt{s}$ from 62.4 to 500 GeV. The two beams, Blue and Yellow, are stored in separate beam pipes for most of their orbit around RHIC. The beam pipes merge, however, at the six interaction regions positioned around the rings. These regions are labeled according to their position, as on a clock. The sPHENIX experiment is at 8 o'clock (see Figure~\ref{fig:rhic_complex}). At each interaction region, the beams are steered into collision at the corresponding interaction point, before being again separated on their way out.

The beams in RHIC are not continuous streams of protons; rather, they are composed of individual packets of protons called \textit{bunches}. RHIC stores 120 bunches per ring, each containing approximately 1.4x10${}^{12}$ protons. Consecutive bunches are separated by 106 ns. However, only 111 bunches are filled with protons. The remaining 9 bunches are left empty, forming the ``abort gap.'' Because the density of protons drops over time as they circulate in the rings, the beams must be periodically dumped and refilled. The time period of continuous stable running between injection and dump is called a \textit{fill}. Dumping is accomplished by a kicker magnet that directs the protons out of orbit and into a beam stop. This magnet has a finite ramp time. Accordingly, turning on the magnet while filled bunches pass through it would result in an uncontrolled dump of protons between the beamline and beam stop. Instead, the magnet is ramped while the unfilled abort gap passes through, such that filled bunches are always safely directed into the beam stop.

RHIC is the first, and presently, only polarized proton collider in the world. Prior to RHIC, most experiments measuring TSSAs relied on unpolarized beams and polarized fixed targets. One drawback of this approach is that the target polarization in such experiments can generally only be alternated between the two antiparallel spin states on the time scale of hours or days\footnote{The HERMES experiment was an exception, reversing its target spin every 60-90 seconds.}. During that time, such experiments are susceptible to time-dependent changes in detector acceptance and efficiency, introducing the possibility of false asymmetries and increasing systematic uncertainties. RHIC solves this problem by setting the spin direction (up or down) of each bunch independently. The 111 filled bunches are organized into a pseudo-random pattern of spin-up and spin-down, known as the \textit{spin pattern}. At the start of each fill, one of four pre-set patterns is chosen for each beam. The patterns are not truly random, but are designed to minimize systematic effects. Thus at RHIC, the polarization direction changes on the time scale of bunch crossing, i.e. 106 ns, thereby drastically reducing time-dependent detector effects.

\subsection{Polarized Proton Considerations}\label{sec:rhic_pol}
Accelerating beams of polarized protons presents two unique challenges compared to the case of unpolarized beams. First, polarization must be maintained, as much as possible, through the multiple acceleration stages. In addition, in order to make measurements with longitudinally (along the direction of beam momentum) or radially (toward the inside or outside of the accelerator ring) polarized protons, a mechanism is needed to rotate the proton spin away from the vertical orientation used throughout the RHIC rings. Second, for meaningful physics measurements involving polarized protons it is necessary to know the magnitude of the beam polarization at the interaction point. We now describe the solutions used at RHIC to address these three problems.

During acceleration in circular accelerators, beam polarization may be lost when proton bunches pass through depolarizing resonances. At RHIC, these are compensated for by systems of magnets called ``Siberian Snakes''~\cite{SiberianSnake}. Each Siberian Snake is a helical configuration of four dipole magnets which rotate the proton spin by a fixed amount. A partial snake (9 degree rotation) is used in the AGS, while two full snakes (180 degree rotation) are used in each RHIC ring, positioned at exactly opposite sides of each ring (see Figure~\ref{fig:rhic_complex}). The axes of rotation of the two snakes are orthogonal, such that one complete rotation around RHIC returns the proton spin to its original orientation.

To allow the detector experiments at RHIC to make measurements with longitudinal or radial beam polarization, a set of spin rotators is placed around the 8 o'clock (sPHENIX) and 6 o'clock (STAR) interaction regions, as shown in Figure~\ref{fig:rhic_complex}. These spin rotators function similarly to the Siberian Snakes, but are designed to rotate the polarization by 90 degrees, and can be configured for either longitudinal or radial polarization. Each interaction region includes two spin rotators per beam: one upstream of the interaction point to rotate to the desired polarization, and one downstream to restore the polarization to vertical via an exact opposite rotation. In the 2024 RHIC run, sPHENIX took data with vertically polarized beams, and so did not use its spin rotators. STAR, however, rotated the beams to radial polarization. Accordingly, it is important for sPHENIX to verify the polarization direction at its interaction point through the use of local polarimetry, described in Section~\ref{sec:localpol}.

RHIC includes two sets of instrumentation to measure the magnitude of beam polarization it provides. The first, located near the 12 o'clock interaction region, is a $p$-carbon Coulomb Nuclear Interference (CNI) polarimeter~\cite{CNI}. This polarimeter measures the left-right asymmetry in the process $p^\uparrow + C \rightarrow p + C$ by using a thin ribbon of carbon inserted into the beam and an array of silicon strip detectors. The cross section for this process is large, allowing a precise, but uncalibrated, measurement of the beam polarization in a relatively short time. Thus, the time dependence of the polarization can be monitored over the course of a fill. 

The second polarimeter, located at 12 o'clock, is a hydrogen jet (H-Jet) polarimeter~\cite{H-jet}. This polarimeter produces a gas jet target of polarized protons, which are periodically introduced into each RHIC beam. The left-right single-spin asymmetry for elastically scattered protons is measured in two ways: once by considering the beam as polarized and averaging over the target polarization, and once taking the target as polarized and averaging over the beam polarization. The measured physics asymmetry $A_N$ for this process must agree in both cases. The measured raw asymmetry $\epsilon$ is related to the polarization $P$ and physics asymmetry by $A_N = \epsilon_\mathrm{beam}/P_\mathrm{beam} = \epsilon_\mathrm{target}/P_\mathrm{target}$ (see Section~\ref{sec:corrections} for more discussion). The polarization of the target is determined using a Breit-Rabi polarimeter. Therefore, the beam polarization can be determined as $P_\mathrm{beam} = P_\mathrm{target}\cdot\epsilon_\mathrm{beam}/\epsilon_\mathrm{target}$. The H-Jet method provides a calibrated absolute polarization, but due to the dilute gas jet used, suffers from limited statistical precision. By combining the calibrated H-Jet measurement with the high-precision CNI measurement, RHIC is able to provide accurate and precise information on the polarization of each beam.

\section{The sPHENIX Detector}
\label{sec:sPHENIX_setup}

\begin{figure}
    \centering
    \includegraphics[totalheight=0.3\textheight]{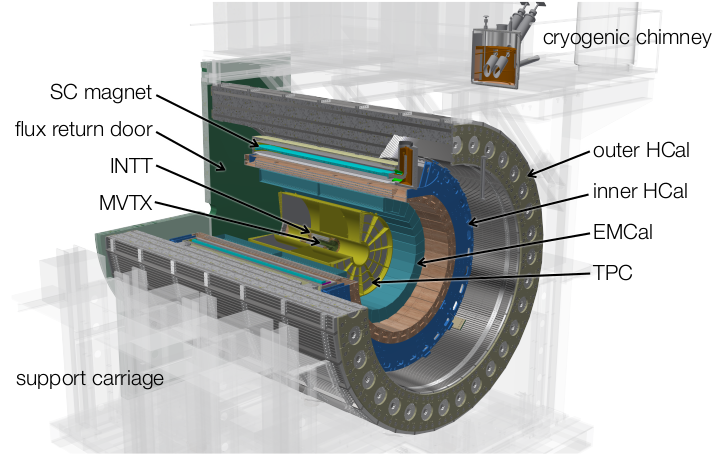}
    \caption{\label{fig:sPHENIX_detector} Overview of the sPHENIX detector. The labeled detector components are described in the following subsections.
    From~\cite{BUP}.}
\end{figure}

The sPHENIX detector is principally composed of a tracking system, electromagnetic and hadronic calorimeters, event characterization detectors, trigger and data acquisition systems, and a superconducting solenoid magnet \cite{TDR}. A schematic sketch of the detector is shown in Figure~\ref{fig:sPHENIX_detector}. These elements are used to detect charged and neutral particles in the ``barrel'' region, given by pseudorapidity\footnote{Pseudorapidity ($\eta$) is a commonly used alternative to the polar angle $\theta$. It is defined as $\eta = -\ln \left[ \tan(\theta/2) \right]$. Thus $\eta = 0$, or ``mid-rapidity,'' corresponds to the direction transverse to the beam line.} $|\eta| < 1.1$, and with full azimuthal $(\phi)$ coverage.

The superconducting solenoid is the same 1.4 T magnet formerly used by the BaBar experiment at SLAC National Accelerator Laboratory. The tracking system comprises, in order of increasing radial position: a silicon pixel micro-vertex detector (MVTX), intermediate silicon strip detector (INTT), and time projection chamber (TPC). An additional TPC Outer Tracker (TPOT) aids in calibration of the TPC. 
Calorimetry is provided by three calorimeters: the electromagnetic (EMCal), inner hadronic (iHCal), and outer hadronic (oHCal) calorimeters. An event plane detector (sEPD) is used for event characterization in heavy-ion collisions.
The sPHENIX trigger system relies on the Minimum Bias Detector (MBD), re-used from the PHENIX detector. A local polarimetry subsystem, comprising the Zero-Degree Calorimeters (ZDC) and Shower Maximum Detectors (SMD), provides information on the local beam polarization and spin orientation at sPHENIX.

The following subsections describe the detector subsystems relevant for polarized $p+p$ collisions in more detail. Exceptionally, the EMCal is described in additional detail in Section~\ref{sec:EMCal}, due to its importance in this analysis.

\subsection{MVTX}
\begin{figure}
    \centering
    \includegraphics[width=0.48\textwidth]{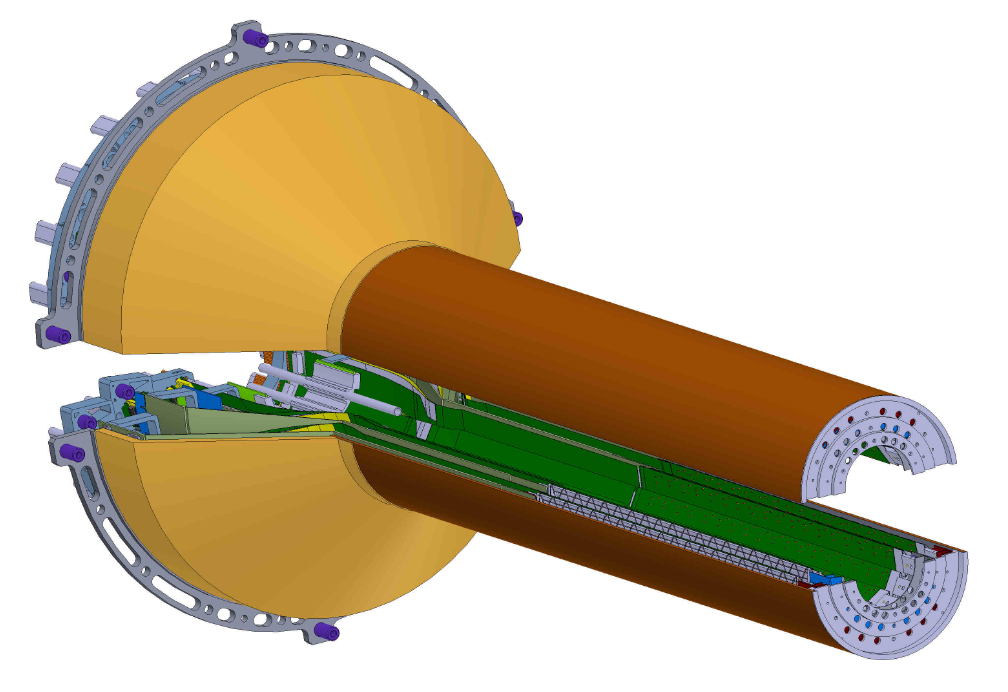}
    \includegraphics[width=0.48\textwidth]{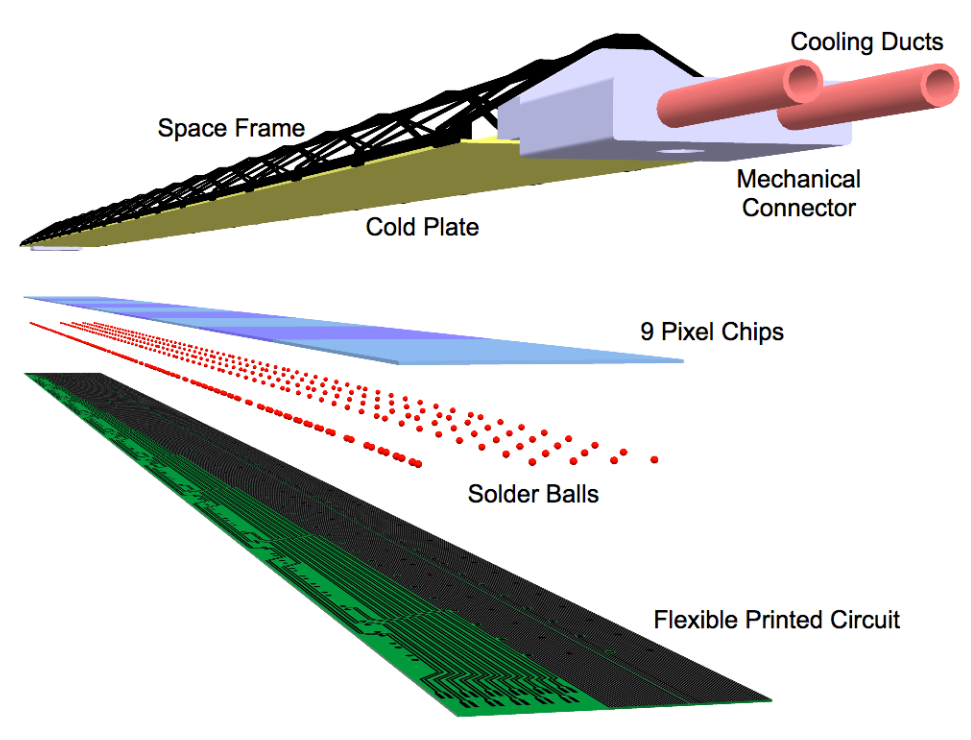}
    \caption{Left: layout of the MVTX. The cylindrical section contains the active detector area, while the conical section on the left-hand side contains cooling and electronic service components. Right: layout of an MVTX stave. From~\cite{MVTX_proposal}.}
    \label{fig:mvtx}
\end{figure}

The MAPS-based Vertex Detector (MVTX), depicted schematically in Figure~\ref{fig:mvtx}, uses Monolithic Active Pixel Sensor (MAPS) technology to provide charged particle tracking and secondary vertex reconstruction~\cite{MVTX_proposal}. It is the innermost sPHENIX detector, spanning the radial distance from 2 to 5 cm from the beam line. The MVTX comprises three cylindrical layers of silicon; each layer is segmented into flat rectangular units called ``staves'' (see Figure~\ref{fig:mvtx} right side). Each stave contains an array of nine pixel sensor chips, a polyimide circuit which provides power and readout for the pixel sensors, a cooling plate, and a carbon fiber support frame.

The pixel sensors are based on those used in the ALICE detector's Inner Tracking System~\cite{ALICE_MAPS}. Each sensor has an active surface area of 15 $\times$ 30 mm and contains 512 $\times$ 1024 individual pixels. As a charged particle passes through the sensor, charge carriers are freed in the p-doped silicon layer. The charge carriers propagate to the nearest pixel, each of which is a n-well signal diode. This results in an effective spatial resolution of about 30 $\mu\mathrm{m}$. The MVTX design also provides high-speed signal readout. Thus, differences in timing between charged particle tracks at different azimuthal angles can be used to infer the position of displaced secondary vertices, as from $D$ and $B$ meson decays.

\subsection{INTT}
\begin{figure}
    \centering
    \includegraphics[width=0.8\linewidth]{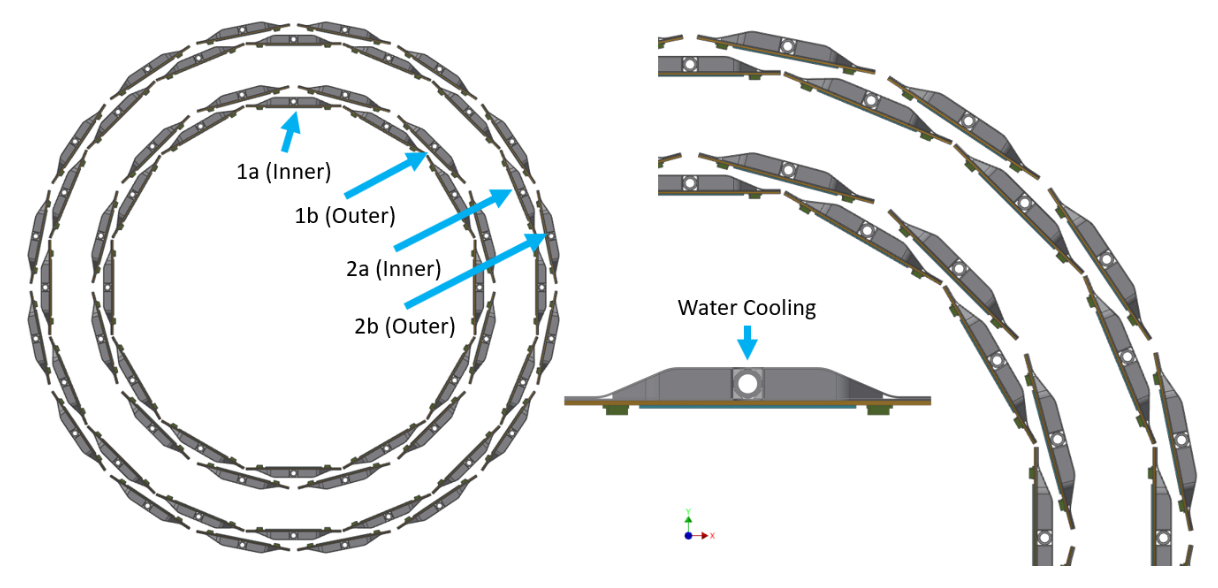}
    \caption{Cross-sectional view of the two INTT layers. The staggered sub-layers provide full azimuthal acceptance. From~\cite{TDR}.}
    \label{fig:intt}
\end{figure}

The Intermediate Tracker (INTT) is a semiconducting silicon strip detector positioned outside the radius of the MVTX~\cite{TDR}. It comprises two cylindrical layers spanning the radial distance from 7 to 10 cm from the beam line. The layers are formed from multiple rectangular units called ``ladders'' which contain the silicon strip sensors, power and readout electronics, water cooling elements, and mechanical support structure. Each layer is composed of two sub-layers which are offset from one another such that alternating ladders overlap, providing full azimuthal coverage (see Figure~\ref{fig:intt}). 

Each ladder contains two silicon modules, each read out from one end of the detector and each comprising two silicon strip sensors and one flexible High Density Interconnect (HDI) circuit board. Two types of sensor are used: Type-A, with eight sets of silicon strips of length 16 mm along the beam axis, and Type-B, with five sets of strips of length 20 mm. In both cases, the strips are separated by a pitch of 78 $\mu\mathrm{m}$. The HDI provides power, bias voltage, and slow control information. Signal readout is provided by 26 128-channel Application-Specific Integrated Circuits (ASICs) mounted on the HDI.

\subsection{TPC}
\begin{figure}
    \centering
    \includegraphics[width=0.8\textwidth]{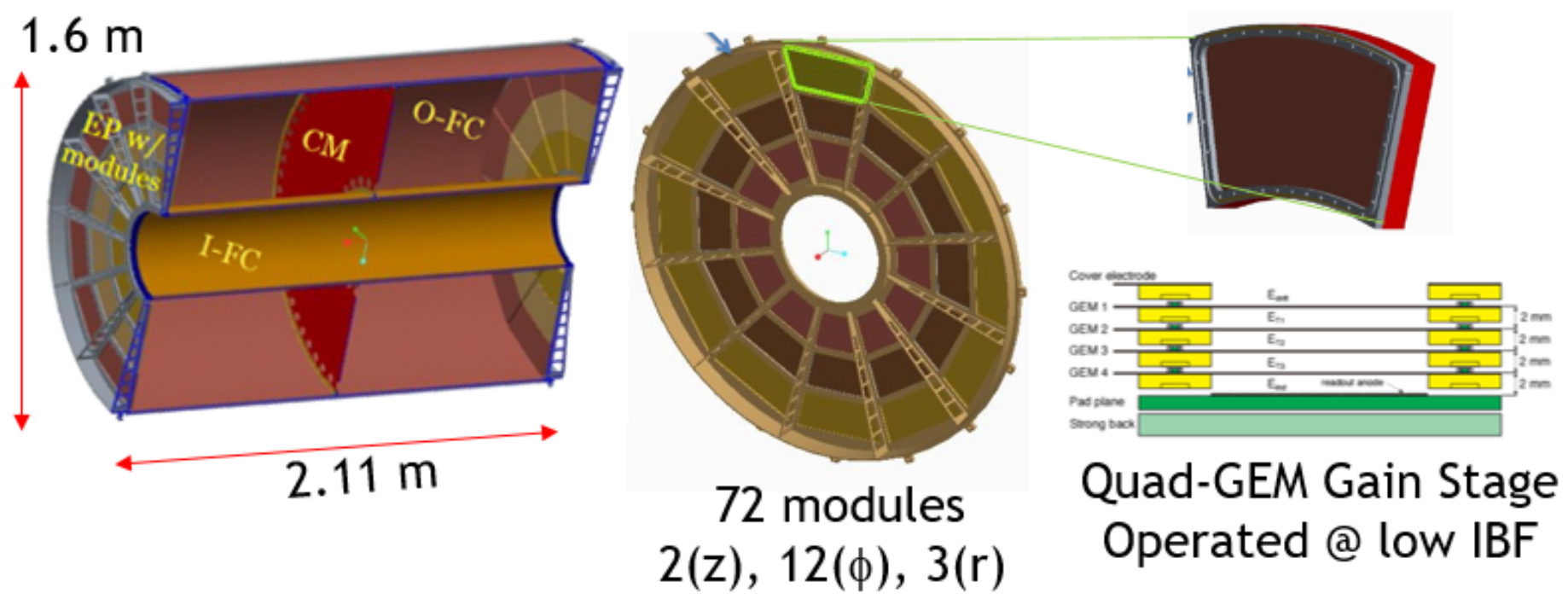}
    \caption{Left: layout of the TPC, indicating the inner field cage (I-FC), outer field cage (O-FC), end plates (EP), and central membrane (CM). Center: segmentation of the end plates into modules for readout. Right: view of the four GEM amplifiers in each module. From~\cite{TDR}.}
    \label{fig:tpc}
\end{figure}

The primary sPHENIX tracking detector is the Time Projection Chamber (TPC)~\cite{TDR}. The TPC is an annular cylinder (see left side of Figure~\ref{fig:tpc}) with inner radius of 20 cm and outer radius of 80 cm. At the center of the cylinder, an electrode called the central membrane is held at high electrical potential. At each end, an end cap (or end plate) is held at ground, creating an electric field along the direction of the beam line. The inner and outer surfaces of the cylinder are composed of field cages, comprising a set of conducting rings held at sequentially decreasing potentials; the cages are designed to maintain a uniform electric field within the TPC volume.

The volume is filled with a gaseous mixture of argon, tetrafluoromethane (CF4) and isobutane (CH(CH${}_3$)${}_3$). When charged particles pass through the detector volume, they ionize the gas molecules. The resulting free electrons drift in the electric field toward the end caps, which serve as readout planes. Each end cap is segmented into 36 modules (3 radial, 12 azimuthal), which each contain a stack of four Gas Electron Multipliers (GEMs). The GEMs produce an avalanche of secondary electrons to provide amplification of the readout signal. Each module includes an array of 32-channel ASICs, resulting in a total of 159,744 readout channels across the entire TPC. Radial and azimuthal coordinates of the drifted electrons are encoded in these channels, while the longitudinal position of the ionization can be inferred from the time delay associated with the drift of the electrons to the readout plane. Thus, a 3-dimensional image of the ionizing particle trajectory can be reconstructed.

An additional TPC Outer Tracker (TPOT) is positioned outside of the TPC and covers a small portion of the tracking system acceptance~\cite{TPOT}. It provides two additional layers of tracking to allow for calibration of the TPC, in particular, to correct for distortions in the drift field due to beam effects and accumulation of ion charge in the TPC volume.

\subsection{EMCal}
The sPHENIX electromagnetic calorimeter (EMCal) is designed to measure the energy and angular position of photons, electrons and positrons~\cite{TDR}. Measurement of photons is crucial because, as neutral particles, they do not produce signatures in the tracking system. Positioned at in inner radius of approximately 95 cm, the EMCal is a sampling spaghetti calorimeter composed of scintillating fiber and tungsten blocks~\cite{EMCalNote}. A photon or $e^{\pm}$ entering the calorimeter deposits energy in the form of an electromagnetic shower, i.e.~a cascade of secondary particles produced by bremsstrahlung and pair production interactions in the dense tungsten.

\begin{figure}
    \centering
    \includegraphics[width=0.48\textwidth]{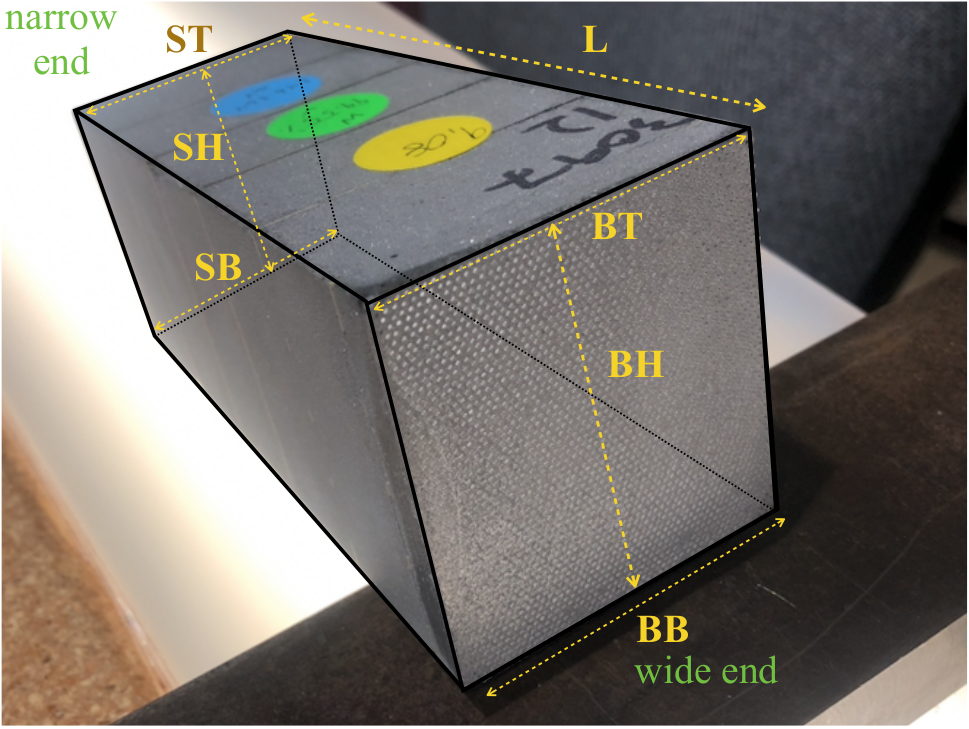}
    \includegraphics[width=0.48\textwidth]{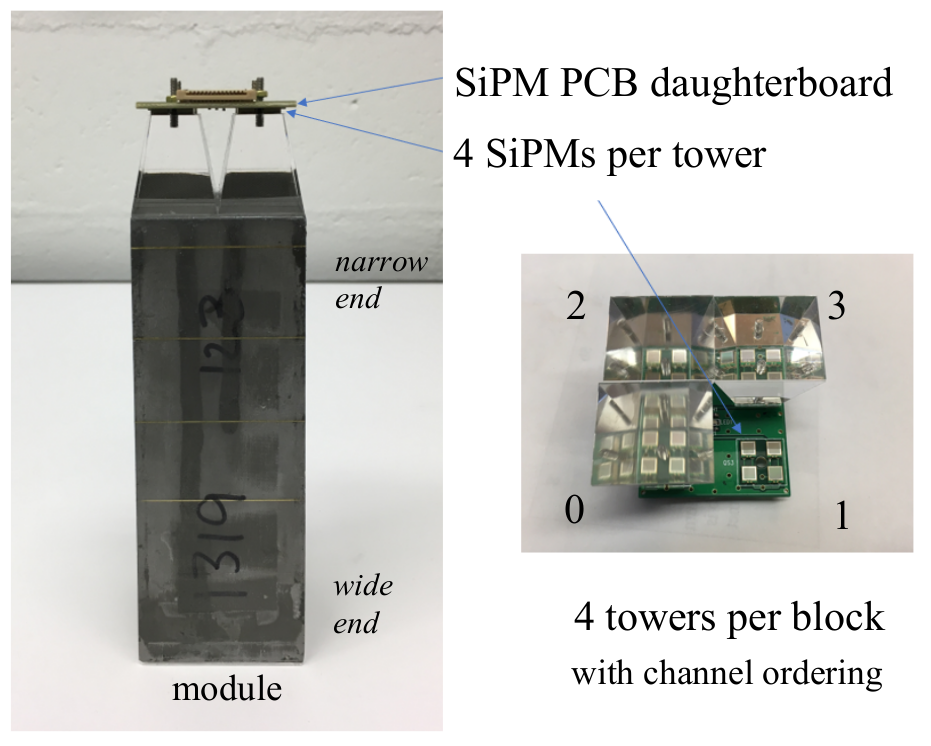}
    \caption{Left: one produced EMCal block, labeled with dimensions and with visible scintillating fibers illuminated from the rear. Right: one complete EMCal module, comprising one block, four light guides, and $4\times2\times2$ SiPMs. From~\cite{EMCalNote}.}
    \label{fig:emcal_block}
\end{figure}

Each calorimeter block contains 2,688 fibers (hence the term ``spaghetti'' calorimeter) of diameter 470 $\mu$m spaced 1 mm apart and held in place by brass meshes. The remaining block volume is filled with tungsten powder and epoxy; the left side of Figure~\ref{fig:emcal_block} shows one block with illuminated fiber ends visible. The calorimeter has a 2-D projective geometry; the blocks are arranged such that the fibers point radially approximately toward the interaction point. To accommodate this geometry blocks at different polar angle have slightly different dimensions and all blocks are tapered with the wider end facing radially outward. The average block size is approximately 16 $\times$ 5 $\times$ 5 cm.

\begin{figure}
    \centering
    \includegraphics[width=0.8\textwidth]{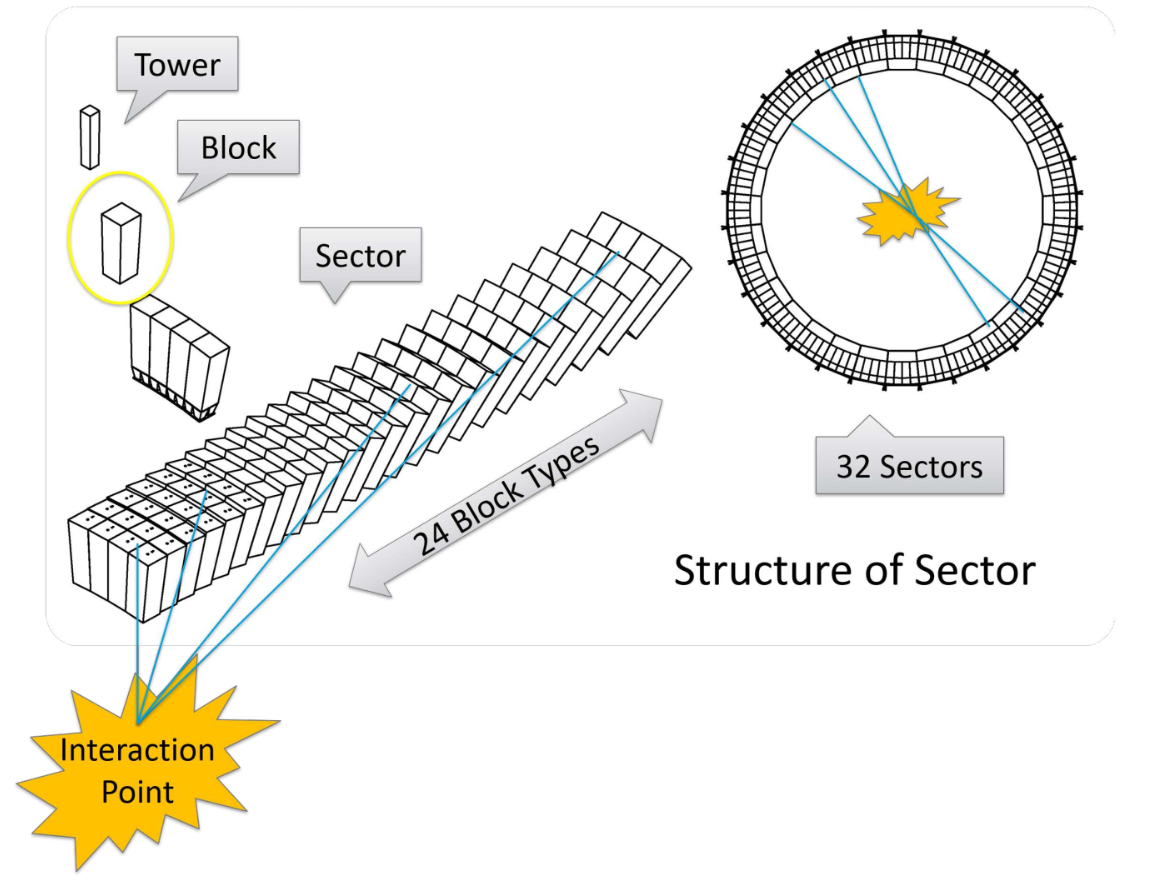}
    \caption{Schematic view of the arrangement of towers within blocks, blocks forming rows, and rows forming sectors. From~\cite{EMCalNote}.}
    \label{fig:emcal_sector}
\end{figure}

The blocks are arranged in 64 azimuthal sectors (32 each on the north and south sides of sPHENIX), each with 24 rows of 4 blocks, as illustrated in Figure~\ref{fig:emcal_sector}. Each block in turn corresponds to four readout channels, also known as towers, for a total of 24,576 channels and angular resolution of $\Delta \eta \times \Delta \phi = 0.024 \times 0.024$. Each block has four acrylic light guides to collect light from the scintillating fibers (one per tower). Each light guide is coupled to a 2$\times$2 array of silicon photomultipliers (SiPMs), as shown in the right side of Figure~\ref{fig:emcal_block}. The analog signals from the four SiPMs are summed into one channel and read out by front-end interface boards.

Prototyping for the EMCal took place at the University of Illinois Urbana-Champaign (UIUC) Nuclear Physics Laboratory (NPL) from 2015-2019. Nearly 5,000 blocks were produced and tested for quality at NPL between 2019 and 2021. UIUC maintains a database with extensive details of each block constructed and shipped to BNL, including each block's position and channel mapping in the fully-assembled calorimeter. Additional blocks were produced at Fudan University in Shanghai, China and the China Institute of Atomic Energy and Peking University in Beijing, China.


\subsection{HCal}
\begin{figure}
    \centering
    \includegraphics[width=0.8\textwidth]{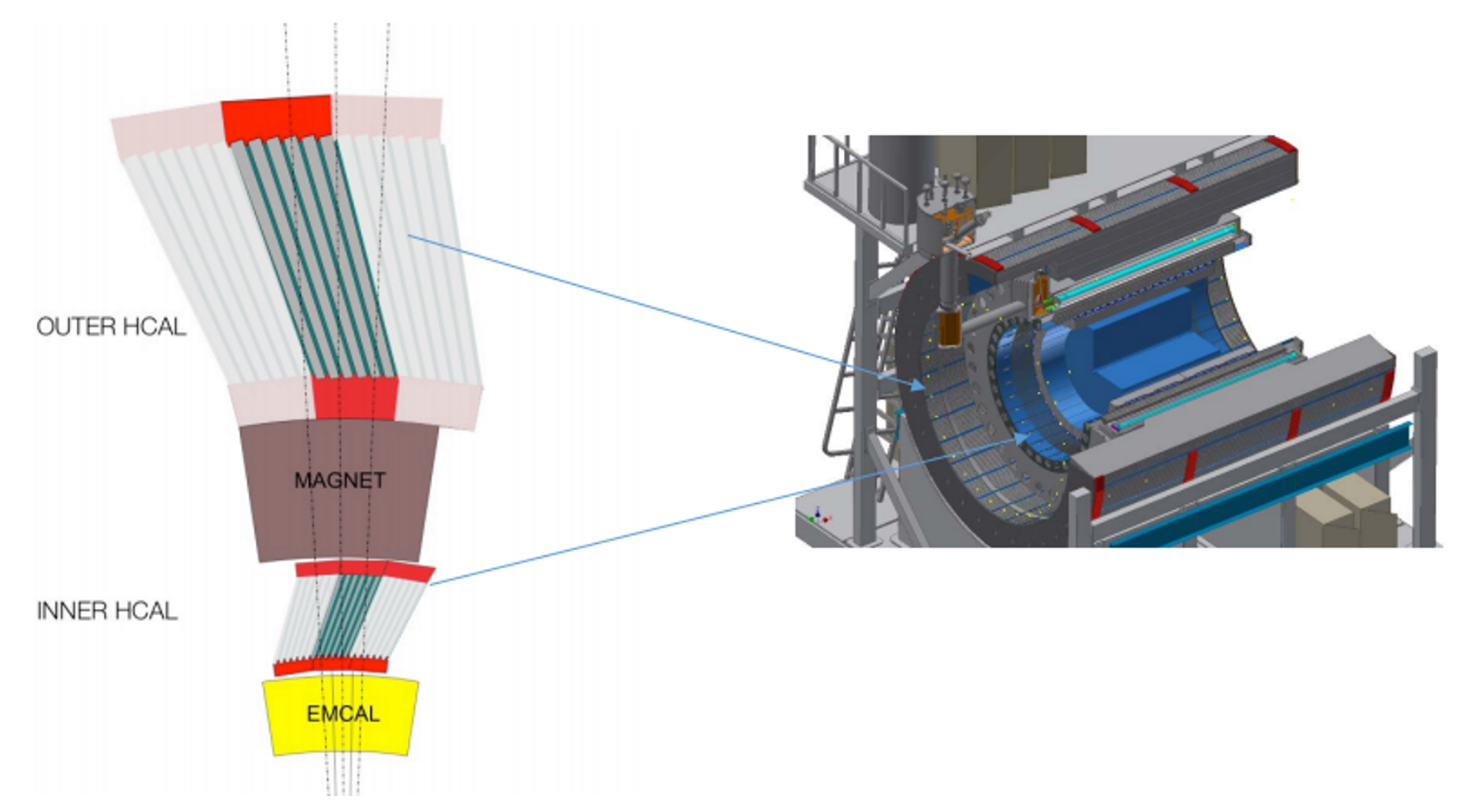}
    \caption{Cross sectional diagram of the sPHENIX calorimeters, showing the radial location of the two HCal compartments and their tilted plate structure. From~\cite{HCalNote}.}
    \label{fig:hcal}
\end{figure}

The sPHENIX Hadronic Calorimeter system (HCal) is designed to measure the energy of hadrons which do not decay within the inner detector systems~\cite{TDR}. As shown in Figure~\ref{fig:hcal}, it is composed of two compartments: the inner calorimeter (iHCal) is located between the EMCal and superconducting solenoid, while the outer calorimeter (oHCal) is positioned outside the magnet. In additional to providing calorimetry, the latter also serves as the flux return for the solenoid.

\begin{figure}
    \centering
    \includegraphics[width=0.8\textwidth]{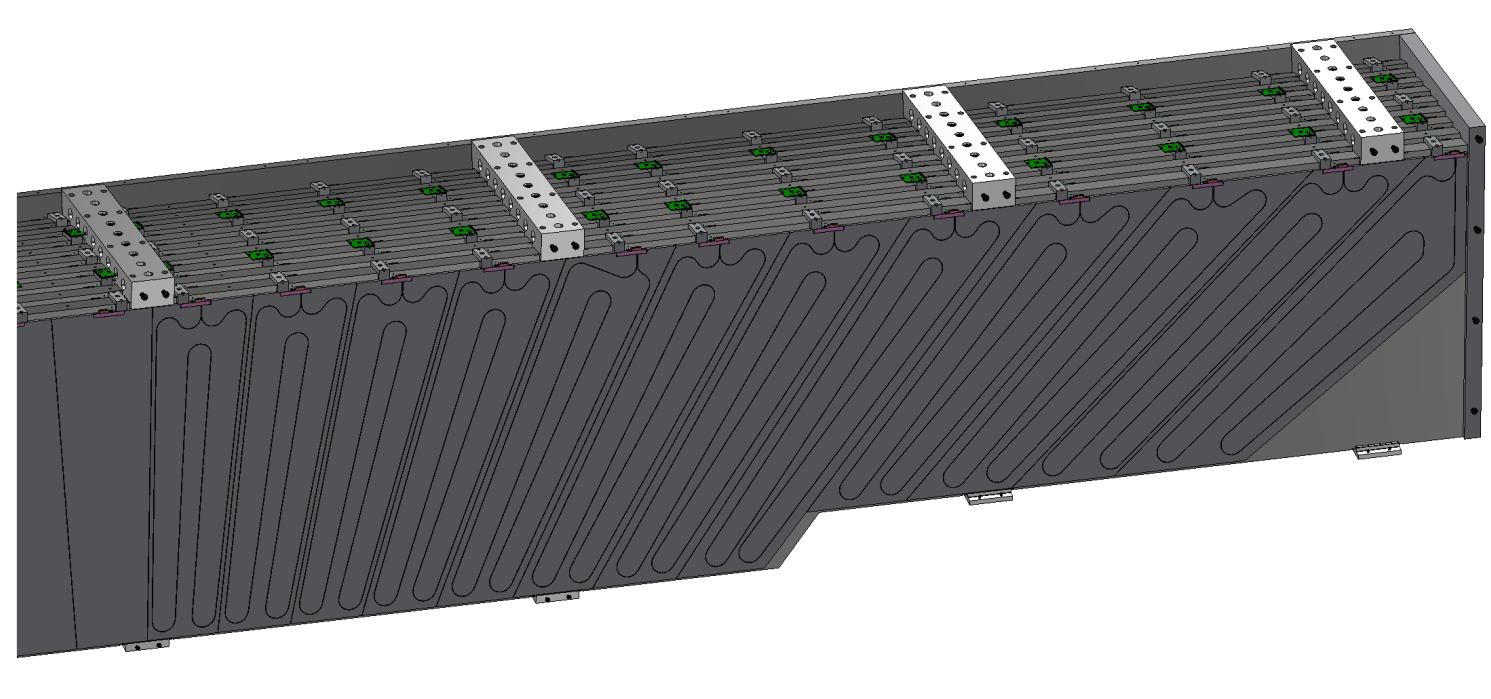}
    \caption{View of one HCal module, showing the tilt of towers along the longitudinal direction. From~\cite{TDR}.}
    \label{fig:hcal2}
\end{figure}

The two HCal compartments share the same basic design, comprising scintillating tiles embedded in thin absorber plates. The absorber material is aluminum for the iHCal and steel for the oHCal. The plates are tilted slightly from the radial direction such that a hadron track passing through the HCal will traverse at least four tiles in each compartment. Each compartment is divided into 32 modules which are further segmented into towers, again corresponding to individual readout channels. Each module includes two towers in the azimuthal direction and 24 towers in the longitudinal direction; the latter are tilted such that each tower is projective in pseudorapidity, as shown in Figure~\ref{fig:hcal2}. The effective readout segmentation is approximately $\Delta\eta\times\Delta\phi = 0.1\times0.1$. Hadrons entering the HCal volume produce showers of secondary particles in the absorber material, which deposit their energy in the scintillating tiles. The light produced is collected by flexible wavelength-shifting fibers embedded in the tiles. The light produces a signal in SiPMs, which is read out by interface boards on the outer radius of each calorimeter compartment.


\subsection{MBD}\label{sec:mbd}
\begin{figure}
    \centering
    \includegraphics[width=0.8\textwidth]{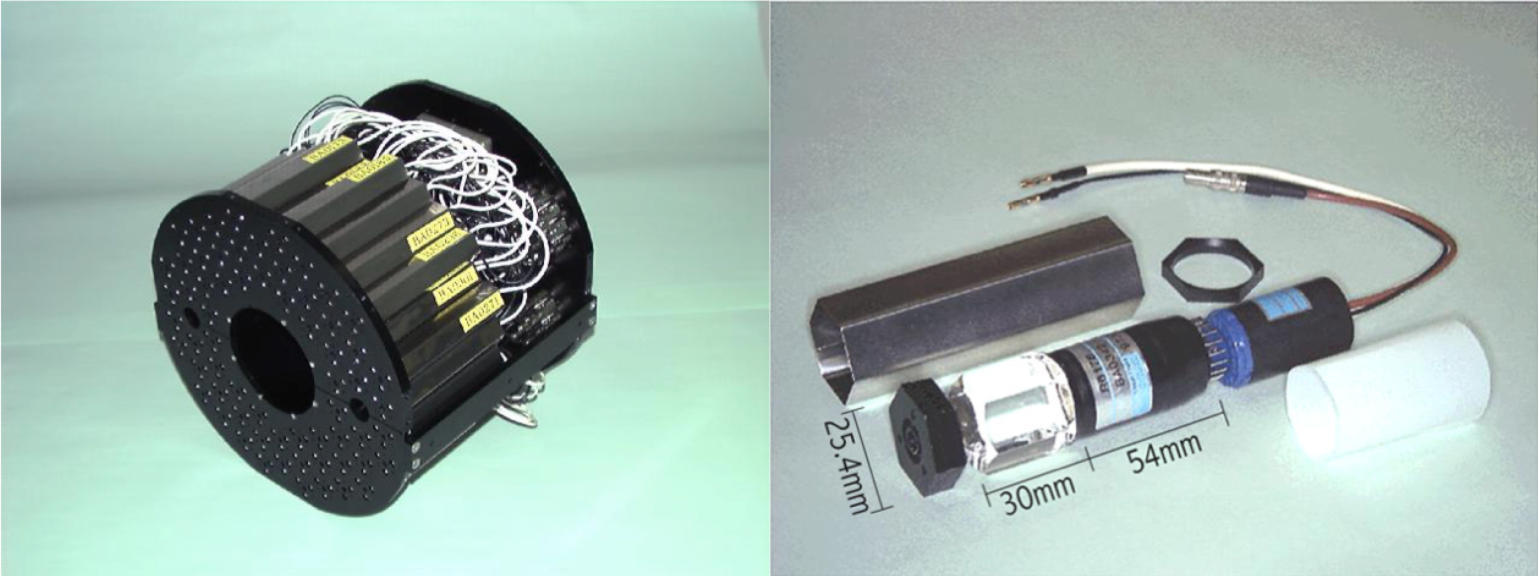}
    \caption{Left: one arm of the MBD, showing the array of 64 modules. Right: one module, with PMT and silica radiator visible. From~\cite{TDR}.}
    \label{fig:mbd}
\end{figure}

The Minimum Bias Detector (MBD) provides the primary trigger and vertex detection for sPHENIX collisions~\cite{TDR}. It comprises two arms positioned along the beam line, one to the north of sPHENIX and one to the south, at a distance 250 cm from the center of the detector. Each arm consists of an array of 64 modules containing photomultiplier tubes (PMTs) coupled to 3 cm length fused-silica Cherenkov radiators, as shown in Figure~\ref{fig:mbd}. The array is arranged azimuthally around a mechanical housing with inner radius 5 cm and outer radius 15 cm; the effective pseudorapidity range of the MBD is therefore $3.5 < |\eta| < 4.6$. Charged particles produced in this far-forward region pass through the Cherenkov radiators and the resulting light produces signals in the PMTs. The signal is passed through a discriminator, yielding a boolean ``yes'' or ``no'' for whether each PMT fired. The MBD arms therefore act as counters, counting the number of PMT hits in each collision. A coincidence between PMT hits in the north and south arms is used to trigger on collisions with minimal bias. Further, the difference in signal timing between the two arms is used to infer the longitudinal ($z$) position of each collision vertex. Finally, the MBD is used as the primary tool to measure the integrated luminosity recorded by sPHENIX: a set of scalers\footnote{A scaler is a simple electronic counter which increments each time it receives a digital signal.}, called the Global Level-1 (GL1) scalers, count the number of north-south coincidence events. Knowledge of the MBD cross section allows one to translate the scaler values to a luminosity. Further, an additional set of 120 independent ``GL1P'' scalers records the luminosity for each of the 120 RHIC bunch crossings, allowing for separation of spin-up and spin-down luminosities.

\subsection{Local Polarimetry}\label{sec:localpol}
\begin{figure}
    \centering
    \includegraphics[width=0.8\textwidth]{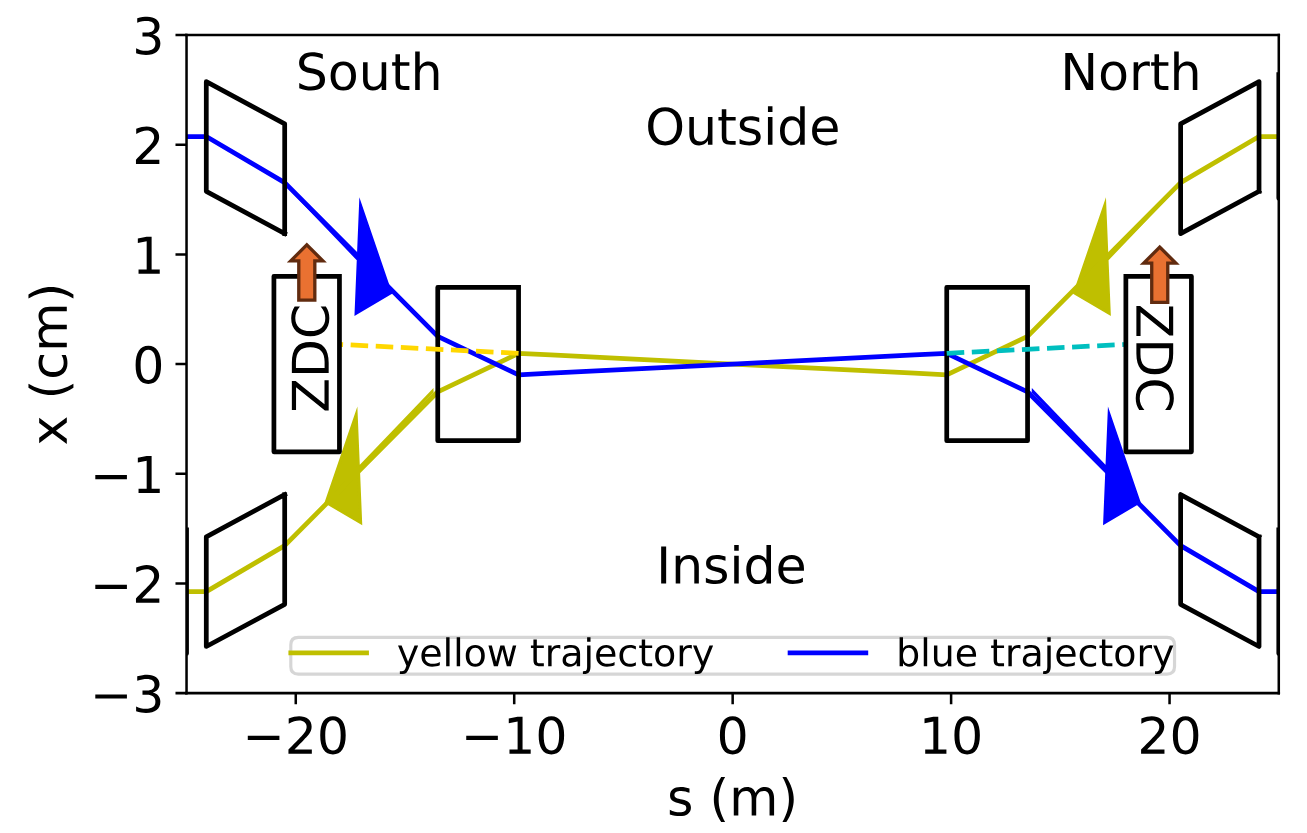}
    \caption{Sketch of the placement of local polarimetry units (here labeled ZDC). Note that this sketch shows the two beams crossing at a nonzero angle, corresponding to the configuration used throughout most of the 2024 RHIC run. The angle is exaggerated for visibility. From~\cite{LocalPol}.}
    \label{fig:localpol}
\end{figure}

Measurement of the proton beam polarization and spin orientation local to sPHENIX is achieved via the Zero Degree Calorimeters (ZDC), Shower Maximum Detectors (SMD), and a set of charged particle veto counters (VCs)\footnote{The author of this thesis participated in the installation and commissioning of the local polarimetry systems at the start of the 2024 RHIC run. This included tests of the ZDC and SMD channel mapping, testing the relative gain of the SMD channels using a radioactive source, installation of the veto counters, and analysis of the forward neutron asymmetries as an independent cross-check.}~\cite{LocalPol}. Similar to the MBD, there are two identical local polarimetry units of ZDC, SMD, and VCs: one to the north of sPHENIX, the other to the south. These units are placed 18 m away from the center of sPHENIX, along the axis of the beams as they pass through sPHENIX. Note the longitudinal placement of the units lies outside the RHIC ``DX'' magnets which bend the beams into and out of collision; see Figure~\ref{fig:localpol}. Therefore they are placed at a polar angle of zero degrees, and, as charged particles are bent away by the DX magnets, detect primarily neutral particles produced at this angle.

The composition of each local polarimetry unit is as follows. The ZDC is a Cherenkov sampling calorimeter composed of tungsten absorber plates and interwoven optical fibers. Particles entering the ZDC shower in the tungsten and produce Cherenkov light, collected by the optical fibers and read out by PMTs. Each ZDC arm is divided longitudinally into three segments. The SMD is sandwiched between the first and second ZDC segments. The SMD is a scintillating strip detector, with 7 vertically-oriented strips to measure the horizontal $x$-coordinate of showers from the ZDC, and 8 horizontally-oriented strips to measure the vertical $y$-coordinate. The scintillation light is collected by light guides and produces signals in a multi-anode PMT. Each local polarimetry unit includes two charged particle veto detectors, placed just upstream and downstream of the ZDC. The veto detectors are simple scintillator paddles read out by PMTs.

This configuration provides polarimetry information based on the known spin-dependent asymmetries of far-forward neutrons~\cite{PHENIX_neutron_AN, RHICf_neutron_AN}. For vertical beam polarization, there is a left-right asymmetry; for horizontal (radial) polarization, the asymmetry is between the vertical up and down directions. Thus by comparing the $x$-$y$ distribution of showers measured by the SMD for spin-up bunches to that for spin-down bunches, one can calculate both the left-right and up-down asymmetries (a similar asymmetry calculation is described in Section~\ref{sec:rawasym}). For vertical beam polarization, the up-down asymmetry vanishes; therefore any measured nonzero up-down asymmetry indicates a deviation from the expected vertical polarization. A dedicated local polarimetry analysis concluded that no such off-vertical polarization is observed in the 2024 $p+p$ data~\cite{SpinQANote}; this observation will be revisited in the context of data quality assurance in Section~\ref{sec:data_QA}, and in the context of the TSSA measurement in Section~\ref{sec:rawresults}. Further, note that in principal, the magnitude of the measured asymmetry can be used to estimate the magnitude of the beam polarization. In practice, however, the limited spatial resolution of the SMD does not allow for a meaningful polarization measurement; the instrumentation described in Section~\ref{sec:rhic_pol} is the definitive source for this information.

\subsection{Trigger and Data Acquisition}
The sPHENIX Data Acquisition (DAQ) system is responsible for aggregating the data recorded by each detector subsystem and transferring it to offline storage for later processing. Saving \textbf{all} of the recorded data would require infeasible transfer rates and storage capacity. Instead, data of interest is selected by a \textit{trigger} system. This system decides, for each bunch crossing, whether to read out and store the detector data. As described in Section~\ref{sec:mbd}, the MBD is used as the primary trigger source in sPHENIX. The simplest trigger is the minimum bias (MB) trigger, which requires only a coincidence between north and south MBD hits. More selective triggers incorporate information on the MBD vertex position, to flag collisions occurring in the ideal $z$ window wherein tracking system acceptance is optimized ($|z|<10$ cm). Other triggers incorporate calorimeter information to flag collisions which likely produce a high-energy photon or jet.

Whenever one of these triggers fires, it defines an \textit{event}. Events are the fundamental unit of data taking. For each event, the fired triggers signal the DAQ to begin the process of reading out the detector data. The DAQ relies on a timing system to ensure data collected by each detector subsystem is aligned temporally. The timing system communicates with each subsystem's front-end electronics and signals them to send the appropriate data forward to dedicated Field-Programmable Gate Arrays (FPGAs) and CPUs. These FPGAs and CPUs collate and compress the data before passing it to so-called Buffer Boxes, which finally send the data to the offline Scientific Data and Computing Center (SDCC) at BNL for storage.

In practice, data in sPHENIX is not collected in a single continuous stream. It is instead segmented into \textit{runs}. Each run corresponds to one continuous group of events recorded under uniform detector conditions, generally ranging from several minutes to one hour. If detector conditions change or a problem develops during data taking, the sPHENIX operators will typically end the current run, fix any significant issues, and start a new run.

\section{EMCal Signal Processing}\label{sec:EMCal}
As the EMCal plays a critical role in this analysis, additional details regarding the transformation of raw SiPM signals into a format useful for physics analysis are given in this section. This includes signal extraction and quality assurance, aggregating the signals from nearby groups of towers into clusters, and calibration of the raw signals.

\subsection{Signal Extraction and Pathologies}\label{sec:EMCal_signal}
The analog signal from an EMCal readout channel is passed by the front-end electronics to an off-detector Analog-to-Digital Converter (ADC). For each event, the ADC takes a sequence of 31 time-samples of the analog signal and returns a 14-bit digital value, called the ADC value, proportional to the signal amplitude at each sample~\cite{Y1Calib}. The resulting sequence of values is known as a waveform. The raw ADC waveform for each calorimeter tower is analyzed using a template-fitting method to extract the waveform peak amplitude, peak timing, pedestal, and waveform fit quality. The peak amplitude minus pedestal is taken as the energy of the tower.

There are two common pathologies which can affect the raw tower waveforms, and, consequently, the extracted peak timing and amplitudes. The first occurs when one bit in the 14-bit ADC value is erroneously flipped, resulting in one time-sample with abnormally large or small ADC value relative to its neighbors. The second occurs when the boundaries of the 14-bit word storing the ADC value for a given channel are incorrectly configured, resulting in a shift in bits and a waveform wrap-around effect. In both of these cases, towers exhibit erroneously large signal amplitudes and noise levels; such towers are termed ``hot.'' Conversely, towers which exhibit abnormally low signals are termed ``cold.'' Finally, towers which uniformly read out zero, due to front-end electronics instrumentation or configuration issues, are called ``dead.'' Hot, cold, and dead towers are identified using a dedicated ``bad tower finder'' algorithm. This algorithm determines the relative frequency with which each EMCal tower records a ``hit,'' i.e. a signal exceeding a nominal noise threshold. The mode and standard deviation of this distribution are calculated. Towers exhibiting a hit frequency more than 2.5 standard deviations above (below) the mode are flagged as hot (cold); towers with a frequency of 0 are flagged as dead. The bad tower finder performs this check for each run and produces a set of run-by-run maps of known bad towers, which can later be masked during data analysis.

Note that the peak amplitude and pedestal determined by the waveform fitting are in ADC units. To arrive at an amplitude in units of energy, the raw ADC values must be calibrated. The calibration procedure, however, requires the reconstruction of groups of towers representing the same electromagnetic shower.

\subsection{Clustering}\label{sec:cluster_merging}
One characteristic parameter of a material is its \textit{Moli\`{e}re radius}, $R_\mathrm{M}$, defined as the radius of a cylinder which contains, on average, 90\% of an electromagnetic shower's energy deposition. Thus $R_\mathrm{M}$ characterizes the lateral size of a shower. For the sPHENIX EMCal this radius is about 2.3 cm~\cite{EMCalNote}. Conversely, each tower has a length of about 2.5 cm. Therefore it is exceedingly unlikely for the shower produced by a photon or $e^\pm$ to be contained entirely within one tower. Rather, a typical shower is spread across a group of neighboring towers.

The process of identifying such groups and reconstructing the kinematics of the shower-producing particle is called \textit{clustering}. Clustering in sPHENIX is performed by a dedicated clustering algorithm or clusterizer. The algorithm first identifies contiguous groups of towers with energy over a predefined threshold (70 MeV after calibration for the 2024 $p+p$ data). This threshold cut is used to filter out electronic noise. Note the formed cluster need not be a rectangular grid of towers, but can have arbitrary shape so long as the towers are contiguous. Next, the algorithm searches for local maxima in the cluster, defined as towers with energy greater than that of the 8 neighboring towers. If more than one local maximum is found, the cluster is divided into sub-clusters and the energy of each tower is redistributed between sub-clusters based on a Monte Carlo simulation-driven template of realistic electromagnetic shower shapes. From here, any sub-clusters are separated and treated as fully independent clusters. The angular position of each cluster is calculated as the energy-weighted average of each constituent tower's $\eta$ and $\phi$ coordinates; a cluster's energy is taken as the sum of tower energies. Finally, a metric called the cluster $\chi^2$ is calculated by comparing the cluster's tower energy distribution to the aforementioned template. This $\chi^2$ is designed to give an indication of how likely it is that the cluster corresponds to a single electromagnetic shower.

\begin{figure}
    \centering
    \includegraphics[width=0.48\textwidth]{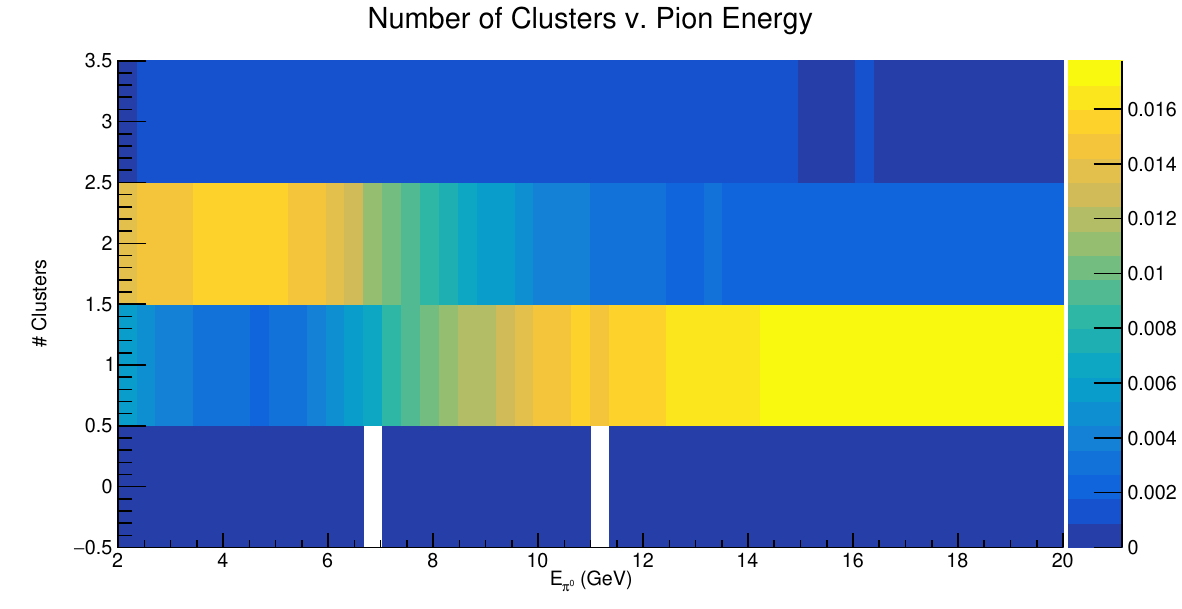}
    \includegraphics[width=0.48\textwidth]{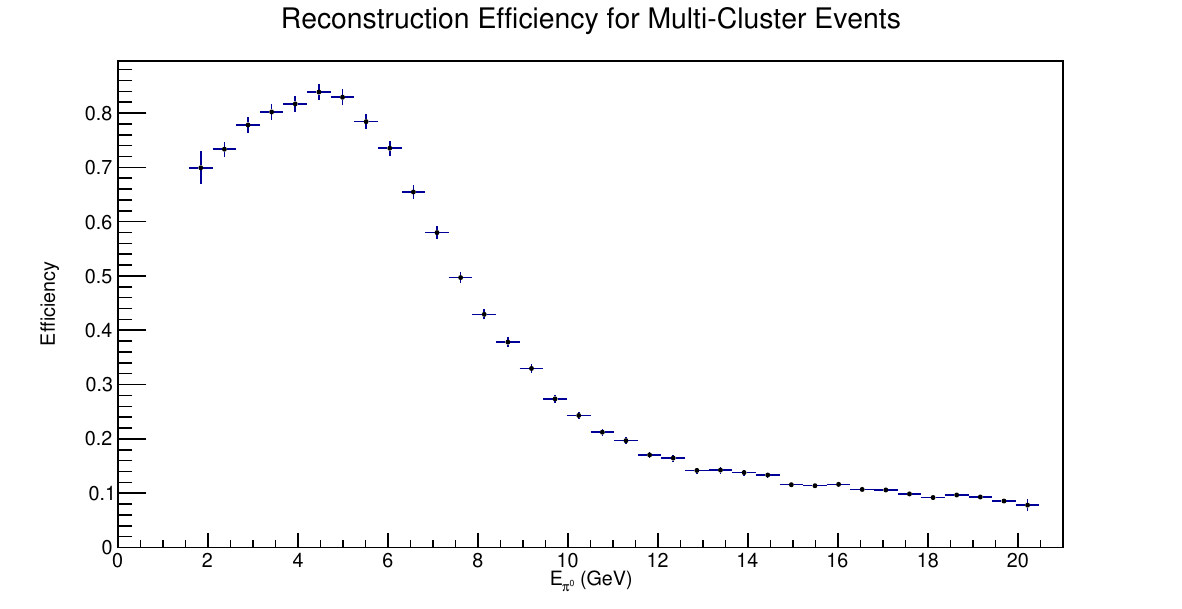}
    \caption{Results from a single-$\pi^0$ ``particle gun'' Monte Carlo simulation of sPHENIX. Left: number of clusters per event with energy above 0.3 GeV, plotted against the initial $\pi^0$ energy. Right: $\pi^0$ reconstruction efficiency, defined as $N_\mathrm{reco}/N_\mathrm{truth}$, plotted against $\pi^0$ energy.}
    \label{fig:cluster_merging}
\end{figure}

For an idealized EMCal and clustering algorithm, each cluster would correspond to exactly one electromagnetic shower, induced by exactly one photon or $e^\pm$. In practice this is not the case, as some clusters are due to background sources. As the sPHENIX EMCal material constitutes approximately one hadronic interaction length, clusters corresponding to showers induced by hadrons are not uncommon. Other background clusters originate not from a shower, but from purely electronic noise exceeding the clusterizer threshold. The most common background case occurs when two or more particles produce showers that overlap in the same group of towers. Consider the primary decay channel of the $\pi^0$ meson, $\pi^0 \rightarrow 2\gamma$. In the pion's rest frame the two photons are emitted back-to-back. However for high-energy $\pi^0$s, the opening angle between the decay photons in the lab frame becomes small, increasing the likelihood that the showers they produce will overlap. At sufficiently high energies the showers overlap completely and cannot be individually resolved. This effect, known as cluster merging, is demonstrated in Figure~\ref{fig:cluster_merging}, based on Monte Carlo simulation of events with a single $\pi^0$ in isolation. The left side of this figure shows the distribution of the number of reconstructed clusters per event. For low pion energies, events with exactly two clusters are by far the most common. However at around 6-8 GeV, there is a transition from two-cluster to single-cluster events, which dominate for energies over about 10 GeV. The effect of cluster merging and limited ability of the clustering algorithm to reconstruct high-energy $\pi^0$s is directly relevant for this analysis, and will be revisited in Section~\ref{sec:corrections}.

\subsection{Calibrations}\label{sec:calibrations}
\begin{figure}
    \centering
    \includegraphics[width=0.8\textwidth]{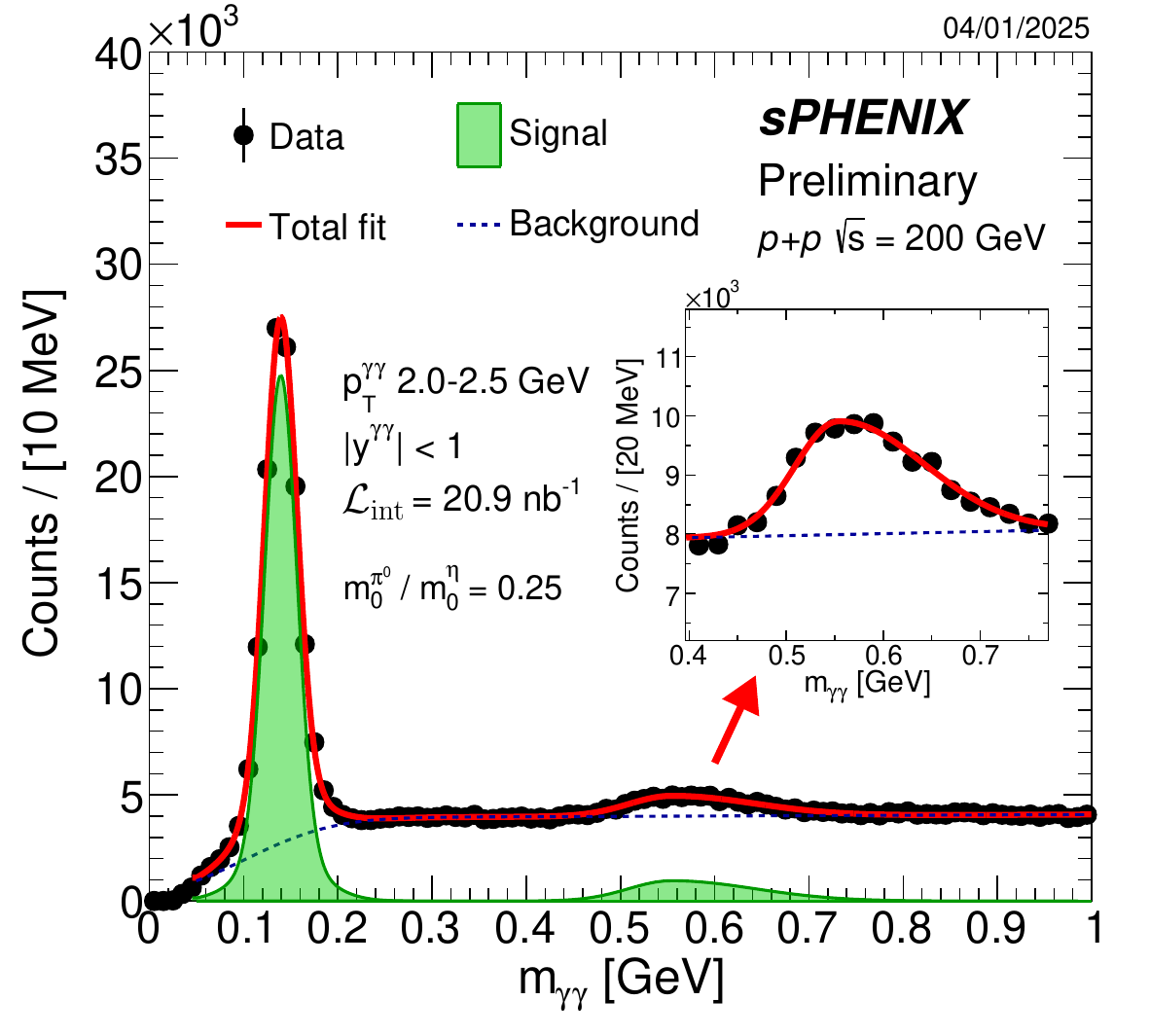}
    \caption{Example of the di-photon invariant mass distribution from 2024 sPHENIX data. The large peak near 0.14 GeV corresponds to the $\pi^0$. sPHENIX Public Result~\cite{PerformancePlots}.}
    \label{fig:calib_mass}
\end{figure}

With the ability to reconstruct clusters encompassing the full energy of a photon established, we can now describe the procedure used to calibrate the EMCal data, i.e. to convert the extracted tower waveform amplitude from units of ADC value to units of energy~\cite{Y1Calib}
. This procedure relies on the reconstruction of di-photons from pairs of clusters; see Section~\ref{sec:diphoton_cuts} for a similar reconstruction procedure. First, an initial calibration factor is assigned to each tower to yield tower energies in units of MeV. Next, the clusterizer is run, producing a list of reconstructed clusters for each event in the data sample. A subset of these clusters is chosen based on a minimum transverse energy ($E_T$) cut and a $\chi^2$ cut to mitigate the influence of background sources. Then, the surviving clusters are combined pair-wise, assuming that each cluster corresponds to one photon, to form a di-photon. Over many events, the invariant mass of the di-photons follows a distribution with a distinct peak corresponding to $\pi^0\rightarrow2\gamma$ decays, as shown in Figure~\ref{fig:calib_mass}.

This distribution is fit with a Gaussian (for the $\pi^0$ peak) plus third-order polynomial (for the background), and the mass position of the Gaussian peak is extracted. The invariant mass distribution is constructed for each tower independently by associating each di-photon with its leading-energy cluster's leading-energy tower. The extracted $\pi^0$ mass position for each tower is compared to a target mass value. Each tower is recalibrated by multiplying its initial calibration factor by $m_\mathrm{target}/m_\mathrm{tower}$. Then the entire procedure, starting from clustering, is repeated iteratively. On the order of 25 iterations are performed such that the multiplicative corrections approach unity and exhibit only statistical fluctuations. The final calibration factors are approximately 3 MeV per ADC unit.

The reader may notice that the fully-calibrated $\pi^0$ mass position in Figure~\ref{fig:calib_mass} and related figures (such as Figure~\ref{fig:integratedDiphotonMass}) differs from the definitive Particle Data Group (PDG) value. This is because the target mass value used in the calibration procedure is not set to the PDG value. The observed mass measured by a detector depends not only on the calibration factors, but also on the momentum distribution of $\pi^0$s, detector energy resolution, and position resolution. Therefore the target mass is chosen based on Monte Carlo simulation of $p+p$ collisions in sPHENIX, and represents the mass an ideally-calibrated sPHENIX EMCal would measure.


\chapter{Data Selection and Meson Reconstruction}\label{ch:data}
\section{Data Production}\label{sec:production}
The process of converting raw detector readout recorded by the DAQ into a form appropriate for higher-level physics analysis, called Data Summary Tables (DTSs), is know as data \textit{production}. In sPHENIX, data is produced centrally for use by all physics analyzers via a production workflow. This begins with event combining: the data packets from independent detector systems are aggregated such that the data in each packet aligns event-by-event. This step also appends the GL1 information, such as the run number and scaler values, to the DST. Next, information from the MBD is used to reconstruct the event vertex as a ``\texttt{GlobalVertex}'' object, also appended to the DST. The following steps depend on the type of data to be produced. Tracking data is produced by a dedicated tracking workflow, due to the complexity of associating TPC tracks with silicon hits and applying TPC distortion corrections.

Directly relevant for this analysis is the calorimeter production workflow, aspects of which are described in detail in Section~\ref{sec:EMCal}. First, raw tower ADC values are converted to units of energy using a set of calibration constants (see Section~\ref{sec:calibrations}) stored in a run-by-run database. This database, called the Conditions Database, includes information about the status of each detector subsystem during each run. The next step is to form clusters from the individual calorimeter towers. However, in order to exclude known hot, cold or dead towers, a run-by-run map of known bad towers (Section~\ref{sec:EMCal_signal}) in the Conditions Database is used to mask these towers. The remaining towers are then used as input to the clusterizer (Section~\ref{sec:cluster_merging}). The result is a DST containing a list of all clusters, including their calibrated energies, $\eta$ and $\phi$ positions, and transverse shower profile information.

\section{Data Selection}
This analysis uses data taken during the 2024 RHIC polarized $p+p$ run. Events are selected in a sequence of run-level and event-level checks as described below.

\subsection{Run Selection}\label{sec:data_QA}
Selection of good physics runs is based on a series of data quality assurance (QA) requirements, summarized in Table~\ref{tab:run_selection}. Selection begins by taking only ``physics'' runs (according to the sPHENIX DAQ Database). This ensures that only physics-quality runs with stable data taking are included; runs used for other purposes, such as calibration, DAQ testing or troubleshooting, and cosmic ray measurements are excluded. From these physics runs, those with less than 500,000 events are omitted. This criteria is chosen to ensure that the EMCal hot, cold and dead tower finder (described in Section~\ref{sec:EMCal_signal}) has enough statistics to correctly identify bad towers. Next, runs are selected from a run-by-run database storing a quality assessment for each detector subsystem, called the Run Triage Database\footnote{The author of this thesis was responsible for implementing the automated calorimeter QA structure during the 2024 RHIC run. This included producing run-by-run QA plots for each calorimeter subsystem; populating a dedicated ``offline QA'' web page with these plots for convenient examination; using the plots to assign a ``good'' or ``bad'' assessment for each run; populating the Run Triage Database; and studying the common pathologies of ``bad'' runs. This workflow was incorporated into the data production process to seamlessly add new runs to the offline QA page and triage database. The author's work represented the first detector subsystem to use this framework, and formed the basis for analogous subsystem QA workflows developed by others.}, where the EMCal status is ``Golden''. This ensures good quality EMCal data by imposing the following QA requirements:
\begin{itemize}
    \item $\leq$ 100 hot towers
    \item $\leq$ 500 cold and dead towers
    \item Timing $|\mu| \leq 1$
    \item Timing $\sigma \leq 2$
    \item Mean MBD vertex $\langle z \rangle \leq$ 30 cm.
\end{itemize}
Here the signal timing is in units of ADC sample, with respect to the expected signal peak time (sample number 6). Each sample is six times the RHIC beam clock frequency, or about 17 ns.

\begin{table}
\centering
\begin{tabular}{l|l|l|l|l}
\underline{Criteria} & \underline{Runs} & \underline{\% Runs} & \underline{Events} & \underline{\% Events} \\\hline
Current production DSTs available & 1679 & 100 & 12945464444 & 100 \\\hline
$>$ 500k events & 1285 & 76.53 & 12898124983 & 99.63 \\\hline
EMCal ``Golden'' & 1165 & 69.39 & 12039268295 & 93.00 \\\hline
FEM clock matching & 1155 & 68.79 & 11928566872 & 92.14 \\\hline
Bad tower map available & 1121 & 66.77 & 11807043808 & 91.21 \\\hline
Spin QA & 927 & 55.21 & 10564239866 & 81.61 \\
\end{tabular}
\caption{Number of runs and corresponding number of events, taken from a central sPHENIX database cataloging the available DST files, which survive each run-level selection criteria.}
\label{tab:run_selection}
\end{table}

Following the EMCal tower-level QA, a Front-End Module (FEM) clock matching requirement ensures that all calorimeter interface boards sending data to the same collating CPU have matching clock values, as read from the DAQ Database. The final requirement for EMCal QA is that a bad tower map be available in the Conditions Database for each run. As mentioned in Section~\ref{sec:production}, this is necessary to mask hot towers before running the clusterizer.

The final QA requirements relate to spin information. Run-by-run spin-related information is stored in a central Spin Database. Runs are excluded if either the Spin Database ``\texttt{BadRunFlag}'' is set, or if the GL1P scalers are uniformly zero. The \texttt{BadRunFlag} encompasses the following requirements:
\begin{itemize}
    \item Spin pattern must match one of the known RHIC Main Control Room  (MCR) patterns
    \item Polarization values for both beams are between 0 and 1 (corresponding to 0\% and 100\%)
    \item $<$ 10 bunches exhibit mismatch between intended and measured spin pattern
\end{itemize}
If any of these checks fails, the 
\texttt{BadRunFlag} is set and the run is excluded from the data sample. More detailed information about the Spin QA, including analysis of the local polarimetry information (see Section~\ref{sec:localpol}), can be found in an analysis note~\cite{SpinQANote} and the spin DB~\cite{SpinDBwiki} and spin run 2024~\cite{Run24wiki} sPHENIX wiki pages.

\subsection{Event Selection}
\label{sec:eventSelection}

\begin{figure}
    \centering
    \includegraphics[page=3, width=0.8\linewidth]{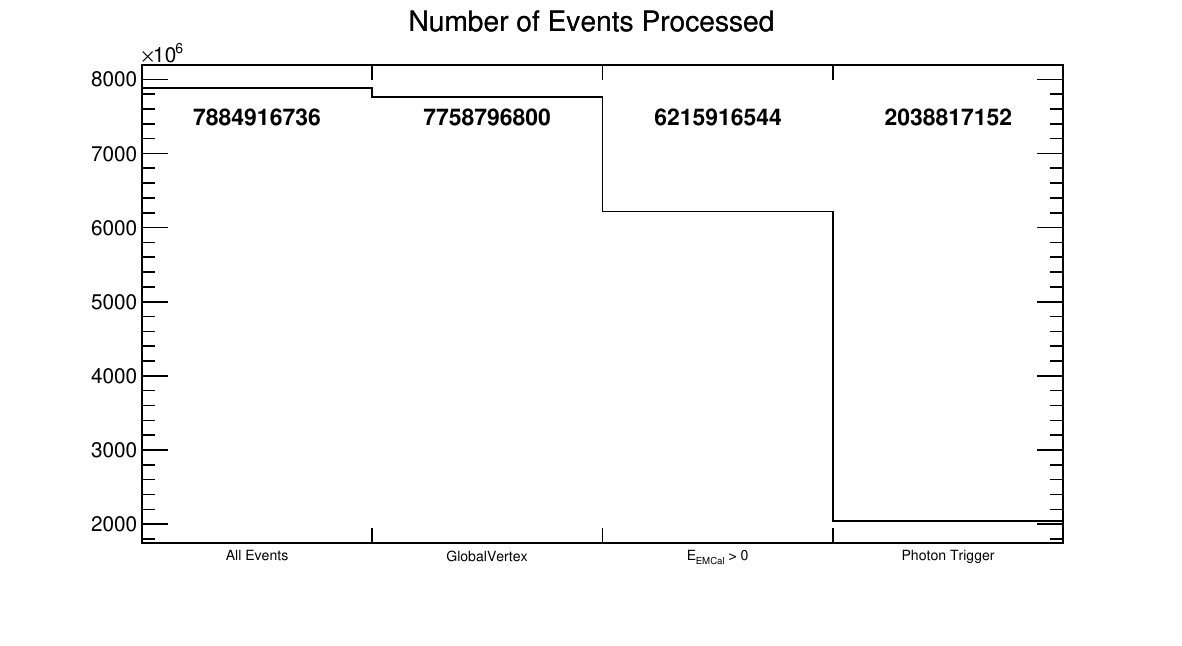}
    \caption{Number of events passing each event-level requirement. Note that the total number of events processed is lower than the number of events in Table~\ref{tab:run_selection} because some DST segments are missing necessary information and are excluded from the analysis.}
    \label{fig:event_selection}
\end{figure}

The event-level data selection begins with a trigger requirement. 
Events are selected which fire either (or both) of the two lowest-energy photon triggers. For the 2024 Run these triggers were set to energy thresholds of 3~GeV and 4~GeV. 
Next, we require that a \texttt{GlobalVertex} be reconstructed for each event. This is necessary for the clustering algorithm to associate the correct pseudorapidity to each cluster. Finally, we require that the total EMCal energy, computed as the sum of the energy in each tower, must be positive. This requirement filters out events where the EMCal does not record any significant energy deposition, and pedestal subtraction in the waveform processing results in a negative total energy. Figure~\ref{fig:event_selection} shows the number of events surviving each event-level cut.

\section{Meson Selection}
\subsection{Cluster-Level Reconstruction and Cuts}
\label{sec:cluster_cuts}
In each event, mesons are reconstructed from pairs of EMCal clusters. 
The cluster's pseudorapidity $\eta$ is calculated taking into account the event's vertex $z$-position. A minimum cluster energy cut of 1.0 GeV is applied to all clusters entering the analysis. In this analysis no cuts are applied on the shower-shape variable $\chi^2$ inherited from the PHENIX clustering algorithm, as this has been found to be a poor metric for separating signal and background in sPHENIX~\cite{ian-ppg11}. A study on the effect of such a cut found that the improvement in signal-to-background ratio is negligible, while the corresponding drop in available statistics results in an increased statistical uncertainty on $A_N$. Note that for the final result for publication, a shower-shape-based method of identifying single-photon clusters will be implemented to improve on the preliminary signal-to-background ratios in this work.

\begin{figure}[ht]
  \centering
  \includegraphics[width=\textwidth]{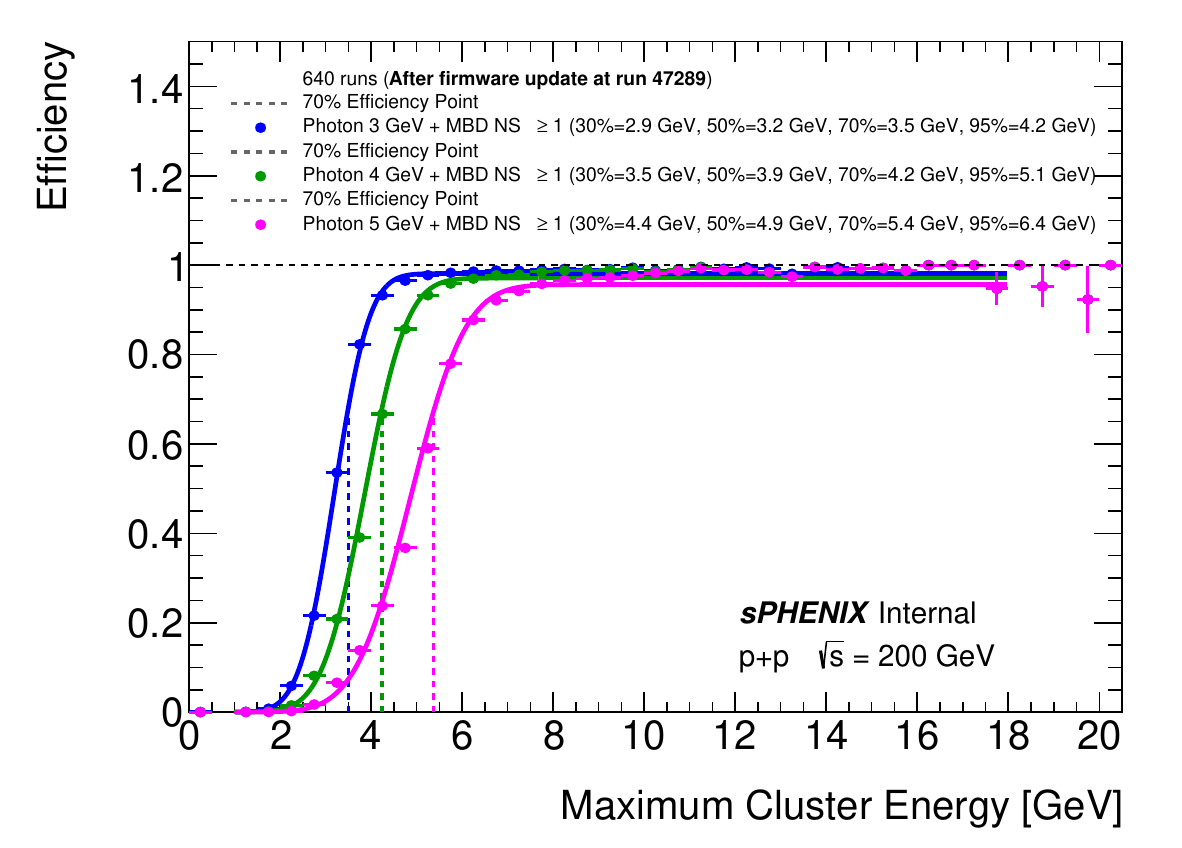}
  \caption{Trigger turn-on curves for the photon triggers. Each set of points shows the efficiency for one of the triggers as a function of the maximum cluster energy in each event. From~\cite{JustinTalk}.} 
  \label{fig:triggercurves}
\end{figure}

To avoid trigger bias in the measurement of low-$p_T$ mesons, it is necessary to select only di-photons which caused the trigger to fire. Information on which trigger tile fired in a given event is not currently available in the sPHENIX data production. Instead, an energy requirement is imposed:
\begin{itemize}
    \item At least one cluster composing the di-photon candidate is required to have sufficient energy to fire the trigger; OR
    \item The combined energy of the two clusters is sufficient to fire the trigger, and the two clusters lie within $\Delta \eta = 0.1$ and $\Delta \phi = 0.1$, such that they may reasonably occupy the same trigger tile.
\end{itemize}
Here ``sufficient energy to fire the trigger'' is defined as the energy threshold at which the trigger becomes 70\% efficient. Figure~\ref{fig:triggercurves} shows the efficiency of each photon trigger, with dotted lines indicating this 70\% threshold energy. For the 3~GeV (4~GeV) trigger, this threshold is 3.5 (4.3) GeV.

\subsection{Di-photon-Level Reconstruction and Cuts}
\label{sec:diphoton_cuts}
\begin{figure}
    \centering
    \includegraphics[page=329,width=0.8\linewidth]{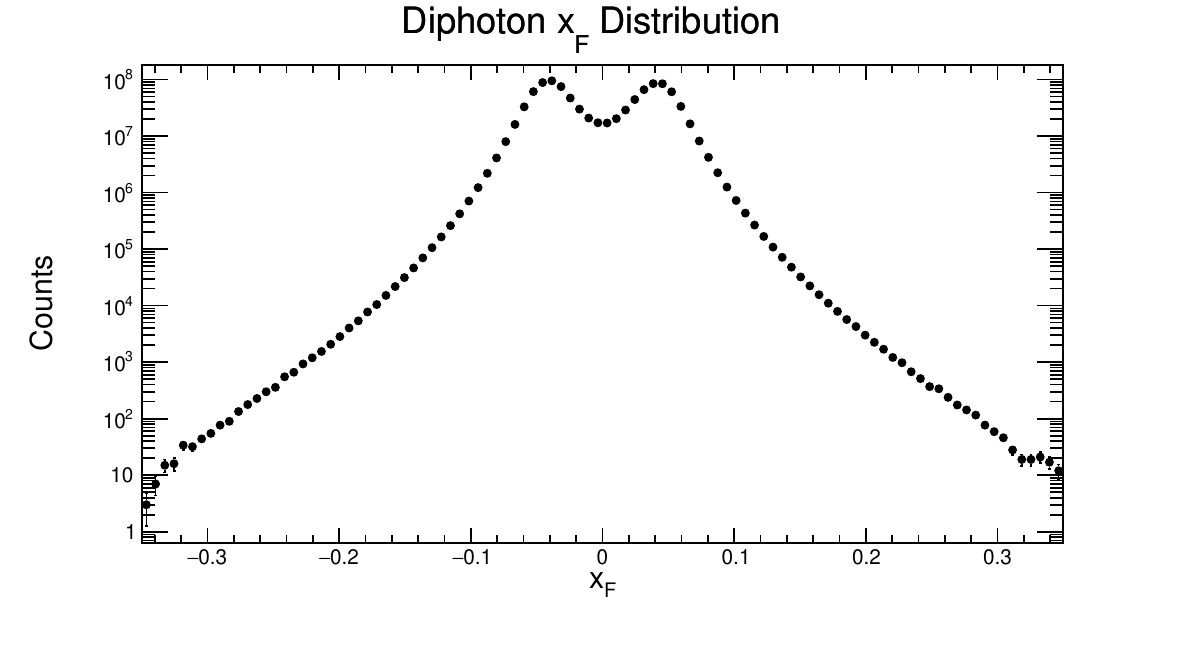}
    \caption{$x_F$ distribution of all di-photons surviving the minimum $p_T$ and maximum $\alpha$ cuts.}
    \label{fig:diphoton_xF}
\end{figure}

\begin{figure}
    \centering
    \includegraphics[page=451,width=0.48\linewidth]{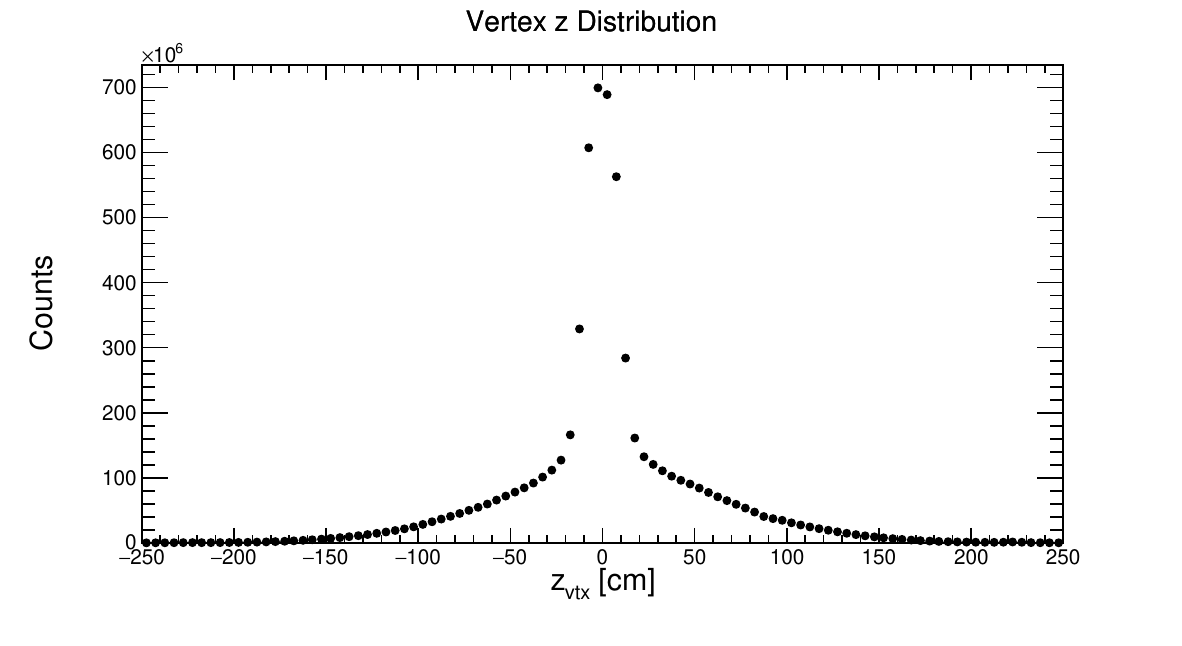}
    \includegraphics[page=8,width=0.48\linewidth]{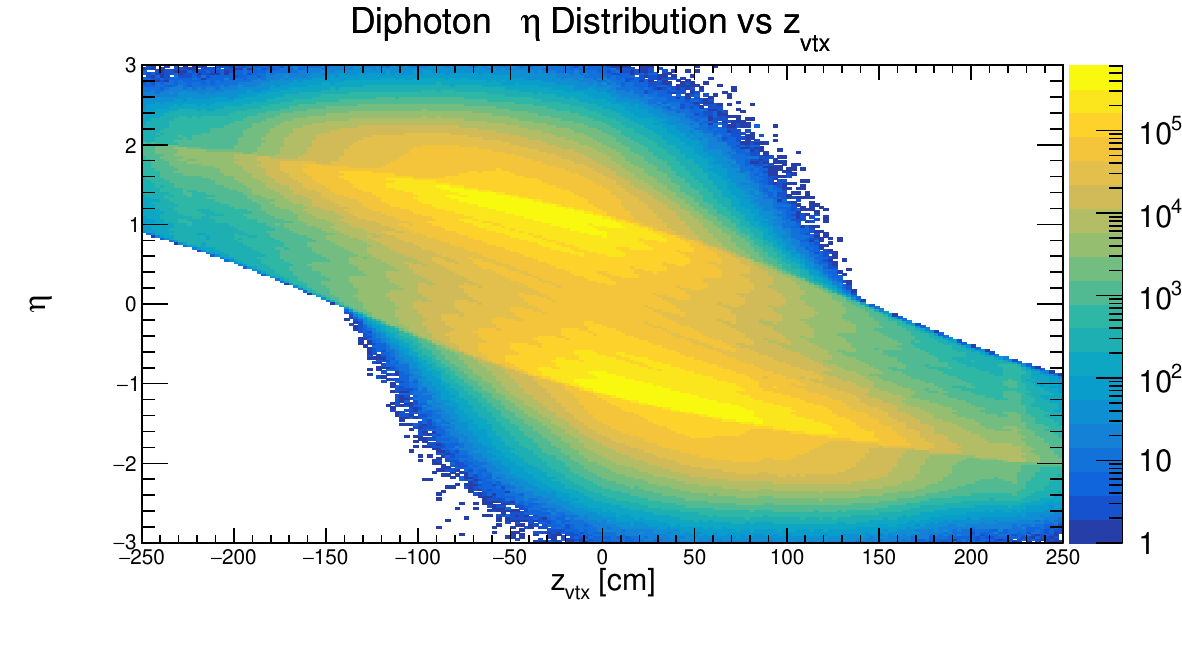}
    \caption{Left: vertex-$z$ distribution for all events entering the analysis. Right: pseudorapidity versus vertex-$z$ distribution of all di-photons surviving the minimum $p_T$ and maximum $\alpha$ cuts.}
    \label{fig:diphoton_eta_vtxz}
\end{figure}

Pairs of clusters surviving the cuts in Section~\ref{sec:cluster_cuts} are used to reconstruct di-photons. First a single photon is reconstructed from a cluster by assuming its mass is zero, or equivalently, assuming its energy and transverse momentum $p_T$ are related by 
\begin{equation}
p_T = \frac{E}{\cosh{\eta}}.
\end{equation}
Next, a di-photon is constructed by adding the four-momenta of two single photons. A minimum di-photon $p_T$ cut of 1.0 GeV is applied, followed by a cut of $\alpha<0.7$ on the pair energy asymmetry $\alpha$ with  
\begin{equation}
\alpha = \left|\frac{E_1 - E_2}{E_1 + E_2}\right|.
\end{equation}
In addition to these cuts, a fiducial di-photon pseudorapidity cut of $|\eta| < 3.0$ and an $x$-Feynman cut of $|x_F| < 0.35$ are applied. The $x_F$ cut is motivated by Figure~\ref{fig:diphoton_xF}, which shows that $\pm 0.35$ is approximately the largest $x_F$ accessible by the sPHENIX acceptance. The $\eta$ cut is more subtle: while the EMCal has a nominal pseudorapidity acceptance of $|\eta| < 1.1$, this assumes an interaction vertex at the nominal interaction point, i.e. at $x = y = z = 0$. For vertices displaced along the $z$-axis, however, the effective $\eta$ coverage can be larger or smaller. To illustrate this point, consider a fixed point on the north (positive-$\eta$) edge of the EMCal, and three different vertices along the beam axis with coordinates $z_1 > z_2 = 0 > z_3$. The vertex $z_2$ will correspond to a pseudorapidity $\eta_2 = 1.1$. The line connecting vertex $z_1$ to our fixed point will make a steeper angle with respect to the beam axis, resulting in $\eta_1 < 1.1$; conversely, the line connecting $z_3$ to the fixed point will make a shallower angle, resulting in $\eta_3 > 1.1$. This effect is shown in the right side of Figure~\ref{fig:diphoton_eta_vtxz}. Here we see that effective pseudorapidity coverage of the sPHENIX data, integrated over all vertex $z$-positions, is approximately $|\eta| < 3.0$.


\begin{figure}[ht]
  \centering
  \includegraphics[width=\textwidth]{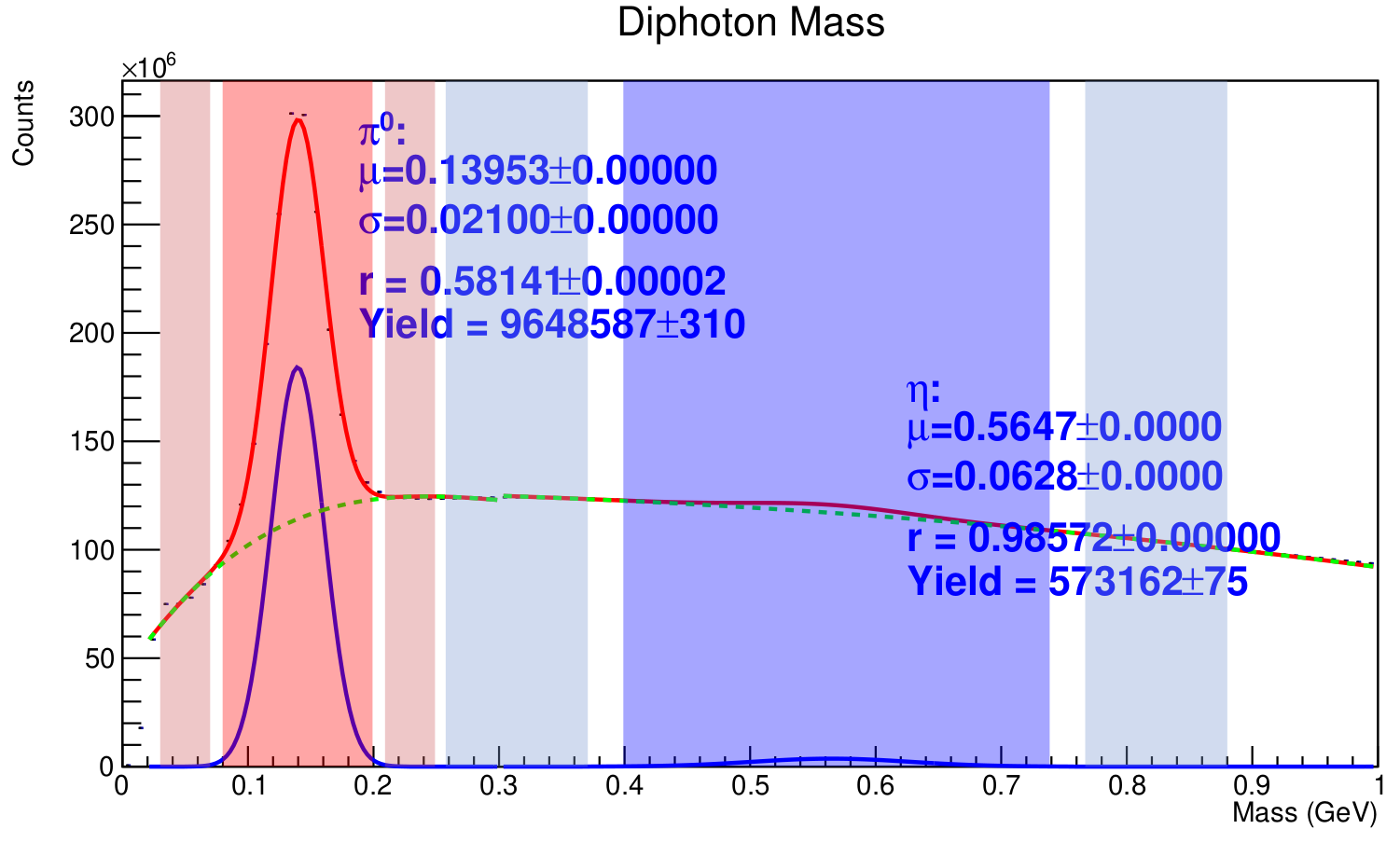}
  \caption{Di-photon invariant mass distribution for di-photons with $p_T > 1.0$ GeV and $\alpha < 0.7$. The fit is a Gaussian (meson signal) plus threshold function with exponential (background) for each meson: $\pi^0$ and $\eta$-meson. The two mesons are fit independently of each other.}
  \label{fig:integratedDiphotonMass}
\end{figure}

Mesons are reconstructed by selecting di-photons based on their invariant mass. For the $\pi^0$, the mass window from 80 to 199 MeV is used. For the $\eta$-meson the chosen mass window is from 399 to 739 MeV. These windows are chosen by defining a 3-$\sigma$ window around the respective meson mass peaks. Figure \ref{fig:integratedDiphotonMass} shows the di-photon mass distribution for all di-photons surviving cuts detailed above. The $\pi^0$ and $\eta$ signal regions are highlighted. The $p_T$-dependent mass distributions will be shown and discussed in detail in Section~\ref{sec:corrections}.

\begin{figure}
    \centering
    \includegraphics[page=4,width=0.48\linewidth]{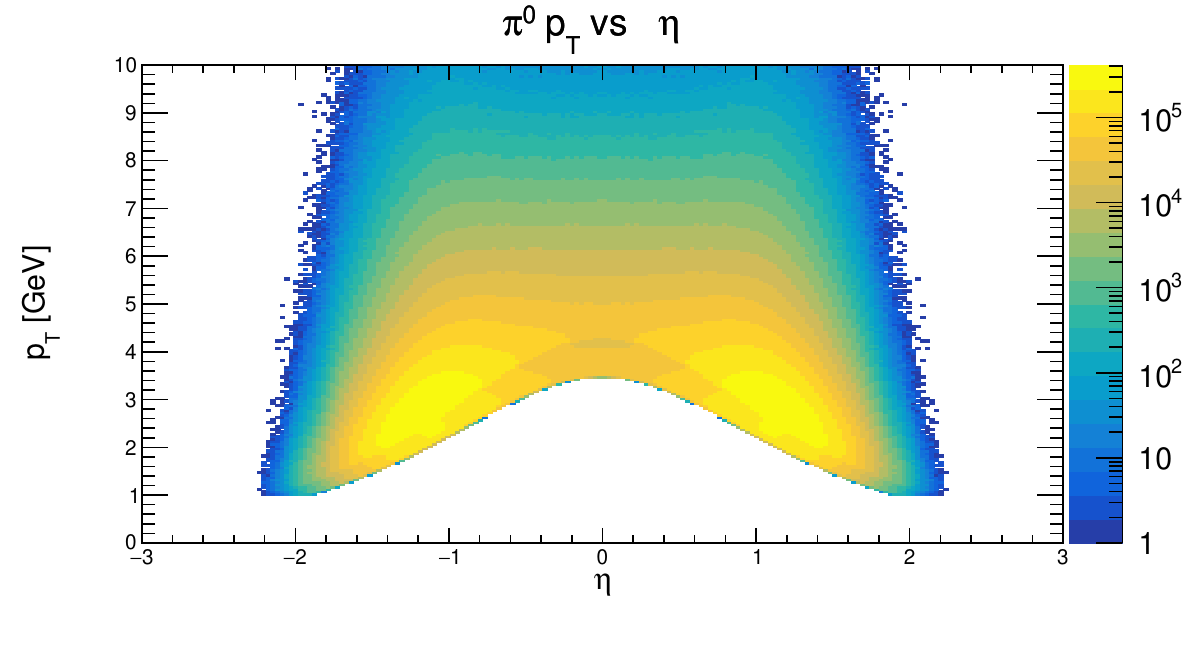}
    \includegraphics[page=10,width=0.48\linewidth]{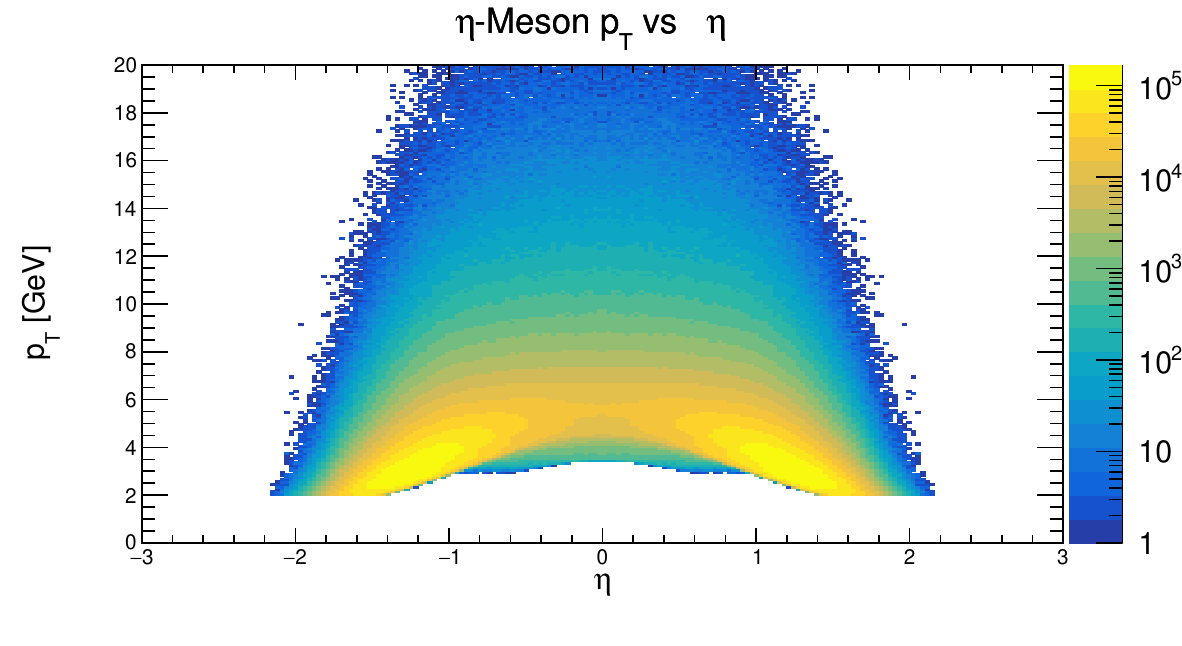}
    \caption{$p_T$ versus pseudorapidity distribution for reconstructed $\pi^0$ (left) and $\eta$-mesons (right).}
    \label{fig:kinematics_pT_vs_eta}
\end{figure}

\begin{figure}
    \centering
    \includegraphics[page=3,width=0.48\linewidth]{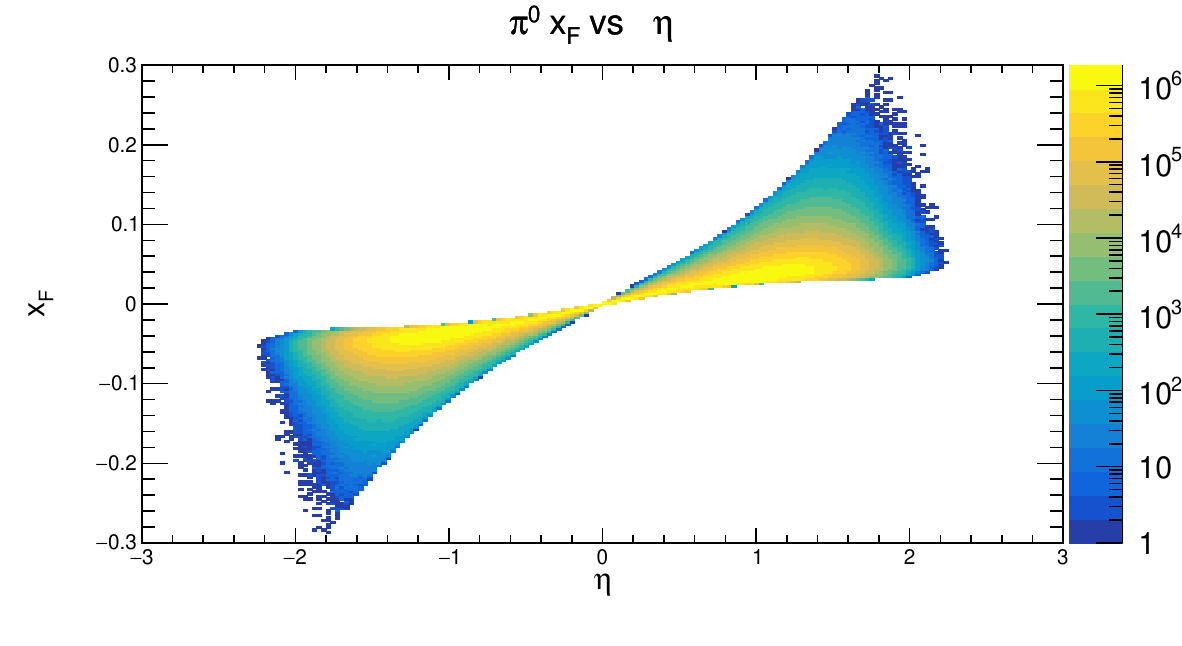}
    \includegraphics[page=9,width=0.48\linewidth]{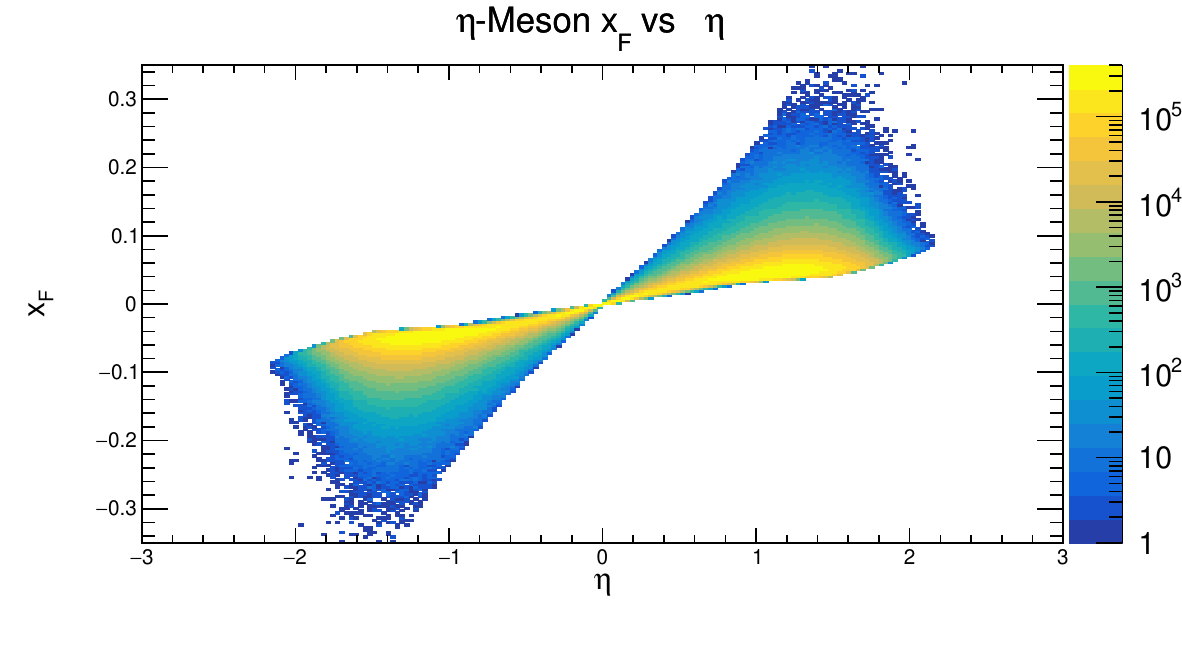}
    \caption{$x_F$ versus pseudorapidity distribution for reconstructed $\pi^0$ (left) and $\eta$-mesons (right).}
    \label{fig:kinematics_xF_vs_eta}
\end{figure}

\begin{figure}
    \centering
    \includegraphics[page=6,width=0.48\linewidth]{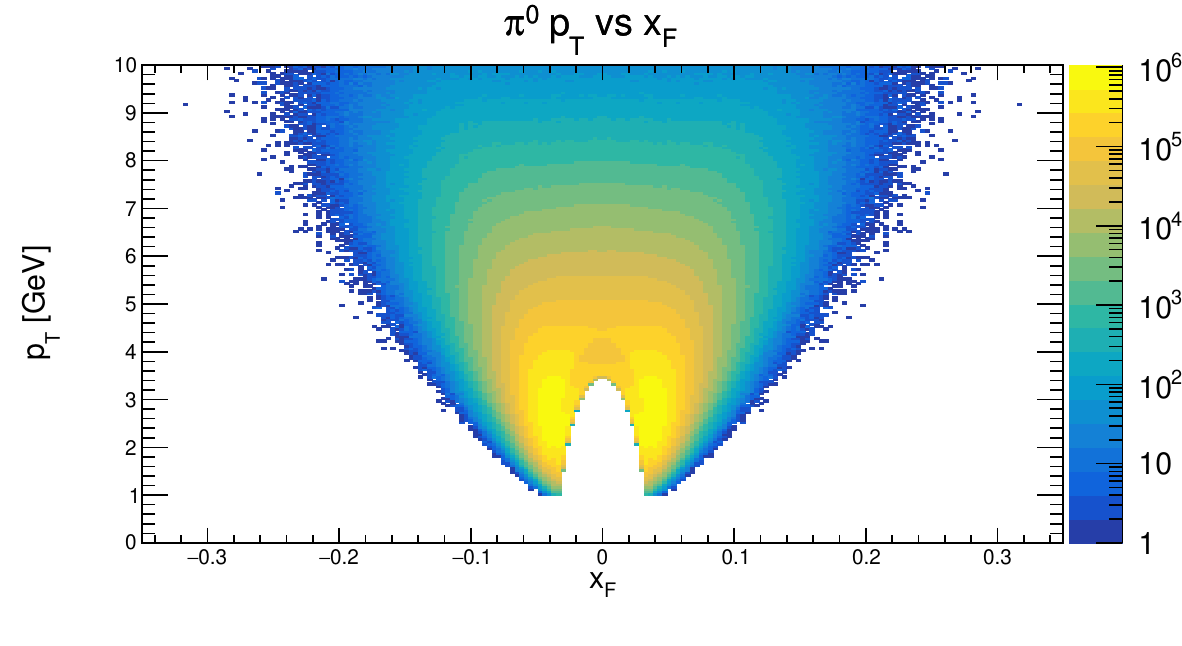}
    \includegraphics[page=12,width=0.48\linewidth]{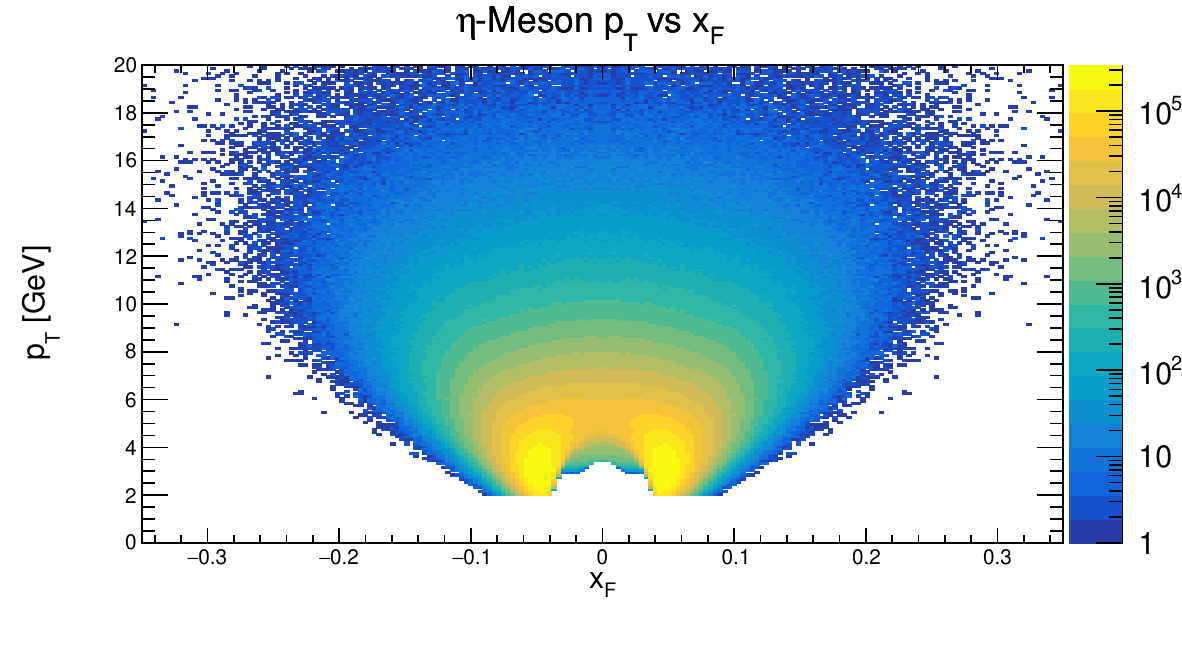}
    \caption{$p_T$ versus $x_F$ distribution for reconstructed $\pi^0$ (left) and $\eta$-mesons (right).}
    \label{fig:kinematics_pT_vs_xF}
\end{figure}

The kinematic region probed by the sPHENIX data is shown in Figures~\ref{fig:kinematics_pT_vs_eta}-\ref{fig:kinematics_pT_vs_xF}. Note that in these plots, $\eta$ and $x_F$ are defined with respect to the Blue beam, such that positive values correspond to the north side of the detector. The implications of the sPHENIX kinematic coverage will be discussed in relation to the existing global TSSA data in Chapter~\ref{ch:results}.


\chapter{Asymmetry Extraction}\label{ch:asym}
This chapter details the extraction of $A_N$ for the $\pi^0$- and $\eta$-mesons, as reconstructed from the sPHENIX data described in Chapter~\ref{ch:data}. This extraction begins with a calculation of the raw asymmetry $\epsilon$ and its statistical uncertainty $\delta \epsilon$ from the spin-dependent azimuthal distributions of di-photons, as detailed in Section~\ref{sec:rawasym}. Section~\ref{sec:pollumi} describes the polarization and relative luminosity calculations needed to measure $\epsilon$. Results for the raw asymmetry are shown in Section~\ref{sec:rawresults}. Section~\ref{sec:corrections} details the corrections to the raw asymmetry needed to obtain the physics asymmetry $A_N$. Results for this corrected $A_N$ are given in Section~\ref{sec:stat-results}. 

\section{Raw Asymmetry Calculation}\label{sec:rawasym}
The spin-dependent asymmetry is calculated in two ways, given by Equations~\ref{eq:AN_rellumi} and \ref{eq:AN_sqrt}. In both formulas, $P$ is the average beam polarization. The relative luminosity formula compares the yield $N$ of mesons on one side ($L$ or $R$ subscript) of the polarized proton beam and takes the difference between spin-up and spin-down yields ($\uparrow$ or $\downarrow$ superscripts):
\begin{equation} \label{eq:AN_rellumi}
A_N = \frac{1}{P} \frac{N^{\uparrow}_L - 
\mathcal{R} N^{\downarrow}_L}{N^{\uparrow}_L + \mathcal{R} N^{\downarrow}_L},
\end{equation}
where $\mathcal{R} = \mathcal{L}^\uparrow/\mathcal{L}^\downarrow$ is the relative luminosity between spin-up and spin-down bunch crossings. The relative luminosity formula is an exact reformulation of Equation~\ref{eq:AN_def}. The square root or geometric formula compares the yields of mesons on opposite sides of the polarized beam:
\begin{equation} \label{eq:AN_sqrt}
A_N = \frac{1}{P} \frac{\sqrt{N^{\uparrow}_{L} N^{\downarrow}_{R}} - \sqrt{N^{\downarrow}_{L} N^{\uparrow}_{R}}}{\sqrt{N^{\uparrow}_{L} N^{\downarrow}_{R}} + \sqrt{N^{\downarrow}_{L} N^{\uparrow}_{R}}}.
\end{equation}
This formula is an approximation of Equation~\ref{eq:AN_def} which, to first order, cancels acceptance effects.

\begin{figure}
    \centering
    \includegraphics[page=7,width=0.8\linewidth]{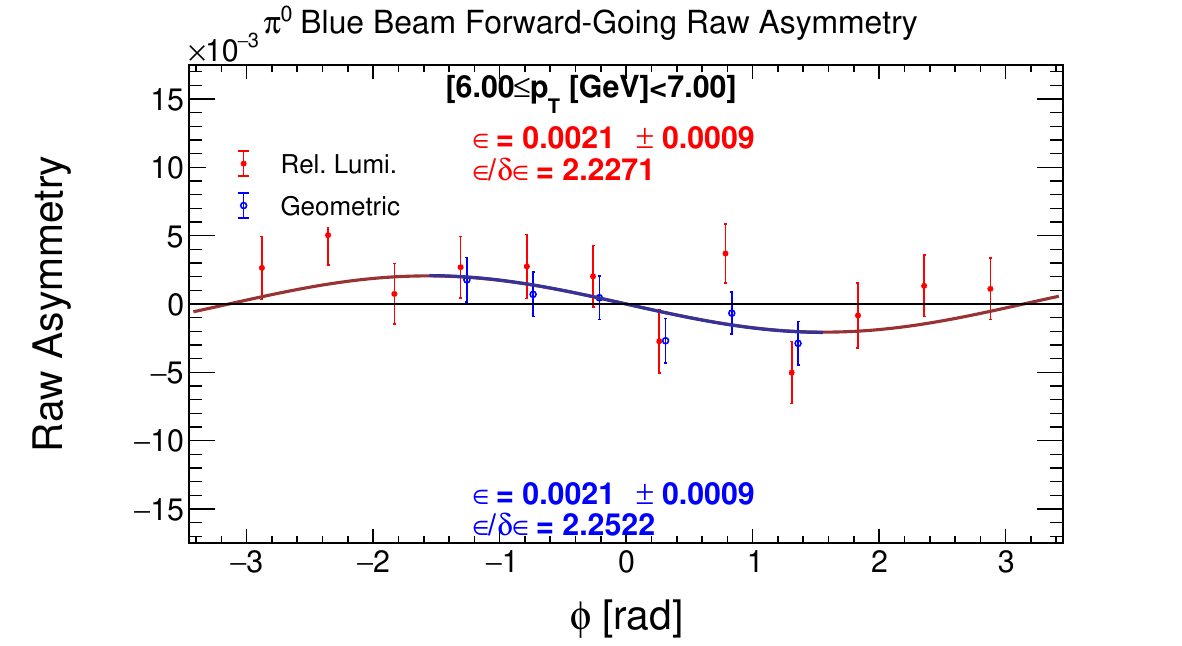}
    \caption{Example plot showing the $\phi$-dependent raw asymmetry and fit to a sinusoid. The extracted amplitude of the fit is taken as the raw asymmetry $\epsilon$.}
    \label{fig:rawasymfit}
\end{figure}

In practice, $A_N$ is determined by first calculating the raw asymmetry $\epsilon$ using the spin-dependent azimuthal ($\phi$) distributions of $\pi^0$ and $\eta$ mesons. The $\phi$-dependent raw asymmetry $\epsilon(\phi)$ is calculated using Equations~\ref{eq:eps_rellumi} and \ref{eq:eps_sqrt}. The relative luminosity formula calculates $\epsilon(\phi)$ over an interval of size $2\pi$: 
\begin{equation} \label{eq:eps_rellumi}
\epsilon(\phi) = \frac{N^{\uparrow}(\phi) - 
\mathcal{R} N^{\downarrow}(\phi)}{N^{\uparrow}(\phi) + \mathcal{R} N^{\downarrow}(\phi)}, \quad \phi \in [0, 2\pi].
\end{equation}
In contrast, because the geometric formula uses information from opposite sides of the detector, it is only defined on an interval of size $\pi$:
\begin{equation} \label{eq:eps_sqrt}
\epsilon(\phi) = \frac{\sqrt{N^{\uparrow}(\phi) N^{\downarrow}(\phi + \pi)} - \sqrt{N^{\downarrow}(\phi) N^{\uparrow}(\phi+\pi)}}{\sqrt{N^{\uparrow}(\phi) N^{\downarrow}(\phi+\pi)} + \sqrt{N^{\downarrow}(\phi) N^{\uparrow}(\phi+\pi)}}, \quad \phi \in [0, \pi].
\end{equation}
To get the raw asymmetry $\epsilon$, $\epsilon(\phi)$ is fit with a sinusoid and the amplitude is taken as $\epsilon$.
An example for the $\pi^0$ asymmetry in one $p_T$ bin is shown in Figure~\ref{fig:rawasymfit}. The uncertainty of the fit parameter is taken as the statistical uncertainty $\delta \epsilon$. To convert this raw asymmetry to a physics asymmetry $A_N$, it must be corrected for polarization and background contamination, as described in Section~\ref{sec:corrections}.

\subsection{Polarization and Relative Luminosity}
\label{sec:pollumi}
\begin{figure}
    \centering
    \includegraphics[width=0.8\linewidth]{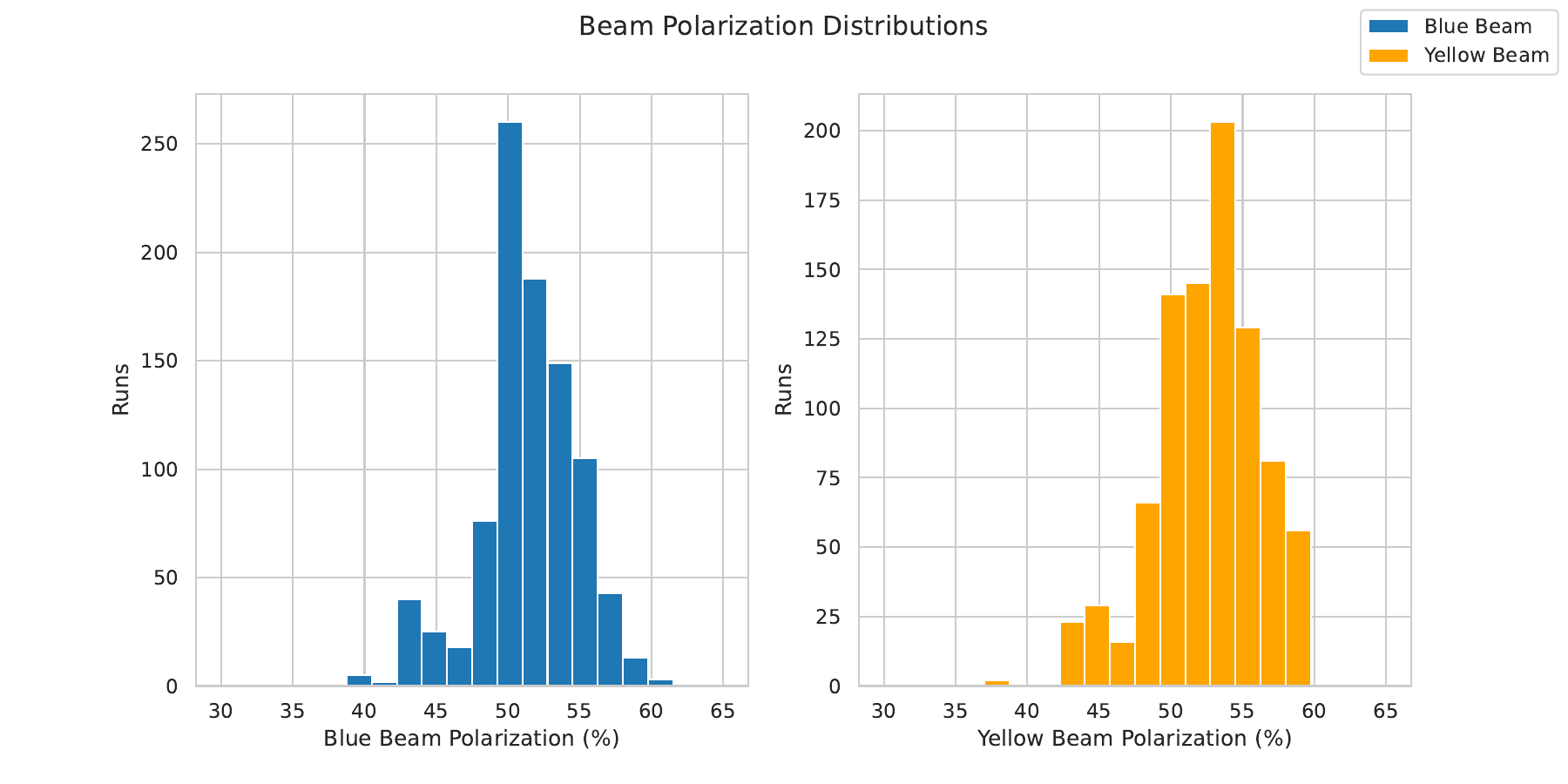}
    \caption{Distribution of beam polarization values for Blue and Yellow beams. Each entry corresponds to one run.}
    \label{fig:polarizations}
\end{figure}

\begin{figure}
    \centering
    \includegraphics[width=0.8\linewidth]{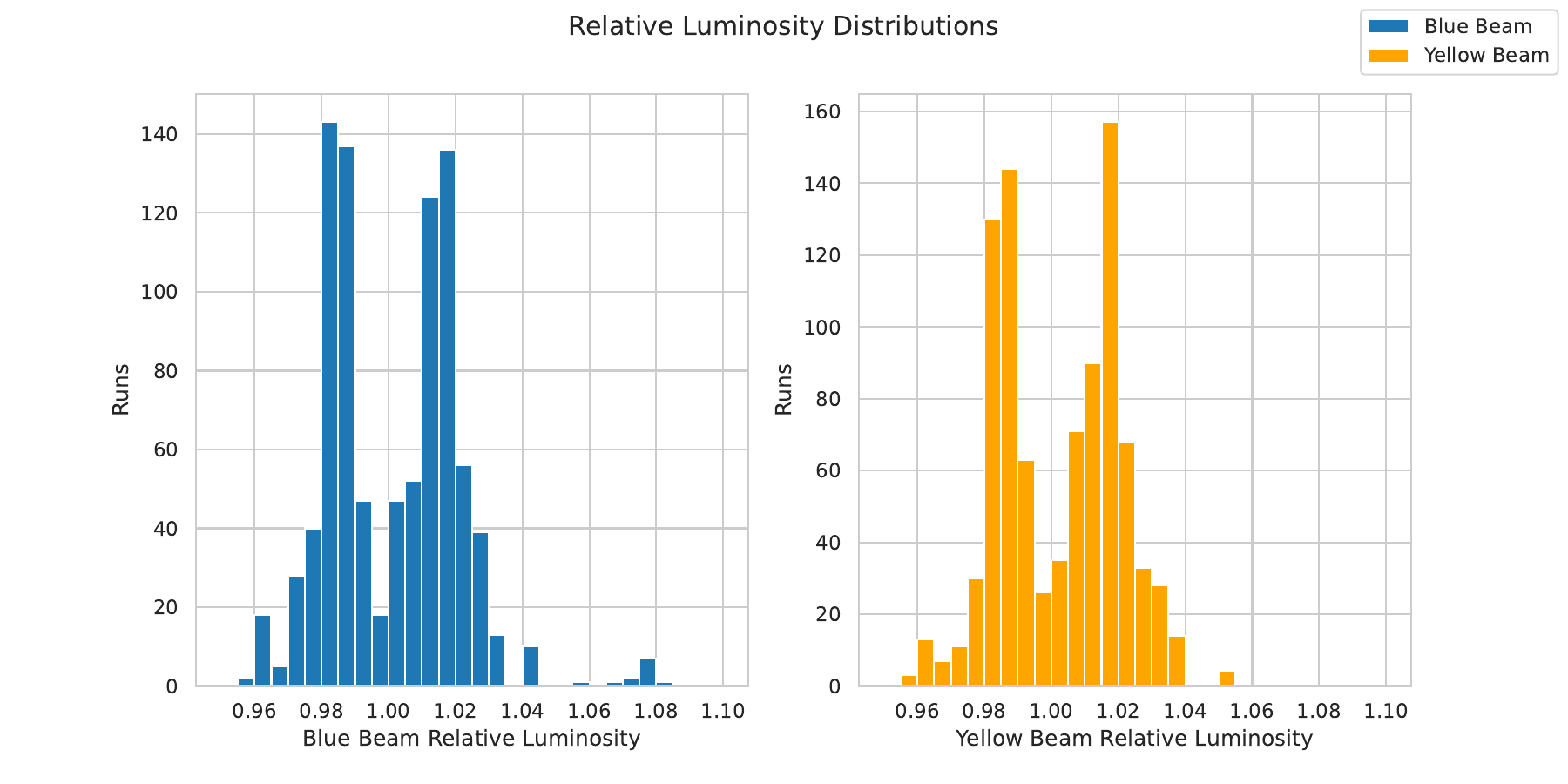}
    \caption{Distribution of relative luminosity values for Blue and Yellow beams. Each entry corresponds to one run.}
    \label{fig:relativelumis}
\end{figure}

The polarization and relative luminosity used in the asymmetry calculations are averaged values for the full 2024 run. The distributions of polarization and relative luminosity are shown in Figures~\ref{fig:polarizations} and \ref{fig:relativelumis}. The averages are weighted by total luminosity:

\begin{equation}
    \label{eq:avgpol}
    \langle P \rangle = \frac{\sum_{run\ i} \mathcal{L}_i P_i}{\sum \mathcal{L}_i},
\end{equation}
\begin{equation}
    \label{eq:avgrellumi}
    \langle \mathcal{R} \rangle = \frac{\sum_{run\ i} \mathcal{L}_i \frac{\mathcal{L}^\uparrow_i}{\mathcal{L}^\downarrow_i}}{\sum \mathcal{L}_i},
\end{equation}
where the sum is taken over all runs used in the analysis. Here $P_i$ is the beam polarization, $\mathcal{L}_i = \mathcal{L}_i^\uparrow + \mathcal{L}_i^\downarrow$ is the total luminosity, and $\mathcal{L}^{\uparrow (\downarrow)}$ is the spin-up (-down) luminosity, each for run $i$. 
The run-by-run polarization values used in Equation~\ref{eq:avgpol} are based on the fill-by-fill CNI polarimeter measurements, with initial scale calibration from the HJet polarimeter, provided by the RHIC polarimetry group \cite{RHICpolarimetry}. Thus, all runs within the same fill share the same polarization values. These values are not included in the sPHENIX DSTs, but rather are read from the Spin Database. The distribution of run-by-run polarization values for each beam is shown in Figure~\ref{fig:polarizations}. The polarization varies from about 40\% to 60\%.

The luminosities are measured using the MBD scalers. Specifically, the spin-dependent luminosities are calculated from the GL1P scalers, described in Section~\ref{sec:mbd}. The spin-up (-down) luminosity is taken as the sum of GL1P scalers for all spin-up (-down) bunch crossings. As with the polarization values, the GL1P scaler values are read from the Spin Database. The distribution of run-by-run relative luminosity values for each beam is shown in Figure~\ref{fig:relativelumis}. They range from roughly 0.96 to 1.08 for both beams.

The average polarization and relative luminosity are calculated independently for each beam. The resulting average polarization values are $\langle P \rangle = 0.51751$ for the Blue beam and $\langle P \rangle = 0.45687$ for the Yellow beam. The corresponding relative luminosities are $\langle \mathcal{R} \rangle = 1.000823$ and $\langle \mathcal{R} \rangle = 1.001796$ for Blue and Yellow beams, respectively. These value are given with six significant digits because the GL1P scalers are integers on the order of $10^6$.

\subsection{Angle Convention}
\begin{figure}
    \centering
    \includegraphics[width=0.8\linewidth]{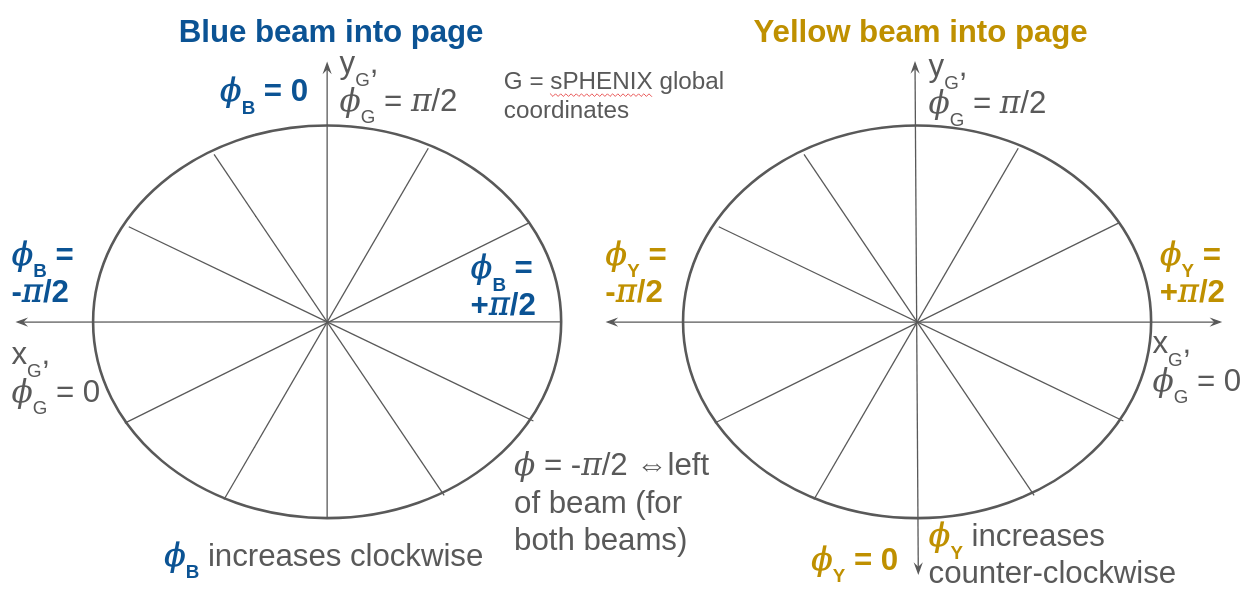}
    \caption{Illustration of the angular convention used in this analysis, showing the definitions of $\phi_B$ and $\phi_Y$.}
    \label{fig:phi_convention}
\end{figure}

Note that the $\phi$ values shown in Figure~\ref{fig:rawasymfit} and throughout the following sections are \textit{not} with respect to the global sPHENIX coordinate system, which we denote $\phi_G$\footnote{The global sPHENIX coordinate system is defined such that the positive $z$-axis lies along the direction of the Blue beam, the positive $y$-axis points vertically up, and the positive $x$-axis points away from the center of the RHIC ring to form a right-handed system.}. Because the global coordinate system is right-handed with respect to the Blue beam but left-handed with respect to the Yellow beam, using the global coordinate leads to the problem that $\phi_G = 0$ corresponds to the direction \textit{left} of the Blue beam and to the direction \textit{right} of the Yellow beam. To ensure consistency between a given $\phi$ value and its direction with respect to the polarized beam, beam-specific coordinates
\begin{equation}
    \phi_B \equiv \phi_G - \frac{\pi}{2}, \quad \phi_B \in [-\pi, \pi]
\end{equation}
and
\begin{equation}
    \phi_Y \equiv \phi_G + \frac{\pi}{2}, \quad \phi_Y \in [-\pi, \pi]
\end{equation}
are defined for the Blue and Yellow beams respectively, as shown in Figure~\ref{fig:phi_convention}. This transformation is made such that:
\begin{itemize}
    \item $\phi_{B,Y} = -\pi/2 \leftrightarrow$ Left of polarized beam
    \item $\phi_{B,Y} = 0 \leftrightarrow$ Vertical (up for Blue, down for Yellow)
    \item $\phi_{B,Y} = +\pi/2 \leftrightarrow$ Right of polarized beam
\end{itemize}

The function used to fit the $\phi$-dependent asymmetries is 
\begin{equation}\label{eq:eps_fit}
    \epsilon(\phi) = -\epsilon \sin(\phi),
\end{equation}
where the amplitude $\epsilon$ is the raw asymmetry and $\phi$ is identified with $\phi_{B(Y)}$ when the Blue (Yellow) beam is taken as polarized. The overall negative sign is chosen such that a positive amplitude corresponds to a positive left-right asymmetry (see Equation~\ref{eq:AN_def}). Note that no phase offset parameter is included in the fit. Such a parameter could account for any off-vertical polarization of the beams. However, local polarimetry analysis concludes that no significant deviation from vertical polarization is observed in the 2024 data~\cite{SpinQANote}. Further, a study of the effects of including a phase offset term in the fit showed that the extracted fit amplitudes are nearly identical with and without a free phase. Therefore we choose to omit the phase term, ensuring consistency with Equation~\ref{eq:AN_def}.

\section{Raw Asymmetry Results}\label{sec:rawresults}
\begin{figure}
    \centering
    \includegraphics[width=0.7\linewidth]{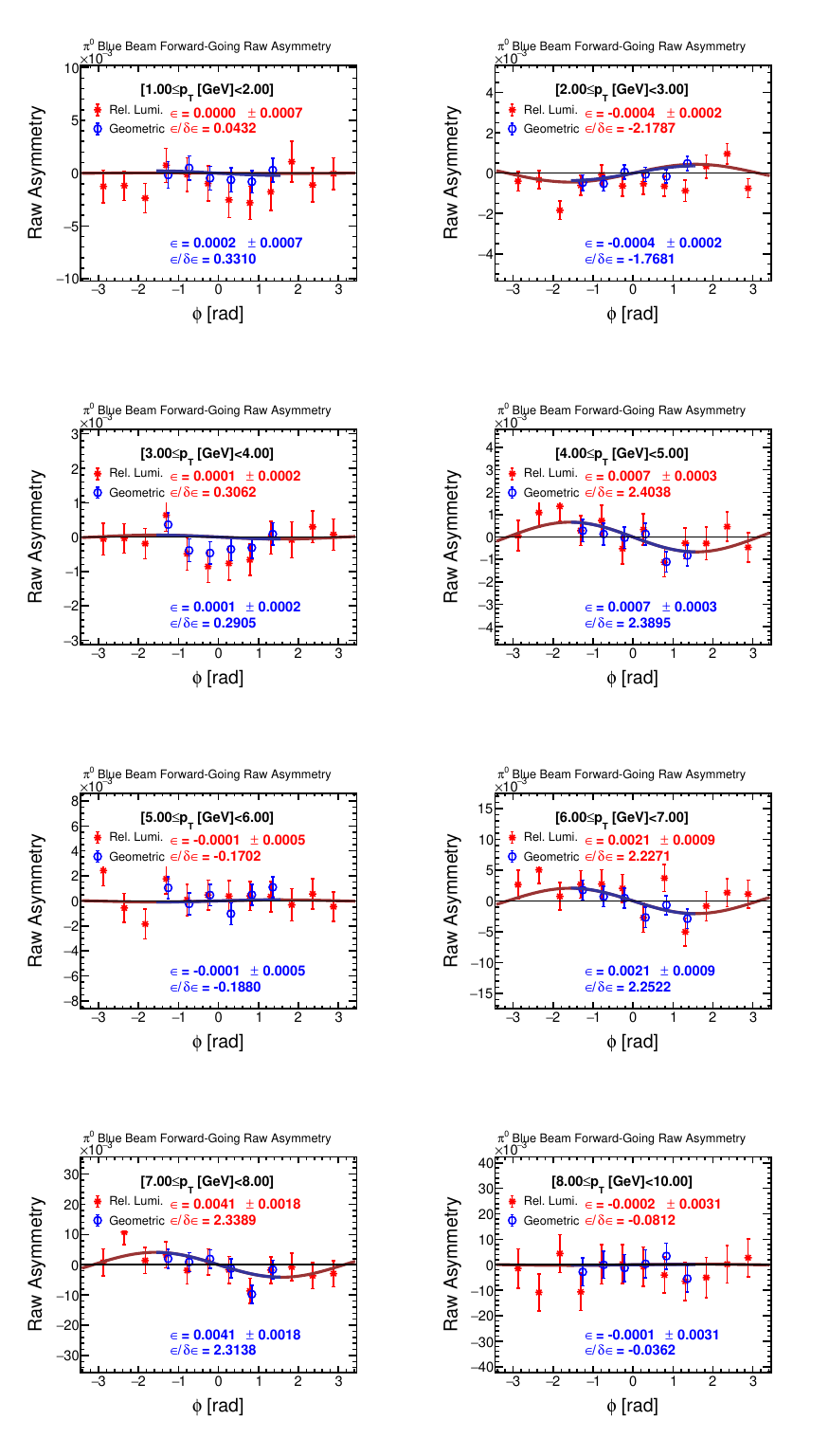}
    \caption{$\phi$-dependent raw asymmetries for the $\pi^0$ signal region with respect to the Blue beam. Each pad corresponds to one $p_T$ bin.}
    \label{fig:pi0_blue_raw}
\end{figure}

\begin{figure}
    \centering
    \includegraphics[width=0.7\linewidth]{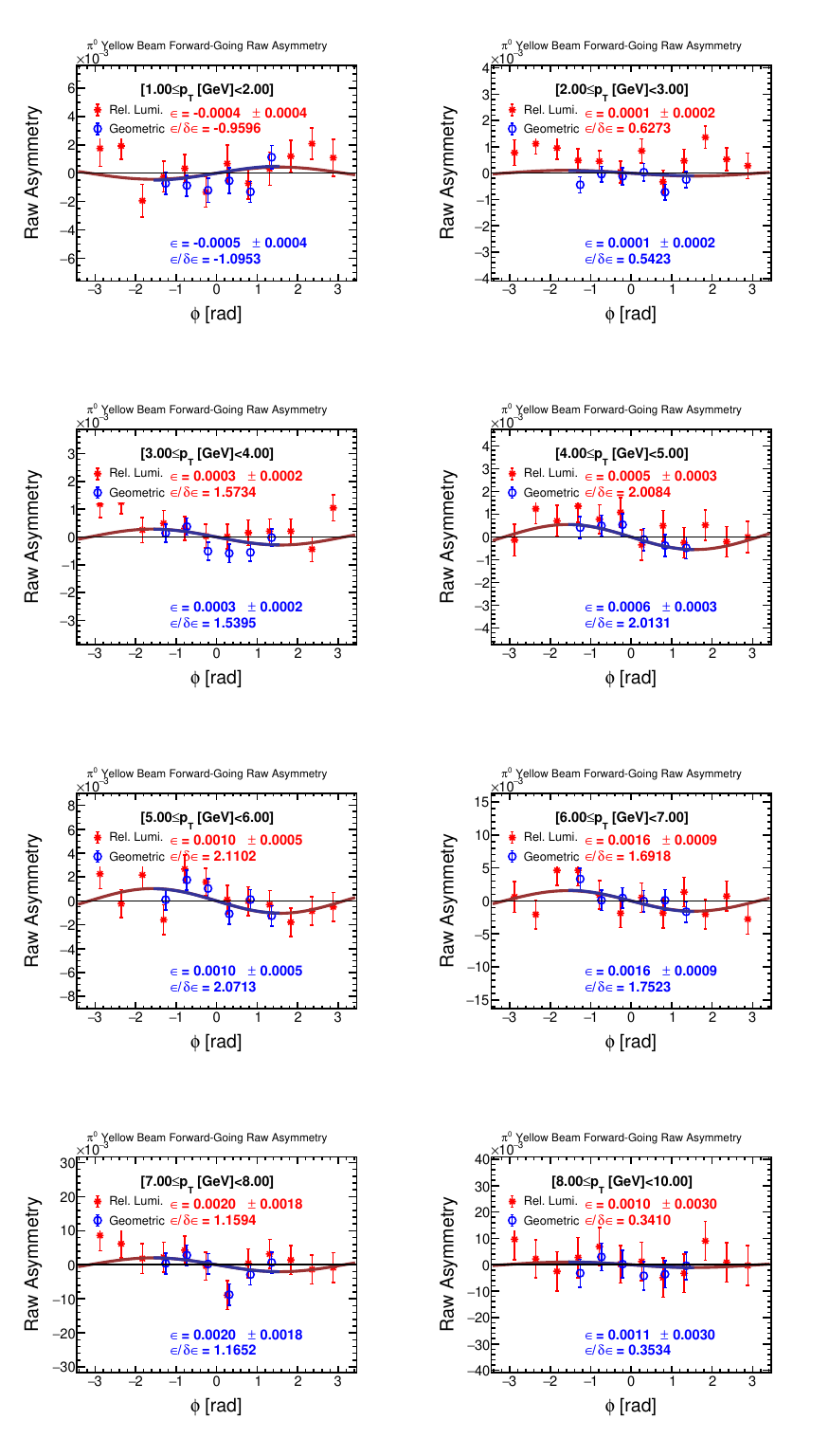}
    \caption{$\phi$-dependent raw asymmetries for the $\pi^0$ signal region with respect to the Yellow beam. Each pad corresponds to one $p_T$ bin.}
    \label{fig:pi0_yellow_raw}
\end{figure}

\begin{figure}
    \centering
    \includegraphics[width=0.7\linewidth]{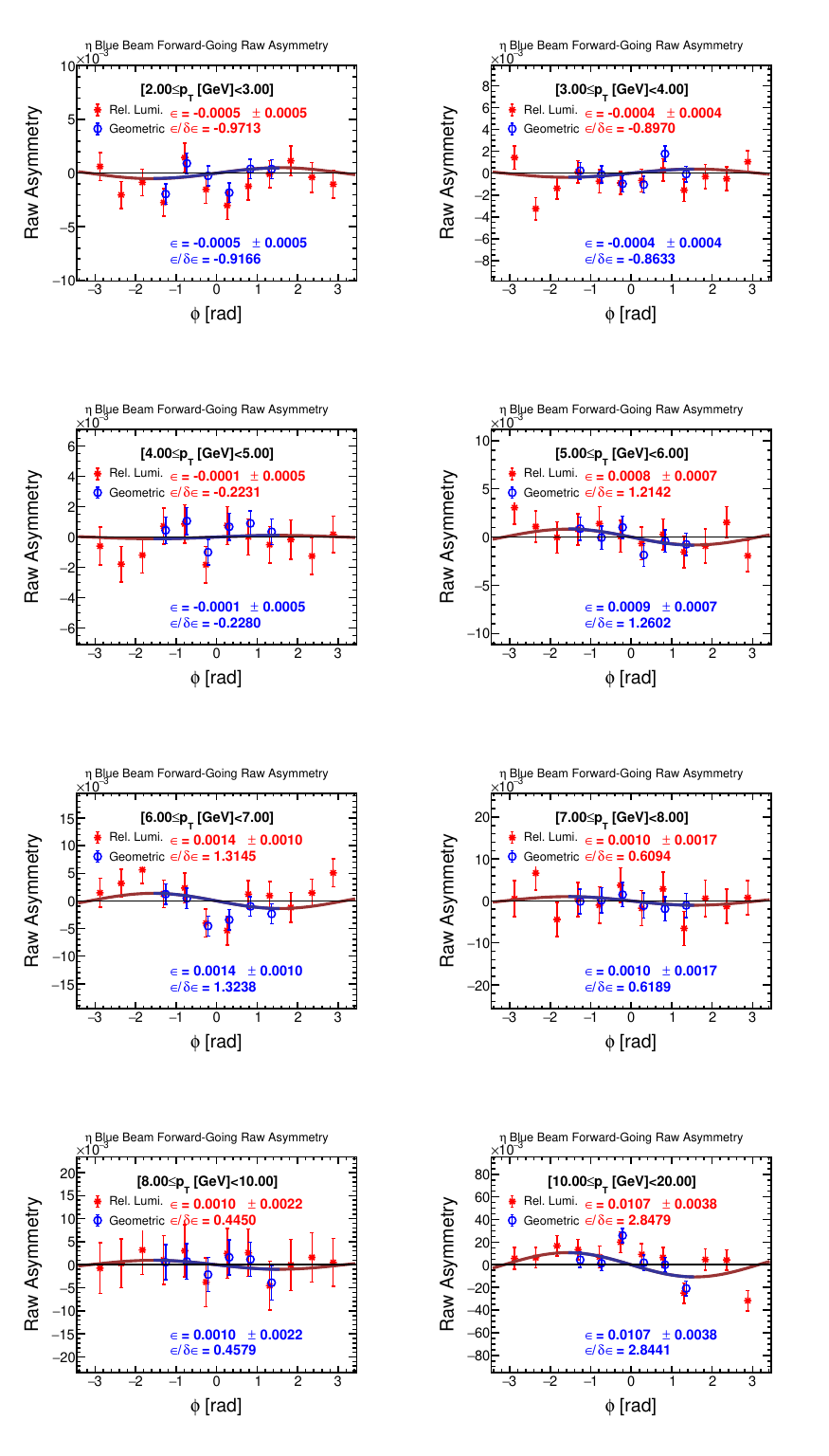}
    \caption{$\phi$-dependent raw asymmetries for the $\eta$ signal region with respect to the Blue beam. Each pad corresponds to one $p_T$ bin.}
    \label{fig:eta_blue_raw}
\end{figure}

\begin{figure}
    \centering
    \includegraphics[width=0.7\linewidth]{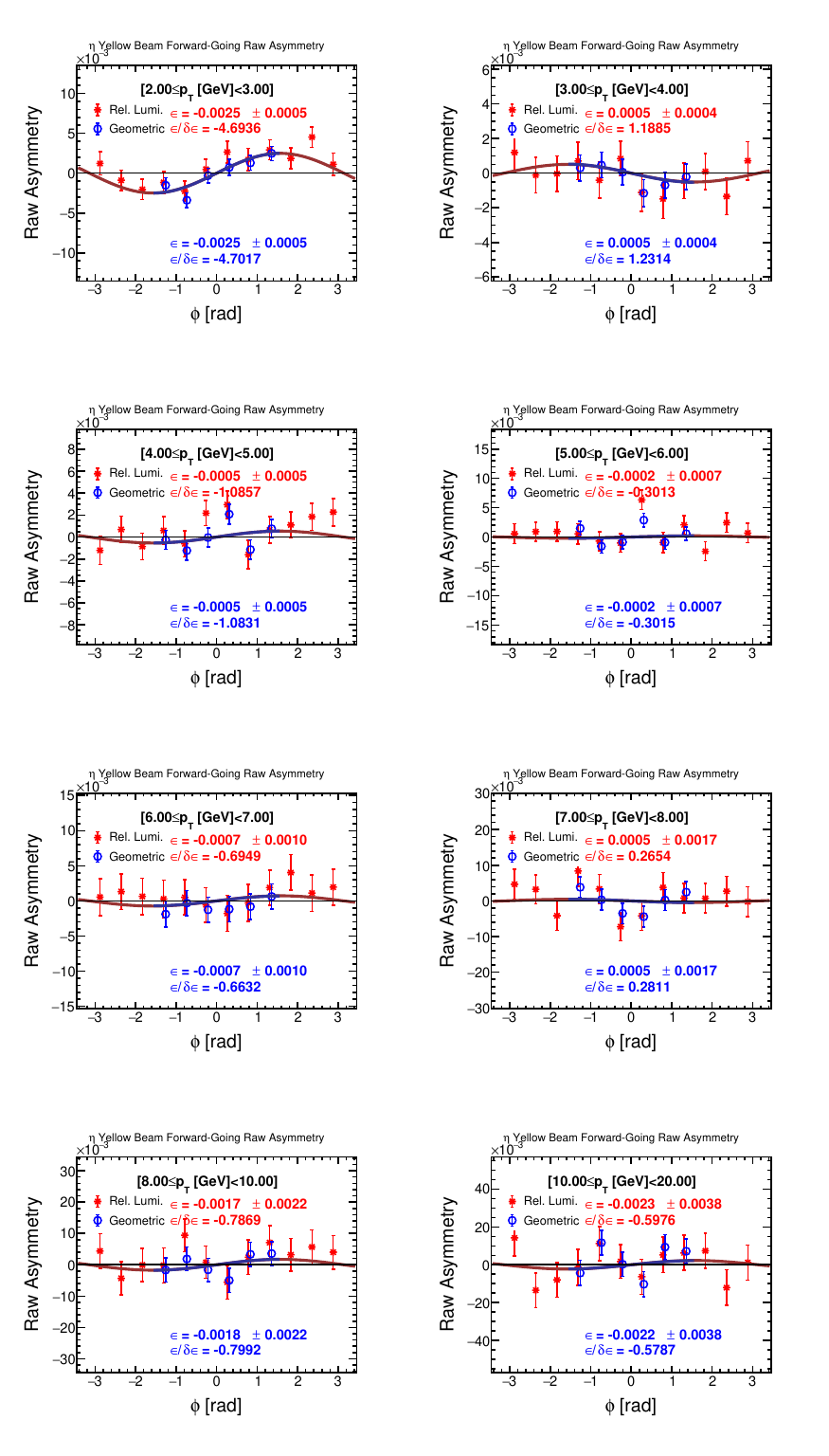}
    \caption{$\phi$-dependent raw asymmetries for the $\eta$ signal region with respect to the Yellow beam. Each pad corresponds to one $p_T$ bin.}
    \label{fig:eta_yellow_raw}
\end{figure}

The $\phi$-dependent raw asymmetries for forward-going mesons with respect to the Blue and Yellow beams are shown in Figures~\ref{fig:pi0_blue_raw}-\ref{fig:eta_yellow_raw}. Here the asymmetries are shown in different bins in meson $p_T$; this kinematic binning anticipates the main result of this analysis, the $p_T$-dependence of $A_N$.


\begin{figure}
    \centering   \includegraphics[page=1,width=0.48\linewidth]{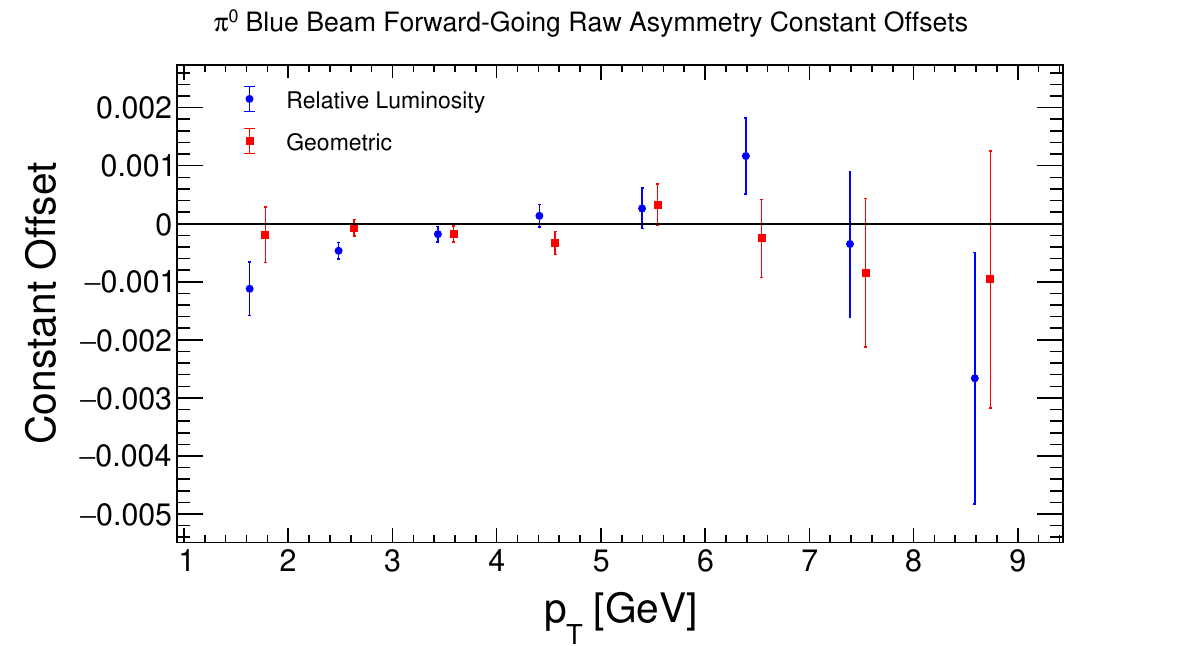}  \includegraphics[page=1,width=0.48\linewidth]{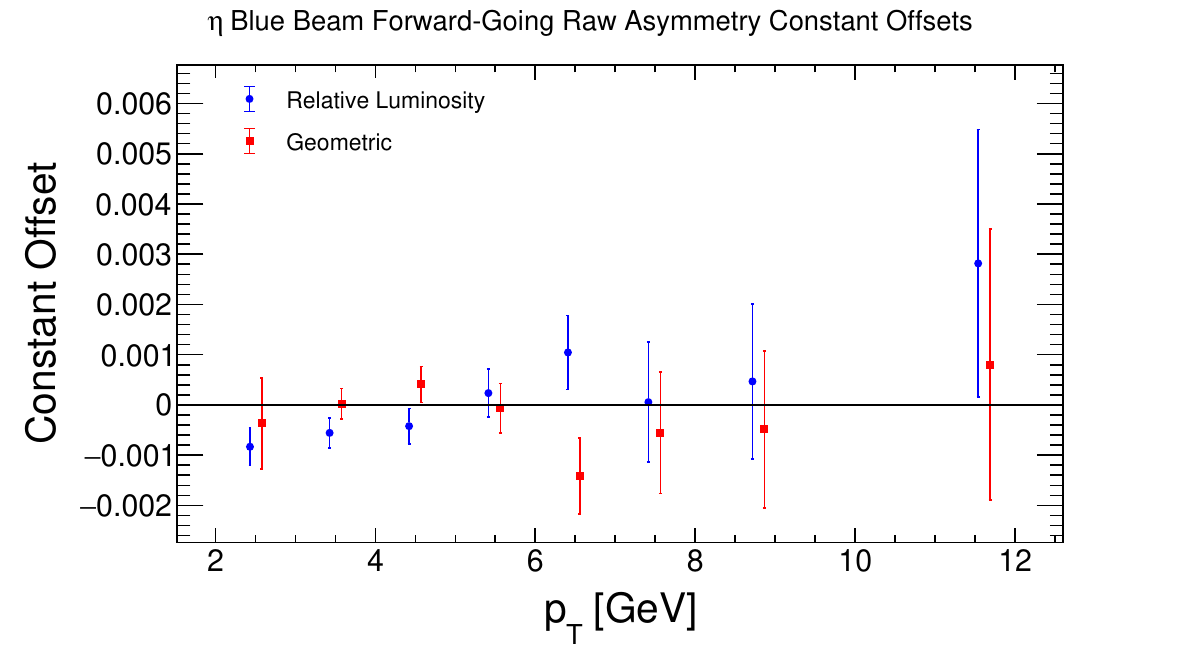}
    \includegraphics[page=2,width=0.48\linewidth]{fig/pi0_offsets}  \includegraphics[page=2,width=0.48\linewidth]
    {fig/eta_offsets.pdf}
    \includegraphics[page=3,width=0.48\linewidth]{fig/pi0_offsets}  \includegraphics[page=3,width=0.48\linewidth]{fig/eta_offsets.pdf}
    \caption{Constant offset in the sinusoidal fit to the $\phi$-dependent asymmetry versus meson transverse momentum $p_T$. The left column shows the $\pi^0$; the right, the $\eta$-meson.}
    \label{fig:constant-term}
\end{figure}

In Figures~\ref{fig:pi0_blue_raw}-\ref{fig:eta_yellow_raw}, some sets of raw asymmetry points are offset from zero by a constant upward or downward shift. Figure~\ref{fig:constant-term} illustrates the meson $p_T$-dependence of this constant offset, comparing both the two asymmetry extraction formulas and the two beams. The offset is most pronounced in the lowest $p_T$ bins and is systematically larger for the relative luminosity formula than for the geometric formula. It also appears to have opposite sign for the Blue and Yellow beams.

The origin, $p_T$- and beam-dependence of this constant offset have not been identified. In principle, a constant offset can be introduced in the relative luminosity asymmetry if an incorrect value is used for $\mathcal{R}$. This was tested by recalculating the relative luminosity asymmetries using artificially large ($\mathcal{R} = 1.1$) and small ($\mathcal{R} = 0.9$) values. Doing so does introduce a large constant shift in all asymmetry values, with opposite sign in the two cases. However, this change in offset due to $\mathcal{R}$ is the same across all $p_T$ bins; the smaller-scale differences in constant offset for different $p_T$ are unaffected. Likewise the opposite behavior with respect to the two beams is unchanged. This is consistent with the mathematical expectation, based on Equation~\ref{eq:eps_rellumi}, that changing the value of $\mathcal{R}$ should not introduce any kinematic dependence.

The effect of including a constant term in the sinusoidal fit was tested. This test showed that the extracted fit amplitudes are similar with and without this constant term, even in the $p_T$ bins for which the offset is largest. We emphasize that the existence of this constant offset should not be interpreted as a failure of the asymmetry extraction method. Rather, it should be regarded as a limitation of the method relevant to the lowest $p_T$ bins considered in this analysis, particularly for the relative luminosity calculation. This limitation will be revisited in Chapter~\ref{ch:summary}. For the remainder of the analysis, we proceed with the fitting procedure given by Equation~\ref{eq:eps_fit} and do not include a constant offset parameter.

\clearpage

\begin{figure}
    \centering   
    \includegraphics[page=198,width=0.48\linewidth]{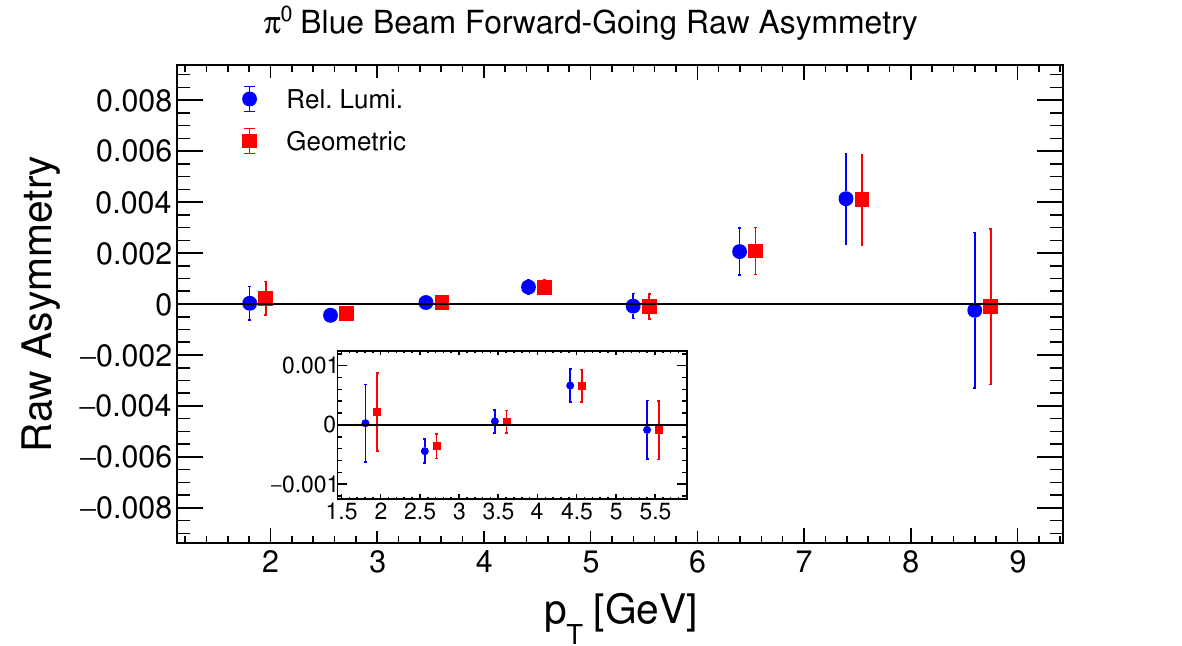}
    \includegraphics[page=203,width=0.48\linewidth]{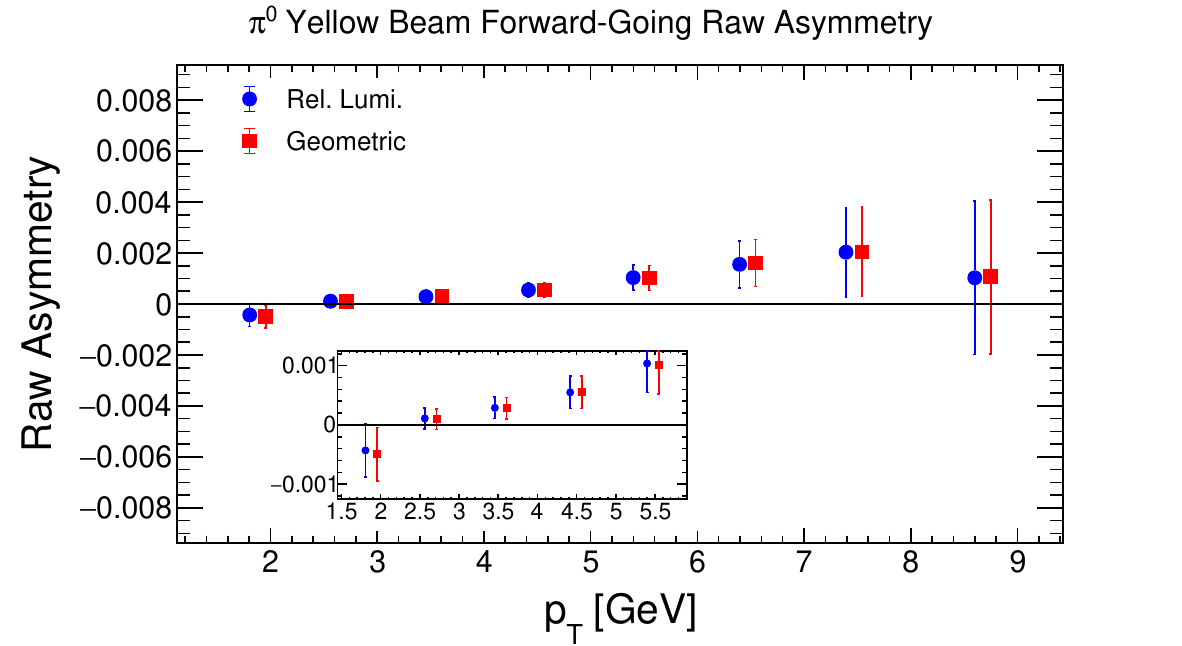}
    \includegraphics[page=209,width=0.48\linewidth]{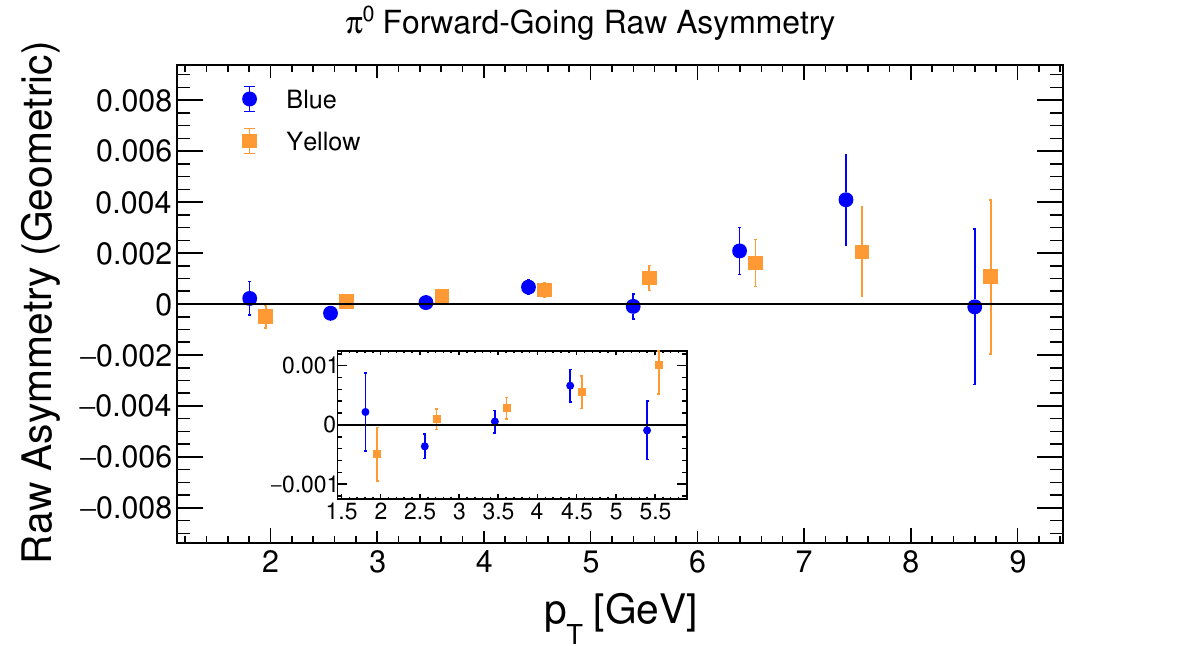}
    \caption{Raw asymmetries for the $\pi^0$ signal region, plotted against di-photon $p_T$. Top: comparison of the two calculation methods, with respect to the Blue (left) and Yellow (right) beam. Bottom: comparison of the Blue and Yellow beam asymmetries using the geometric calculation.}
    \label{fig:pi0_raw_pT}
\end{figure}

\begin{figure}
    \centering
    \includegraphics[page=264,width=0.48\linewidth]{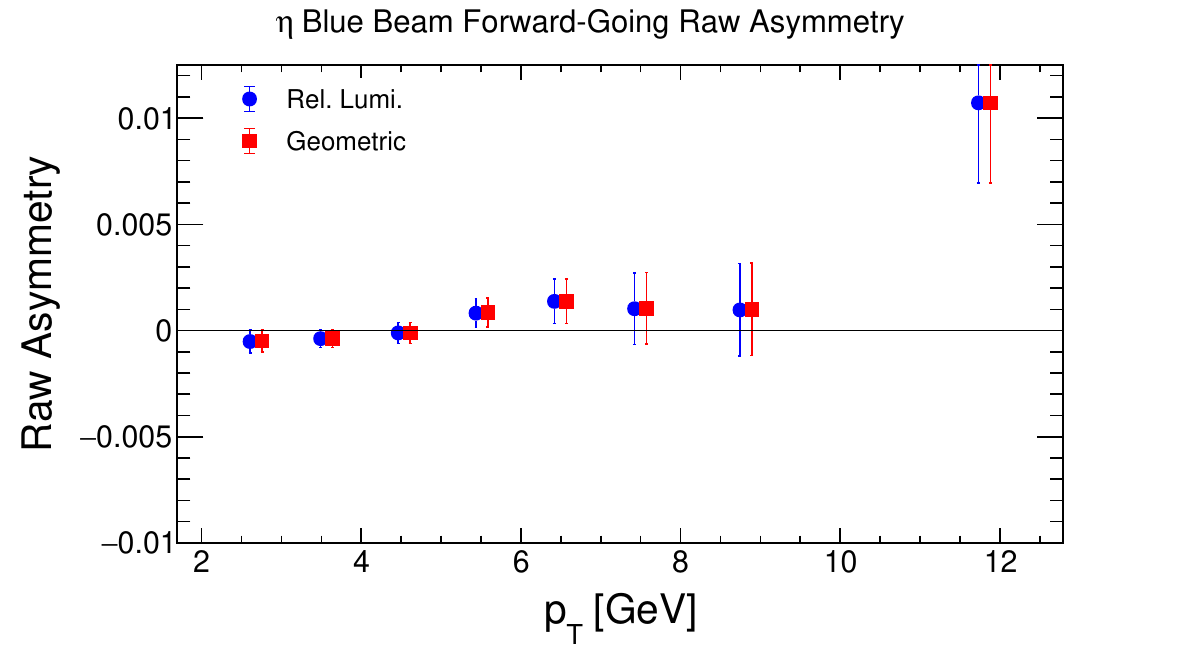}    
    \includegraphics[page=269,width=0.48\linewidth]{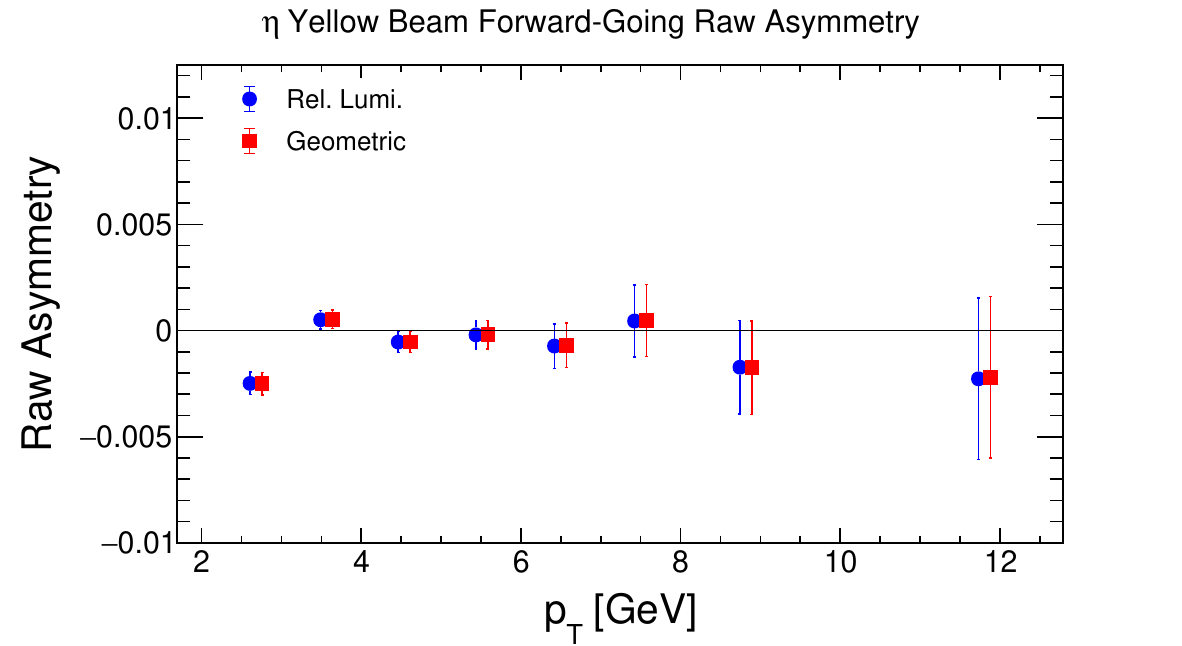}
    \includegraphics[page=275,width=0.48\linewidth]{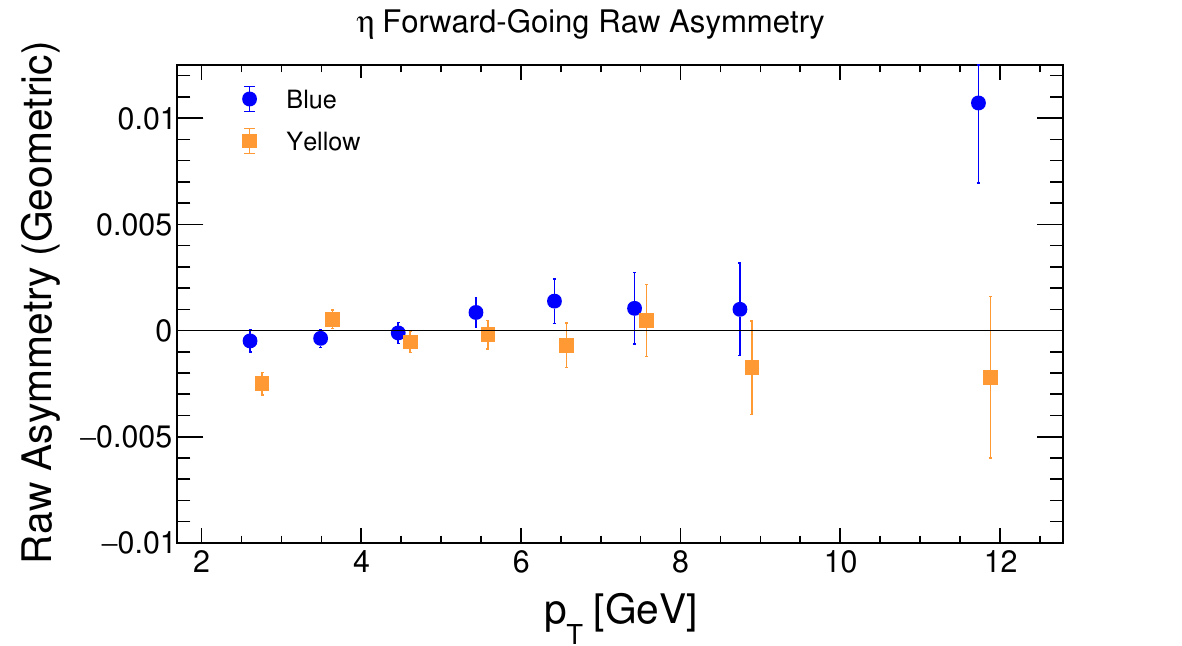}
    \caption{Raw asymmetries for the $\eta$-meson signal region, plotted against di-photon $p_T$. Top: comparison of the two calculation methods, with respect to the Blue (left) and Yellow (right) beam. Bottom: comparison of the Blue and Yellow beam asymmetries using the geometric calculation.}
    \label{fig:eta_raw_pT}
\end{figure}

The raw asymmetries extracted in each $p_T$ bin are shown in Figure~\ref{fig:pi0_raw_pT} for the $\pi^0$ and in Figure~\ref{fig:eta_raw_pT} for the $\eta$-meson. These figures show there is excellent agreement between the relative luminosity and geometric asymmetries. This is expected, as the two calculations are performed on the same data and are therefore not statistically independent. Agreement between Blue and Yellow beam asymmetries is less pronounced, because in this case the two data samples \textit{are} statistically independent. Nonetheless, results from the two beams are generally compatible at the 1-$\sigma$ level. The compatibility of relative luminosity with geometric and Blue beam with Yellow beam results will be discussed in more detail in Section~\ref{sec:stat-results}. The corresponding raw asymmetries for the background side band regions are shown in the following subsection, wherein their importance will be described.

\section{Asymmetry Corrections} 
\label{sec:corrections}
To calculate the physics asymmetry $A_N$, the raw asymmetry $\epsilon$ must be corrected for both beam polarization and background contamination. The polarization correction simply scales $\epsilon$ and its statistical uncertainty $\delta\epsilon$ by the proton beam polarization, $1/P$:
\begin{eqnarray}
\epsilon \longrightarrow A_N^{sig} &=& \frac{1}{P} \epsilon,\\
\delta\epsilon \longrightarrow \delta A_N^{sig} &=& \frac{1}{P} \delta\epsilon,
\end{eqnarray}
where the value of $P$ is taken from Section~\ref{sec:pollumi} and the $sig$ superscript signifies that this asymmetry is calculated from all di-photons within the signal mass region. Accordingly, $A_N^{sig}$ contains effects from both true mesons and combinatorial background:
\begin{equation}\label{eq:invertedbkgrcorr}
    A_N^{sig} = (1-r)A_N^{meson} + r A_N^{bkgr},
\end{equation}
where $r$ is the background fraction and $A_N^{bkgr}$ is the polarization-corrected asymmetry for background (i.e. non-meson) di-photons. To find the physics asymmetry $A_N$ we invert Equation~\ref{eq:invertedbkgrcorr}:
\begin{equation}
    A_N \equiv A_N^{meson} = \frac{A_N^{sig} - r A_N^{bkgr}}{1-r}.
    \label{eq:bkgrcorr}
\end{equation}

The background fraction $r$ is calculated by fitting the di-photon invariant mass distribution to a Gaussian signal plus threshold with exponential background function; the fitting procedure is detailed in the following subsection. The true-meson and background yields, $N_{meson}$ and $N_{bkgr}$ respectively, are taken as the integrals of the resulting fit functions within the signal mass regions. The background fraction is then calculated as
\begin{equation}
r = \frac{N_{bkgr}}{N_{meson}+N_{bkgr}}.
\label{eq:bgfrac}
\end{equation}
The corresponding statistical uncertainty $\delta r$ is obtained by propagating the covariance matrix of the fit parameters through to the function integrals. The $p_T$-dependent invariant mass distributions, with corresponding background fractions, are shown in the following subsection.

\begin{figure}
    \centering
    \includegraphics[page=213, width=0.48\linewidth]{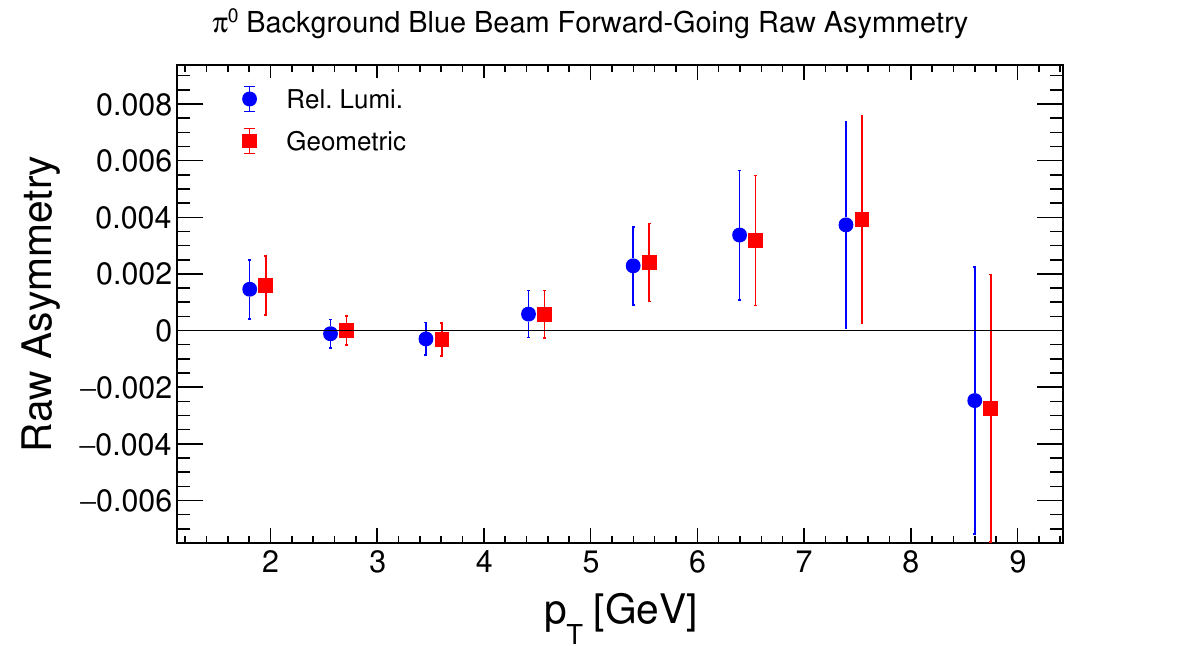}
    \includegraphics[page=214, width=0.48\linewidth]{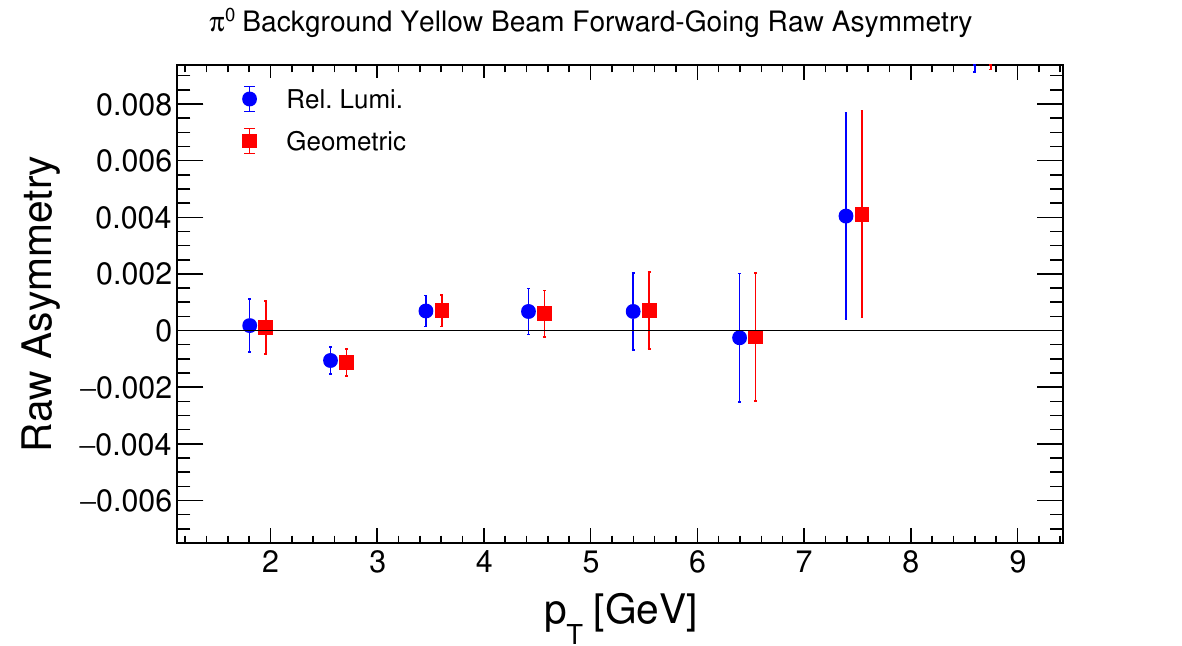}
    \caption{Raw asymmetries for di-photons in the $\pi^0$ background side band regions, for (left) Blue beam and (right) Yellow beam.}
    \label{fig:pi0_bkgr}
\end{figure}

\begin{figure}
    \centering
    \includegraphics[page=279, width=0.48\linewidth]{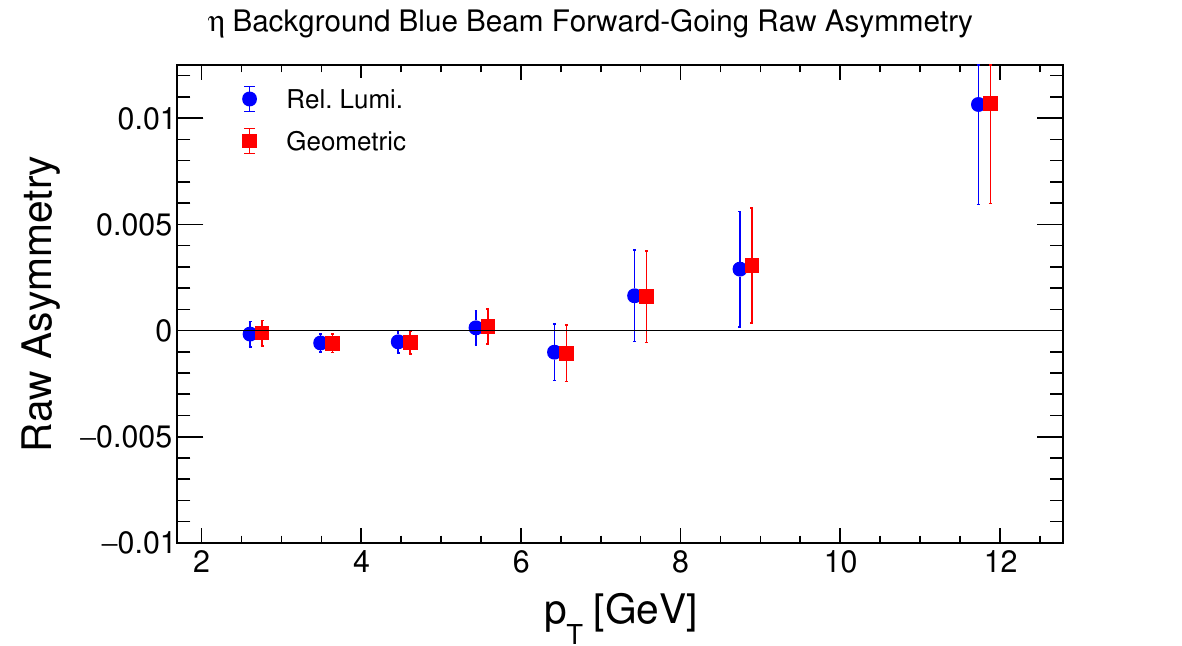}
    \includegraphics[page=280, width=0.48\linewidth]{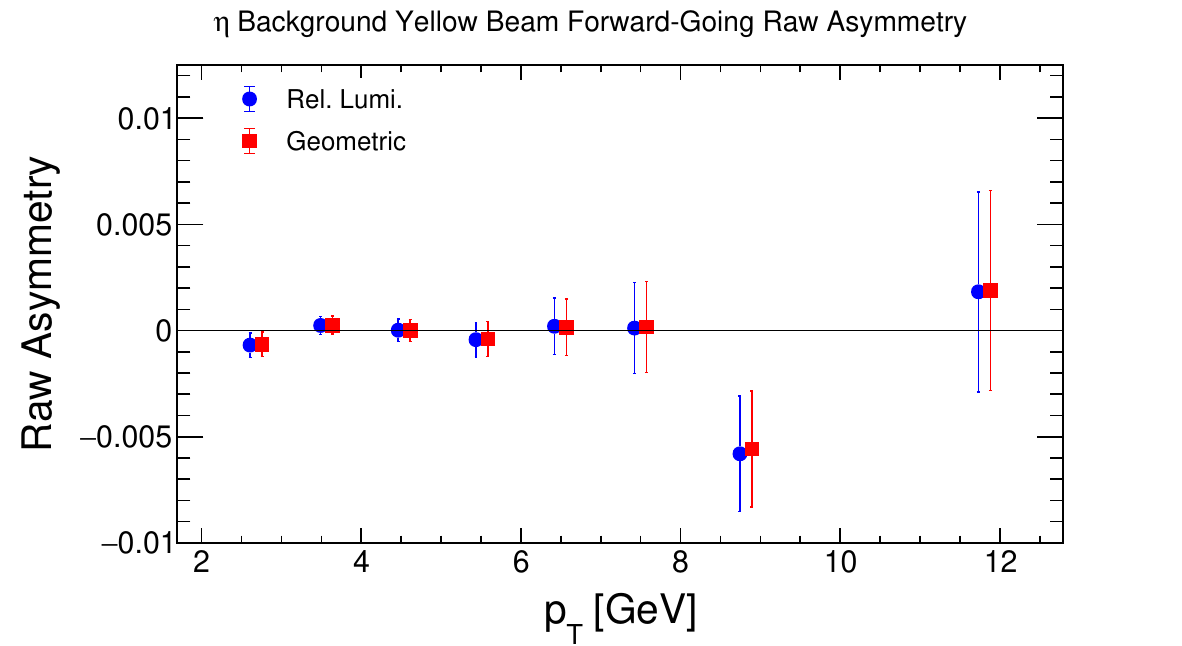}
    \caption{Raw asymmetries for di-photons in the $\eta$-meson background side band regions, for (left) Blue beam and (right) Yellow beam.}
    \label{fig:eta_bkgr}
\end{figure}

The background asymmetry $A_N^{bkgr}$ cannot be calculated directly. Instead, it is estimated by calculating the asymmetry for di-photons in the \textit{background side bands}, which correspond to the lighter shaded regions in Figure~\ref{fig:integratedDiphotonMass}. These regions are chosen by selecting a $3.5-5.5 \sigma$ window away from the meson mass peaks. The resulting mass ranges are $[30, 70]$ MeV and $[209, 249]$ MeV for the $\pi^0$, and $[257, 371]$ MeV and $[767, 880]$ MeV for the $\eta$-meson. By estimating the background asymmetry in this way, we implicitly assume the asymmetry for combinatorial background in the signal region is the same as that for the background side bands. The raw asymmetries for di-photons in the background side bands are shown in Figure~\ref{fig:pi0_bkgr} for the $\pi^0$ and Figure~\ref{fig:eta_bkgr} for the $\eta$-meson. These 
asymmetries are generally consistent with zero asymmetry at the 1-$\sigma$ level. Out of 32 points, 10 deviate from zero by more than 1-$\sigma$ and only one point deviates by more than 2-$\sigma$. This is consistent with statistical noise on a zero measurement. We therefore conclude that no nonzero background asymmetry is observed.


The statistical uncertainty $\delta A_N$ on the physics asymmetry is determined by propagation of error using the statistical uncertainties on the signal-window asymmetry ($\delta A_N^{\mathrm{sig}}$), the background asymmetry ($\delta A_N^{\mathrm{bkgd}}$), and the background fraction ($\delta r$):
\begin{equation}
\delta A_N\equiv\delta A_N^{\mathrm{meson}} = \sqrt{ \left(\frac{1}{1 - r} \delta A_N^{\mathrm{sig}} \right)^2 + \left(\frac{r}{1 - r} \delta A_N^{\mathrm{bkgr}} \right)^2 + \left(\frac{A_N^{\mathrm{sig}} - A_N^{\mathrm{bkgr}}}{(1 - r)^2} \delta r \right)^2 }.
\label{eq:deltaAbg}
\end{equation}
This procedure increases the statistical uncertainties because background contributions are removed.


\clearpage

\subsection{Di-photon Invariant Mass Fitting}\label{sec:fitting}
As described above, the background fraction $r$ in each kinematic bin is determined by fitting separate signal and background contributions to the di-photon invariant mass distributions. The $\pi^0$ and $\eta$-meson signals are modeled as Gaussians; the background is modeled using a threshold function with exponential. The implications of this choice for assessing the background contribution are discussed in Section~\ref{sec:sys-bg}.

To fit the background, we first define a threshold mass $m_\mathrm{thresh}$ as the lowest mass value for which the distribution reaches 0.1\% of its maximum, to approximate the mass below which no di-photons are produced. For masses above $m_\mathrm{thresh}$, we fit the distribution using the following function:
\begin{equation}
  f(m) = A \exp \left( q \log \left( \frac{m - m_{\mathrm{thresh}}}{m_{\mathrm{max}} - m_{\mathrm{thresh}}} \right) + \frac{q}{p} \left( 1 - \left( \frac{m - m_{\mathrm{thresh}}}{m_{\mathrm{max}} - m_{\mathrm{thresh}}} \right)^p \right)\right),
\end{equation}
with free parameters $A, m_\mathrm{max}, q$ and $p$. Here $A$ is the overall amplitude, $m_\mathrm{max}$ is the mass at which the background reaches its maximum, and $q$ and $p$ control the steepness of the curve.

The fit is performed twice: once for the $\pi^0$ region (from $m_\mathrm{thresh}$ to 0.27 GeV), and separately for the $\eta$-meson region (from 0.32 to 1 GeV). For each region, the fit proceeds in three steps. First, the background function is fit in the respective meson region, but excluding the meson peak. For the $\pi^0$ this is the region $[m_\mathrm{thresh}, \max(m_\mathrm{offset}, 0.1)] \cup [0.18, 0.27]$, where numerical values are in units of GeV and $m_\mathrm{offset}$ is defined as the lowest mass for which the distribution reaches 10\% of its maximum. Thus $m_\mathrm{offset}$ approximates the low-mass edge of the $\pi^0$ peak. For the $\eta$-meson, the background fit region is $[0.35, 0.45] \cup [0.75, 1.0]$, again in units of GeV. Next, the background parameters are fixed at their fitted values, and the full background-plus-signal model is fit to the full meson region. Finally, the background parameters are released, and the full background-plus-signal model is fit again.

\begin{figure}
    \centering
    \includegraphics[page=14, width=0.48\linewidth]{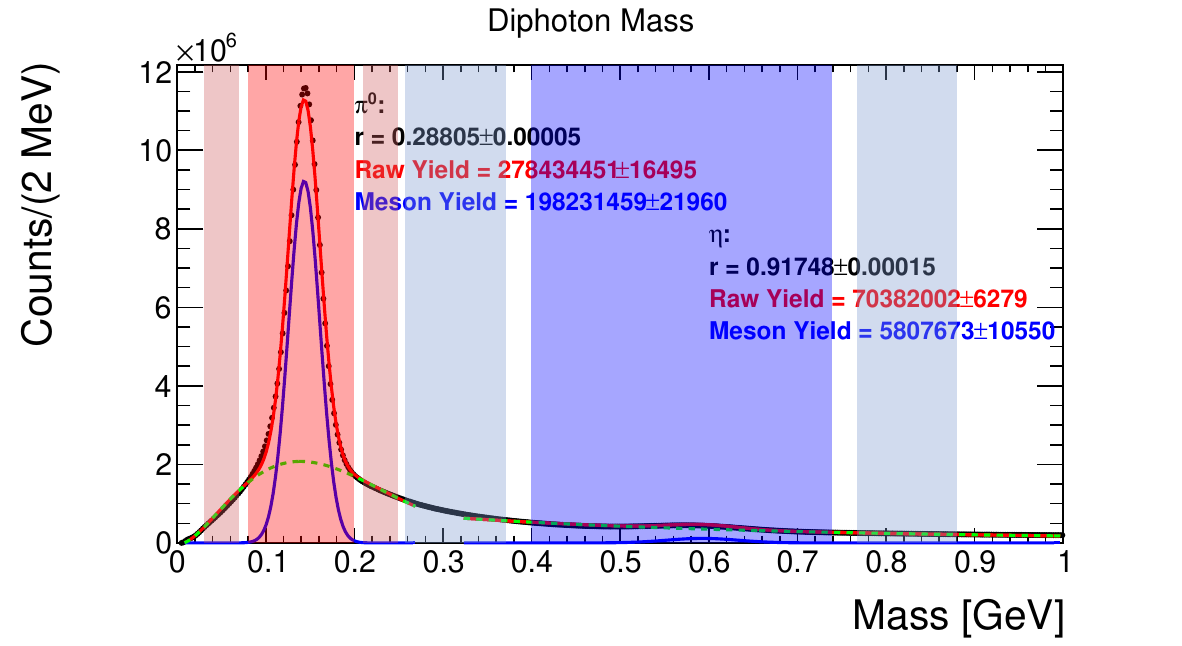}
    \includegraphics[page=15, width=0.48\linewidth]{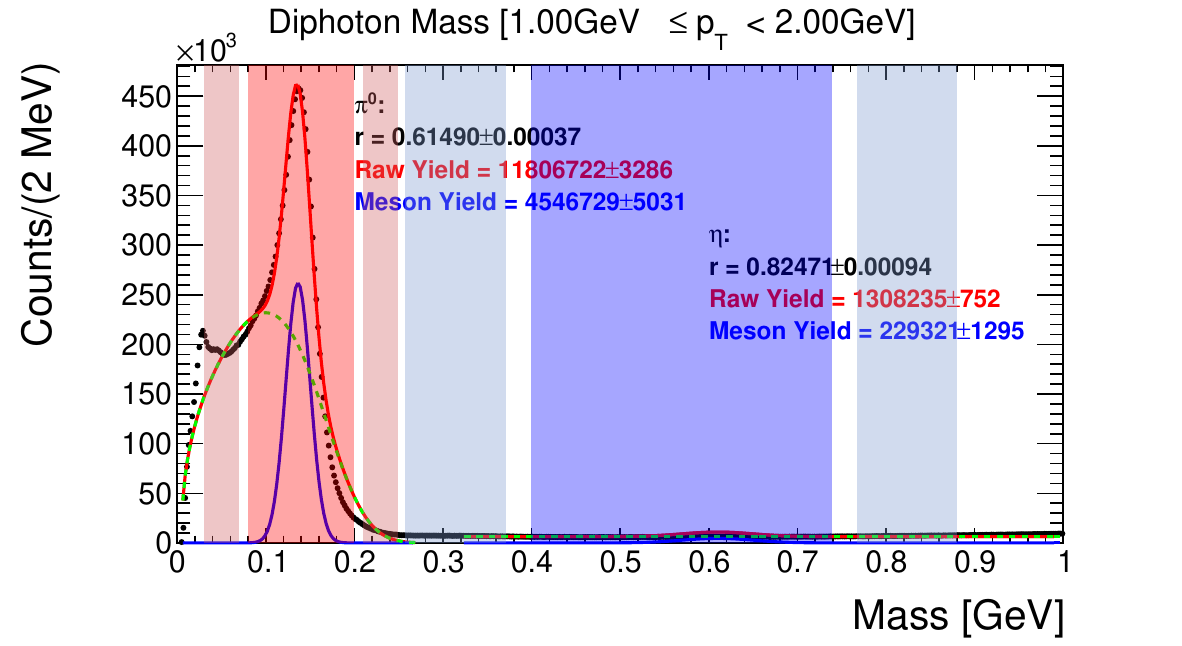}
    \includegraphics[page=16, width=0.48\linewidth]{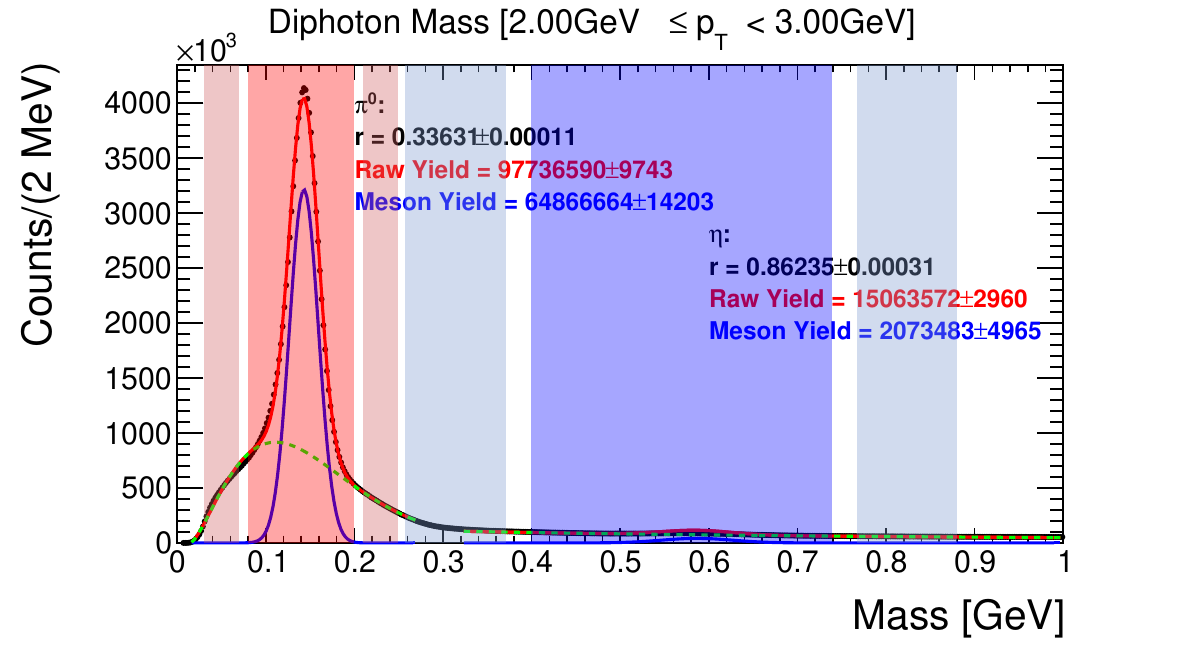}
    \includegraphics[page=17, width=0.48\linewidth]{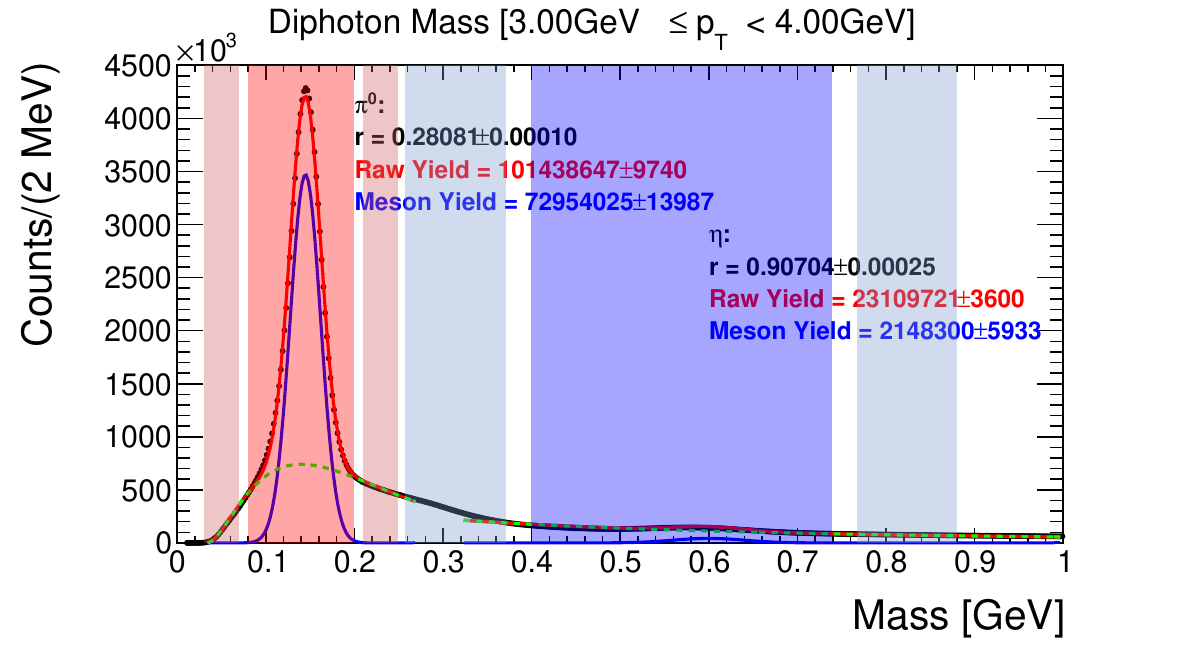}
    \includegraphics[page=18, width=0.48\linewidth]{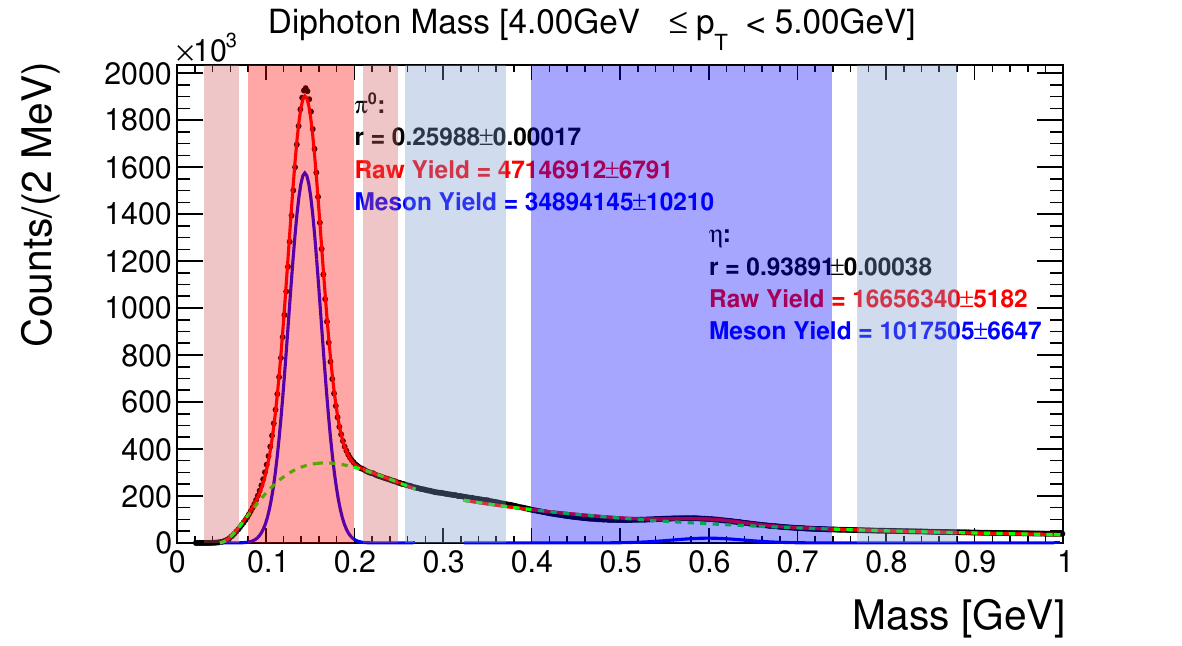}
    \includegraphics[page=19, width=0.48\linewidth]{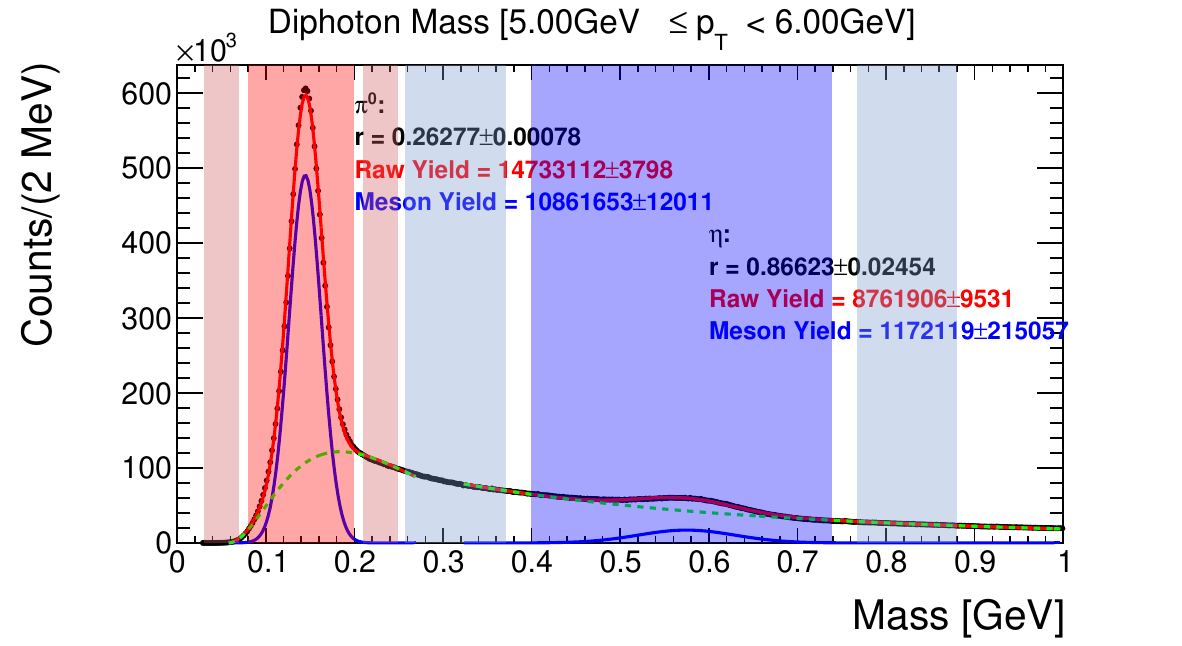}
    \includegraphics[page=20, width=0.48\linewidth]{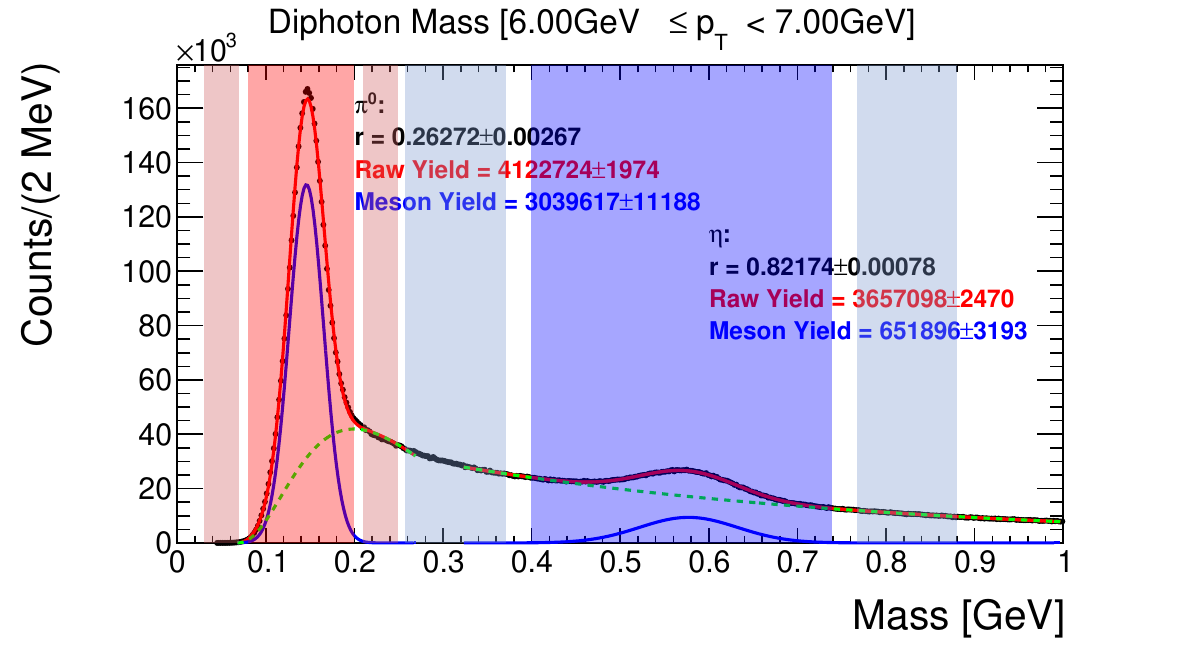}
    \includegraphics[page=21, width=0.48\linewidth]{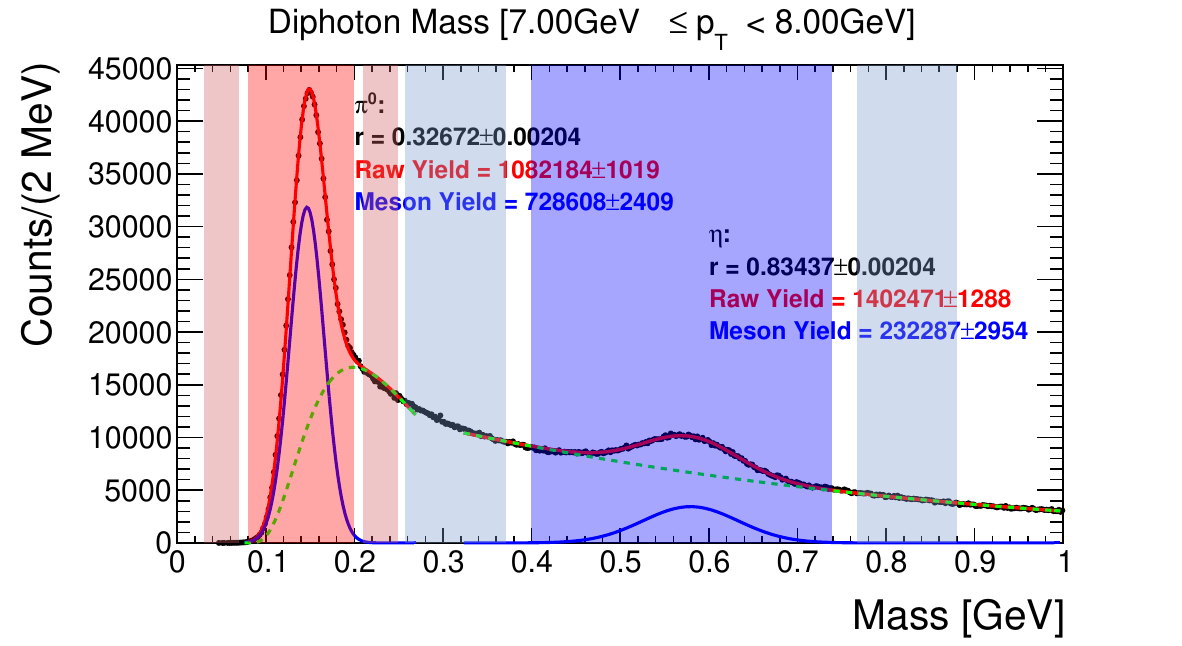}
    \includegraphics[page=22, width=0.48\linewidth]{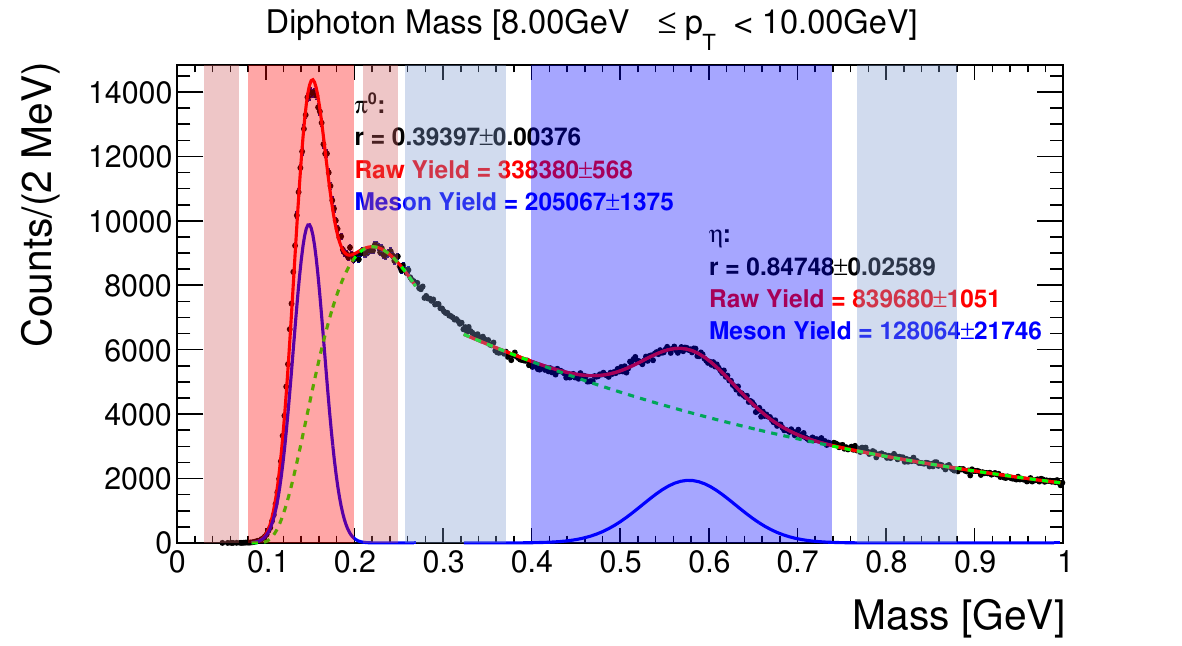}
    \includegraphics[page=23, width=0.48\linewidth]{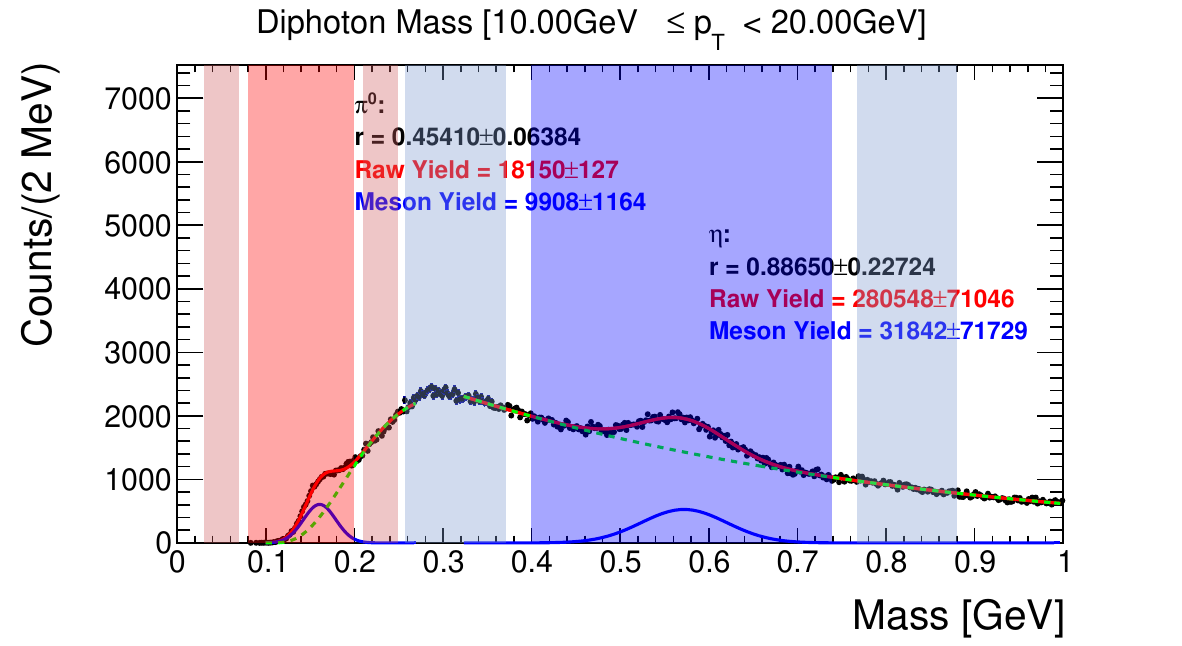}
    \caption{Di-photon invariant mass distributions and fits with meson signal and background side bands highlighted. The top left pad shows the $p_T$-integrated distribution. Each remaining pad corresponds to one $p_T$ bin.}
    \label{fig:mass_dists}
\end{figure}

The $p_T$-dependent invariant mass distributions, fit with the above procedure and labeled with corresponding background fractions, are shown in Figure~\ref{fig:mass_dists}. This fit procedure yields good agreement with data in most $p_T$ bins. One notable exception is the lowest-$p_T$ bin (from 1 to 2 GeV). In this bin, the background function fails to fit the regions above and below $\pi^0$ peak. This result and its consequences are examined in more detail in Section~\ref{sec:sys-bg}.

Figure~\ref{fig:mass_dists} provides two important insights which guide the remainder of the analysis. First, note that in the lowest $p_T$ bin (1-2 GeV), no true $\eta$-meson peak can be resolved from the combinatorial background. Therefore, we limit the analysis of $\eta$-meson asymmetries to di-photons with $p_T > 2$ GeV. Second, we see in the highest $p_T$ bin (10-20 GeV) that no clear $\pi^0$ peak can be resolved. This is due to the cluster merging effect (discussed in Section~\ref{sec:cluster_merging}); at such high energies, effectively all $\pi^0$ decay photons overlap in the EMCal and therefore cannot be reconstructed via their di-photon decay channel. As such, we limit the analysis of $\pi^0$ asymmetries to di-photons with $p_T < 10$ GeV. 

\section{Corrected Asymmetry Results}
\label{sec:stat-results}
Table~\ref{tab:meson_yields} shows the background-subtracted meson yields and combinatorial background fractions $r$, as explained in Section~\ref{sec:corrections}, for each $p_T$ bin considered in the analysis. This information is taken from the $p_T$-dependent di-photon mass distributions and fits in Figure~\ref{fig:mass_dists}. While the signal region and background asymmetries are calculated from the raw yields (i.e. without background subtraction), these values give an indication of the statistical power of the background-corrected asymmetry.

\begin{table}[h]
\centering
\begin{tabular}{c|rr|rr}
$p_T$ [GeV] & $\pi^0$ Yield & $r_{\pi^0}$ & $\eta$ Yield & $r_{\eta}$ \\\hline
1-2  & 5,152,967 & 0.602  & -- & -- \\
2-3   &  67,156,991  & 0.313 & 2,169,696  & 0.957  \\
3-4   & 73,056,941  & 0.280  & 4,558,106 & 0.900  \\
4-5   & 34,894,145  & 0.260  & 3,243,292   & 0.866  \\
5-6   & 10,861,653  & 0.263   & 1,474,037  & 0.852  \\
6-7   & 3,039,617  & 0.263  & 601,243  & 0.839   \\
7-8   & 728,608  & 0.327  & 225,360  & 0.840  \\
8-10  & 205,067 & 0.394  & 128,214 & 0.847   \\
10-20 & -- & -- & 31,842  & 0.887 \\
\end{tabular}
\caption{$\pi^0$ and $\eta$-meson background-subtracted yields and background fractions $r$ for each bin in $p_T$.}
\label{tab:meson_yields}
\end{table}

\begin{figure}
    \centering   \includegraphics[page=220,width=0.48\linewidth]{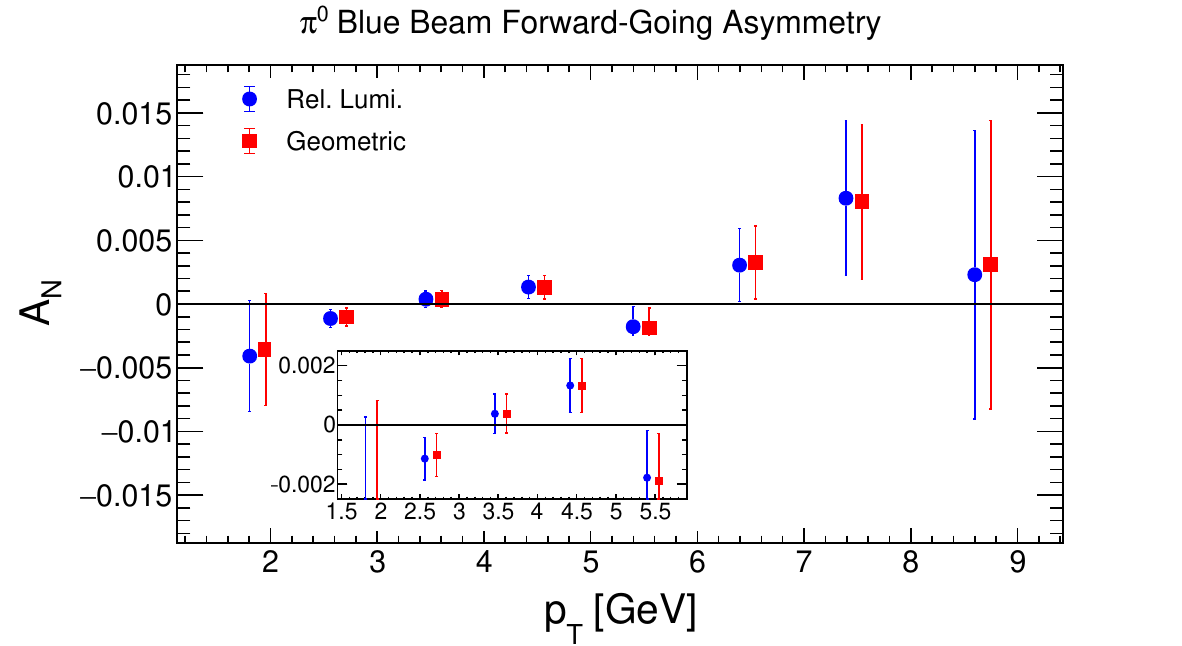}  \includegraphics[page=223,width=0.48\linewidth]{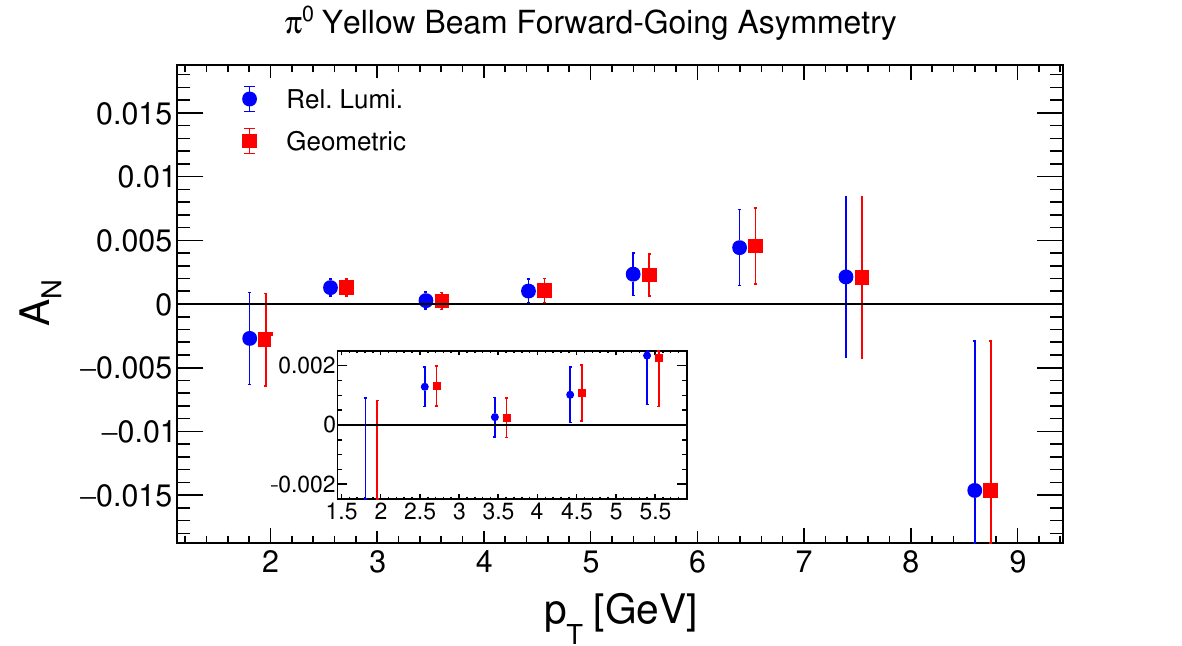}
    \caption{Corrected $A_N$ for the $\pi^0$ with respect to the Blue (left) and Yellow (right) beam, plotted against $p_T$.}
    \label{fig:pi0_pT_blueyellow}
\end{figure}

\begin{figure}
    \centering    \includegraphics[page=286,width=0.48\linewidth]{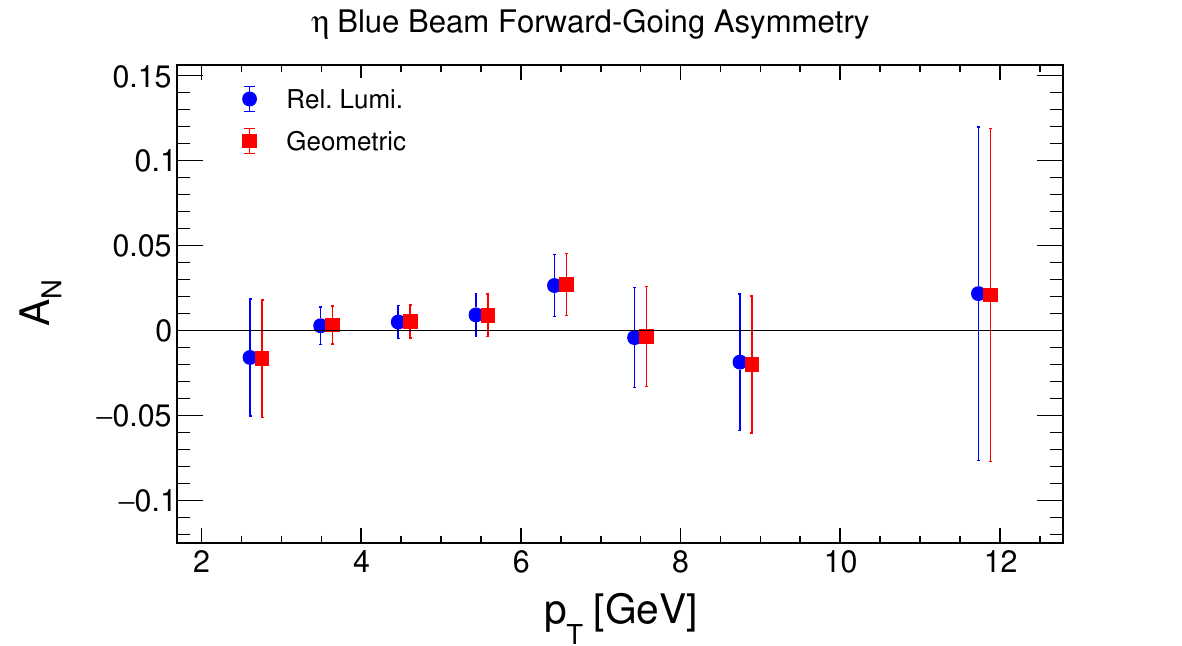}    \includegraphics[page=289,width=0.48\linewidth]{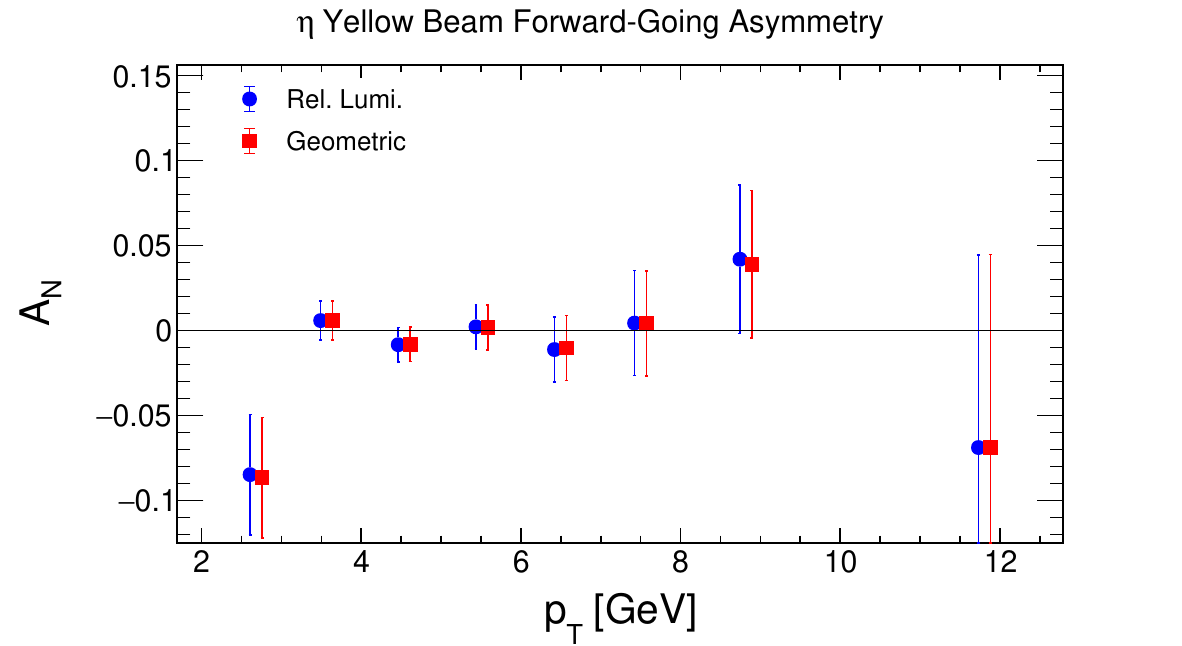}
    \caption{Corrected $A_N$ for the $\eta$-meson with respect to the Blue (left) and Yellow (right) beam, plotted against $p_T$. }
    \label{fig:eta_pT_blueyellow}
\end{figure}

The corrected asymmetries, as described in Section~\ref{sec:corrections}, are shown in Figure~\ref{fig:pi0_pT_blueyellow} for the $\pi^0$ and Figure~\ref{fig:eta_pT_blueyellow} for the $\eta$-meson. We note that in these figures, only mesons produced in the forward direction relative to the polarized beam are included, i.e. $x_F>0$. Results for mesons produced in the backward direction, where the asymmetry is expected to vanish~\cite{PHENIX_backward_AN}, are included in Appendix~\ref{app_backward}.

\begin{figure}
    \centering   \includegraphics[page=228,width=0.48\linewidth]{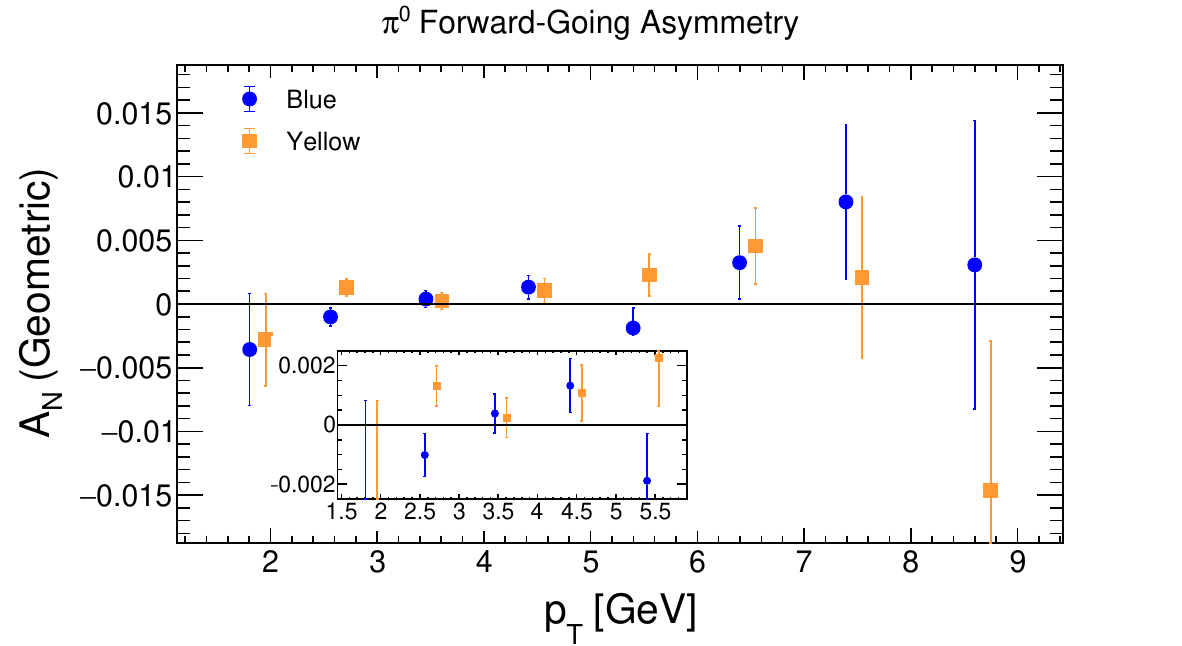}  \includegraphics[page=229,width=0.48\linewidth]{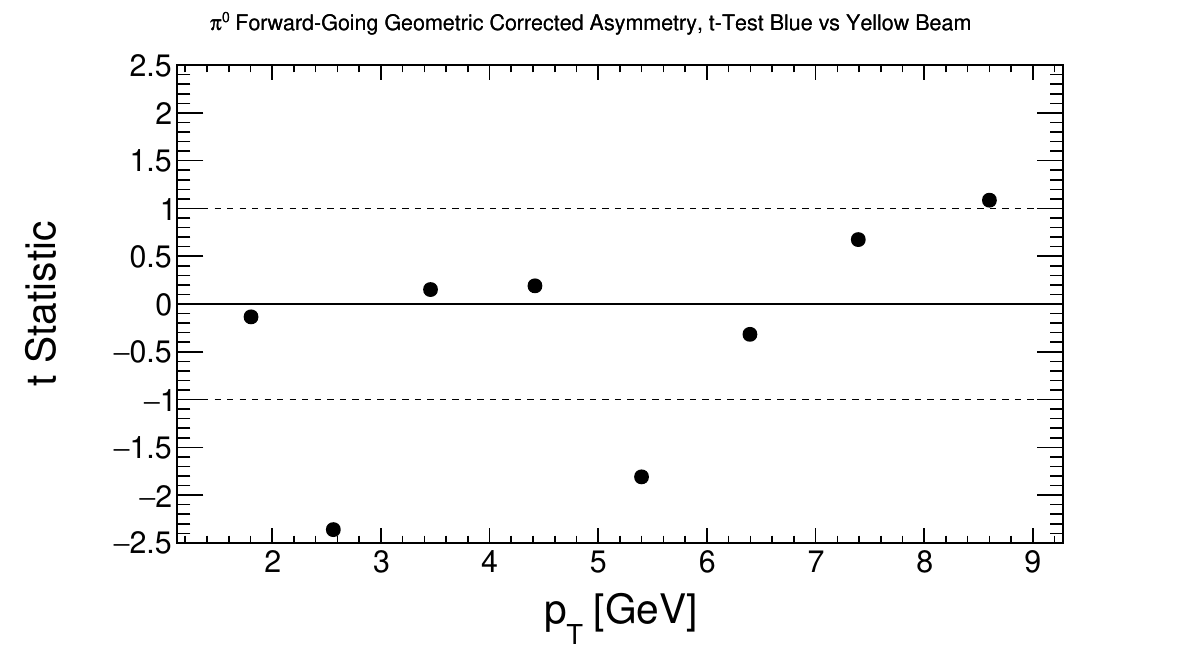}
    \caption{Left: $A_N$ for the $\pi^0$ with respect to the Blue and Yellow beams, plotted against $p_T$. Right: $t$-test for compatibility between Blue and Yellow beam $A_N$.}
    \label{fig:pi0_pT_beam_comparison}
\end{figure}

\begin{figure}
    \centering   \includegraphics[page=294,width=0.48\linewidth]{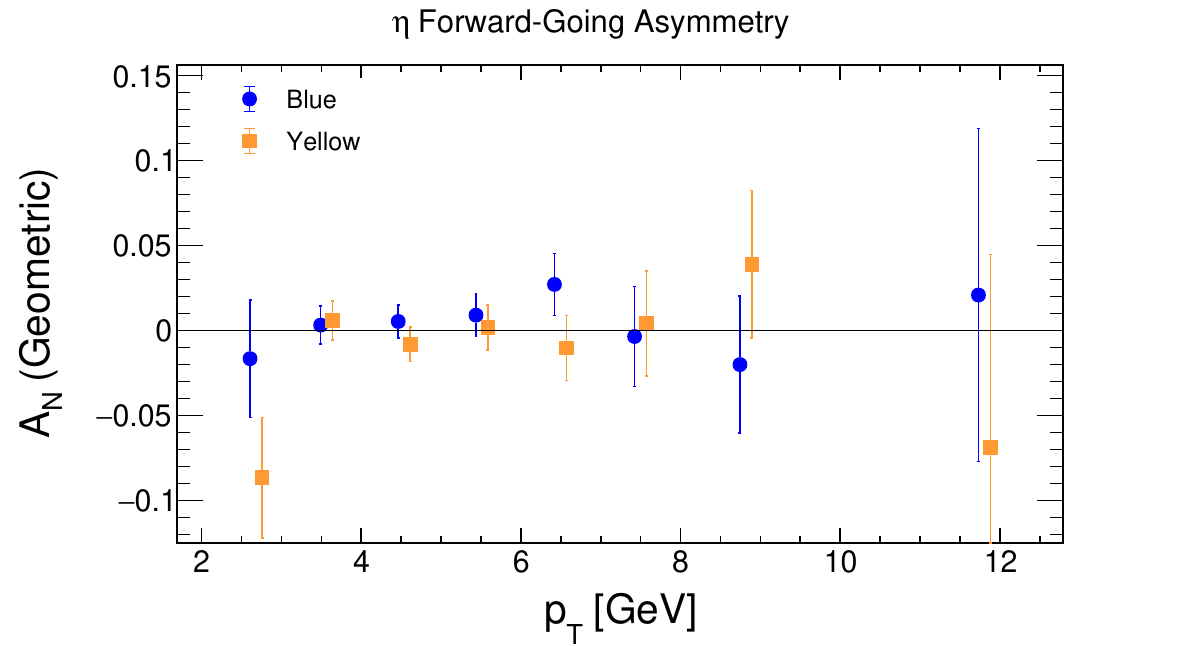}  \includegraphics[page=295,width=0.48\linewidth]{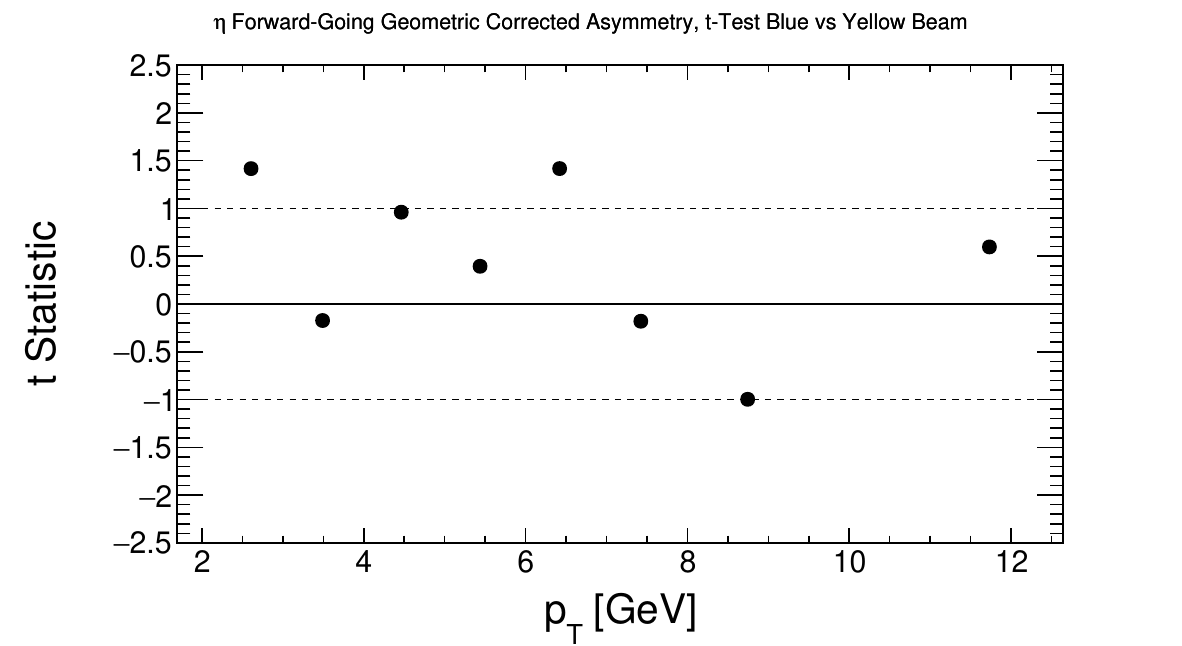}
    \caption{Left: $A_N$ for the $\eta$ with respect to the Blue and Yellow beams, plotted against $p_T$. Right: $t$-test for compatibility between Blue and Yellow beam $A_N$.}
    \label{fig:eta_pT_beam_comparison}
\end{figure}

A comparison between $A_N$ calculated with the geometric formula for the Blue and Yellow beams is shown on the left side of Figure~\ref{fig:pi0_pT_beam_comparison} for the $\pi^0$ and Figure~\ref{fig:eta_pT_beam_comparison} for the $\eta$-meson. The right side of these figures shows a $t$-test for compatibility between results from the two beams. For independent samples, as in the case of the independent Blue and Yellow beam measurements, the $t$-statistic is calculated as
\begin{equation}
    t = \frac{A_N^1 - A_N^2}{\sqrt{(\delta A_N^1)^2 + (\delta A_N^2)^2}}.
\end{equation}
The expected distribution of the $t$-statistic is a Gaussian distribution with mean 0 and variance 1. Thus, on average, 2/3 of the sampled $t$ values should lie within $\pm 1$ and 95\% should lie within $\pm 2$. While difficult to judge a distribution by only eight data points, we see in Figures~\ref{fig:pi0_pT_beam_comparison} and \ref{fig:eta_pT_beam_comparison} a mix of positive and negative values, with most points lying within $\pm 1$ and the remaining points not far outside this range. This implies the independent results from the Blue and Yellow beams are consistent within statistical uncertainties. Results for $A_N$ calculated with the relative luminosity formula are similarly consistent.


\begin{figure}
    \centering   \includegraphics[page=233,width=0.48\linewidth]{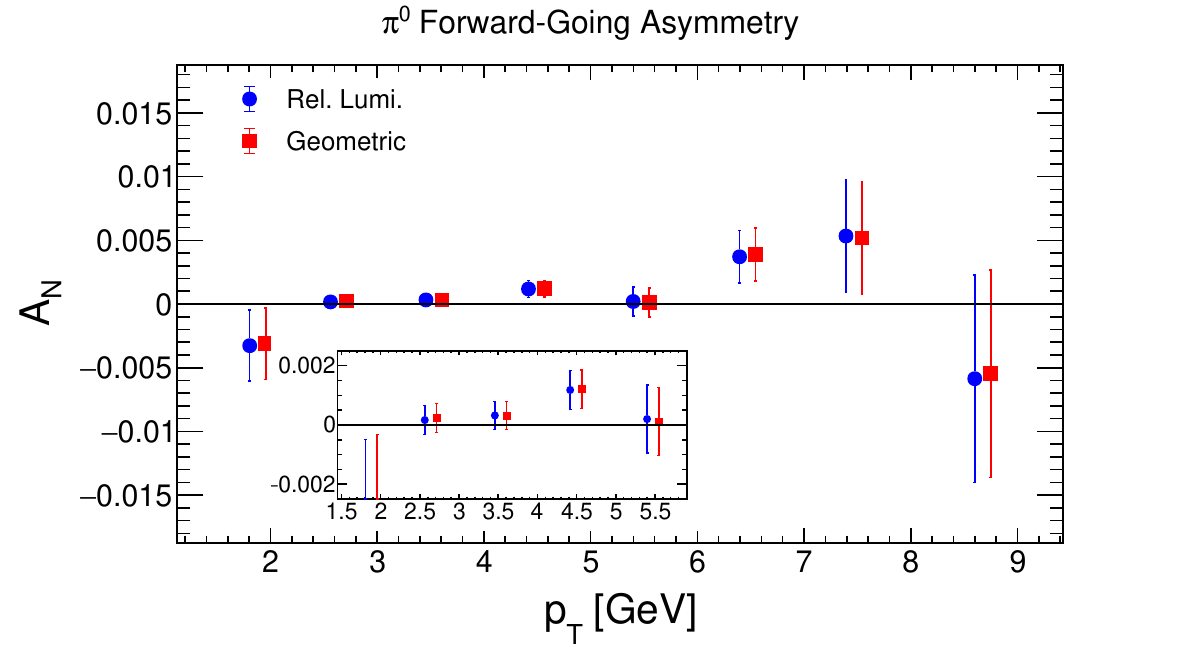}  \includegraphics[page=235,width=0.48\linewidth]{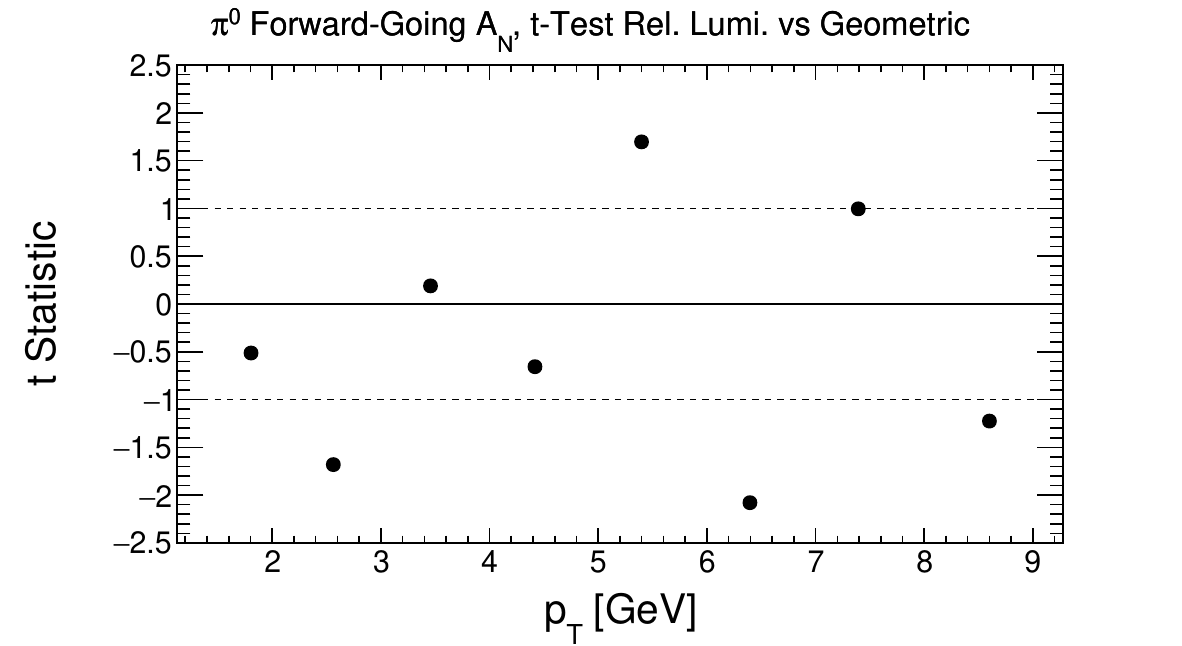}
    \caption{Left: Beam-averaged $A_N$ for the $\pi^0$ calculated using the relative luminosity and geometric formulas, plotted against $p_T$. Right: $t$-test for compatibility between the two calculation methods.}
    \label{fig:pi0_pT_avg}
\end{figure}

\begin{figure}
    \centering   \includegraphics[page=299,width=0.48\linewidth]{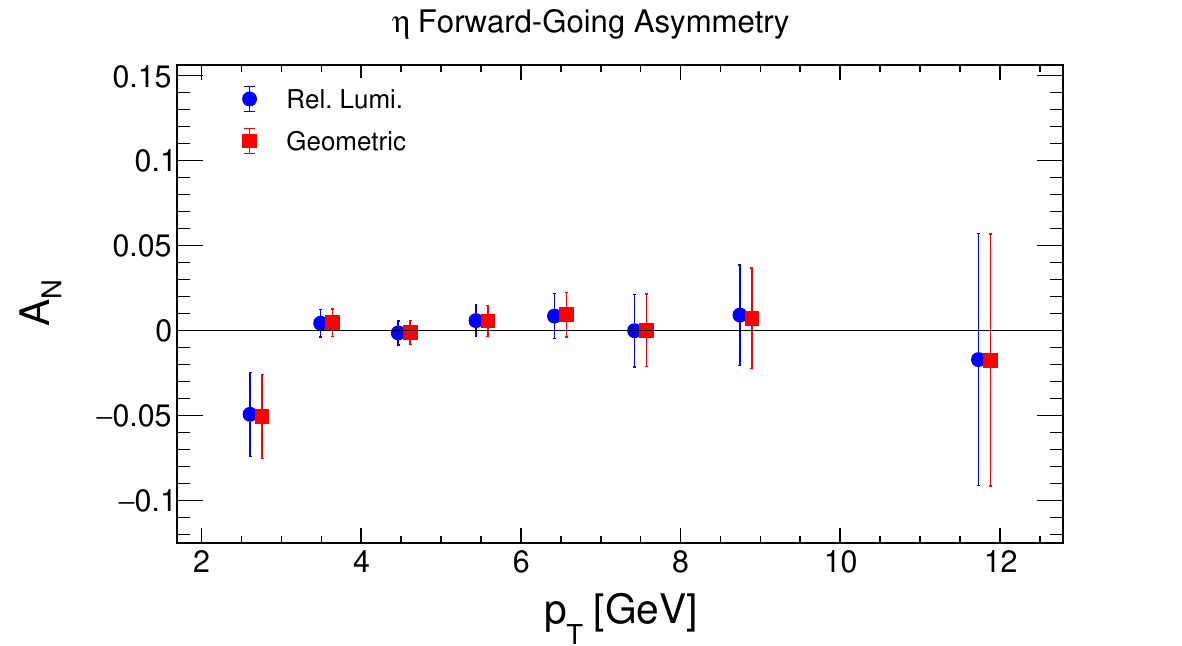}  \includegraphics[page=301,width=0.48\linewidth]{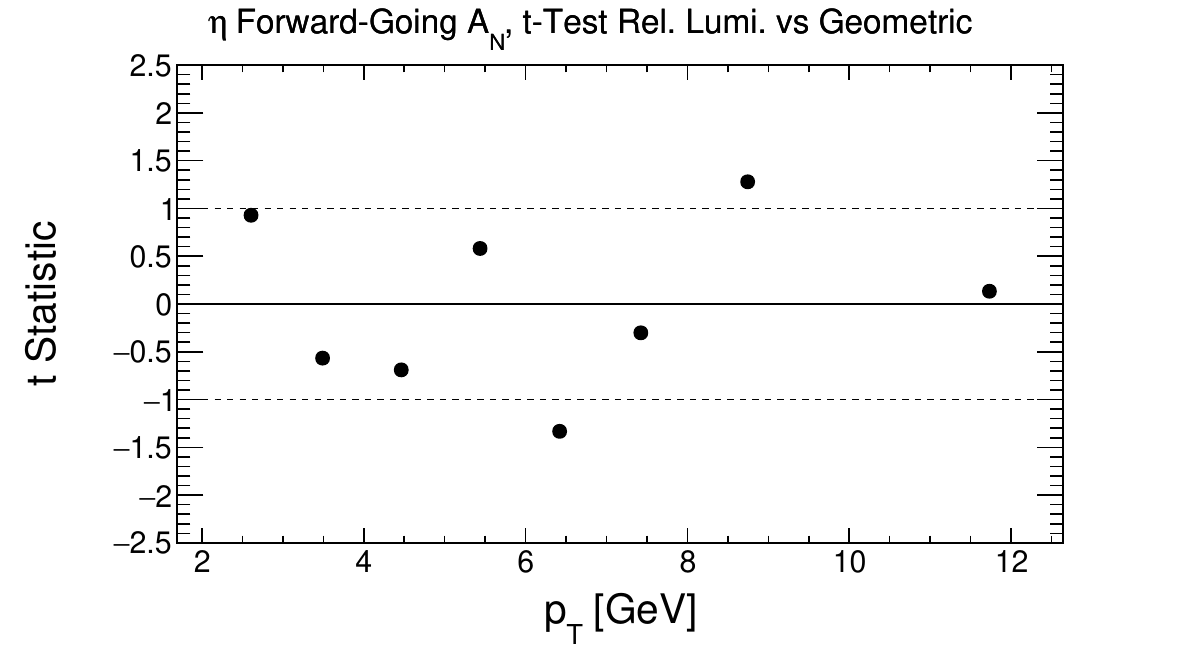}
    \caption{Left: Beam-averaged $A_N$ for the $\eta$-meson calculated using the relative luminosity and geometric formulas, plotted against $p_T$. Right: $t$-test for compatibility between the two calculation methods.}
    \label{fig:eta_pT_avg}
\end{figure}

To arrive at a final value for $A_N$, results from the Blue and Yellow beams are averaged, weighted by their respective uncertainties:
\begin{equation}
A_N=\frac{A_N^B/(\delta A_N^B)^2+A_N^Y/(\delta A_N^Y)^2}{1/(\delta A_N^B)^2+1/(\delta A_N^Y)^2}
\end{equation}
\begin{equation}
\delta A_N=\frac{1}{\sqrt{1/(\delta A_N^B)^2+1/(\delta A_N^Y)^2}}
\end{equation}
The beam-averaged asymmetries obtained from both the relative luminosity and geometric formulas are shown on the left side of Figure~\ref{fig:pi0_pT_avg} for the $\pi^0$ and Figure~\ref{fig:eta_pT_avg} for the $\eta$-meson. The right side of these figures shows a $t$-test for compatibility between the two calculation methods. Because the two calculations use the same data, they are not independent like the Blue and Yellow beam results, but are fully correlated. Accordingly the $t$-statistic is calculated as
\begin{equation}
    t = \frac{A_N^1 - A_N^2}{\sqrt{\left|(\delta A_N^1)^2 - (\delta A_N^2)^2\right|}}.
\end{equation}
Although calculated differently, the $t$-statistic is still expected to follow a Gaussian distribution with mean 0 and variance 1. Figure~\ref{fig:eta_pT_avg} suggests reasonable agreement between the two calculations for the $\eta$-meson. In Figure~\ref{fig:pi0_pT_avg} not all $t$-values are within the $\pm 1$ window.
This discrepancy will be accounted for as a systematic uncertainty (see Section~\ref{sec:sys-calc}).


\begin{figure}
    \centering   \includegraphics[page=245,width=0.48\linewidth]{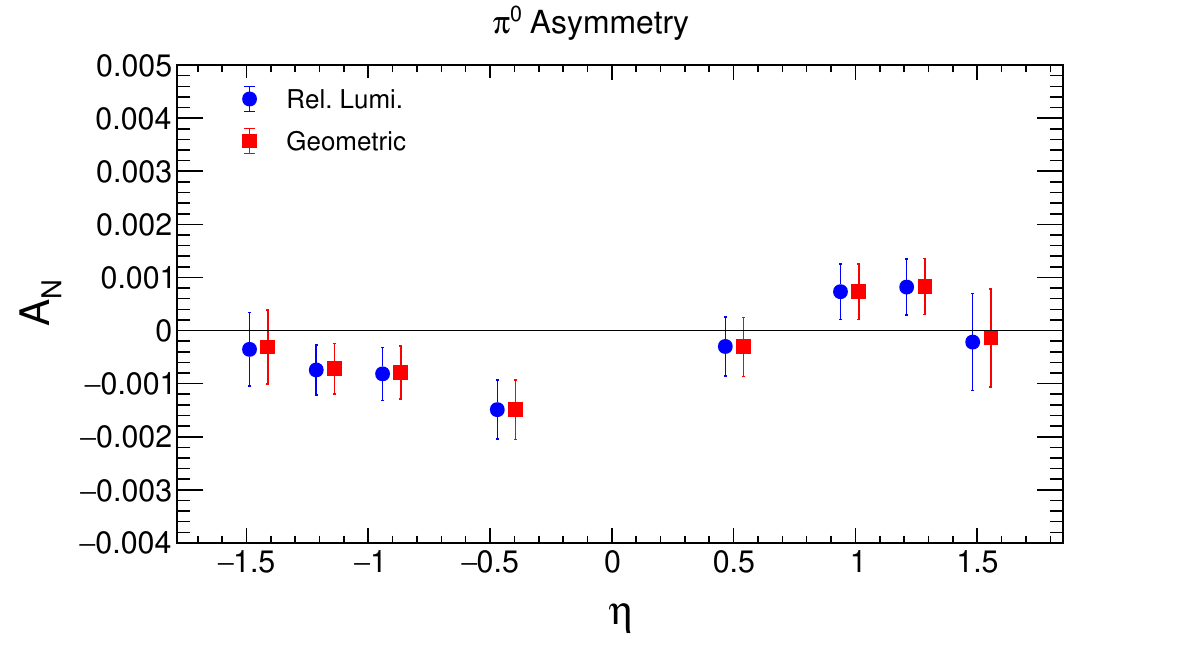} \includegraphics[page=252,width=0.48\linewidth]{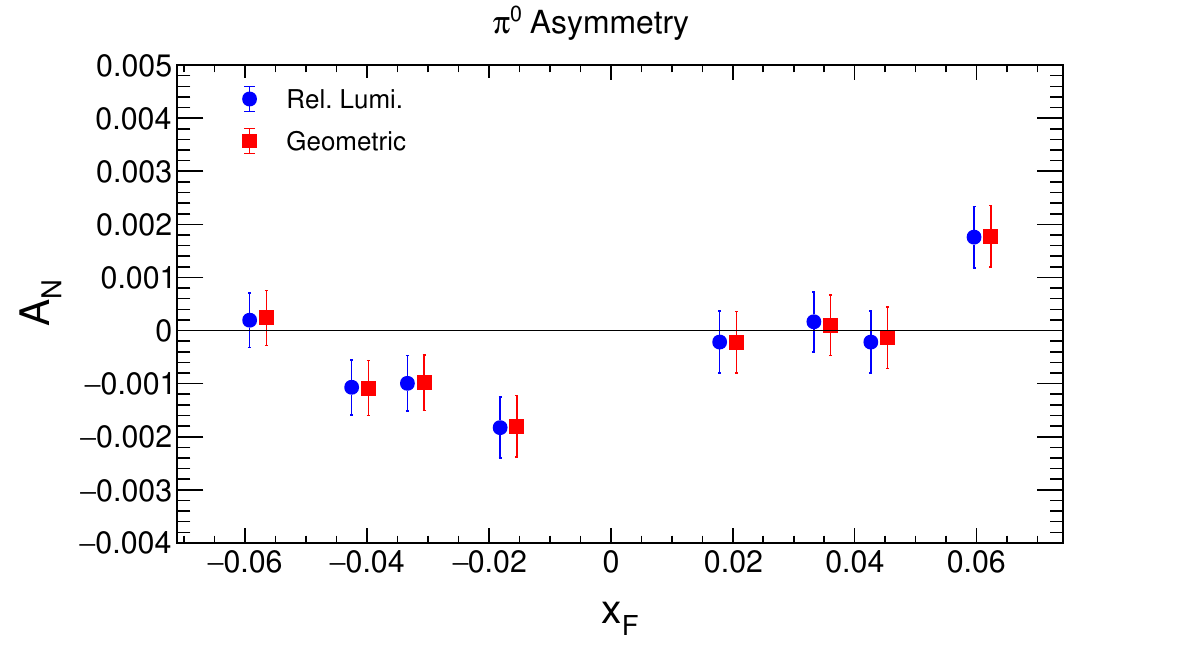}
    \caption{Beam-averaged $A_N$ for the $\pi^0$, plotted against $\eta$ (left) and $x_F$ (right). Positive (negative) $\eta$- and $x_F$-values correspond to the forward (backward) direction. }
    \label{fig:pi0_eta_xF}
\end{figure}

\begin{figure}
    \centering   \includegraphics[page=311,width=0.48\linewidth]{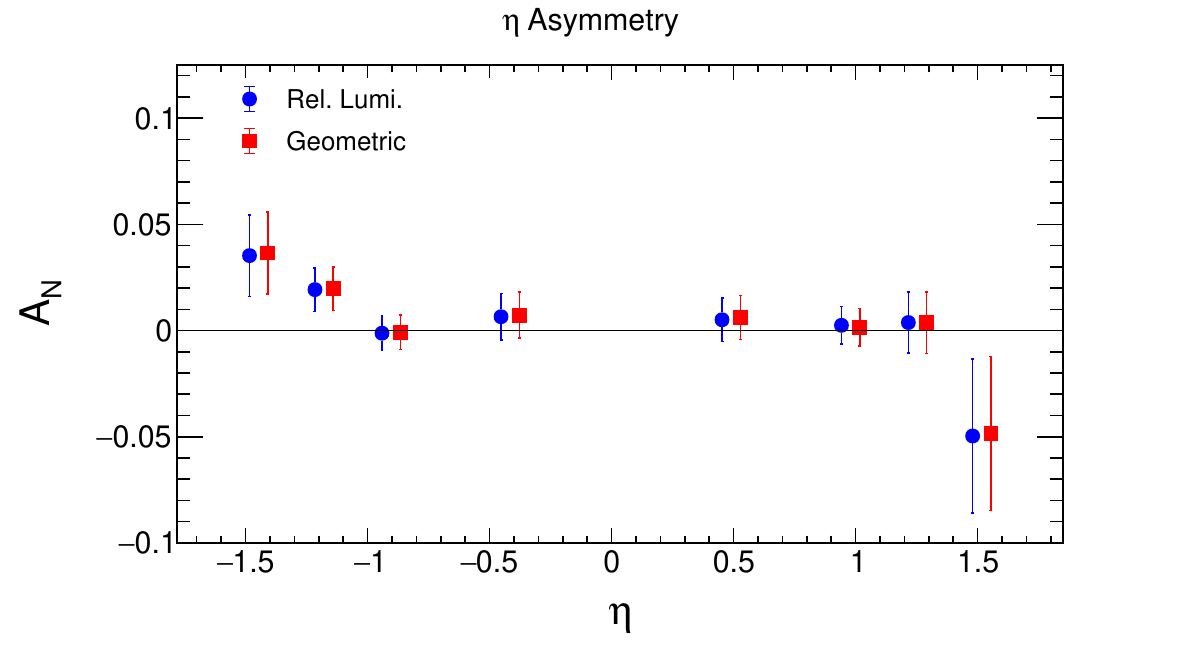} \includegraphics[page=318,width=0.48\linewidth]{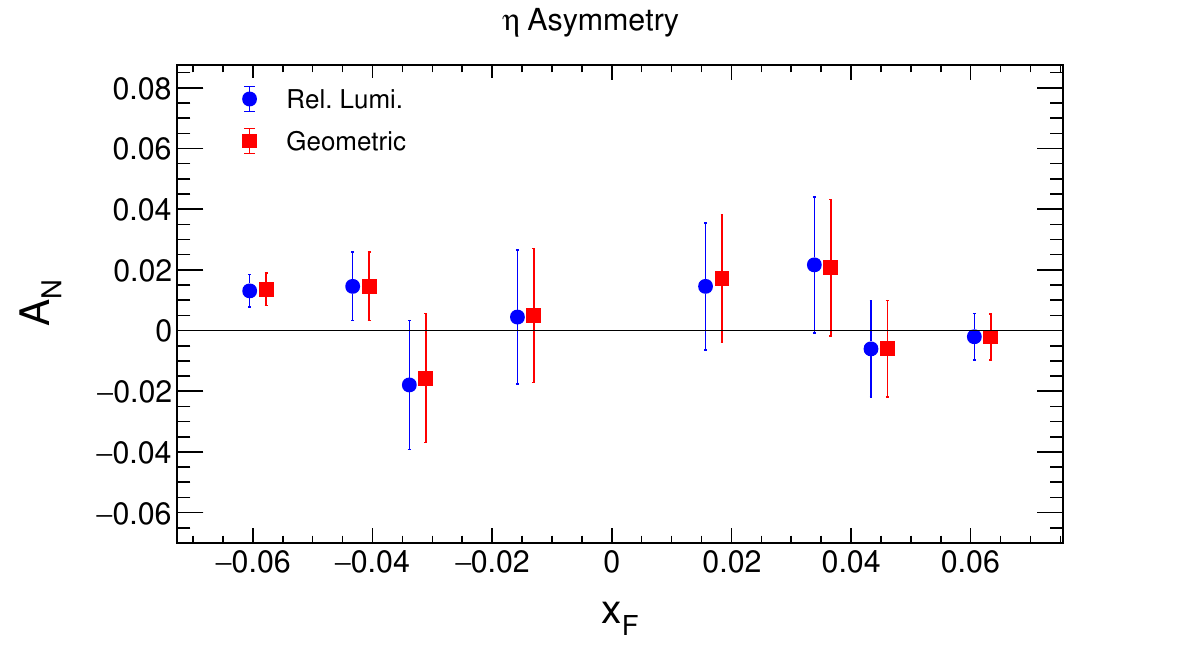}
    \caption{Beam-averaged $A_N$ for the $\eta$-meson, plotted against $\eta$ (left) and $x_F$ (right). Positive (negative) $\eta$- and $x_F$-values correspond to the forward (backward) direction. }
    \label{fig:eta_eta_xF}
\end{figure}

The beam-averaged asymmetries are shown plotted against pseudorapidity $\eta$ and $x$-Feynman $x_F = 2p_z/\sqrt{s}$ in Figure~\ref{fig:pi0_eta_xF} for the $\pi^0$ and Figure~\ref{fig:eta_eta_xF} for the $\eta$-meson. In these figures, positive (negative) $\eta$ and $x_F$ correspond to the direction forward (backward) of the polarized beam, following the standard convention in the literature~\cite{RHICf_pi0_AN}.


The results shown in this section include only statistical uncertainties on $A_N$. Chapter~\ref{ch:syst} discusses the determination of systematic uncertainties. Detailed interpretation of the results, including both statistical and systematic uncertainties, is discussed in Chapter~\ref{ch:results}.


\chapter{Systematic Uncertainties}\label{ch:syst}
This chapter details the determination of systematic uncertainties associated with measuring $A_N$. Three sources of systematic uncertainty are considered in this analysis: the method of calculating the asymmetry, in Section~\ref{sec:sys-calc}; the method of determining the background fraction, in Section~\ref{sec:sys-bg}; and bunch shuffling, in Section~\ref{sec:sys-shuffle}. These considerations result in systematic uncertainties $\delta A_N^\mathrm{sys,calc}$, $\delta A_N^\mathrm{sys,bg}$, and $\delta A_N^\mathrm{sys,shuffle}$, respectively. Section~\ref{sec:sys-tot} summarizes the results for each source of systematic uncertainty as well as the total systematic uncertainty $\delta A_N^\mathrm{sys,tot}$. As with $A_N$ and its statistical uncertainty $A_N^\mathrm{stat}$, each systematic uncertainty is determined independently for each kinematic bin in which the asymmetry is extracted.

In addition to the sources of systematic uncertainty detailed in this chapter, there is a kinematic-independent uncertainty associated with the beam polarization measurement. This analysis uses the online RHIC polarization values, as the calibrated values are not yet available for Run-2024. The RHIC polarimetry group has estimated a 7\% scale uncertainty on these online measurements \cite{RHICpolarimetry}. However, such an uncertainty scales both the measured $A_N$ and its statistical uncertainty in the same way. Therefore it does not change the statistical significance of results of this analysis relative to $A_N = 0$.

\section{Calculation Method}\label{sec:sys-calc}
As detailed in Section~\ref{sec:rawasym}, the asymmetries are calculated using two different formulas (Equations~\ref{eq:AN_rellumi} and \ref{eq:AN_sqrt}). As discussed in Sections~\ref{sec:rawresults} and \ref{sec:stat-results}, the two formulas generally give compatible results within statistical uncertainties. Nonetheless, we conservatively assign the difference between these calculations as a systematic uncertainty:
\begin{equation}
	\delta A_N^\mathrm{sys,calc} = \left| A_N^\mathrm{rellumi} - A_N^\mathrm{geometric} \right|.
\end{equation}
Table~\ref{tab:sys-calc} gives the values for both asymmetry calculations and the corresponding $\delta A_N^\mathrm{sys,calc}$ in each $p_T$ bin.

\begin{table}[h]
\centering
\begin{tabular}{c|rrr|rrr}
& \multicolumn{3}{|c|}{$\pi^0$} & \multicolumn{3}{|c}{$\eta$} \\
$p_T$ [GeV] & $A_N^\mathrm{rellumi}$ & $A_N^\mathrm{geometric}$ & $\delta A_N^\mathrm{sys,calc}$  & $A_N^\mathrm{rellumi}$ & $A_N^\mathrm{geometric}$ & $\delta A_N^\mathrm{sys,calc}$ \\\hline
1-2 & -0.00123 & -0.00107 & 0.00016 & -- & -- & -- \\
2-3 & 0.00026 & 0.00032 & 0.00006 & -0.02549 & -0.02644 & 0.00095 \\
3-4 & 0.00031 & 0.00031 & 0.00001 & -0.00577 & -0.00470 & 0.00107 \\
4-5 & 0.00118 & 0.00120 & 0.00002 & -0.00237 & -0.00187 & 0.00050 \\
5-6 & 0.00020 & 0.00011 & 0.00009 & 0.01297 & 0.01250 & 0.00047 \\
6-7 & 0.00371 & 0.00388 & 0.00017 & 0.01669 & 0.01833 & 0.00164 \\
7-8 & 0.00534 & 0.00517 & 0.00016 & -0.00088 & -0.00056 & 0.00032 \\
8-10 & -0.00586 & -0.00547 & 0.00039 & 0.02798 & 0.02223 & 0.00575 \\
10-20 & -- & -- & -- & -0.10527 & -0.10677 & 0.00150 \\
\end{tabular}
\caption{Values for $A_N$ computed with the relative luminosity and geometric methods, with corresponding $\delta A_N^\mathrm{sys,calc}$, binned in $p_T$.}
\label{tab:sys-calc}
\end{table}

\section{Background Fraction Determination}\label{sec:sys-bg}
The background correction described in Section~\ref{sec:corrections} depends crucially on the background fraction $r$ calculated by fitting signal and background contributions to the di-photon invariant mass distribution. The resulting values of $r$ accordingly depend on the specific procedure used for the fits. To account for this sensitivity, two different fitting methods are used: one for the main analysis, and an alternative method used for comparison. The background fraction in each kinematic bin is calculated using both methods, resulting in fractions $r^\mathrm{main~analysis}$ and $r^\mathrm{alternative}$ respectively. The corresponding asymmetries $A_N^\mathrm{main~analysis}$ and $A_N^\mathrm{alternative}$, respectively, are then calculated using these independent background fractions. In analogy with $\delta A_N^\mathrm{sys,calc}$, the difference between these two calculations is assigned as a systematic uncertainty:
\begin{equation}
	\delta A_N^\mathrm{sys,bg} = \left| A_N^\mathrm{main~analysis} - A_N^\mathrm{alternative} \right|.
\end{equation}

The main analysis uses the threshold function with exponential described in Section~\ref{sec:fitting}. The alternative fitting method uses Gaussian Process Regression (GPR). GPR is a machine learning technique which fits data with a non-parametric model \cite{GPR_tutorial, GPR_signal_extraction}. No functional form is assumed for the background contribution. Instead, GPR models the data using a joint Gaussian distribution. Data is fit by assuming each point $i$ corresponds to a single-variable Gaussian distribution with mean $\mu_i$ and standard deviation $\sigma_i$. Information on the correlation between data points is encoded in the joint covariance matrix. In practice, one gives as input to the GPR model a \textit{kernel} function $k_{ij}$, representing the covariance between points $i$ and $j$. This analysis uses the Radial Basis Function (RBF) kernel, defined by
\begin{equation}
    k_\mathrm{RBF}(x_i, x_j) = e^{-\frac{\left| x_i - x_j \right|}{2\sigma^2}},
\end{equation}
where $\sigma$ is a parameter (``hyperparameter'' in machine learning parlance) controlling the length scale of correlations. In the ``training'' step, the GPR model then infers optimal values for the means $\mu_i$ and widths $\sigma_i$ such that the joint likelihood over the data is maximized. The $\sigma_i$ are naturally interpreted as uncertainties on the predicted means, which can be propagated to the yield $N_\mathrm{bkgd}$.

If trained on the full di-photon mass distribution, the GPR model will attempt to recreate the distribution exactly, including meson signal peaks. To separate the signal and background contributions, the model is instead trained only on data with the signal regions excluded. Thus the model receives no information about the signal peaks and extrapolates from the background side bands to infer values in the signal regions. As in the main analysis, the meson signal peaks are modeled as Gaussian distributions. These Gaussians are fit separately for each meson by subtracting the GPR background predictions from the data in the respective meson signal regions. For the $\pi^0$, the signal region is the same as for the main analysis, i.e. a 3-$\sigma$ window around the $\pi^0$ peak. For the $\eta$-meson, a 2-$\sigma$ window is used to limit the range over which the GPR model must extrapolate. For consistency, these windows are defined using the same $\mu$ and $\sigma$ values used in the main analysis, i.e. those determined from the threshold function with exponential fit to the $p_T$-integrated mass distribution.

\begin{figure}
    \centering
    \includegraphics[height=0.8\textheight]{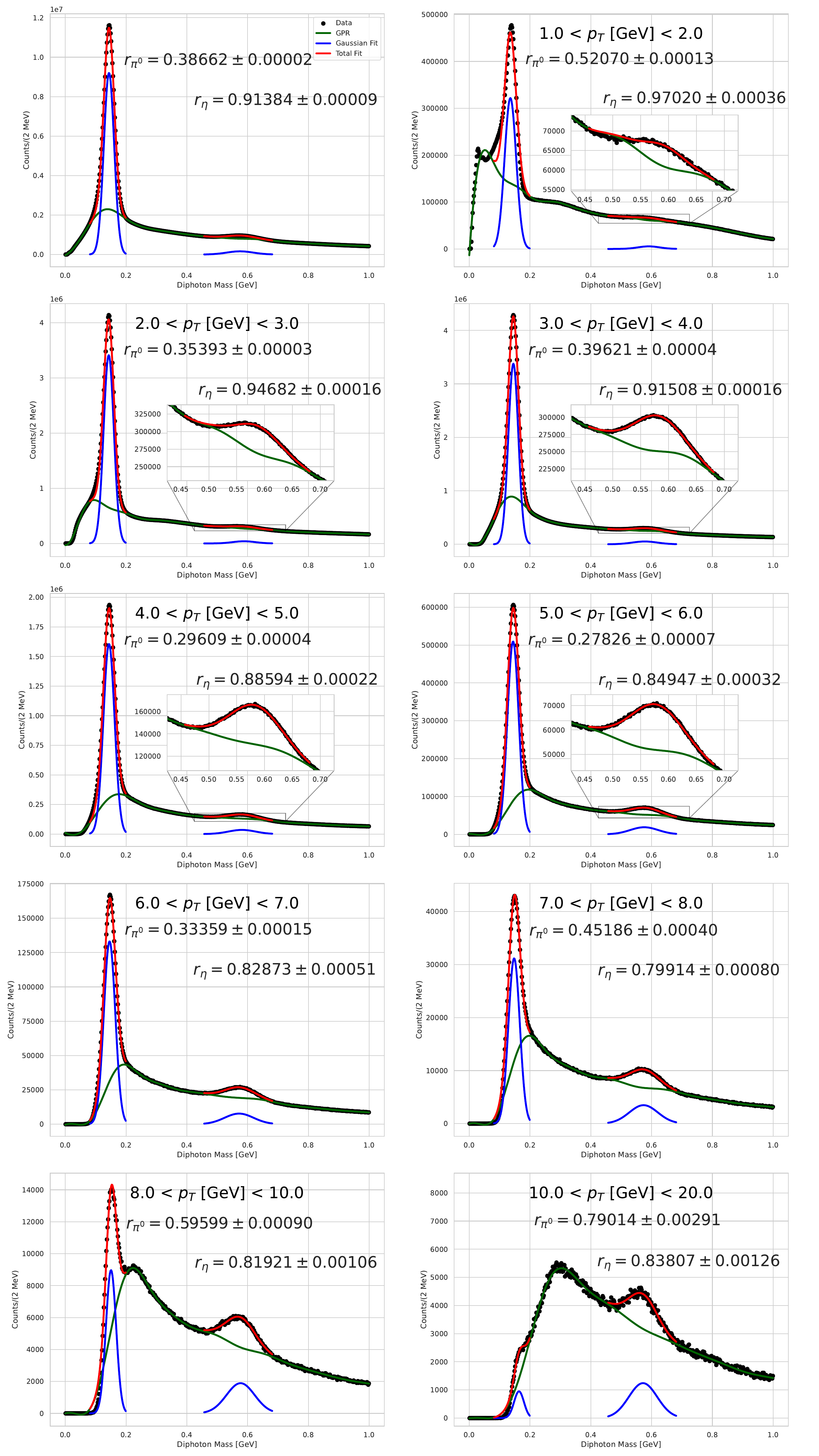}
    \caption{Di-photon invariant mass distributions fit using Gaussian Process Regression and labeled with background fractions. The top left pad shows the $p_T$-integrated distribution. Each remaining pad corresponds to one $p_T$ bin.}
    \label{fig:mass_gpr}
\end{figure}

The $p_T$-dependent invariant mass distributions, fit with the above procedure and labeled with corresponding background fractions, are shown in Figure~\ref{fig:mass_gpr}. The background fractions and their uncertainties are calculated in analogy with those of the main analysis, with one minor difference. Instead of calculating the true-meson yield $N_\mathrm{meson}$ from the Gaussian fits, we calculate directly the total yield
\begin{equation}
    N_\mathrm{total} = \sum_i{y_i}
\end{equation}
and its uncertainty
\begin{equation}
    \delta N_\mathrm{total} = \sqrt{\sum_i{(\delta y_i)^2}},
\end{equation}
where $y_i$ are the raw counts at mass $m_i$. The counts are assumed to follow Poisson statistics, such that $\delta y_i = \sqrt{y_i}$. Similarly, the background yield $N_\mathrm{bkgd}$ is taken as the sum of the GPR predicted means $\mu_i$, and its uncertainty $\delta N_\mathrm{bkgd}$ calculated by propagating the predicted uncertainties $\sigma_i$. The background fraction is then
\begin{equation}
    r = \frac{N_\mathrm{bkgd}}{N_\mathrm{total}}
\end{equation}
with uncertainty
\begin{equation}
    \delta r = r \sqrt{\left(\frac{\delta N_\mathrm{bkgd}}{N_\mathrm{bkgd}}\right)^2 + \left(\frac{\delta N_\mathrm{total}}{N_\mathrm{total}}\right)^2}.
\end{equation}

The background fractions calculated in both the main analysis and the alternative analysis are summarized in Tables~\ref{tab:bgfracs_pi0} and \ref{tab:bgfracs_eta} for the $\pi^0$ and $\eta$-mesons, respectively, together with the values of $A_N$ determined using each method. The right column of these tables gives the resulting systematic uncertainty $\delta A_N^\mathrm{sys,bg}$.

\begin{table}[h]
\centering
\begin{tabular}{c|rr|rr|r}
$p_T$ [GeV] & $r^\mathrm{main~analysis}$ & $r^\mathrm{alternative}$ & $A_N^\mathrm{main~analysis}$ & $A_N^\mathrm{alternative}$ & $\delta A_N^\mathrm{sys,bg}$ \\\hline
1-2 & 0.60159 $\pm$ 0.00033 & 0.52070 $\pm$ 0.00013 & -0.00107 & -0.00090 & 0.00017 \\
2-3 & 0.31259 $\pm$ 0.00009 & 0.35393 $\pm$ 0.00003 & 0.00032 & 0.00042 & 0.00010 \\
3-4 & 0.27980 $\pm$ 0.00010 & 0.39621 $\pm$ 0.00004 & 0.00031 & 0.00029 & 0.00002 \\
4-5 & 0.25988 $\pm$ 0.00017 & 0.29609 $\pm$ 0.00004 & 0.00120 & 0.00121 & 0.00000 \\
5-6 & 0.26277 $\pm$ 0.00078 & 0.27826 $\pm$ 0.00007 & 0.00011 & 0.00005 & 0.00006 \\
6-7 & 0.26272 $\pm$ 0.00267 & 0.33359 $\pm$ 0.00015 & 0.00388 & 0.00397 & 0.00010 \\
7-8 & 0.32672 $\pm$ 0.00204 & 0.45186 $\pm$ 0.00040 & 0.00517 & 0.00456 & 0.00061 \\
8-10 & 0.39397 $\pm$ 0.00376 & 0.59599 $\pm$ 0.00090 & -0.00547 & -0.01356 & 0.00809 \\
10-20 & -- & -- & -- & -- & -- \\
\end{tabular}
\caption{$\pi^0$-meson background fractions $r$ and corresponding $A_N$ for both background fitting methods, binned in $p_T$.}
\label{tab:bgfracs_pi0}
\end{table}

\begin{table}[h]
\centering
\begin{tabular}{c|rr|rr|r}
$p_T$ [GeV] & $r^\mathrm{main~analysis}$ & $r^\mathrm{alternative}$ & $A_N^\mathrm{main~analysis}$ & $A_N^\mathrm{alternative}$ & $\delta A_N^\mathrm{sys,bg}$ \\\hline
1-2 & -- & -- & -- & -- & -- \\
2-3 & 0.95655 $\pm$ 0.00019 & 0.94682 $\pm$ 0.00016 & -0.02644 & -0.02492 & 0.00152 \\
3-4 & 0.90003 $\pm$ 0.00033 & 0.91508 $\pm$ 0.00016 & -0.00470 & -0.00479 & 0.00008 \\
4-5 & 0.86599 $\pm$ 0.00448 & 0.88594 $\pm$ 0.00022 & -0.00187 & -0.00181 & 0.00006 \\
5-6 & 0.85221 $\pm$ 0.00079 & 0.84947 $\pm$ 0.00032 & 0.01250 & 0.00924 & 0.00326 \\
6-7 & 0.83868 $\pm$ 0.00174 & 0.82873 $\pm$ 0.00051 & 0.01833 & 0.01532 & 0.00301 \\
7-8 & 0.83949 $\pm$ 0.01732 & 0.79914 $\pm$ 0.00080 & -0.00056 & -0.00029 & 0.00028 \\
8-10 & 0.84730 $\pm$ 0.02642 & 0.81921 $\pm$ 0.00106 & 0.02223 & 0.01470 & 0.00753 \\
10-20 & 0.88649 $\pm$ 0.04509 & 0.83807 $\pm$ 0.00103 & -0.10677 & -0.05252 & 0.05424 \\
\end{tabular}
\caption{$\eta$-meson background fractions $r$ and corresponding $A_N$ for both background fitting methods, binned in $p_T$.}
\label{tab:bgfracs_eta}
\end{table}

\section{Bunch Shuffling}\label{sec:sys-shuffle}
Bunch shuffling is a method for investigating potential sources of biases and uncertainty related to instrumentation effects. This technique can help identify any sensitivity to false asymmetries, as well as any over- or under-estimation of statistical uncertainties on the measured asymmetries. In bunch shuffling, the true spin pattern of the 111 filled RHIC bunches (see Section~\ref{sec:rhic_beam}) is replaced with a randomized pattern, and the raw asymmetry $\epsilon$ and its statistical uncertainty $\delta \epsilon$ are recalculated. This process is repeated many times; this analysis uses 10,000 iterations. On average, the polarization of the beams cancels out and the physics asymmetry vanishes. The raw asymmetries are calculated using the geometric formula to avoid the need to recalculate the relative luminosity for each randomized spin pattern.

The spin pattern randomization is performed carefully to preserve an equal number of spin-up and spin-down bunches. Rather than generating a truly random pattern with variable number of up and down spins, the true physics spin pattern is shuffled into a random re-ordering. To account for the fact that there are an odd number of filled bunches, the spin direction of all 111 shuffled bunches is then simultaneously flipped at random, with 50\% probability. This ensures that, on average, the shuffled patterns have no net polarization.

Studying the distribution of the quantity $\epsilon/\delta \epsilon$ over many iterations gives insight into systematic effects. In the absence of any such effects, $\epsilon/\delta \epsilon$ will be distributed as a Gaussian with mean zero and unit variance, corresponding to statistical noise on a zero asymmetry measurement. A non-zero mean of this distribution would indicate sensitivity to a false asymmetry, while a variance different from one would indicate the shuffled asymmetries vary more than expected due to pure statistical fluctuation. 
In particular, if one fits the distribution of $\epsilon/\delta \epsilon$ with a Gaussian and finds a fitted width $\sigma_\mathrm{fit} > 1$, then the true statistical uncertainty which should be assigned to $\epsilon$ is not simply $\delta \epsilon$, but rather $\sigma_\mathrm{fit} \delta \epsilon$. This can be seen by considering the distribution from which $\epsilon$ is drawn:
\begin{align}
    \epsilon/\delta \epsilon & \sim \mathcal{N}(0, \sigma_\mathrm{fit}) \\
    \Rightarrow \epsilon & \sim \mathcal{N}(0, \sigma_\mathrm{fit} \delta \epsilon).
\end{align}
If $\sigma_\mathrm{fit}$ is found to be larger than one, then $\delta \epsilon$ is an underestimation of the true uncertainty $\sigma_\mathrm{fit} \delta \epsilon$. Likewise, the statistical uncertainty on the physics asymmetry scales in the same way: $\delta A_N^\mathrm{stat} \rightarrow \sigma_\mathrm{fit} \delta A_N^\mathrm{stat}$. In this case, a systematic uncertainty $\delta A_N^\mathrm{sys,shuffle}$ is determined as the additional uncertainty which must be added to $\delta A_N^\mathrm{stat}$ in quadrature to recover the true statistical uncertainty:
\begin{align}
    \sigma_\mathrm{fit} \delta A_N^\mathrm{stat} &= \sqrt{(\delta A_N^\mathrm{stat})^2 + (\delta A_N^\mathrm{sys,shuffle})^2} \\
    \Rightarrow \delta A_N^\mathrm{sys,shuffle} &= \delta A_N^\mathrm{stat} \sqrt{\sigma_\mathrm{fit}^2 - 1}.
\end{align}
If $\sigma_\mathrm{fit}$ is found to be consistent with 1 within statistical uncertainty, no such systematic uncertainty is assigned.

\begin{figure}
    \centering
    \includegraphics[page=1,height=0.9\textheight]{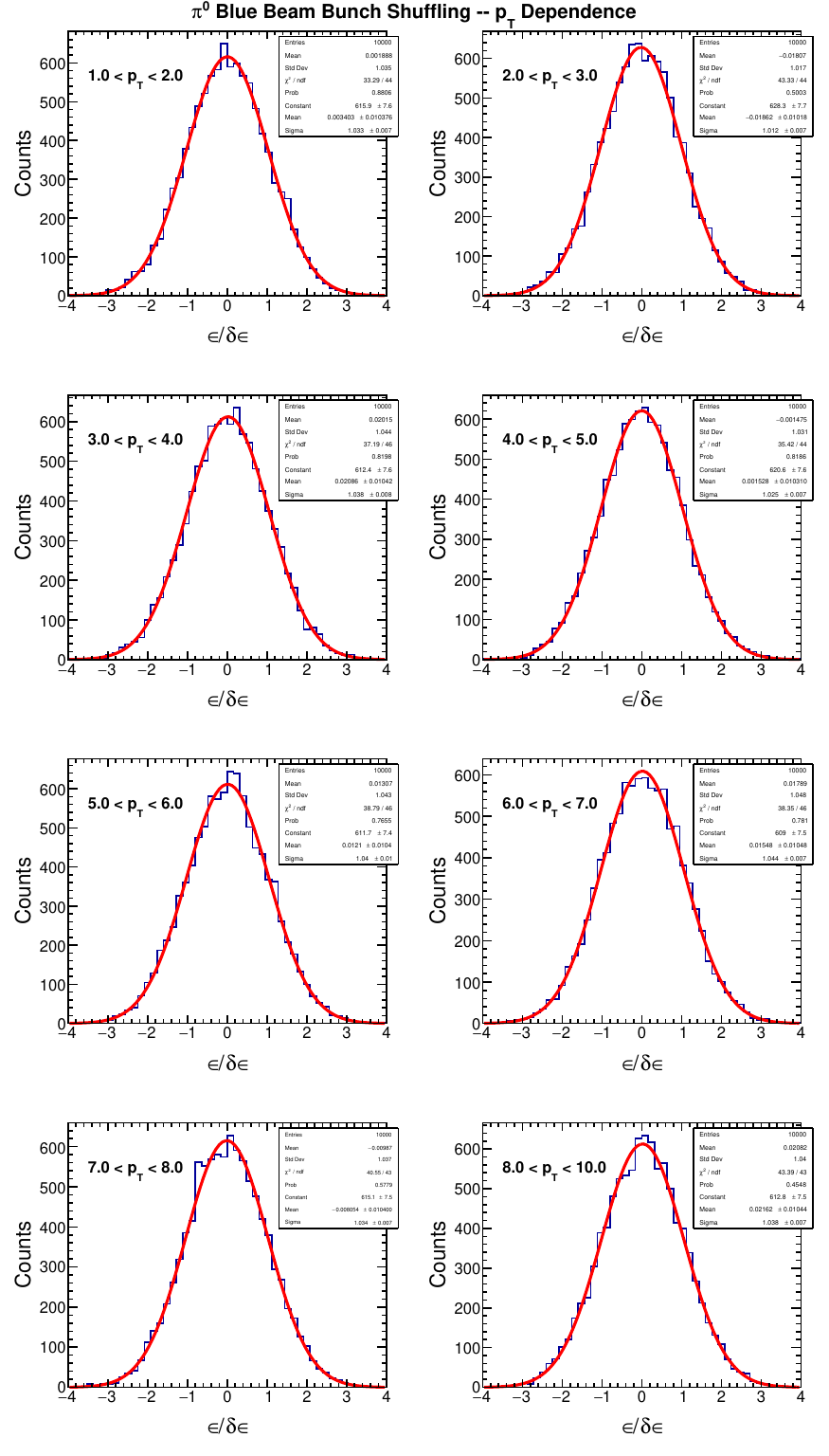}
    \caption{Bunch shuffling results for di-photons in the $\pi^0$ signal region, with the Blue beam taken as polarized. Each pad corresponds to one $p_T$ bin.}
    \label{fig:bs1}
\end{figure}

\begin{figure}
    \centering
    \includegraphics[page=2,height=0.9\textheight]{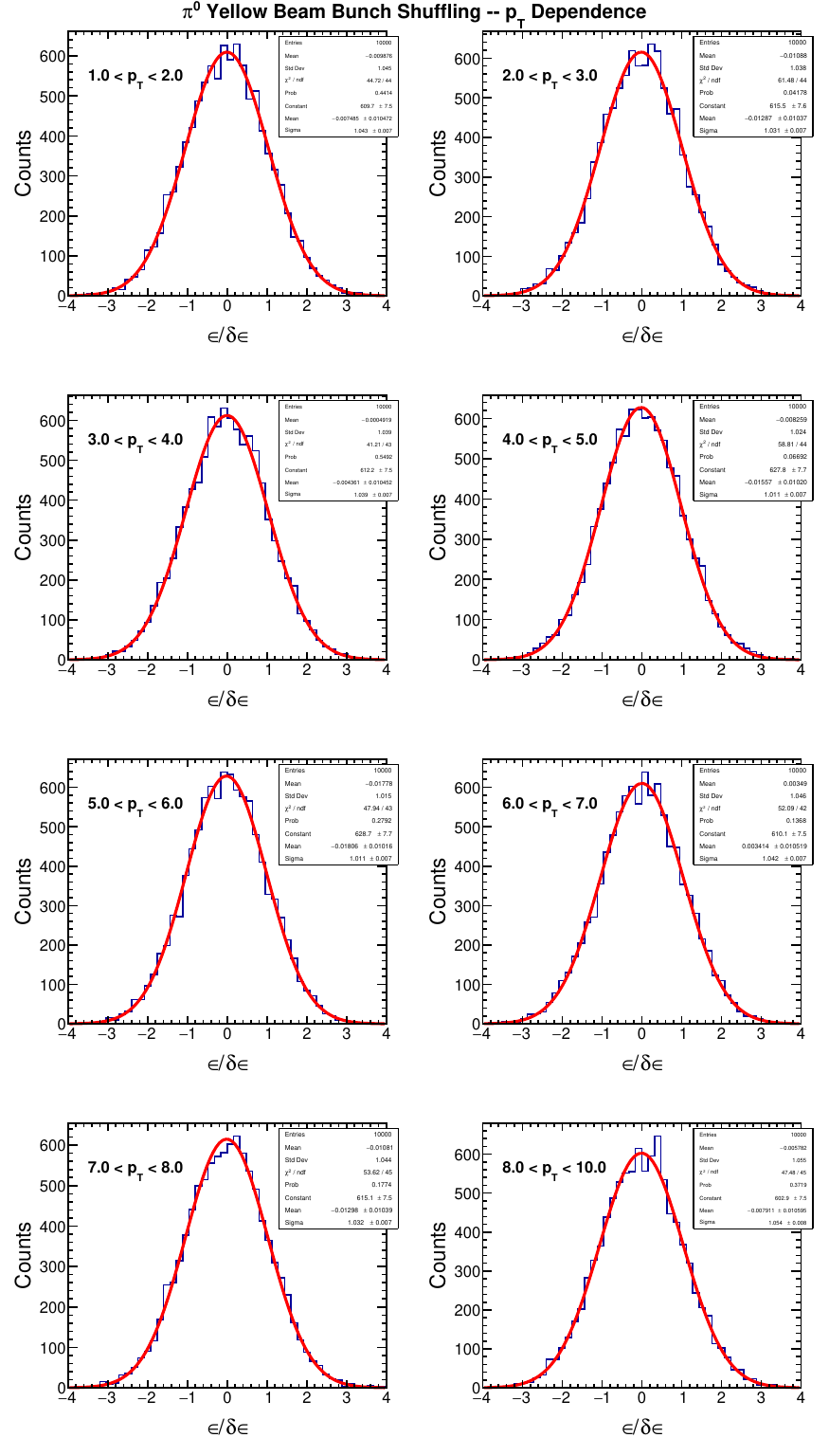}
    \caption{Bunch shuffling results for di-photons in the $\pi^0$ signal region, with the Yellow beam taken as polarized. Each pad corresponds to one $p_T$ bin.}
    \label{fig:bs2}
\end{figure}

\begin{figure}
    \centering
    \includegraphics[page=3,height=0.9\textheight]{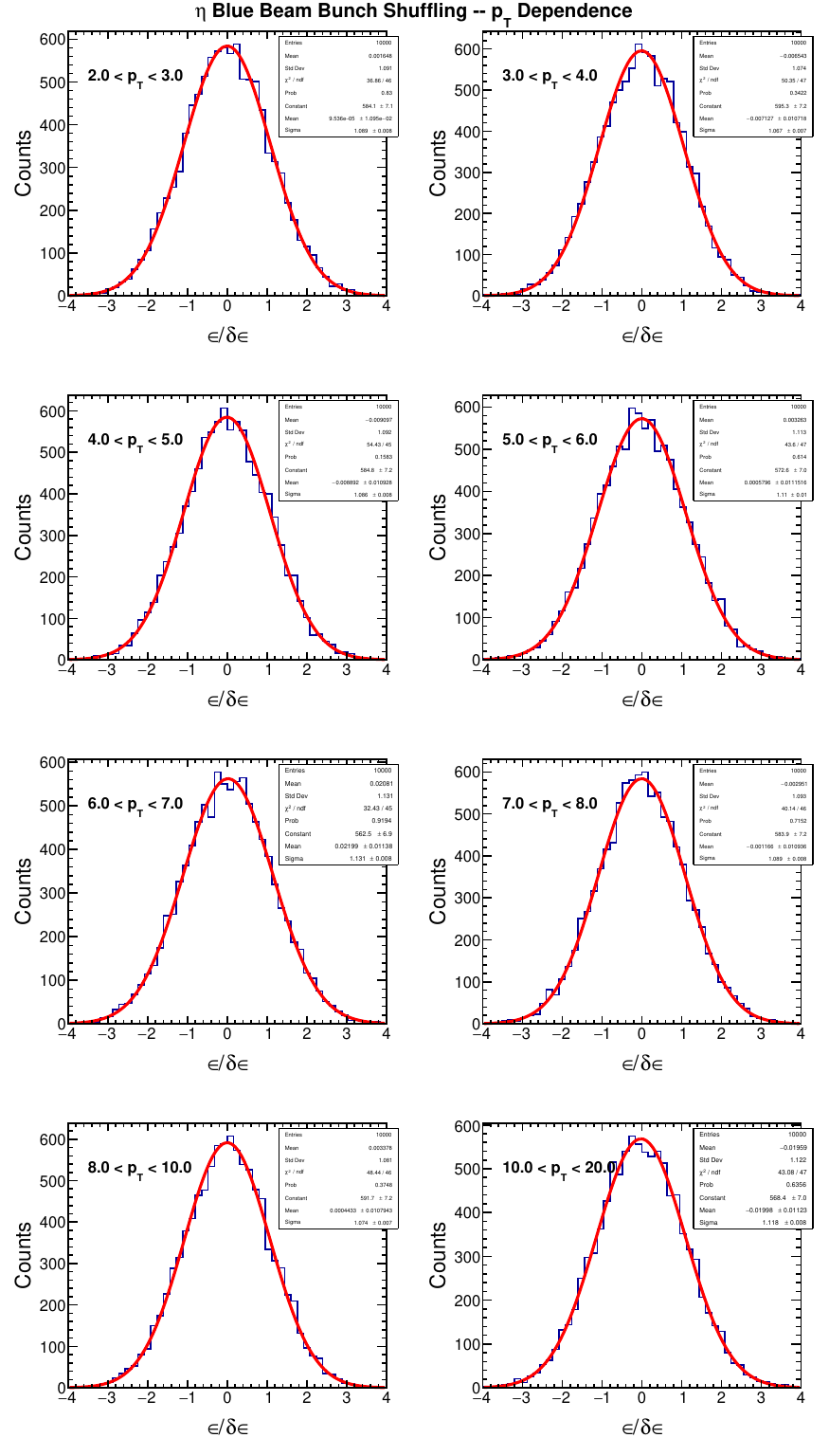}
    \caption{Bunch shuffling results for di-photons in the $\eta$-meson signal region, with the Blue beam taken as polarized. Each pad corresponds to one $p_T$ bin.}
    \label{fig:bs3}
\end{figure}

\begin{figure}
    \centering
    \includegraphics[page=4,height=0.9\textheight]{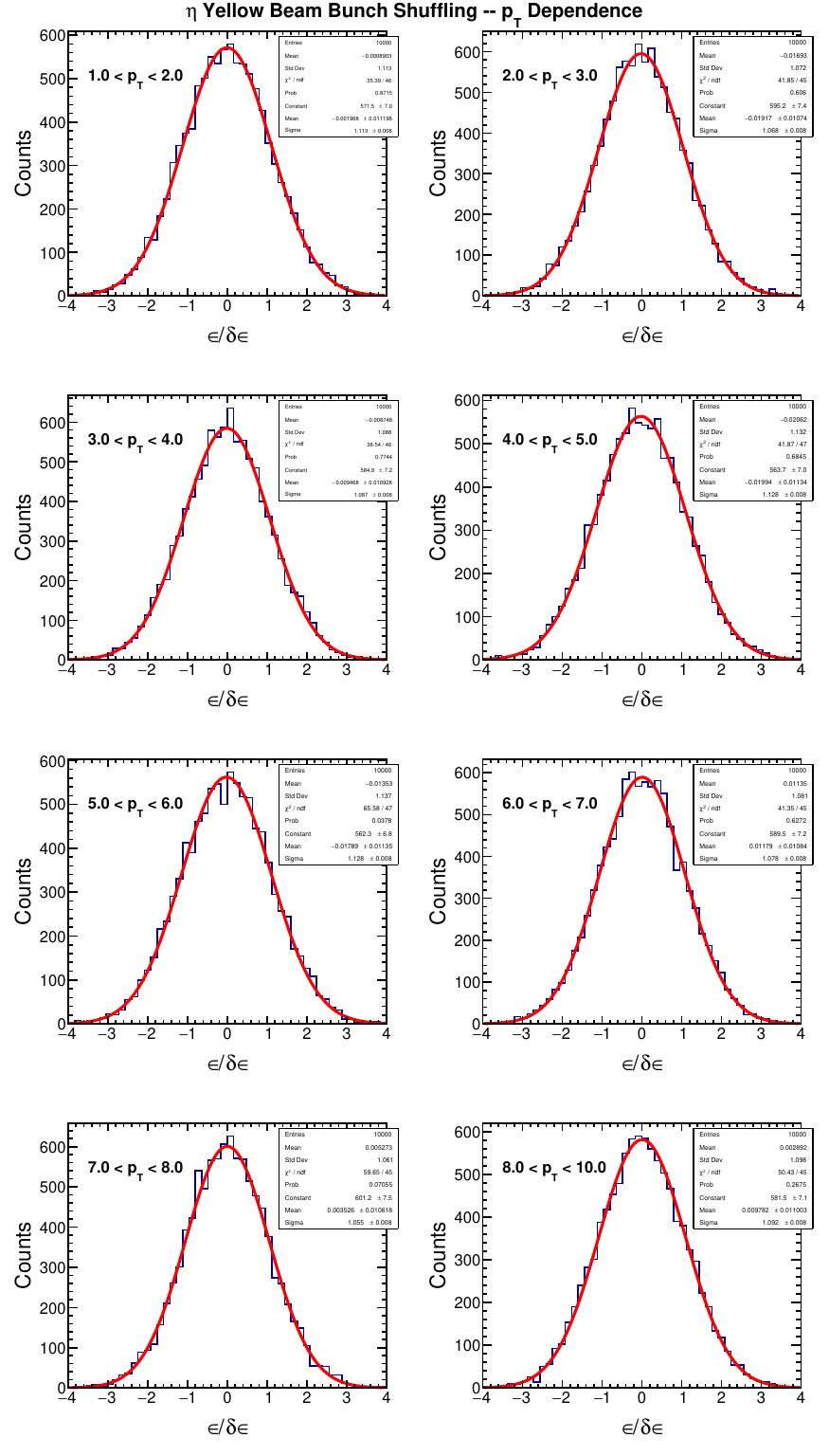}
    \caption{Bunch shuffling results for di-photons in the $\eta$-meson signal region, with the Yellow beam taken as polarized. Each pad corresponds to one $p_T$ bin.}
    \label{fig:bs4}
\end{figure}

\begin{figure}
    \centering
    \includegraphics[page=5,width=0.48\textwidth]{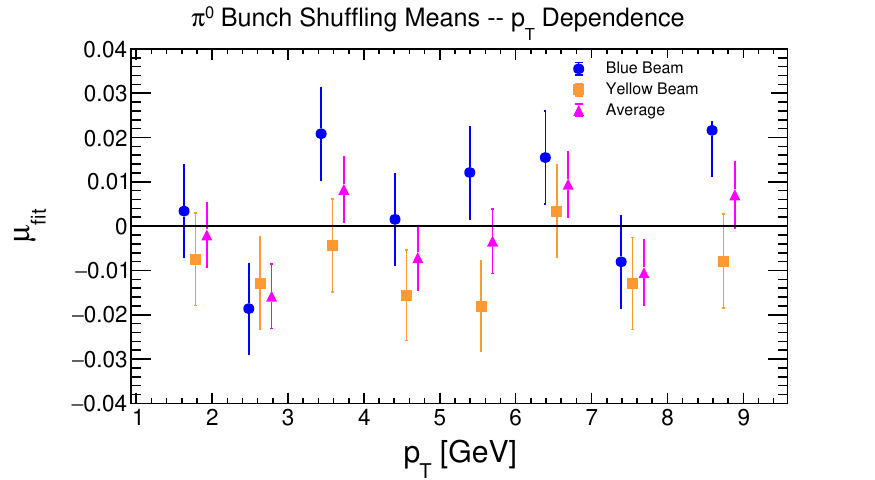}
    \includegraphics[page=6,width=0.48\textwidth]{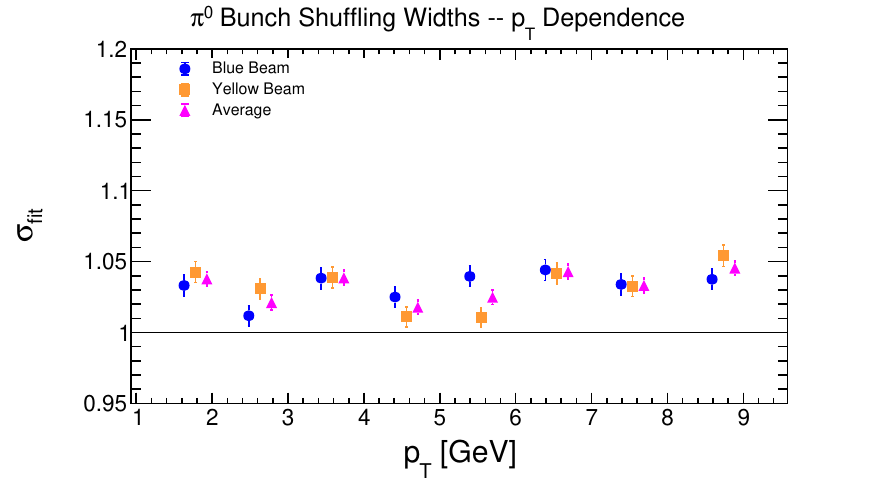}
    \caption{Bunch shuffling fitted means (left) and widths (right) for di-photons in the $\pi^0$ signal region, as a function of $p_T$. As the fitted widths are greater than unity, an additional uncertainty $\delta A_N^\mathrm{sys,shuffle}$ is assigned in all bins.}
    \label{fig:bs5}
\end{figure}

\begin{figure}
    \centering
    \includegraphics[page=7,width=0.48\textwidth]{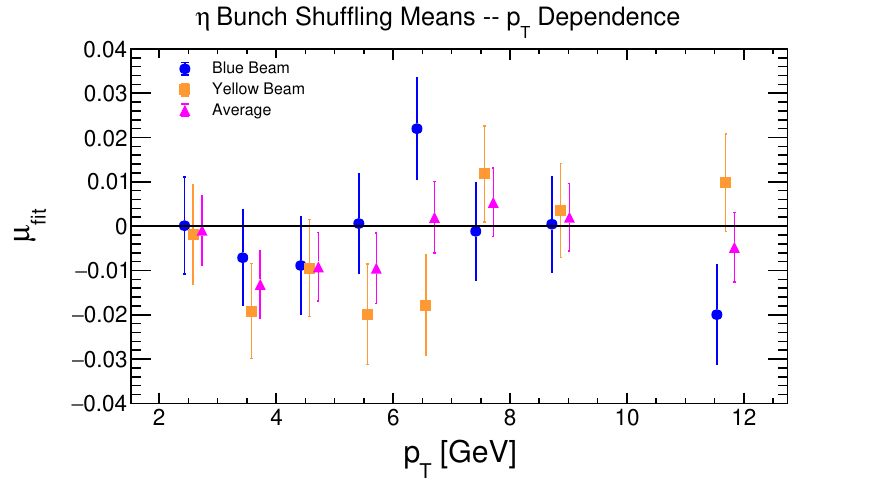}
    \includegraphics[page=8,width=0.48\textwidth]{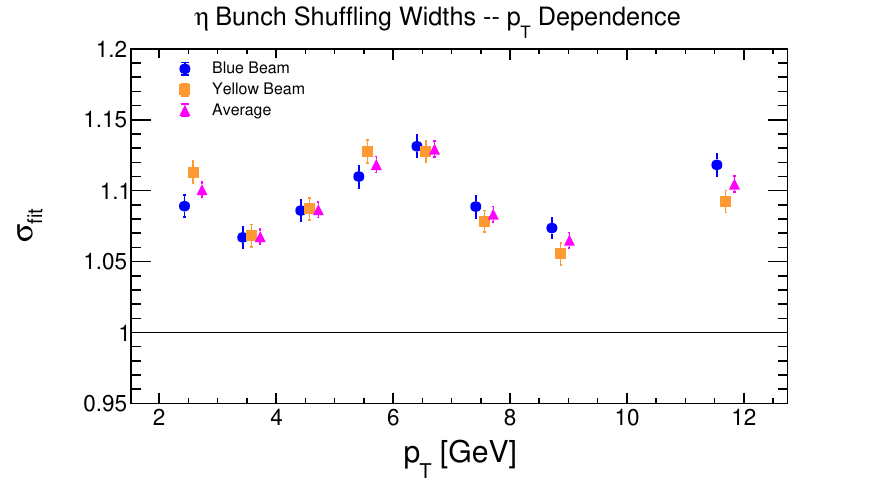}
    \caption{Bunch shuffling fitted means (left) and widths (right) for di-photons in the $\eta$-meson signal region, as a function of $p_T$. As the fitted widths are greater than unity, an additional uncertainty $\delta A_N^\mathrm{sys,shuffle}$ is assigned in all bins.}
    \label{fig:bs6}
\end{figure}

The distributions of $\epsilon/\delta \epsilon$ for 10,000 bunch shuffling iterations, with Gaussian fits, are shown in Figures~\ref{fig:bs1}-\ref{fig:bs4}. In all cases, the distributions are Gaussian and the fitted means $\mu_\mathrm{fit}$ are consistent with zero at or slightly above the 1-$\sigma$ level, indicating that no false asymmetry is observed. However, the fitted widths $\sigma_\mathrm{fit}$ are greater than one in all $p_T$ bins for both mesons. The means and widths extracted from the fits are shown as a function of $p_T$ in Figure~\ref{fig:bs5} for the $\pi^0$ and Figure~\ref{fig:bs6} for the $\eta$-meson. The bunch shuffling is performed independently for the Blue and Yellow beams. Results from the two beams are generally consistent within statistical uncertainties. With this consistency established, the $\sigma_\mathrm{fit}$ from both beams are averaged to determine a single value for the systematic uncertainty in each kinematic bin. The averaged $\sigma_\mathrm{fit}$ and corresponding $\delta A_N^\mathrm{sys,shuffle}$ are given in Table~\ref{tab:sys-shuffle}.


\begin{table}
\centering
\begin{tabular}{c|rr|rr}
& \multicolumn{2}{|c|}{$\pi^0$} & \multicolumn{2}{|c}{$\eta$} \\
$p_T$ [GeV] & $\sigma_\mathrm{fit}$ & $\delta A_N^\mathrm{sys,shuffle}$ & $\sigma_\mathrm{fit}$ & $\delta A_N^\mathrm{sys,shuffle}$ \\\hline
1-2 & 1.03770 $\pm$ 0.00523 & 0.00068 & -- & -- \\
2-3 & 1.02071 $\pm$ 0.00517 & 0.00010 & 1.10074 $\pm$ 0.00551 & 0.00453 \\
3-4 & 1.03907 $\pm$ 0.00532 & 0.00013 & 1.06768 $\pm$ 0.00540 & 0.00267 \\
4-5 & 1.01783 $\pm$ 0.00507 & 0.00012 & 1.08626 $\pm$ 0.00548 & 0.00320 \\
5-6 & 1.02485 $\pm$ 0.00504 & 0.00026 & 1.11846 $\pm$ 0.00567 & 0.00786 \\
6-7 & 1.04287 $\pm$ 0.00525 & 0.00061 & 1.12934 $\pm$ 0.00556 & 0.01288 \\
7-8 & 1.03348 $\pm$ 0.00513 & 0.00114 & 1.08334 $\pm$ 0.00549 & 0.01458 \\
8-10 & 1.04585 $\pm$ 0.00532 & 0.00250 & 1.06521 $\pm$ 0.00538 & 0.02721 \\
10-20 & -- & -- & 1.10471 $\pm$ 0.00557 & 0.11669 \\
\end{tabular}
\caption{Values for $\sigma_\mathrm{fit}$ with corresponding $\delta A_N^\mathrm{sys,shuffle}$, binned in $p_T$.}
\label{tab:sys-shuffle}
\end{table}

\clearpage

\section{Total Systematic Uncertainty}\label{sec:sys-tot}
The three systematic uncertainties detailed above are added in quadrature to calculate the total systematic uncertainty $\delta A_N^\mathrm{sys,tot}$:
\begin{equation}
	\delta A_N^\mathrm{sys,tot} = \sqrt{(\delta A_N^\mathrm{sys,calc})^2 + (\delta A_N^\mathrm{sys,bg})^2 + (\delta A_N^\mathrm{sys,shuffle})^2}.
	\label{eq:sys-total}
\end{equation}
The values of each systematic uncertainty, and the corresponding total systematic uncertainty, are summarized in Table~\ref{tab:sys}.

\begin{table}[h]
\centering
\begin{tabular}{r|c|c|c|c||r|r|r||r|r|r}
\multicolumn{11}{c}{}\\
\multicolumn{11}{c}{\Large $p^\uparrow p\rightarrow\pi^0 X$}\\
\multicolumn{11}{c}{}\\\hline
Bin & $p_T$ & $\langle p_T \rangle$ & $\langle \eta \rangle$ & $\langle x_F \rangle$ & $A_N$ & $\delta A_N^\mathrm{stat}$ & $\delta A_N^\mathrm{sys,tot}$ & $\delta A_N^\mathrm{sys,calc}$ & $\delta A_N^\mathrm{sys,bg}$ & $\delta A_N^\mathrm{sys,shuffle}$ \\
\# & [GeV] & [GeV] & & & & & & & & \\
\hline\hline
1 & 1-2 & 1.786 & 1.496 & 0.038 & -0.00107 & 0.00246 & 0.00072 & 0.00016 & 0.00017 & 0.00068 \\
2 & 2-3 & 2.562 & 1.213 & 0.040 & 0.00032 & 0.00048 & 0.00015 & 0.00006 & 0.00010 & 0.00010 \\
3 & 3-4 & 3.457 & 0.932 & 0.038 & 0.00031 & 0.00047 & 0.00013 & 0.00001 & 0.00002 & 0.00013 \\
4 & 4-5 & 4.418 & 0.707 & 0.036 & 0.00120 & 0.00066 & 0.00013 & 0.00002 & 0.00000 & 0.00012 \\
5 & 5-6 & 5.399 & 0.630 & 0.039 & 0.00011 & 0.00115 & 0.00028 & 0.00009 & 0.00006 & 0.00026 \\
6 & 6-7 & 6.396 & 0.617 & 0.045 & 0.00388 & 0.00207 & 0.00064 & 0.00017 & 0.00010 & 0.00061 \\
7 & 7-8 & 7.393 & 0.634 & 0.054 & 0.00517 & 0.00439 & 0.00131 & 0.00016 & 0.00061 & 0.00114 \\
8 & 8-10 & 8.600 & 0.674 & 0.067 & -0.00547 & 0.00815 & 0.00847 & 0.00039 & 0.00809 & 0.00250 \\
\multicolumn{11}{c}{}\\
\multicolumn{11}{c}{\Large $p^\uparrow p\rightarrow\eta X$}\\
\multicolumn{11}{c}{}\\\hline
Bin & $p_T$ & $\langle p_T \rangle$ & $\langle \eta \rangle$ & $\langle x_F \rangle$ & $A_N$ & $\delta A_N^\mathrm{stat}$ & $\delta A_N^\mathrm{sys,tot}$ & $\delta A_N^\mathrm{sys,calc}$ & $\delta A_N^\mathrm{sys,bg}$ & $\delta A_N^\mathrm{sys,shuffle}$ \\
\# & [GeV] & [GeV] & & & & & & & & \\
\hline\hline
1 & 2-3 & 2.537 & 1.237 & 0.040 & -0.02644 & 0.00985 & 0.00487 & 0.00095 & 0.00152 & 0.00453 \\
2 & 3-4 & 3.460 & 0.982 & 0.041 & -0.00470 & 0.00713 & 0.00288 & 0.00107 & 0.00008 & 0.00267 \\
3 & 4-5 & 4.435 & 0.756 & 0.039 & -0.00187 & 0.00755 & 0.00324 & 0.00050 & 0.00006 & 0.00320 \\
4 & 5-6 & 5.421 & 0.620 & 0.038 & 0.01250 & 0.01568 & 0.00852 & 0.00047 & 0.00326 & 0.00786 \\
5 & 6-7 & 6.418 & 0.559 & 0.040 & 0.01833 & 0.02455 & 0.01333 & 0.00164 & 0.00301 & 0.01288 \\
6 & 7-8 & 7.424 & 0.540 & 0.045 & -0.00056 & 0.03499 & 0.01458 & 0.00032 & 0.00028 & 0.01458 \\
7 & 8-10 & 8.744 & 0.530 & 0.052 & 0.02223 & 0.07414 & 0.02881 & 0.00575 & 0.00753 & 0.02721 \\
8 & 10-20 & 11.732 & 0.511 & 0.066 & -0.10480 & 0.24856 & 0.12831 & 0.00139 & 0.05334 & 0.11669 \\
\hline
\end{tabular}
\caption{Statistical and systematic uncertainties on $A_N$ for both $\pi^0$ and $\eta$-mesons, extracted in different bins in $p_T$. Uncertainty values shown as 0.00000 are less than 5x10${}^{-6}$.} 
\label{tab:sys}
\end{table}

\begin{figure}
    \centering
    \includegraphics[page=7,width=0.48\linewidth]{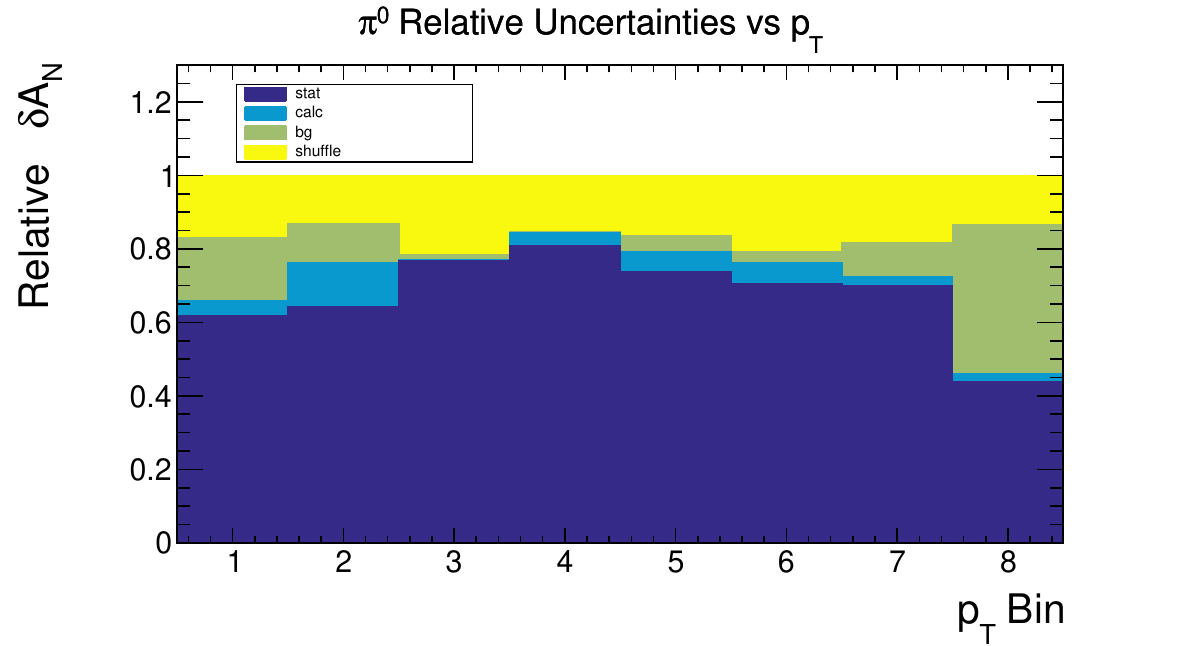}
    \includegraphics[page=10,width=0.48\linewidth]{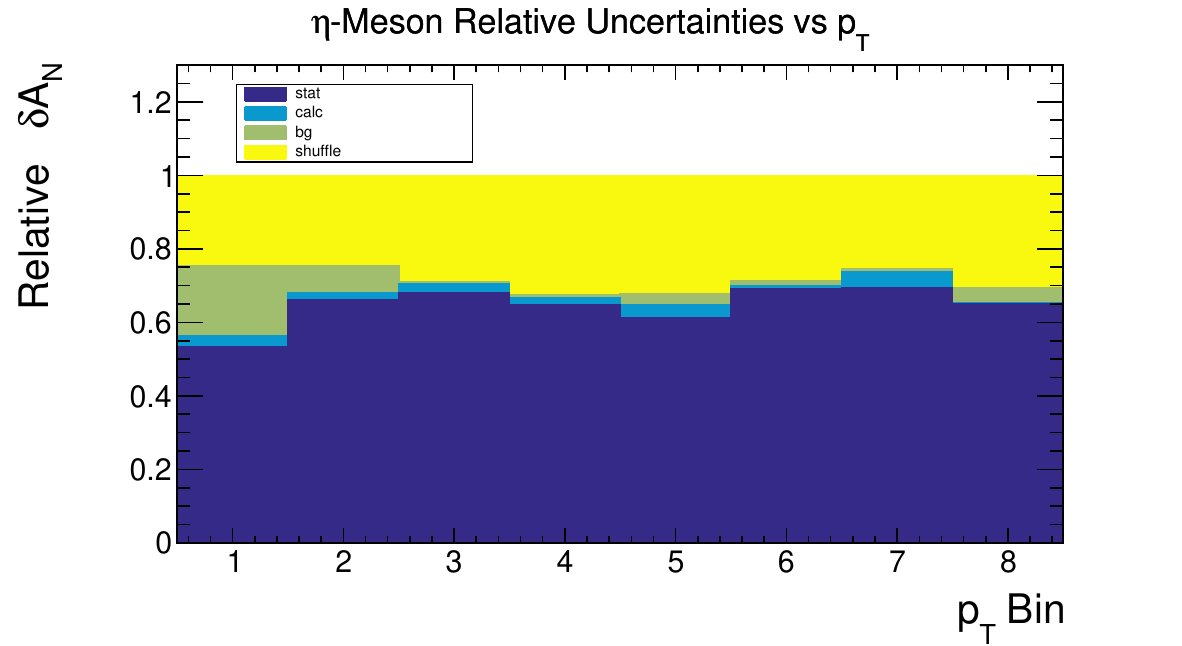}
    \caption{Comparison of statistical and systematic uncertainties for $\pi^0$ (left) and $\eta$-meson (right) as a function of $p_T$.}
    \label{fig:systematics_pT}
\end{figure}

The relative contributions of each source of uncertainty (statistical and systematic) are shown graphically in Figure~\ref{fig:systematics_pT}. In this figure, the four sources of uncertainty are shown stacked one above another. The bin number shown on the $x$-axis corresponds to the left-most column in Table~\ref{tab:sys}. Figure~\ref{fig:systematics_pT} demonstrates that the systematic uncertainties are generally smaller than the statistical uncertainties. No single systematic source dominates for the $\pi^0$, while $A_N^\mathrm{sys,shuffle}$ is generally the largest for the $\eta$-meson. Note that the contribution from $\delta A_N^\mathrm{sys,calc}$ is generally the smallest and is negligible in many $p_T$-bins, especially for the $\eta$-meson.

\clearpage

\section{$\eta$ and $x_F$ Binning}\label{sec:sys-eta-xF}
\begin{figure}
    \centering
    \includegraphics[page=8,width=0.48\linewidth]{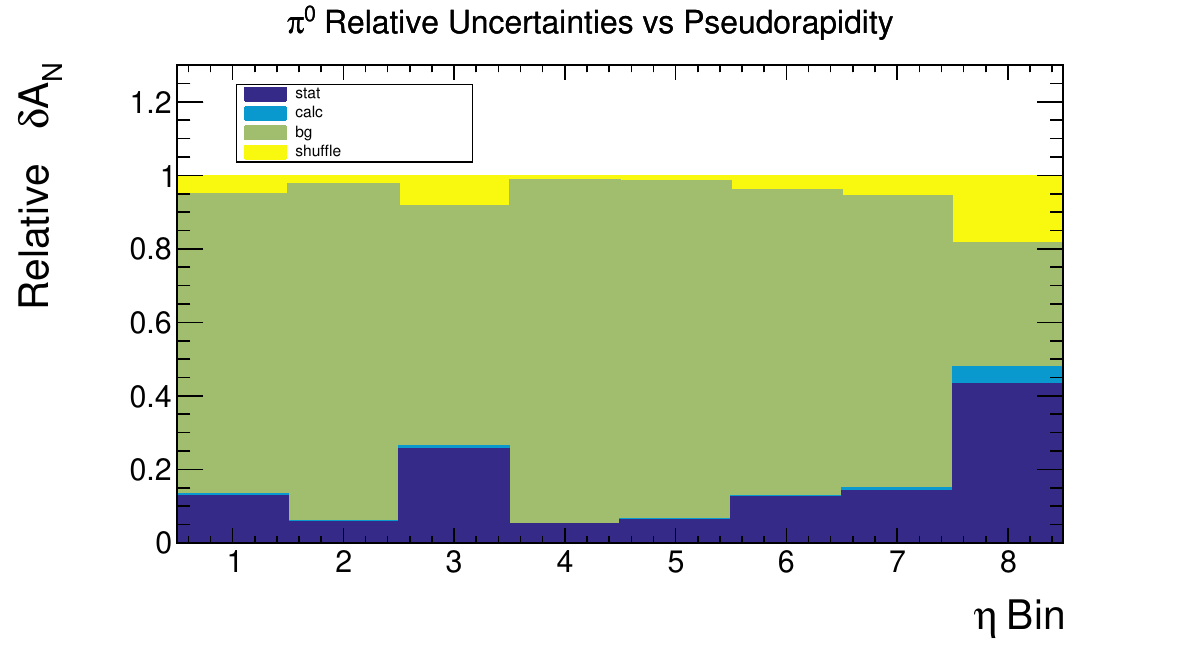}
    \includegraphics[page=11,width=0.48\linewidth]{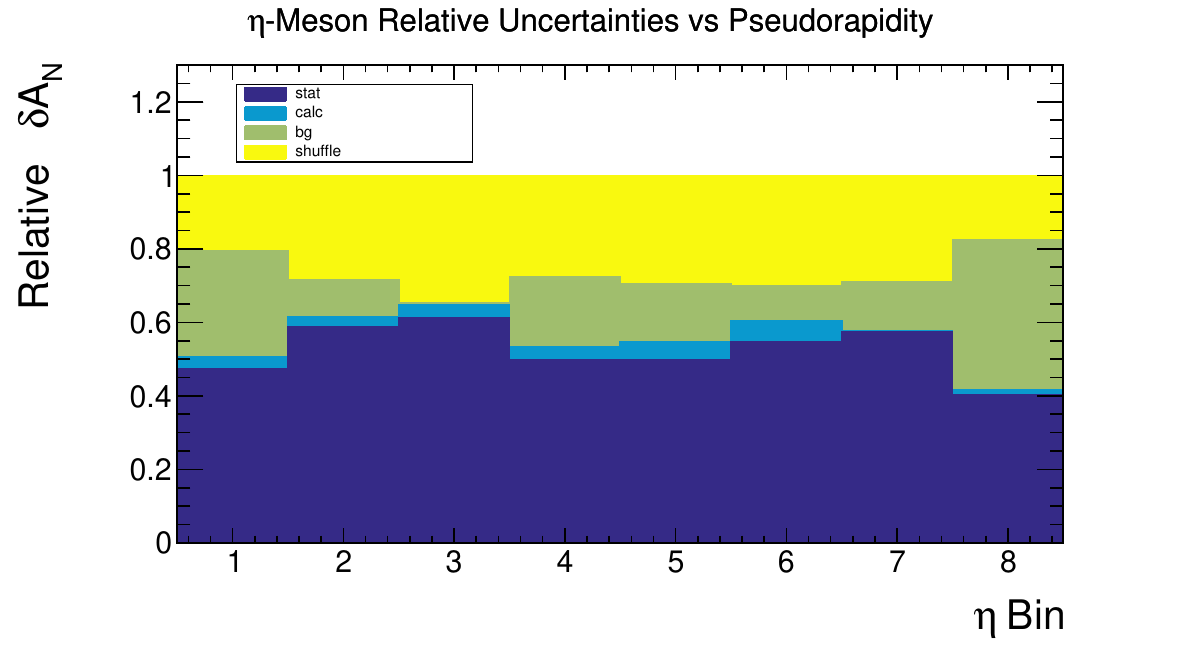}
    \caption{Comparison of statistical and systematic uncertainties for $\pi^0$ (left) and $\eta$-meson (right) as a function of $\eta$.}
    \label{fig:systematics_eta}
\end{figure}

\begin{figure}
    \centering
    \includegraphics[page=9,width=0.48\linewidth]{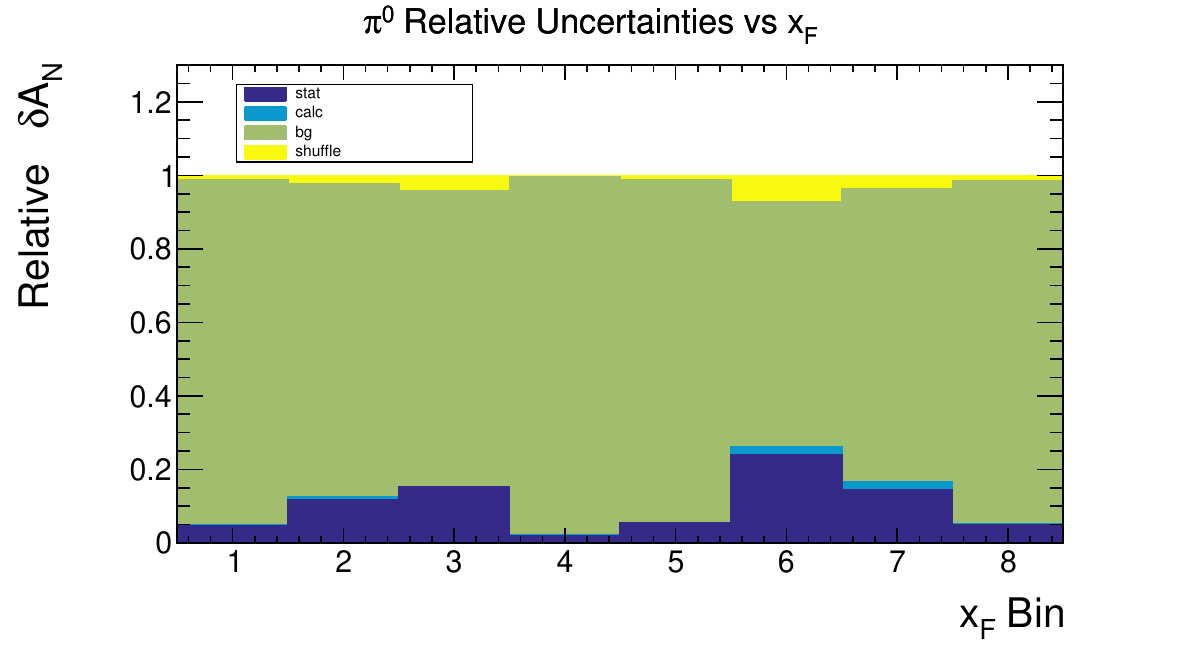}
    \includegraphics[page=12,width=0.48\linewidth]{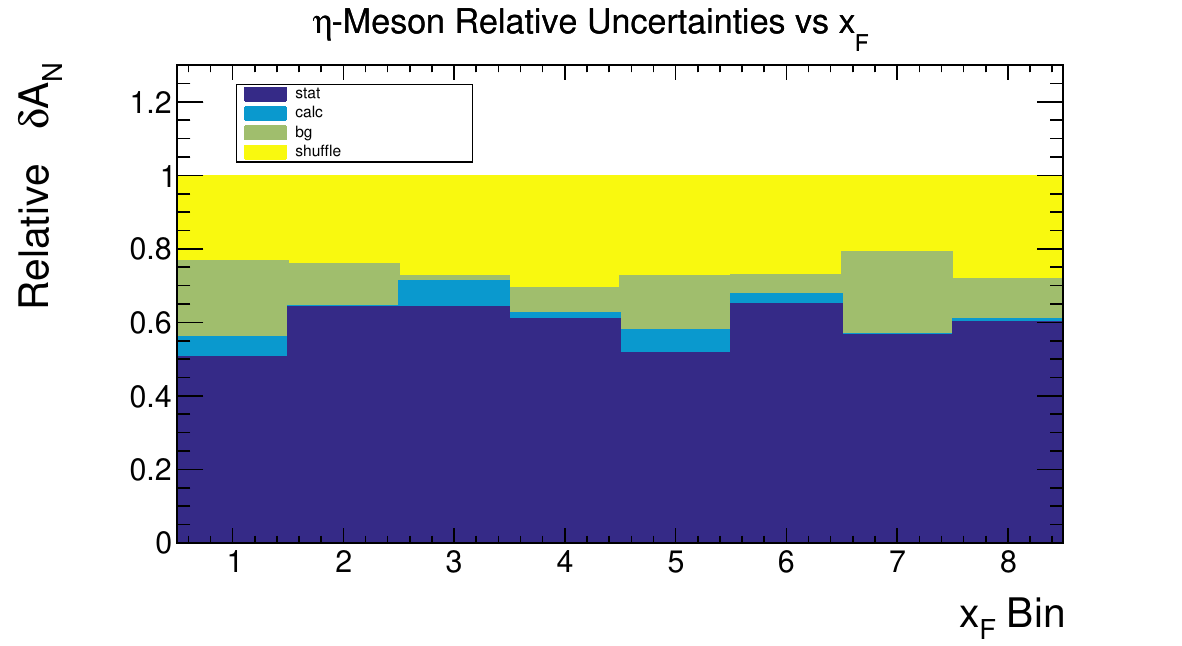}
    \caption{Comparison of statistical and systematic uncertainties for $\pi^0$ (left) and $\eta$-meson (right) as a function of $x_F$.}
    \label{fig:systematics_xF}
\end{figure}

The three sources of systematic uncertainty detailed above for the $p_T$ binning are calculated in parallel for the pseudorapidity and $x$-Feynman binnings. The resulting uncertainties are summarized in Tables~\ref{tab:sys_eta} and \ref{tab:sys_xF} and Figures~\ref{fig:systematics_eta} and \ref{fig:systematics_xF}. For the $\pi^0$, the dominating contribution in both binnings is $\delta A_N^\mathrm{sys,bg}$. For the $\eta$-meson, no one source of uncertainty dominates, but $\delta A_N^\mathrm{sys,shuffle}$ is generally the largest.

\begin{figure}
    \centering
    \includegraphics[page=36,width=0.48\textwidth]{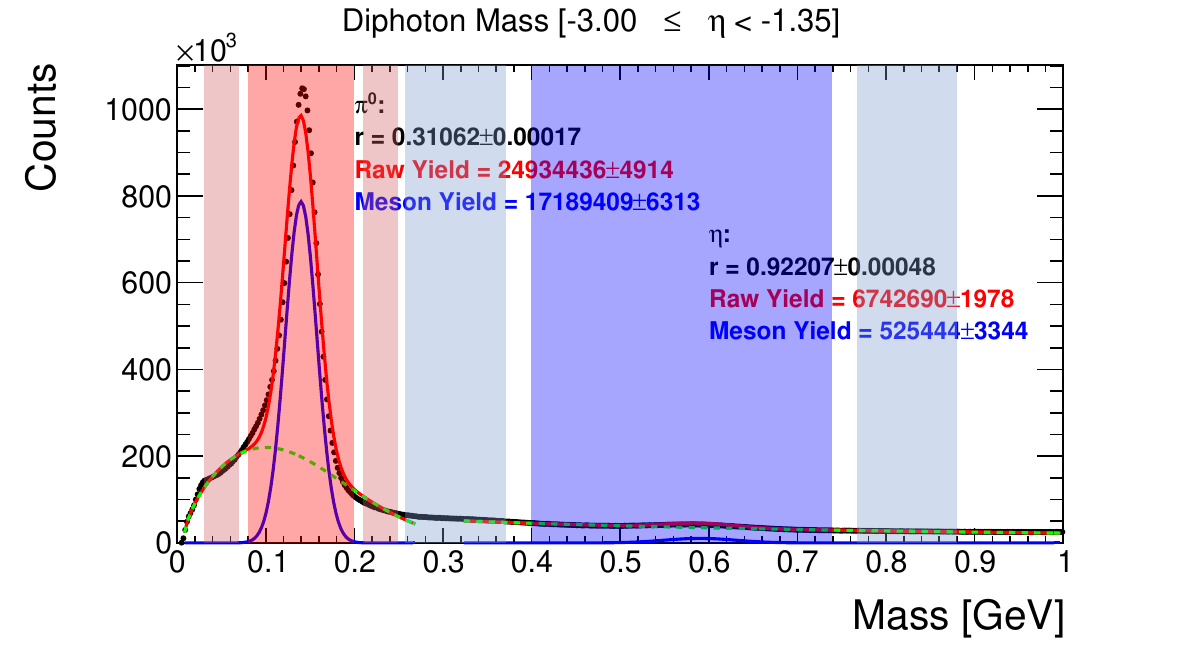}
    \includegraphics[page=37,width=0.48\textwidth]{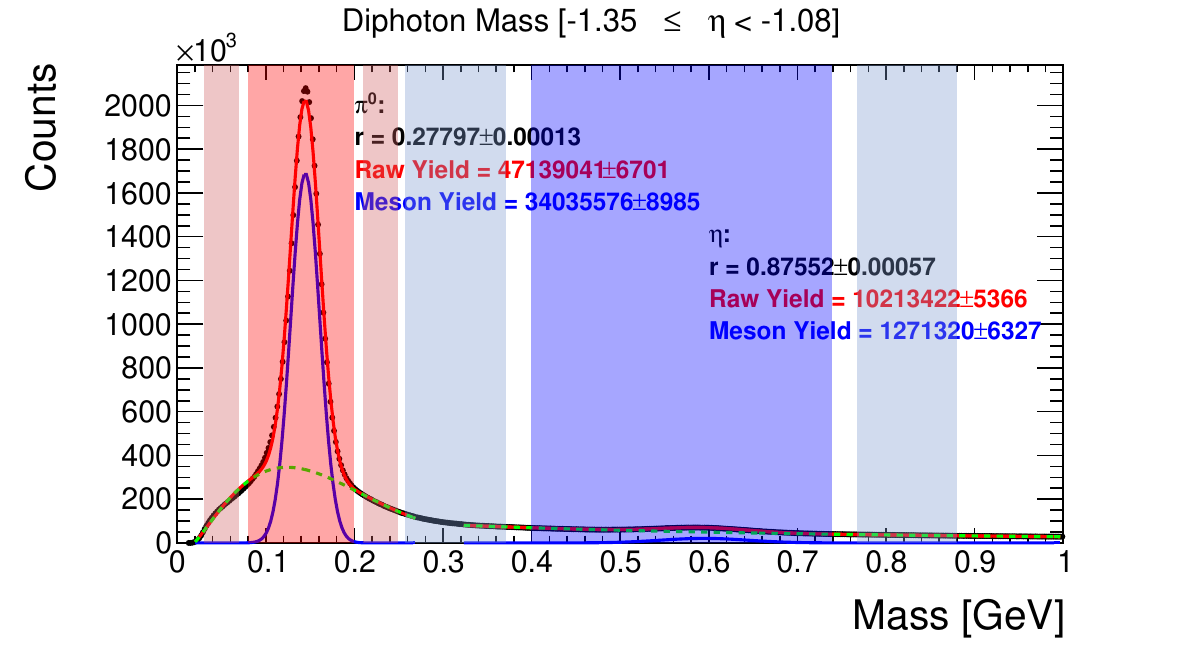}
    \includegraphics[page=38,width=0.48\textwidth]{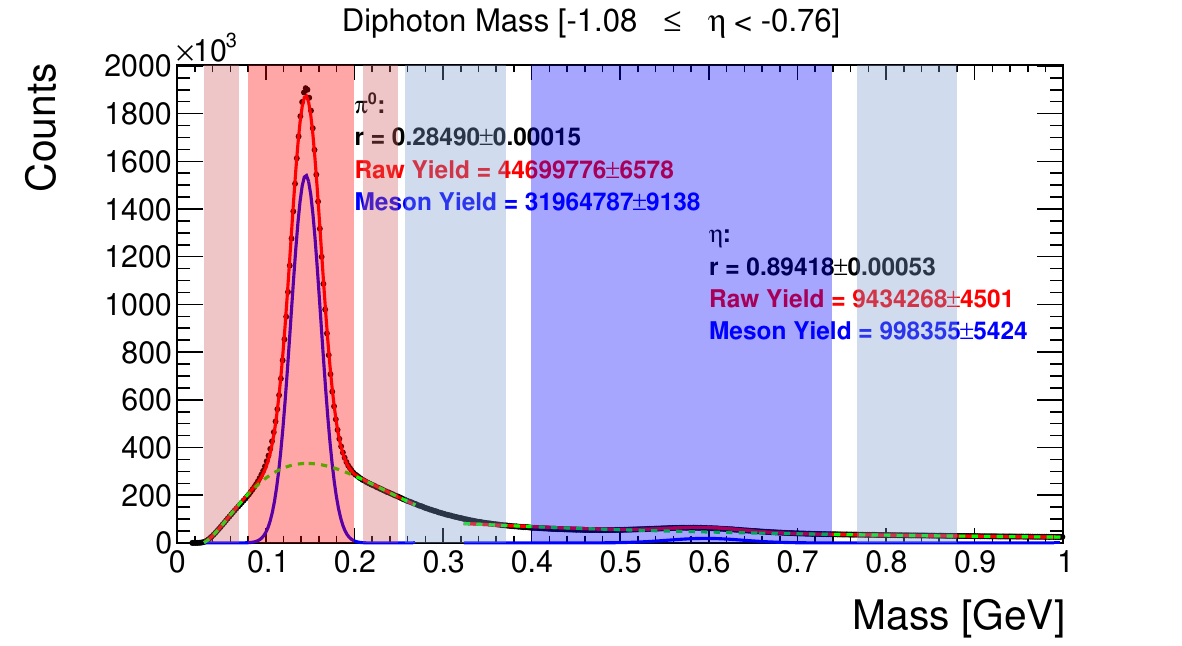}
    \includegraphics[page=39,width=0.48\textwidth]{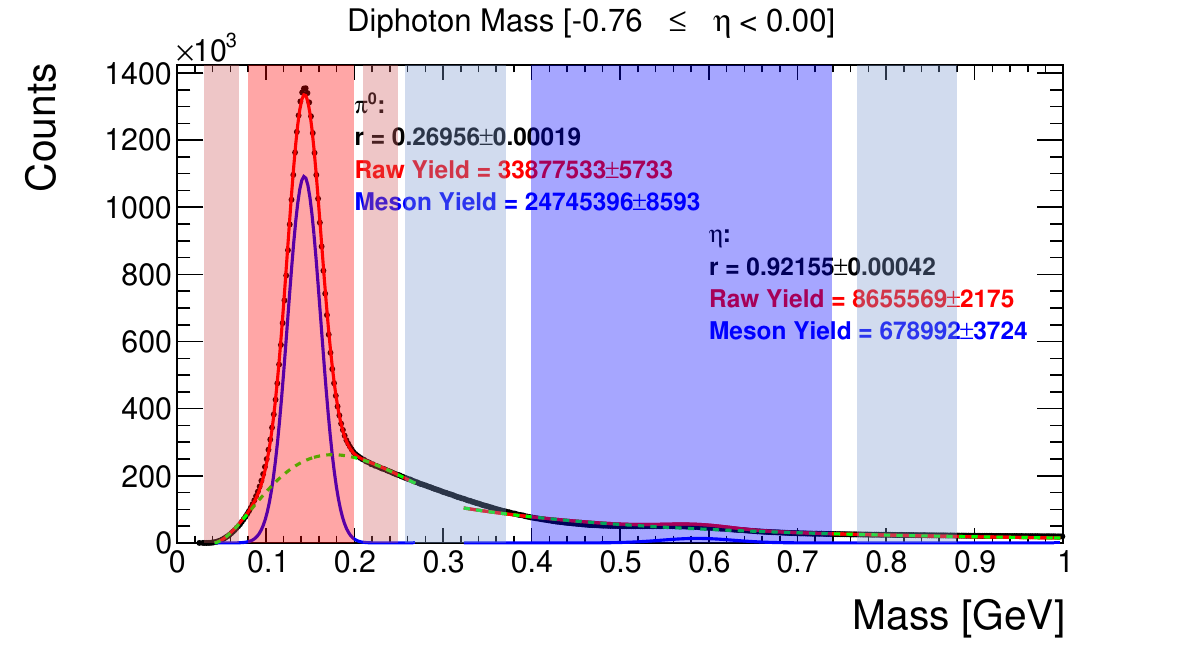}
    \includegraphics[page=40,width=0.48\textwidth]{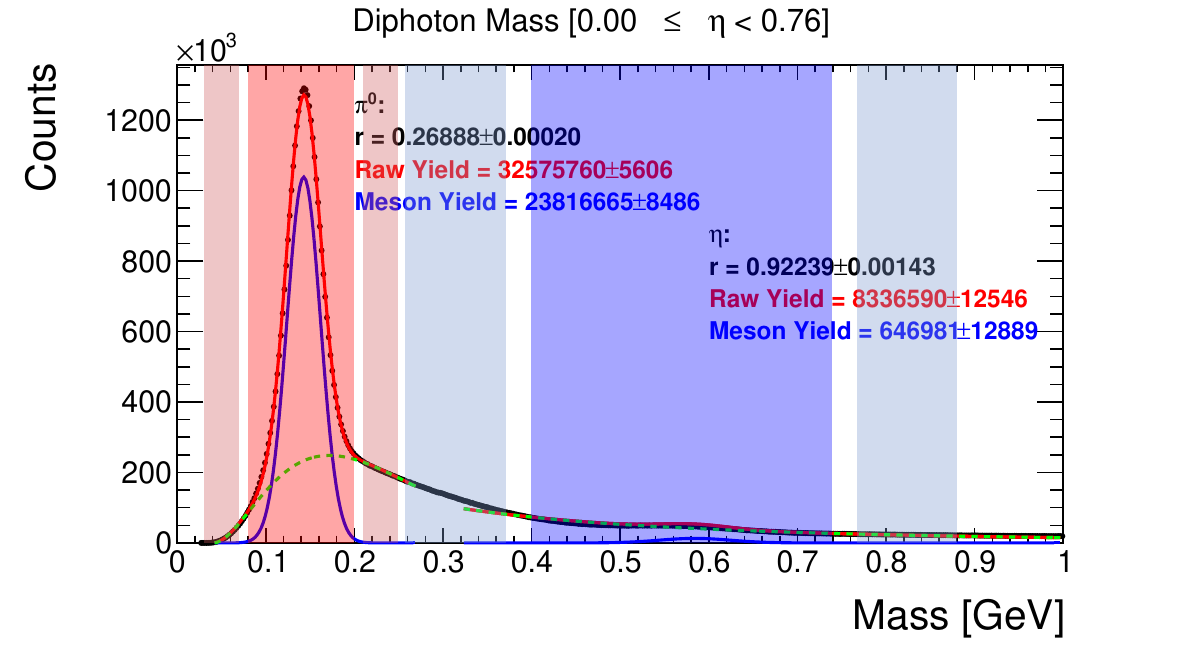}
    \includegraphics[page=41,width=0.48\textwidth]{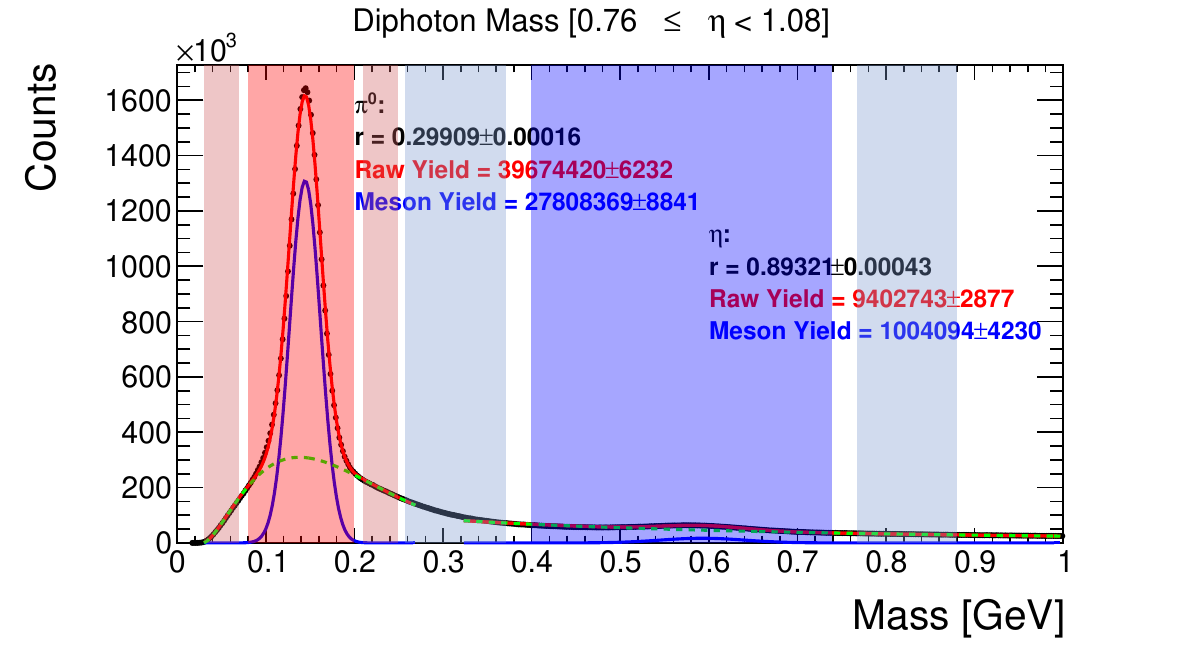}
    \includegraphics[page=42,width=0.48\textwidth]{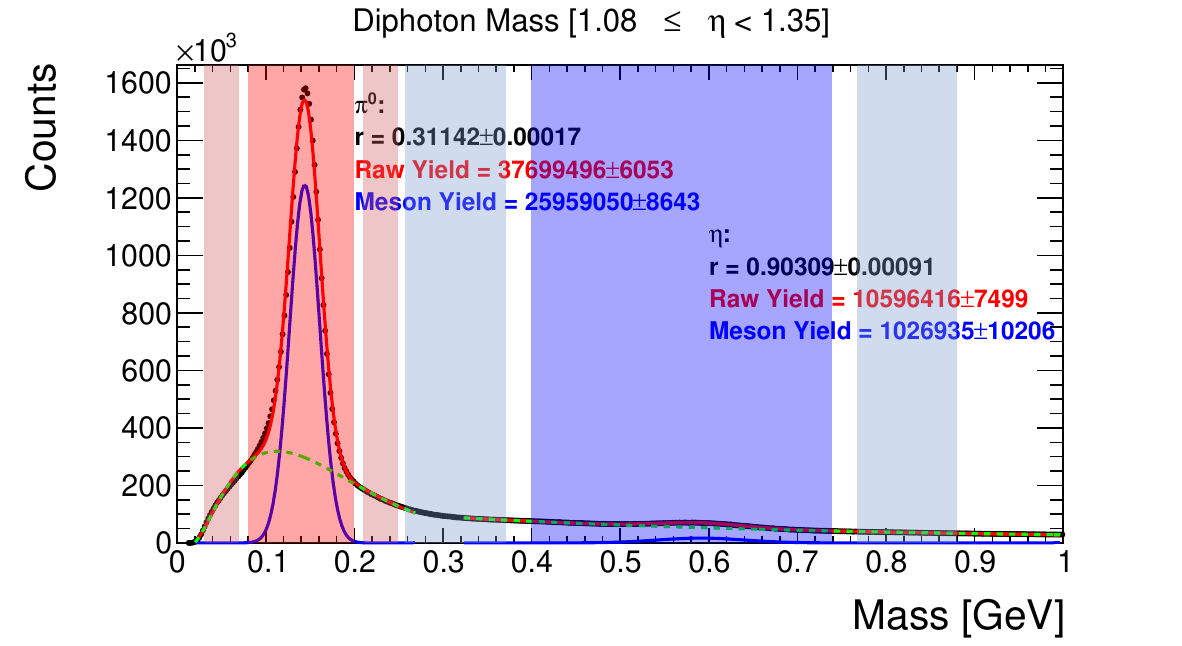}
    \includegraphics[page=43,width=0.48\textwidth]{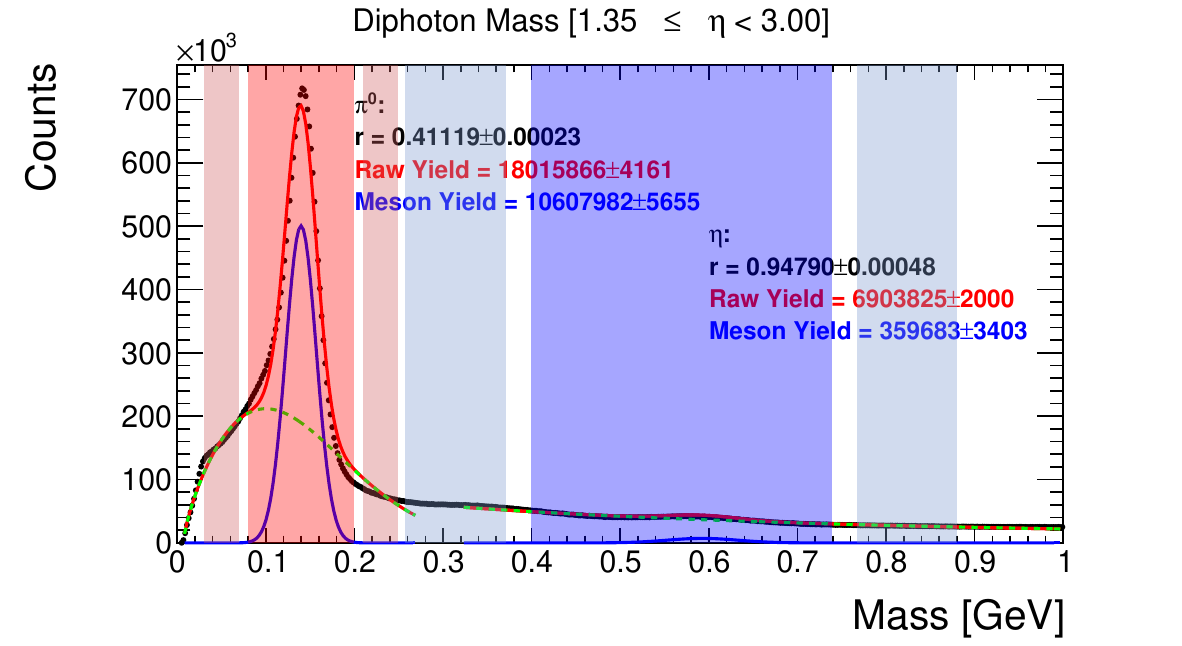}
    \caption{Di-photon mass distributions with main analysis (threshold function) fits, binned in $\eta$.}
    \label{fig:mass_thr_eta}
\end{figure}

\begin{figure}
    \centering
    \includegraphics[page=1,height=0.9\textheight]{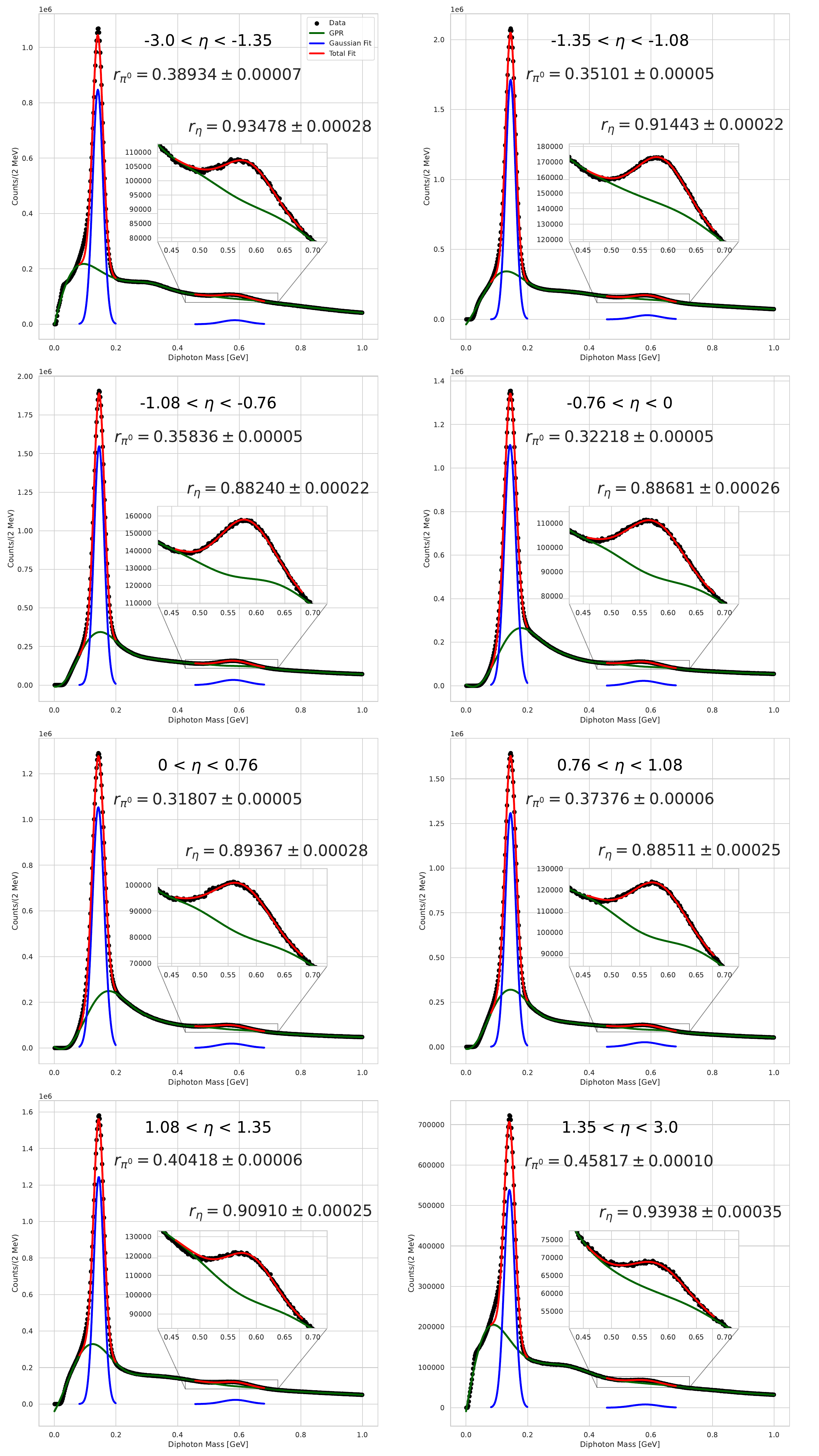}
    \caption{Di-photon mass distributions with alternative (GPR) fits, binned in $\eta$.}
    \label{fig:mass_gpr_eta}
\end{figure}

\begin{figure}
    \centering
    \includegraphics[page=56,width=0.48\textwidth]{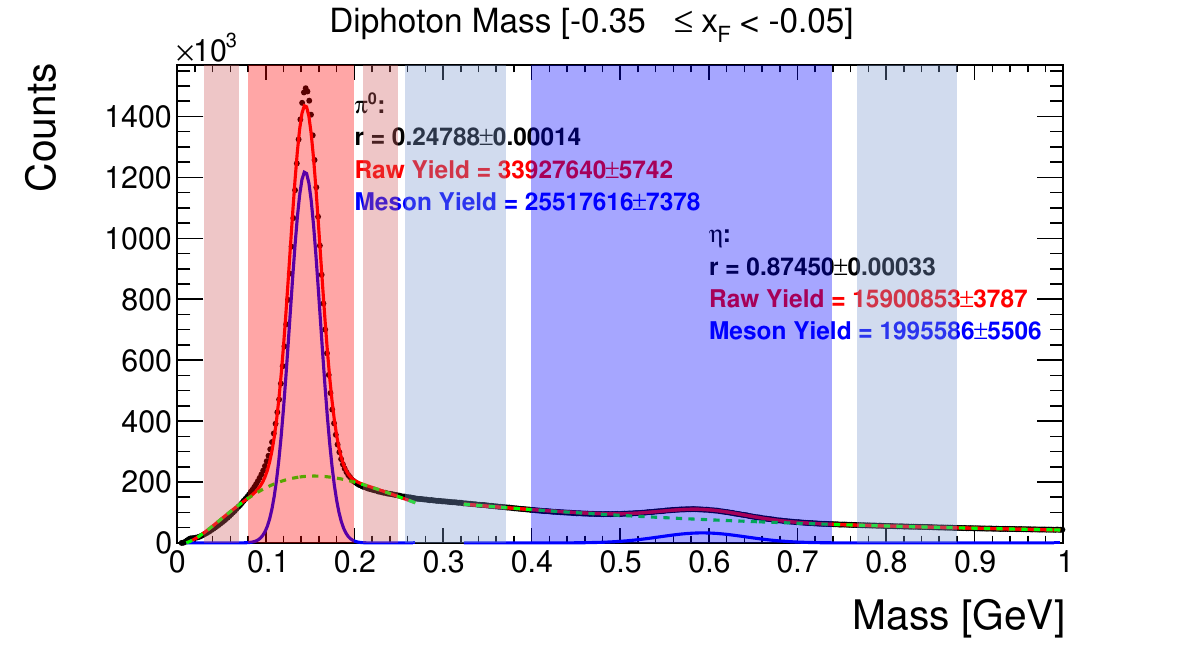}
    \includegraphics[page=57,width=0.48\textwidth]{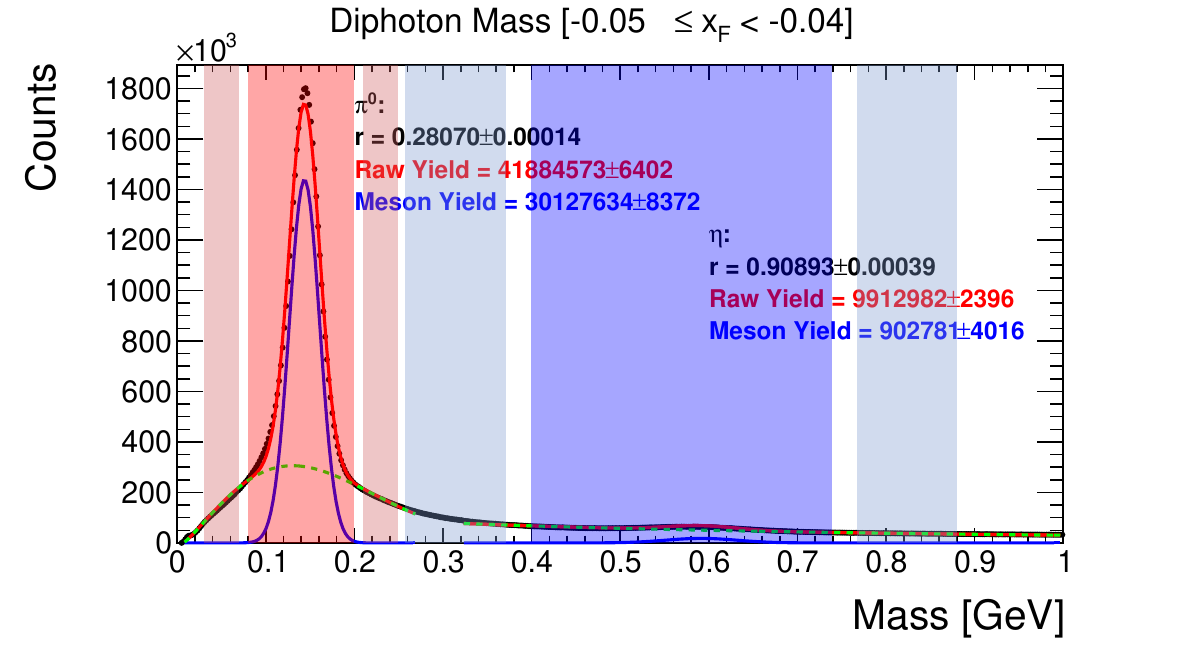}
    \includegraphics[page=58,width=0.48\textwidth]{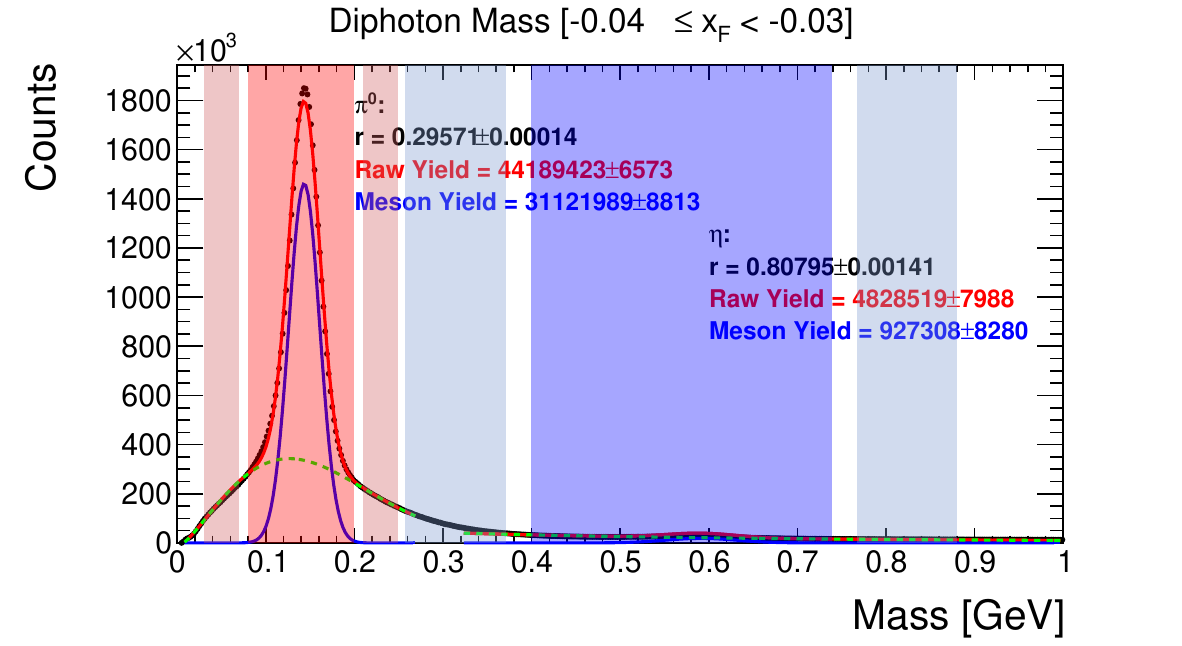}
    \includegraphics[page=59,width=0.48\textwidth]{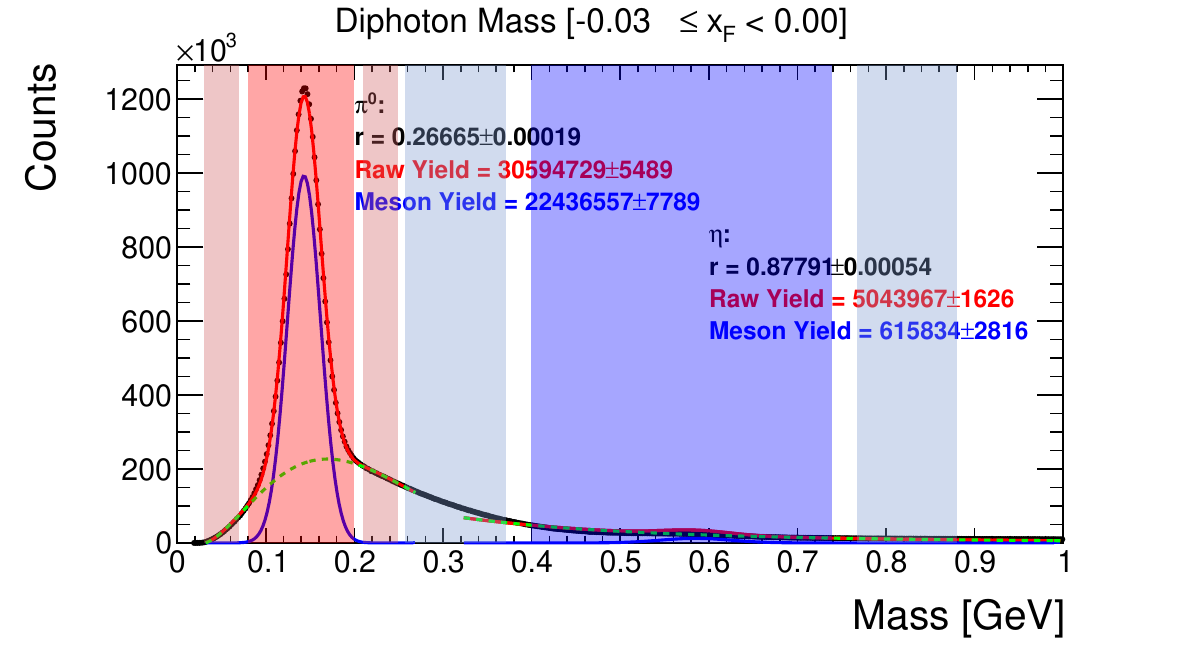}
    \includegraphics[page=60,width=0.48\textwidth]{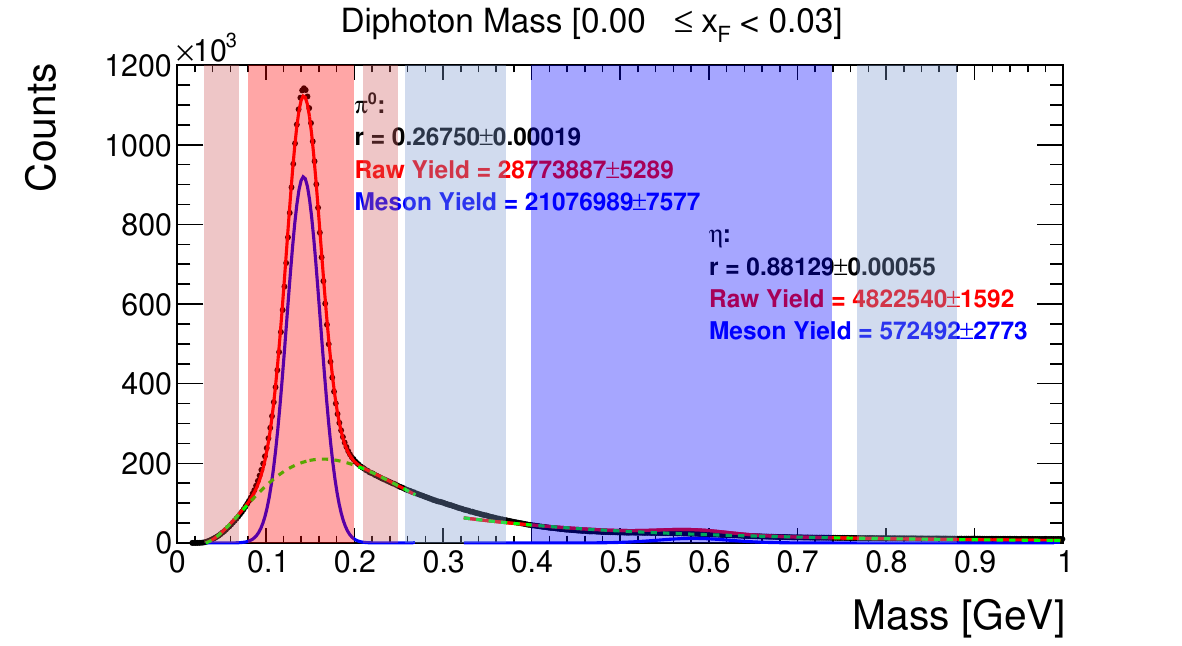}
    \includegraphics[page=61,width=0.48\textwidth]{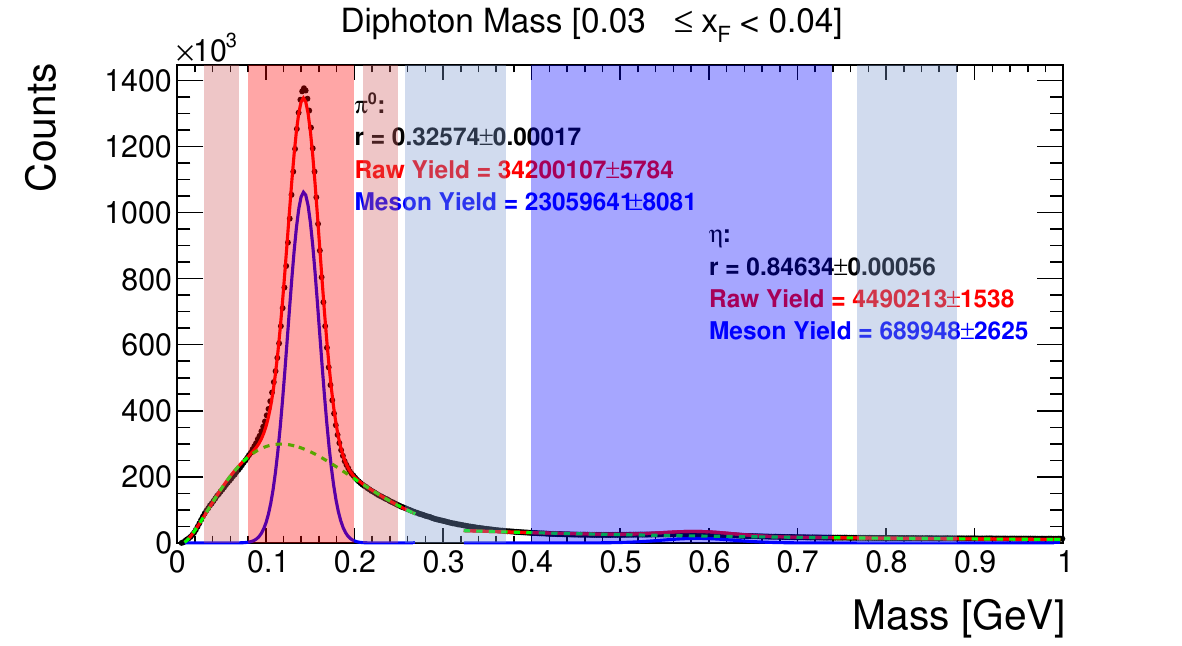}
    \includegraphics[page=62,width=0.48\textwidth]{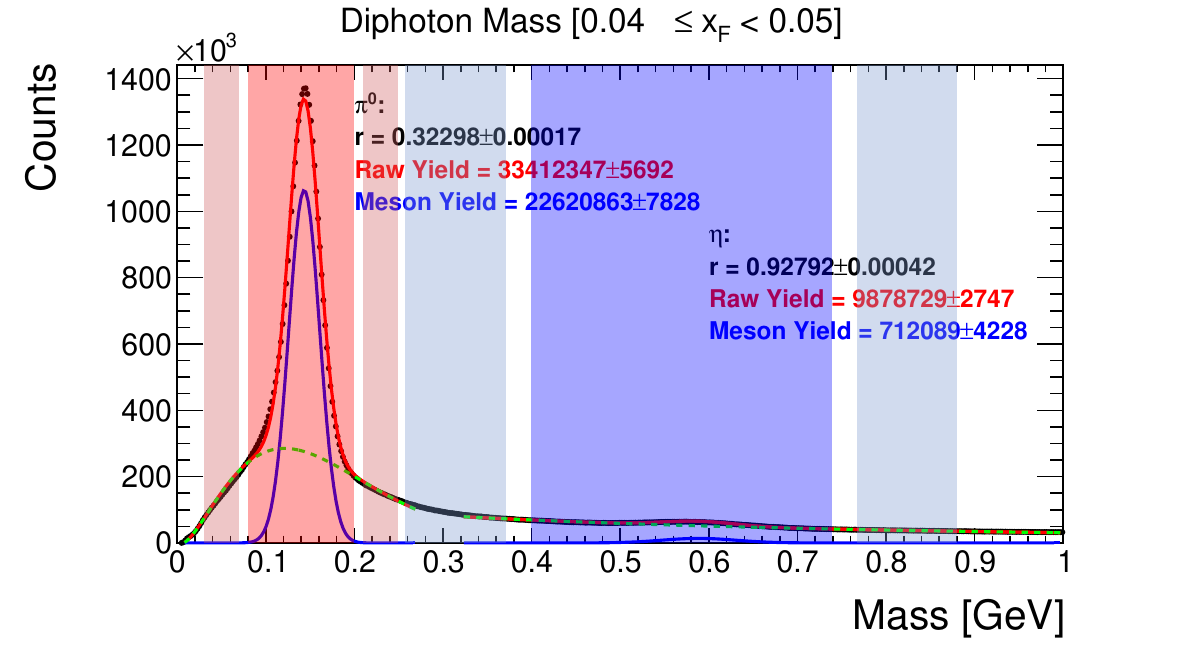}
    \includegraphics[page=63,width=0.48\textwidth]{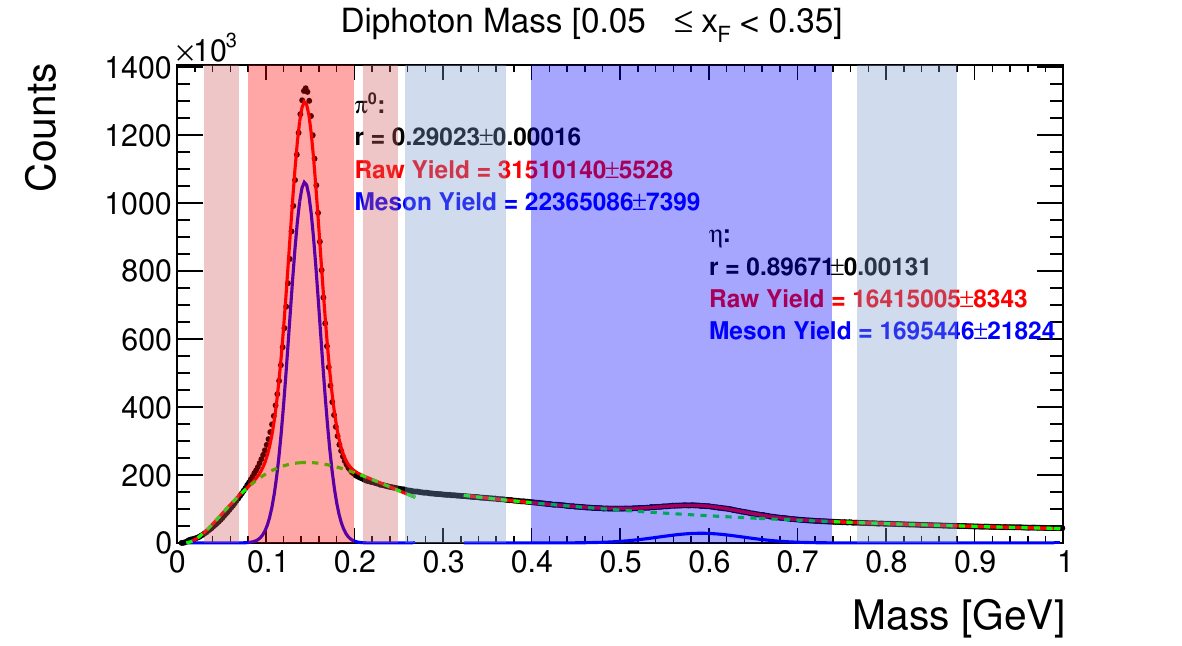}
    \caption{Di-photon mass distributions with main analysis (threshold function) fits, binned in $x_F$.}
    \label{fig:mass_thr_xF}
\end{figure}

\begin{figure}
    \centering
    \includegraphics[page=1,height=0.9\textheight]{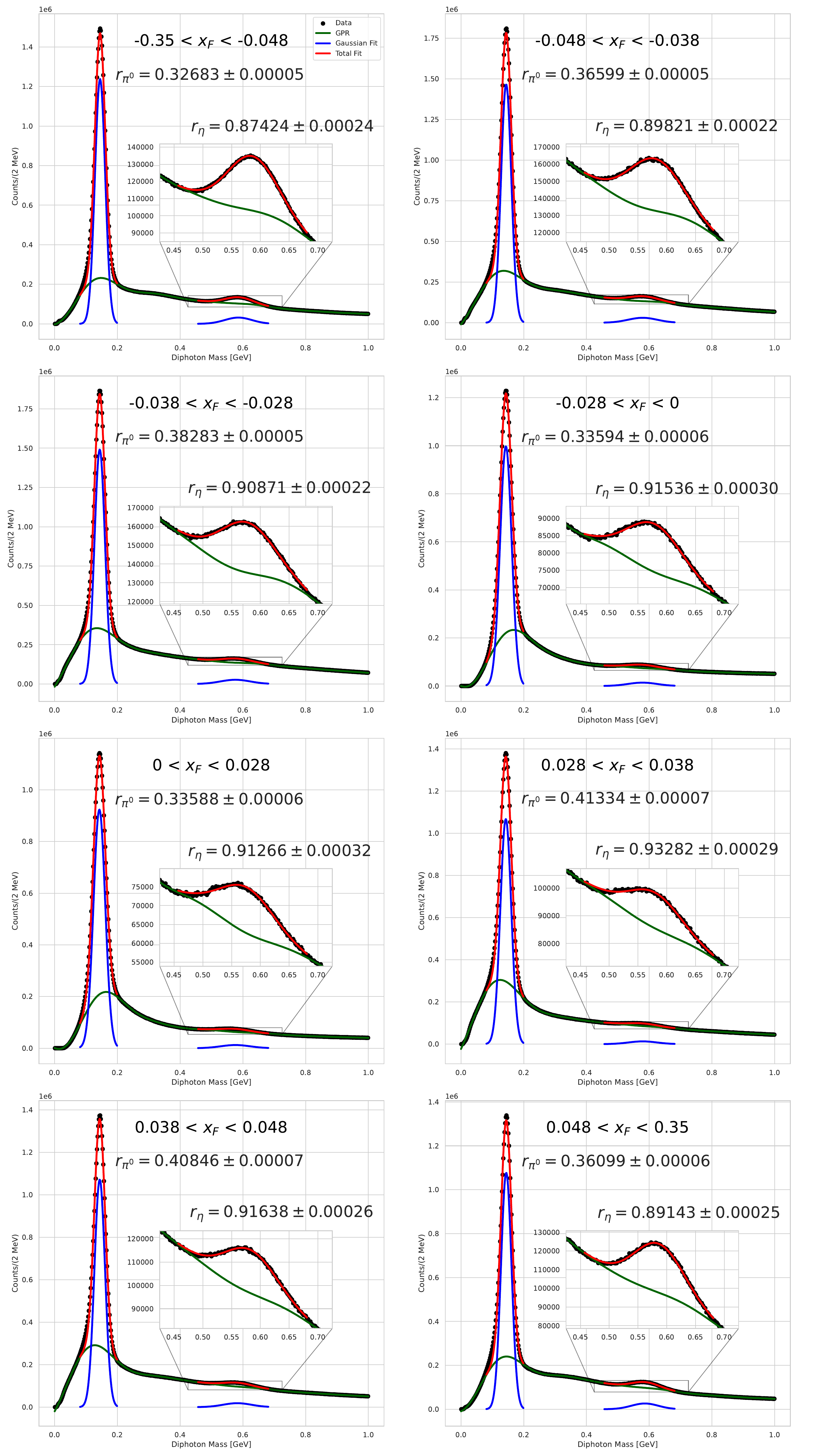}
    \caption{Di-photon mass distributions with alternative (GPR) fits, binned in $x_F$.}
    \label{fig:mass_gpr_xF}
\end{figure}

\begin{figure}
    \centering
    \includegraphics[page=9,height=0.9\textheight]{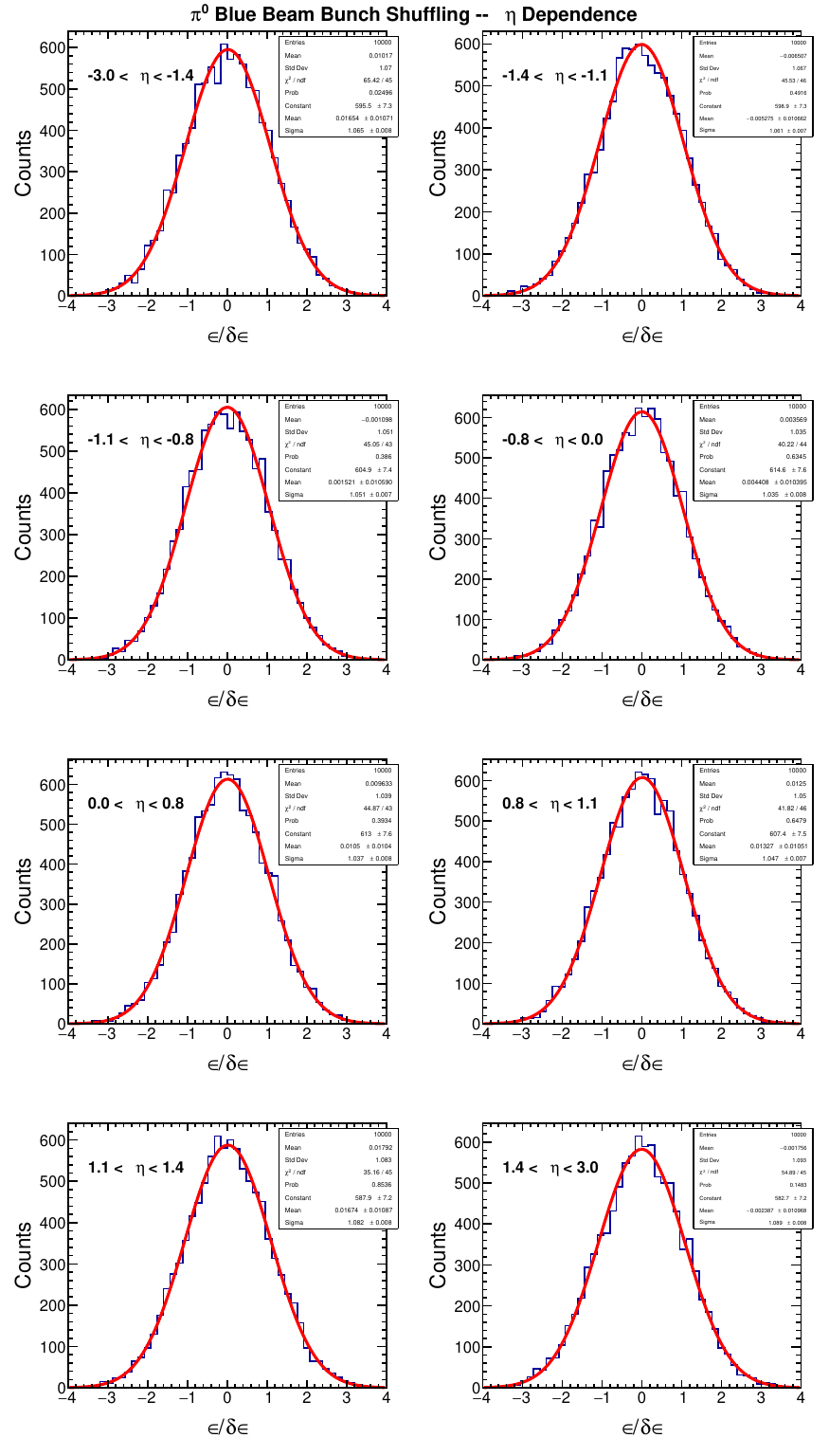}
    \caption{Bunch shuffling results for di-photons in the $\pi^0$ signal region, with the Blue beam taken as polarized. Each pad corresponds to one pseudorapidity bin.}
    \label{fig:bs1_eta}
\end{figure}

\begin{figure}
    \centering
    \includegraphics[page=10,height=0.9\textheight]{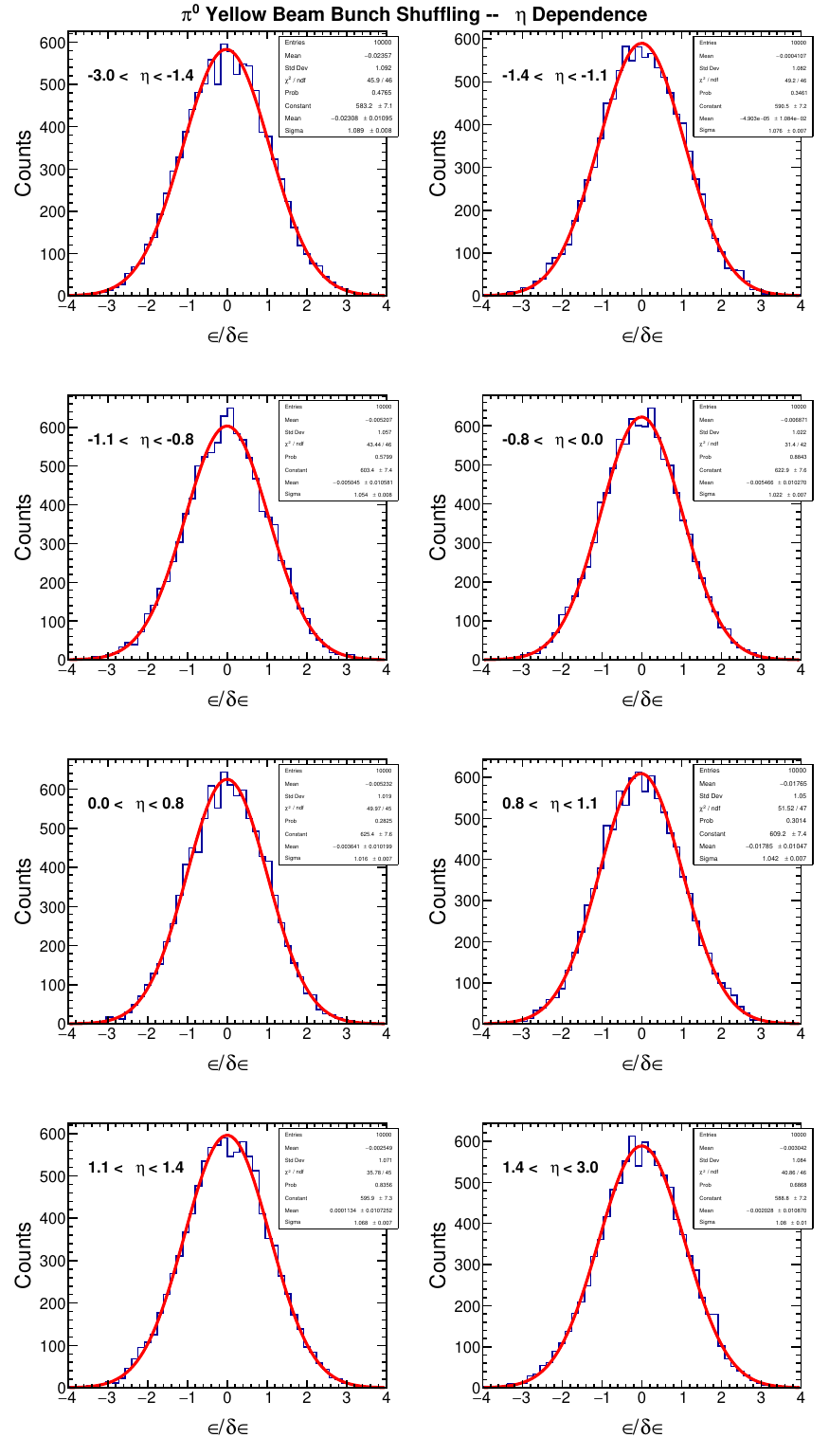}
    \caption{Bunch shuffling results for di-photons in the $\pi^0$ signal region, with the Yellow beam taken as polarized. Each pad corresponds to one pseudorapidity bin.}
    \label{fig:bs2_eta}
\end{figure}

\begin{figure}
    \centering
    \includegraphics[page=11,height=0.9\textheight]{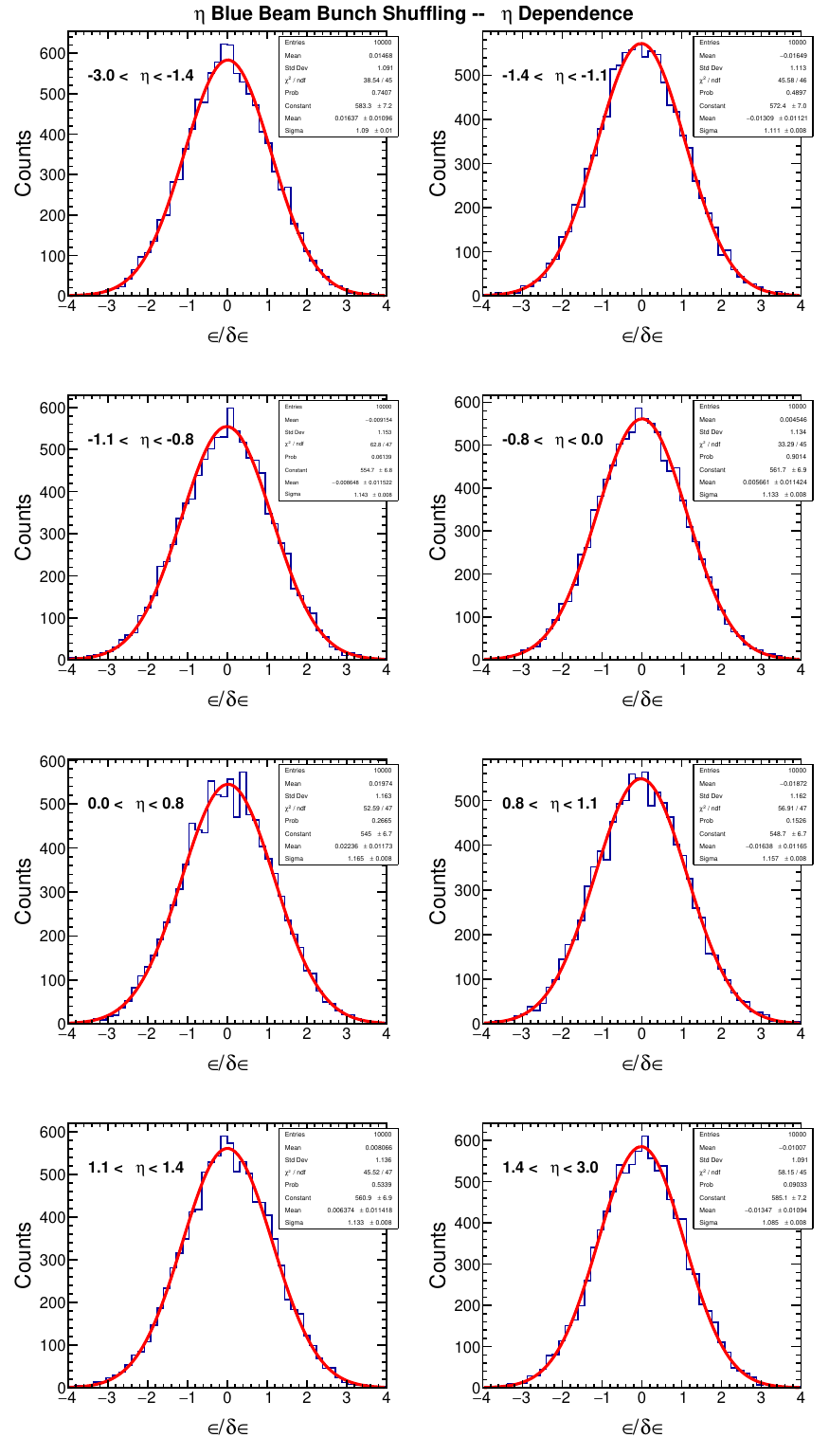}
    \caption{Bunch shuffling results for di-photons in the $\eta$-meson signal region, with the Blue beam taken as polarized. Each pad corresponds to one pseudorapidity bin.}
    \label{fig:bs3_eta}
\end{figure}

\begin{figure}
    \centering
    \includegraphics[page=12,height=0.9\textheight]{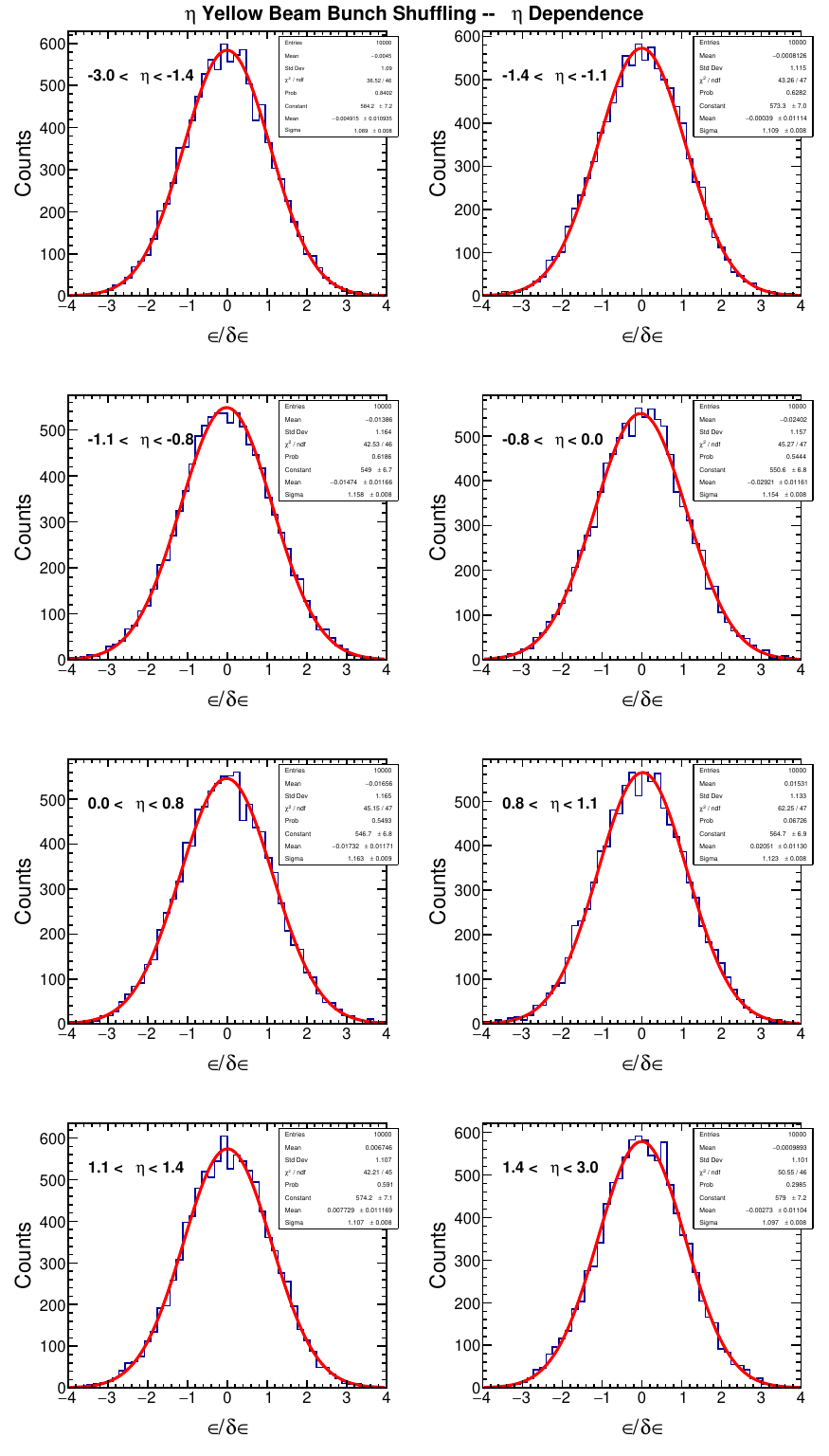}
    \caption{Bunch shuffling results for di-photons in the $\eta$-meson signal region, with the Yellow beam taken as polarized. Each pad corresponds to one pseudorapidity bin.}
    \label{fig:bs4_eta}
\end{figure}

\begin{figure}
    \centering
    \includegraphics[page=13,width=0.48\textwidth]{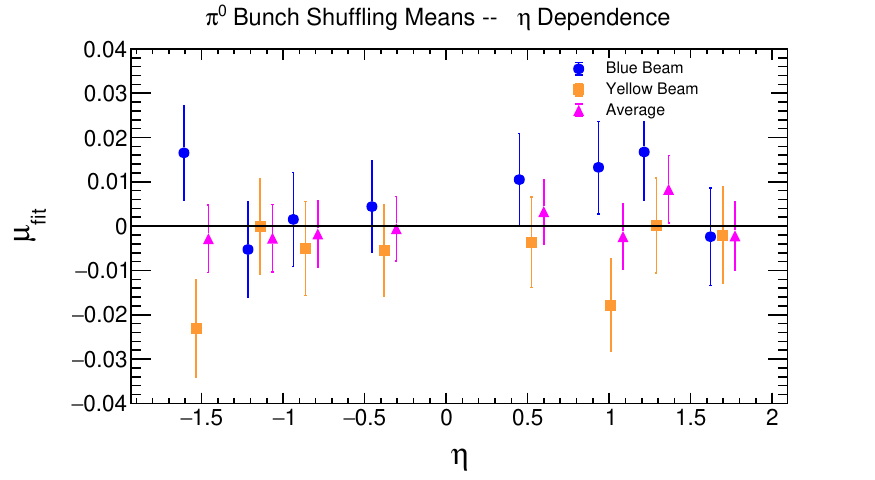}
    \includegraphics[page=14,width=0.48\textwidth]{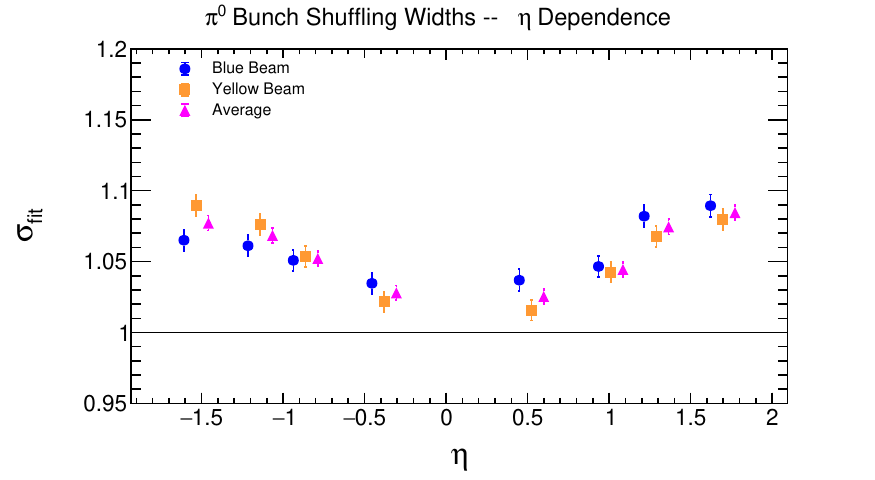}
    \caption{Bunch shuffling fitted means (left) and widths (right) for di-photons in the $\pi^0$ signal region, as a function of pseudorapidity. As the fitted widths are greater than unity, an additional uncertainty $\delta A_N^\mathrm{sys,shuffle}$ is assigned in all bins.}
    \label{fig:bs5_eta}
\end{figure}

\begin{figure}
    \centering
    \includegraphics[page=15,width=0.48\textwidth]{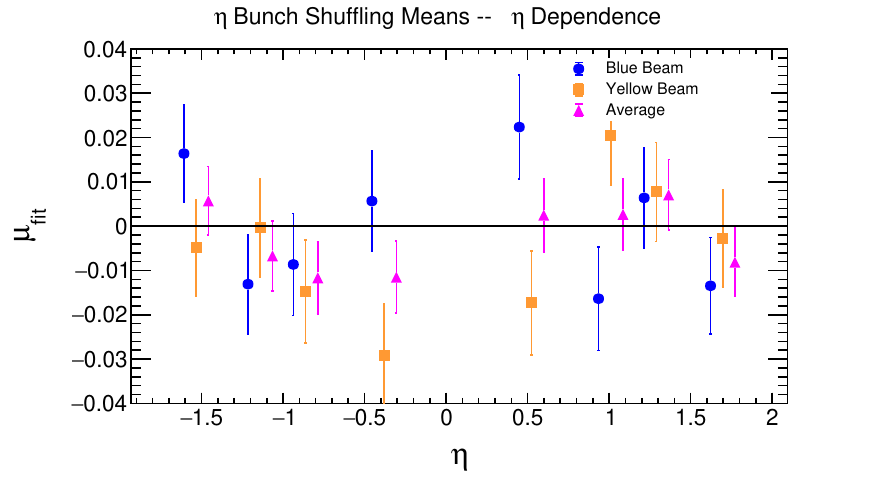}
    \includegraphics[page=16,width=0.48\textwidth]{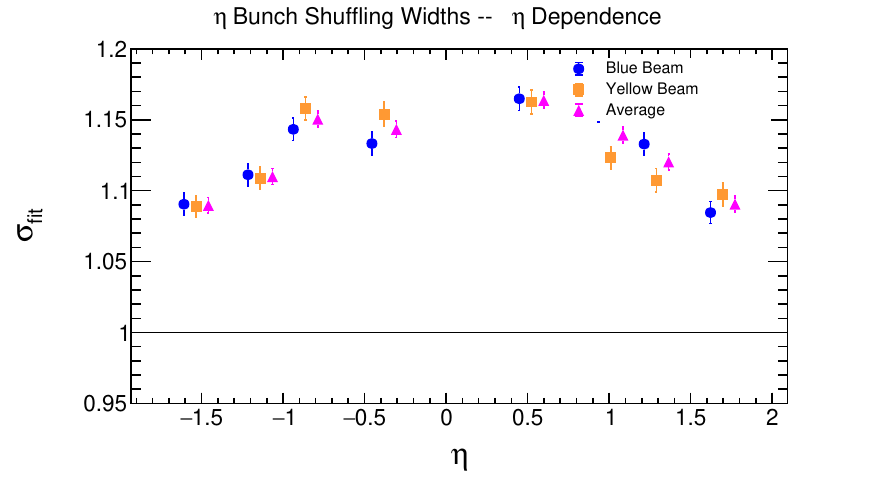}
    \caption{Bunch shuffling fitted means (left) and widths (right) for di-photons in the $\eta$-meson signal region, as a function of pseudorapidity. As the fitted widths are greater than unity, an additional uncertainty $\delta A_N^\mathrm{sys,shuffle}$ is assigned in all bins.}
    \label{fig:bs6_eta}
\end{figure}

\begin{figure}
    \centering
    \includegraphics[page=17,height=0.9\textheight]{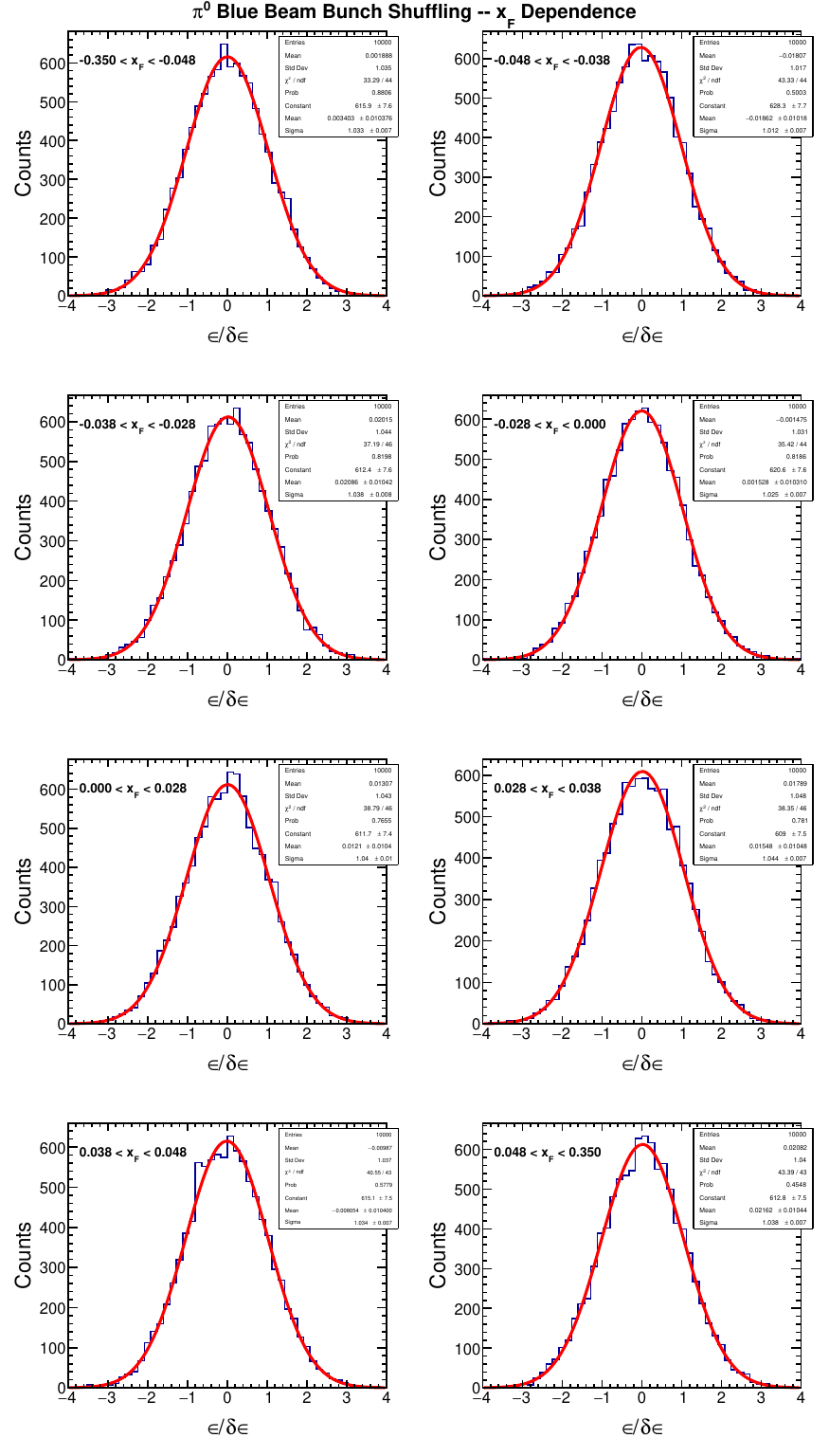}
    \caption{Bunch shuffling results for di-photons in the $\pi^0$ signal region, with the Blue beam taken as polarized. Each pad corresponds to one $x$-Feynman bin.}
    \label{fig:bs1_xF}
\end{figure}

\begin{figure}
    \centering
    \includegraphics[page=18,height=0.9\textheight]{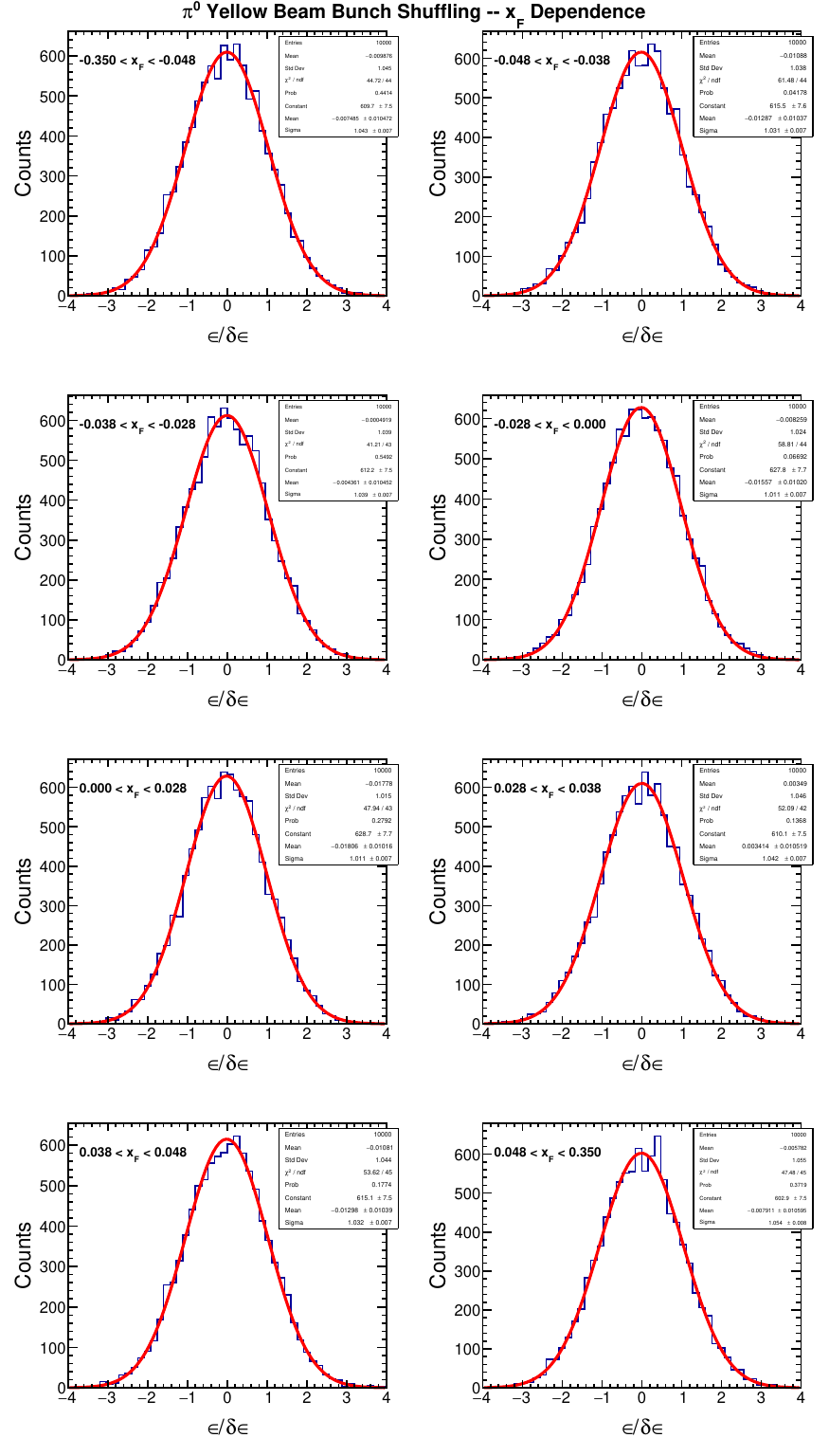}
    \caption{Bunch shuffling results for di-photons in the $\pi^0$ signal region, with the Yellow beam taken as polarized. Each pad corresponds to one $x$-Feynman bin.}
    \label{fig:bs2_xF}
\end{figure}

\begin{figure}
    \centering
    \includegraphics[page=19,height=0.9\textheight]{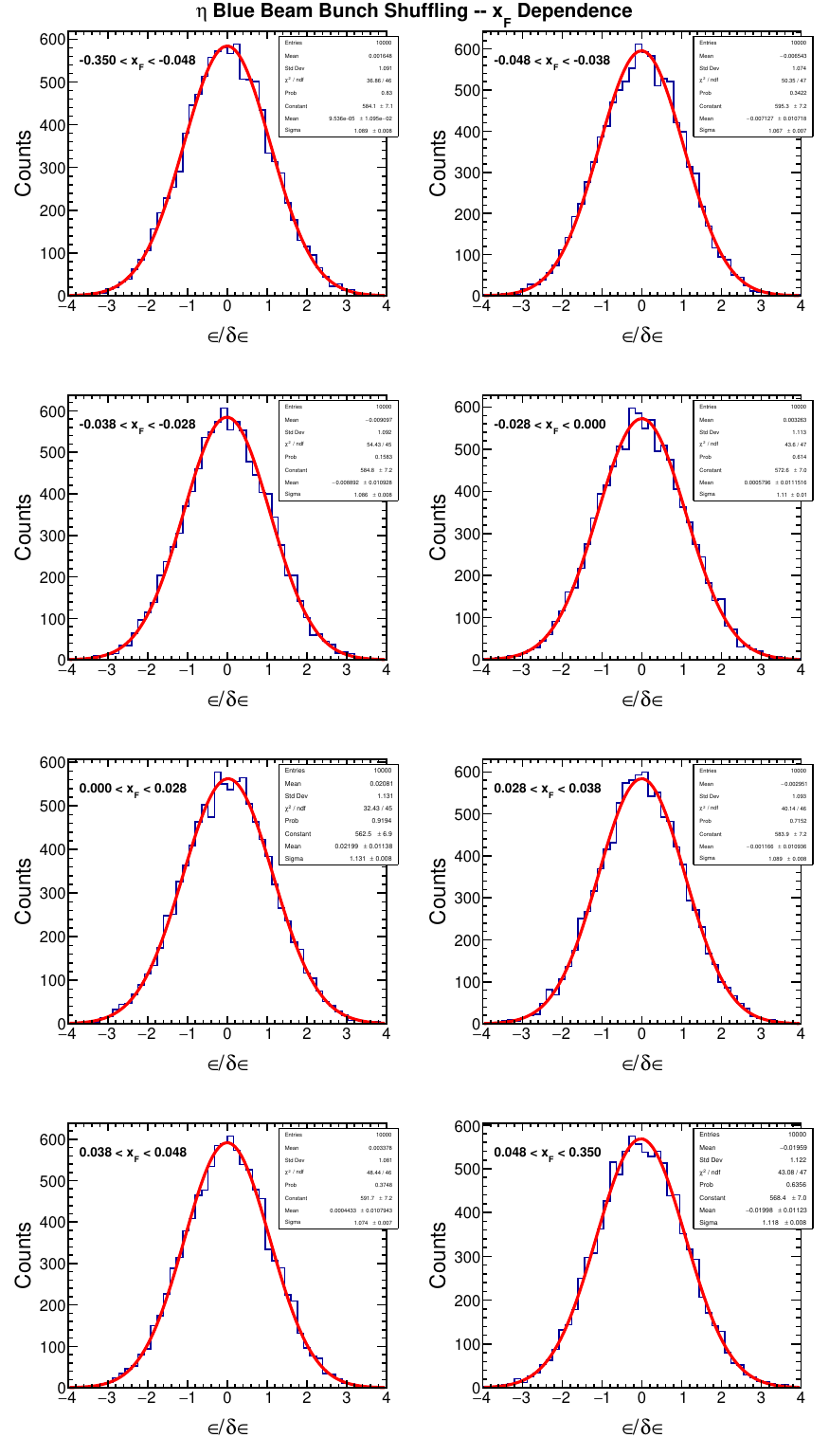}
    \caption{Bunch shuffling results for di-photons in the $\eta$-meson signal region, with the Blue beam taken as polarized. Each pad corresponds to one $x$-Feynman bin.}
    \label{fig:bs3_xF}
\end{figure}

\begin{figure}
    \centering
    \includegraphics[page=20,height=0.9\textheight]{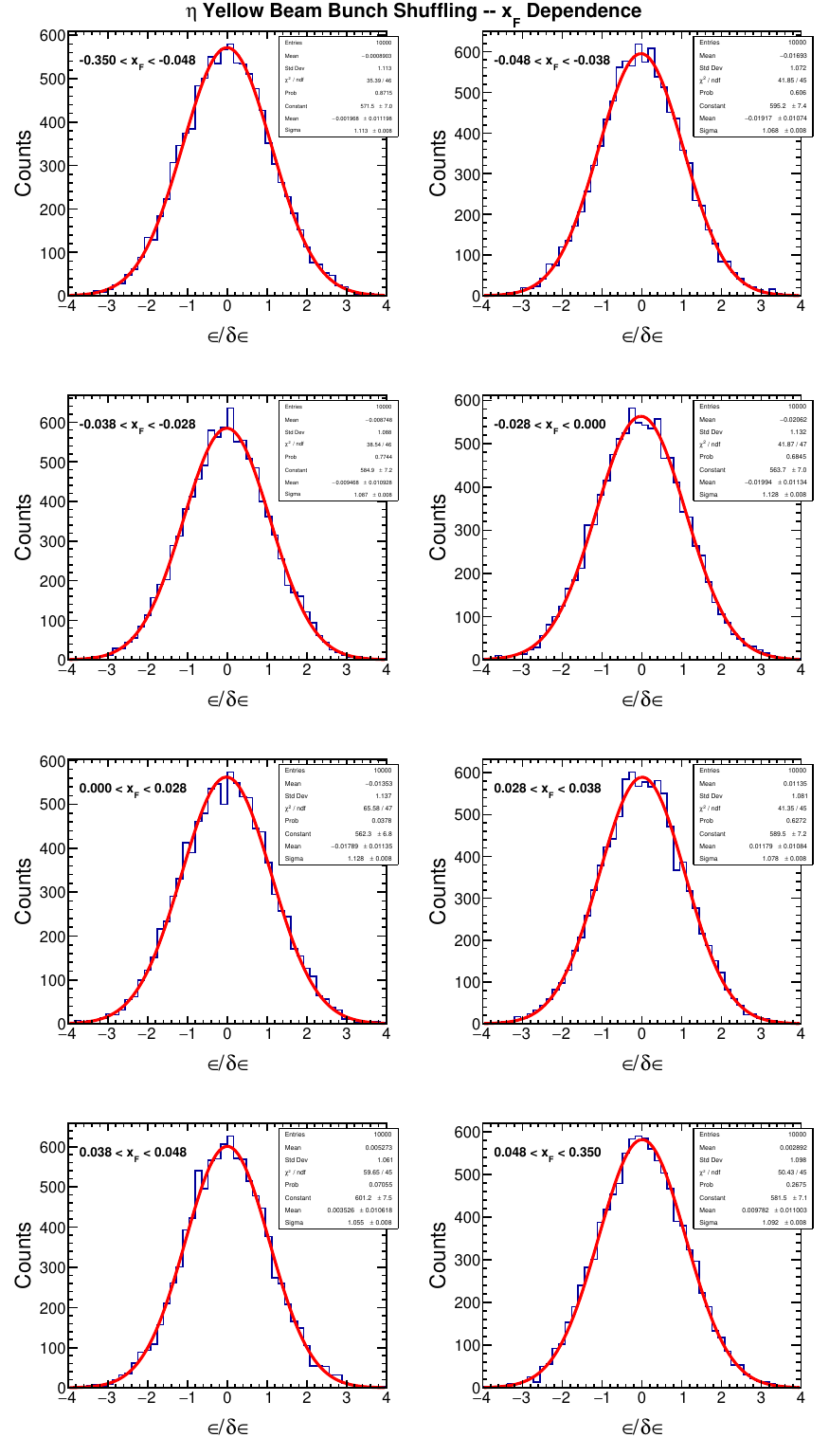}
    \caption{Bunch shuffling results for di-photons in the $\eta$-meson signal region, with the Yellow beam taken as polarized. Each pad corresponds to one $x$-Feynman bin.}
    \label{fig:bs4_xF}
\end{figure}

\begin{figure}
    \centering
    \includegraphics[page=21,width=0.48\textwidth]{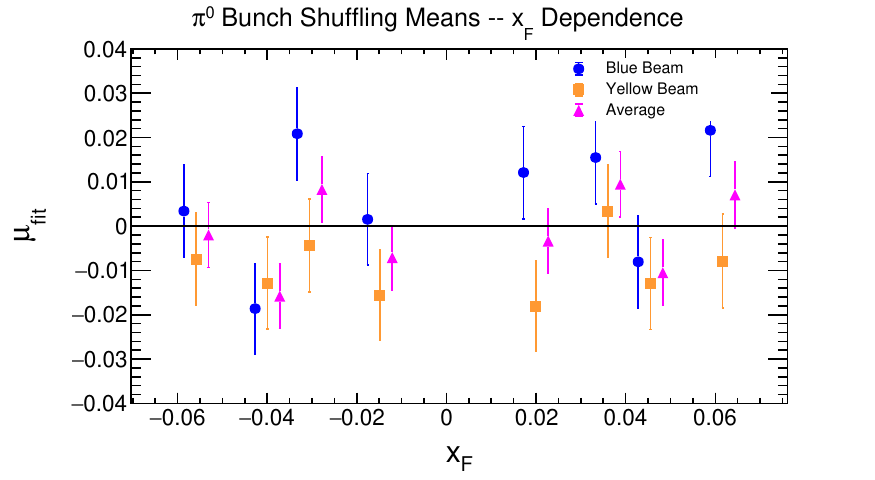}
    \includegraphics[page=22,width=0.48\textwidth]{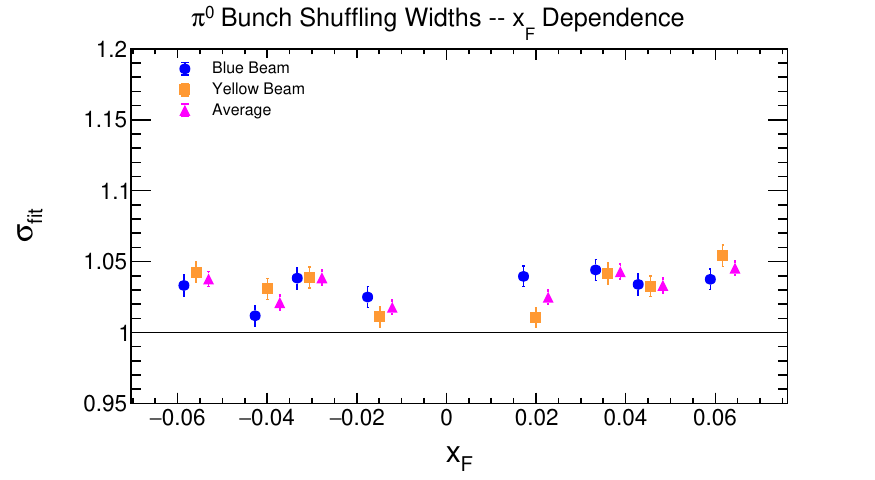}
    \caption{Bunch shuffling fitted means (left) and widths (right) for di-photons in the $\pi^0$ signal region, as a function of $x$-Feynman. As the fitted widths are greater than unity, an additional uncertainty $\delta A_N^\mathrm{sys,shuffle}$ is assigned in all bins.}
    \label{fig:bs5_xF}
\end{figure}

\begin{figure}
    \centering
    \includegraphics[page=23,width=0.48\textwidth]{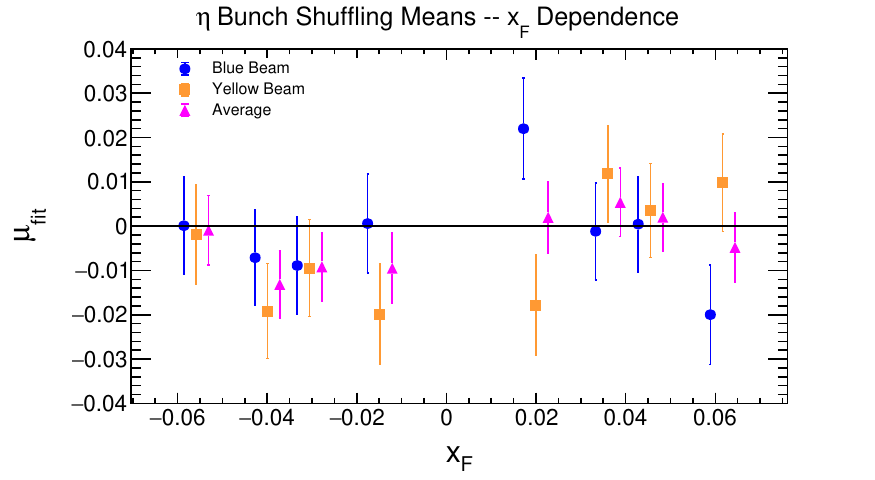}
    \includegraphics[page=24,width=0.48\textwidth]{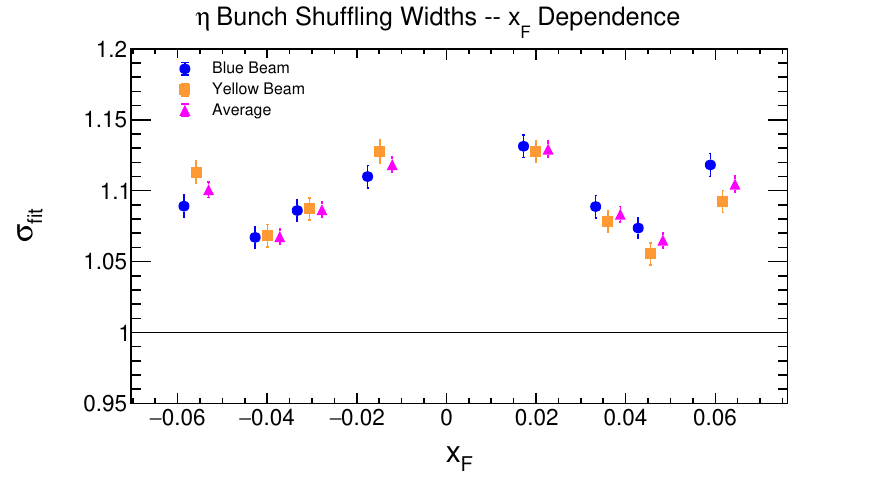}
    \caption{Bunch shuffling fitted means (left) and widths (right) for di-photons in the $\eta$-meson signal region, as a function of $x$-Feynman. As the fitted widths are greater than unity, an additional uncertainty $\delta A_N^\mathrm{sys,shuffle}$ is assigned in all bins.}
    \label{fig:bs6_xF}
\end{figure}

Due to its simplicity, additional details on $\delta A_N^\mathrm{sys,calc}$ in analogy with Table~\ref{tab:sys-calc} are omitted. However, $\delta A_N^\mathrm{sys,bg}$ and $\delta A_N^\mathrm{sys,shuffle}$ are given further context in Figures~\ref{fig:mass_thr_eta}-\ref{fig:bs6_xF}. Figures~\ref{fig:mass_thr_eta} and \ref{fig:mass_gpr_eta} show the di-photon mass distribution with main analysis and alternative fits, respectively, for the $\eta$ binning; Figures~\ref{fig:mass_thr_xF} and \ref{fig:mass_gpr_xF} show the same for the $x_F$ binning. As noted above, $\delta A_N^\mathrm{sys,bg}$ is the dominating uncertainty for the $\pi^0$. This is due in part to the exaggerated contribution from low-$p_T$ di-photons, where the discrepancy between main analysis and alternative fits is most pronounced. Another factor contributing to the discrepancy is that the alternative GPR fits were tuned primarily to produce smooth, reasonable background predictions in the $\eta$-meson signal region. Predictions for the $\pi^0$ region are therefore less robust. Further tuning of the GPR model may yield results more consistent with the main analysis fit, in turn reducing $\delta A_N^\mathrm{sys,bg}$.

Figures~\ref{fig:bs1_eta}-\ref{fig:bs6_eta} show the bunch shuffling results for the $\eta$ binning; Figures~\ref{fig:bs1_xF}-\ref{fig:bs6_xF}, for the $x_F$ binning. As with the $p_T$ binning, the bunch shuffling distributions $\epsilon/\delta \epsilon$ are Gaussians. The means $\mu_\mathrm{fit}$ are generally consistent with zero at the 1-$\sigma$ level. Means for the individual beams differ from zero by 2-$\sigma$ or more in 3 out of 64 data points, consistent with expected statistical fluctuations. We therefore conclude that no false asymmetry is observed in the $\eta$ and $x_F$ binnings. However as the fitted widths $\sigma_\mathrm{fit}$ are uniformly greater than one, a bunch shuffling systematic uncertainty is assigned in all bins.

\begin{table}[h]
\centering
\begin{tabular}{r|c|c|c|c||r|r|r||r|r|r}
\multicolumn{11}{c}{}\\
\multicolumn{11}{c}{\Large $p^\uparrow p\rightarrow\pi^0 X$}\\
\multicolumn{11}{c}{}\\\hline
Bin & $\eta$ & $\langle \eta \rangle$ & $\langle p_T \rangle$ & $\langle x_F \rangle$ & $A_N$ & $\delta A_N^\mathrm{stat}$ & $\delta A_N^\mathrm{sys,tot}$ & $\delta A_N^\mathrm{sys,calc}$ & $\delta A_N^\mathrm{sys,bg}$ & $\delta A_N^\mathrm{sys,shuffle}$ \\
\# & Range & & [GeV] & & & & & & & \\
\hline\hline
\multirow{2}*{1} & \multirow{2}{3em}{[-3.0, -1.35]} & \multirow{2}*{-1.492} & \multirow{2}*{2.416} & \multirow{2}*{-0.051} & \multirow{2}*{-0.00023} & \multirow{2}*{0.00066} & \multirow{2}*{0.00239} & \multirow{2}*{0.00006} & \multirow{2}*{0.00237} & \multirow{2}*{0.00027} \\
& & & & & & & & & \\
\multirow{2}*{2} & \multirow{2}{3em}{[-1.35, -1.08]} & \multirow{2}*{-1.214} & \multirow{2}*{2.961} & \multirow{2}*{-0.045} & \multirow{2}*{-0.00087} & \multirow{2}*{0.00046} & \multirow{2}*{0.00261} & \multirow{2}*{0.00002} & \multirow{2}*{0.00261} & \multirow{2}*{0.00017} \\ & & & & & & & & & \\
\multirow{2}*{3} & \multirow{2}{3em}{[-1.08, -0.76]} & \multirow{2}*{-0.941} & \multirow{2}*{3.495} & \multirow{2}*{-0.038} & \multirow{2}*{-0.00073} & \multirow{2}*{0.00050} & \multirow{2}*{0.00019} & \multirow{2}*{0.00003} & \multirow{2}*{0.00010} & \multirow{2}*{0.00016} \\ & & & & & & & & & \\
\multirow{2}*{4} & \multirow{2}{3em}{[-0.76, 0.0]} & \multirow{2}*{-0.470} & \multirow{2}*{4.290} & \multirow{2}*{-0.021} & \multirow{2}*{-0.00149} & \multirow{2}*{0.00056} & \multirow{2}*{0.01360} & \multirow{2}*{0.00000} & \multirow{2}*{0.01360} & \multirow{2}*{0.00013} \\ & & & & & & & & & \\
\multirow{2}*{5} & \multirow{2}{3em}{[0.0, 0.76]} & \multirow{2}*{0.466} & \multirow{2}*{4.306} & \multirow{2}*{0.021} & \multirow{2}*{-0.00031} & \multirow{2}*{0.00056} & \multirow{2}*{0.01373} & \multirow{2}*{0.00001} & \multirow{2}*{0.01373} & \multirow{2}*{0.00013} \\ & & & & & & & & & \\
\multirow{2}*{6} & \multirow{2}{3em}{[0.76, 1.08]} & \multirow{2}*{0.939} & \multirow{2}*{3.554} & \multirow{2}*{0.038} & \multirow{2}*{0.00070} & \multirow{2}*{0.00052} & \multirow{2}*{0.00622} & \multirow{2}*{0.00000} & \multirow{2}*{0.00621} & \multirow{2}*{0.00016} \\ & & & & & & & & & \\
\multirow{2}*{7} & \multirow{2}{3em}{[1.08, 1.35]} & \multirow{2}*{1.210} & \multirow{2}*{3.076} & \multirow{2}*{0.047} & \multirow{2}*{0.00108} & \multirow{2}*{0.00052} & \multirow{2}*{0.01213} & \multirow{2}*{0.00000} & \multirow{2}*{0.01212} & \multirow{2}*{0.00020} \\ & & & & & & & & & \\
\multirow{2}*{8} & \multirow{2}{3em}{[1.35, 3.0]} & \multirow{2}*{1.483} & \multirow{2}*{2.581} & \multirow{2}*{0.054} & \multirow{2}*{0.00005} & \multirow{2}*{0.00087} & \multirow{2}*{0.00250} & \multirow{2}*{0.00006} & \multirow{2}*{0.00247} & \multirow{2}*{0.00036} \\ & & & & & & & & & \\
\multicolumn{11}{c}{}\\
\multicolumn{11}{c}{\Large $p^\uparrow p\rightarrow\eta X$}\\
\multicolumn{11}{c}{}\\\hline
Bin & $\eta$ & $\langle \eta \rangle$ & $\langle p_T \rangle$ & $\langle x_F \rangle$ & $A_N$ & $\delta A_N^\mathrm{stat}$ & $\delta A_N^\mathrm{sys,tot}$ & $\delta A_N^\mathrm{sys,calc}$ & $\delta A_N^\mathrm{sys,bg}$ & $\delta A_N^\mathrm{sys,shuffle}$ \\
\# & Range & & [GeV] & & & & & & & \\
\hline\hline
\multirow{2}*{1} & \multirow{2}{3em}{[-3.0, -1.35]} & \multirow{2}*{-1.467} & \multirow{2}*{2.634} & \multirow{2}*{-0.054} & \multirow{2}*{0.02119} & \multirow{2}*{0.01549} & \multirow{2}*{0.00732} & \multirow{2}*{0.00289} & \multirow{2}*{0.00049} & \multirow{2}*{0.00670} \\ & & & & & & & & & \\
\multirow{2}*{2} & \multirow{2}{3em}{[-1.35, -1.08]} & \multirow{2}*{-1.211} & \multirow{2}*{3.005} & \multirow{2}*{-0.046} & \multirow{2}*{0.01334} & \multirow{2}*{0.00738} & \multirow{2}*{0.00364} & \multirow{2}*{0.00049} & \multirow{2}*{0.00056} & \multirow{2}*{0.00356} \\ & & & & & & & & & \\
\multirow{2}*{3} & \multirow{2}{3em}{[-1.08, -0.76]} & \multirow{2}*{-0.942} & \multirow{2}*{3.620} & \multirow{2}*{-0.039} & \multirow{2}*{-0.00099} & \multirow{2}*{0.00582} & \multirow{2}*{0.00330} & \multirow{2}*{0.00009} & \multirow{2}*{0.00015} & \multirow{2}*{0.00329} \\ & & & & & & & & & \\
\multirow{2}*{4} & \multirow{2}{3em}{[-0.76, 0.0]} & \multirow{2}*{-0.472} & \multirow{2}*{4.727} & \multirow{2}*{-0.023} & \multirow{2}*{0.00177} & \multirow{2}*{0.00852} & \multirow{2}*{0.00504} & \multirow{2}*{0.00140} & \multirow{2}*{0.00109} & \multirow{2}*{0.00472} \\ & & & & & & & & & \\
\multirow{2}*{5} & \multirow{2}{3em}{[0.0, 0.76]} & \multirow{2}*{0.462} & \multirow{2}*{4.810} & \multirow{2}*{0.023} & \multirow{2}*{-0.00127} & \multirow{2}*{0.00810} & \multirow{2}*{0.00502} & \multirow{2}*{0.00138} & \multirow{2}*{0.00002} & \multirow{2}*{0.00482} \\ & & & & & & & & & \\
\multirow{2}*{6} & \multirow{2}{3em}{[0.76, 1.08]} & \multirow{2}*{0.939} & \multirow{2}*{3.787} & \multirow{2}*{0.041} & \multirow{2}*{-0.00851} & \multirow{2}*{0.00644} & \multirow{2}*{0.00365} & \multirow{2}*{0.00066} & \multirow{2}*{0.00070} & \multirow{2}*{0.00353} \\ & & & & & & & & & \\
\multirow{2}*{7} & \multirow{2}{3em}{[1.08, 1.35]} & \multirow{2}*{1.212} & \multirow{2}*{3.188} & \multirow{2}*{0.049} & \multirow{2}*{-0.00231} & \multirow{2}*{0.01045} & \multirow{2}*{0.00530} & \multirow{2}*{0.00033} & \multirow{2}*{0.00033} & \multirow{2}*{0.00528} \\ & & & & & & & & & \\
\multirow{2}*{8} & \multirow{2}{3em}{[1.35, 3.0]} & \multirow{2}*{1.470} & \multirow{2}*{2.740} & \multirow{2}*{0.056} & \multirow{2}*{-0.05733} & \multirow{2}*{0.02942} & \multirow{2}*{0.02875} & \multirow{2}*{0.00339} & \multirow{2}*{0.02551} & \multirow{2}*{0.01280} \\ & & & & & & & & & \\
\hline
\end{tabular}
\caption{Statistical and systematic uncertainties on $A_N$ for both $\pi^0$ and $\eta$-mesons, extracted in different bins in pseudorapidity $\eta$. Uncertainty values shown as 0.00000 are less than 5x10${}^{-6}$.} 
\label{tab:sys_eta}
\end{table}

\begin{table}[h]
\centering
\begin{tabular}{r|c|r|r|r||r|r|r||r|r|r}
\multicolumn{11}{c}{}\\
\multicolumn{11}{c}{\Large $p^\uparrow p\rightarrow\pi^0 X$}\\
\multicolumn{11}{c}{}\\\hline
Bin & $x_F$ & $\langle x_F \rangle$ & $\langle p_T \rangle$ & $\langle \eta \rangle$ & $A_N$ & $\delta A_N^\mathrm{stat}$ & $\delta A_N^\mathrm{sys,tot}$ & $\delta A_N^\mathrm{sys,calc}$ & $\delta A_N^\mathrm{sys,bg}$ & $\delta A_N^\mathrm{sys,shuffle}$ \\
\# & Range & & [GeV] & & & & & & & \\
\hline\hline
\multirow{2}*{1} & \multirow{2}{3.25em}{[-0.350, -0.048]} & \multirow{2}*{-0.059} & \multirow{2}*{3.664} & \multirow{2}*{-1.287} & \multirow{2}*{0.00021} & \multirow{2}*{0.00052} & \multirow{2}*{0.00992} & \multirow{2}*{0.00004} & \multirow{2}*{0.00992} & \multirow{2}*{0.00014} \\ & & & & & & & & & \\
\multirow{2}*{2} & \multirow{2}{3.25em}{[-0.048, -0.038]} & \multirow{2}*{-0.043} & \multirow{2}*{2.989} & \multirow{2}*{-1.191} & \multirow{2}*{-0.00101} & \multirow{2}*{0.00050} & \multirow{2}*{0.00236} & \multirow{2}*{0.00002} & \multirow{2}*{0.00236} & \multirow{2}*{0.00010} \\ & & & & & & & & & \\
\multirow{2}*{3} & \multirow{2}{3.25em}{[-0.038, -0.028]} & \multirow{2}*{-0.033} & \multirow{2}*{2.958} & \multirow{2}*{-1.015} & \multirow{2}*{-0.00106} & \multirow{2}*{0.00051} & \multirow{2}*{0.00335} & \multirow{2}*{0.00002} & \multirow{2}*{0.00335} & \multirow{2}*{0.00014} \\ & & & & & & & & & \\
\multirow{2}*{4} & \multirow{2}{3.25em}{[-0.028, 0.000]} & \multirow{2}*{-0.018} & \multirow{2}*{3.928} & \multirow{2}*{-0.480} & \multirow{2}*{-0.00177} & \multirow{2}*{0.00058} & \multirow{2}*{0.02572} & \multirow{2}*{0.00003} & \multirow{2}*{0.02572} & \multirow{2}*{0.00011} \\ & & & & & & & & & \\
\multirow{2}*{5} & \multirow{2}{3.25em}{[0.000, 0.028]} & \multirow{2}*{0.018} & \multirow{2}*{3.973} & \multirow{2}*{0.465} & \multirow{2}*{-0.00023} & \multirow{2}*{0.00058} & \multirow{2}*{0.00955} & \multirow{2}*{0.00000} & \multirow{2}*{0.00955} & \multirow{2}*{0.00013} \\ & & & & & & & & & \\
\multirow{2}*{6} & \multirow{2}{3.25em}{[0.028, 0.038]} & \multirow{2}*{0.033} & \multirow{2}*{3.119} & \multirow{2}*{0.969} & \multirow{2}*{0.00031} & \multirow{2}*{0.00055} & \multirow{2}*{0.00835} & \multirow{2}*{0.00006} & \multirow{2}*{0.00835} & \multirow{2}*{0.00016} \\ & & & & & & & & & \\
\multirow{2}*{7} & \multirow{2}{3.25em}{[0.038, 0.048]} & \multirow{2}*{0.043} & \multirow{2}*{3.125} & \multirow{2}*{1.155} & \multirow{2}*{-0.00001} & \multirow{2}*{0.00057} & \multirow{2}*{0.00040} & \multirow{2}*{0.00005} & \multirow{2}*{0.00037} & \multirow{2}*{0.00015} \\ & & & & & & & & & \\
\multirow{2}*{8} & \multirow{2}{3.25em}{[0.048, 0.350]} & \multirow{2}*{0.060} & \multirow{2}*{3.721} & \multirow{2}*{1.276} & \multirow{2}*{0.00182} & \multirow{2}*{0.00058} & \multirow{2}*{0.01099} & \multirow{2}*{0.00001} & \multirow{2}*{0.01099} & \multirow{2}*{0.00018} \\ & & & & & & & & & \\
\multicolumn{11}{c}{}\\
\multicolumn{11}{c}{\Large $p^\uparrow p\rightarrow\eta X$}\\
\multicolumn{11}{c}{}\\\hline
Bin & $x_F$ & $\langle x_F \rangle$ & $\langle p_T \rangle$ & $\langle \eta \rangle$ & $A_N$ & $\delta A_N^\mathrm{stat}$ & $\delta A_N^\mathrm{sys,tot}$ & $\delta A_N^\mathrm{sys,calc}$ & $\delta A_N^\mathrm{sys,bg}$ & $\delta A_N^\mathrm{sys,shuffle}$ \\
\# & Range & & [GeV] & & & & & & & \\
\hline\hline
\multirow{2}*{1} & \multirow{2}{3.25em}{[-0.350, -0.048]} & \multirow{2}*{-0.059} & \multirow{2}*{3.664} & \multirow{2}*{-1.287} & \multirow{2}*{0.00021} & \multirow{2}*{0.00052} & \multirow{2}*{0.00992} & \multirow{2}*{0.00004} & \multirow{2}*{0.00992} & \multirow{2}*{0.00014} \\ & & & & & & & & & \\
\multirow{2}*{2} & \multirow{2}{3.25em}{[-0.048, -0.038]} & \multirow{2}*{-0.043} & \multirow{2}*{2.989} & \multirow{2}*{-1.191} & \multirow{2}*{-0.00101} & \multirow{2}*{0.00050} & \multirow{2}*{0.00236} & \multirow{2}*{0.00002} & \multirow{2}*{0.00236} & \multirow{2}*{0.00010} \\ & & & & & & & & & \\
\multirow{2}*{3} & \multirow{2}{3.25em}{[-0.038, -0.028]} & \multirow{2}*{-0.033} & \multirow{2}*{2.958} & \multirow{2}*{-1.015} & \multirow{2}*{-0.00106} & \multirow{2}*{0.00051} & \multirow{2}*{0.00335} & \multirow{2}*{0.00002} & \multirow{2}*{0.00335} & \multirow{2}*{0.00014} \\ & & & & & & & & & \\
\multirow{2}*{4} & \multirow{2}{3.25em}{[-0.028, 0.000]} & \multirow{2}*{-0.018} & \multirow{2}*{3.928} & \multirow{2}*{-0.480} & \multirow{2}*{-0.00177} & \multirow{2}*{0.00058} & \multirow{2}*{0.02572} & \multirow{2}*{0.00003} & \multirow{2}*{0.02572} & \multirow{2}*{0.00011} \\ & & & & & & & & & \\
\multirow{2}*{5} & \multirow{2}{3.25em}{[0.000, 0.028]} & \multirow{2}*{0.018} & \multirow{2}*{3.973} & \multirow{2}*{0.465} & \multirow{2}*{-0.00023} & \multirow{2}*{0.00058} & \multirow{2}*{0.00955} & \multirow{2}*{0.00000} & \multirow{2}*{0.00955} & \multirow{2}*{0.00013} \\ & & & & & & & & & \\
\multirow{2}*{6} & \multirow{2}{3.25em}{[0.028, 0.038]} & \multirow{2}*{0.033} & \multirow{2}*{3.119} & \multirow{2}*{0.969} & \multirow{2}*{0.00031} & \multirow{2}*{0.00055} & \multirow{2}*{0.00835} & \multirow{2}*{0.00006} & \multirow{2}*{0.00835} & \multirow{2}*{0.00016} \\ & & & & & & & & & \\
\multirow{2}*{7} & \multirow{2}{3.25em}{[0.038, 0.048]} & \multirow{2}*{0.043} & \multirow{2}*{3.125} & \multirow{2}*{1.155} & \multirow{2}*{-0.00001} & \multirow{2}*{0.00057} & \multirow{2}*{0.00040} & \multirow{2}*{0.00005} & \multirow{2}*{0.00037} & \multirow{2}*{0.00015} \\ & & & & & & & & & \\
\multirow{2}*{8} & \multirow{2}{3.25em}{[0.048, 0.350]} & \multirow{2}*{0.060} & \multirow{2}*{3.721} & \multirow{2}*{1.276} & \multirow{2}*{0.00182} & \multirow{2}*{0.00058} & \multirow{2}*{0.01099} & \multirow{2}*{0.00001} & \multirow{2}*{0.01099} & \multirow{2}*{0.00018} \\ & & & & & & & & & \\
\hline
\end{tabular}
\caption{Statistical and systematic uncertainties on $A_N$ for both $\pi^0$ and $\eta$-mesons, extracted in different bins in $x_F$. Uncertainty values shown as 0.00000 are less than 5x10${}^{-6}$.}
\label{tab:sys_xF}
\end{table}


\clearpage





\chapter{Results}\label{ch:results}
\section{Main Results}
\begin{figure}[h]
    \centering
    \includegraphics[page=2, width=0.8\textwidth]{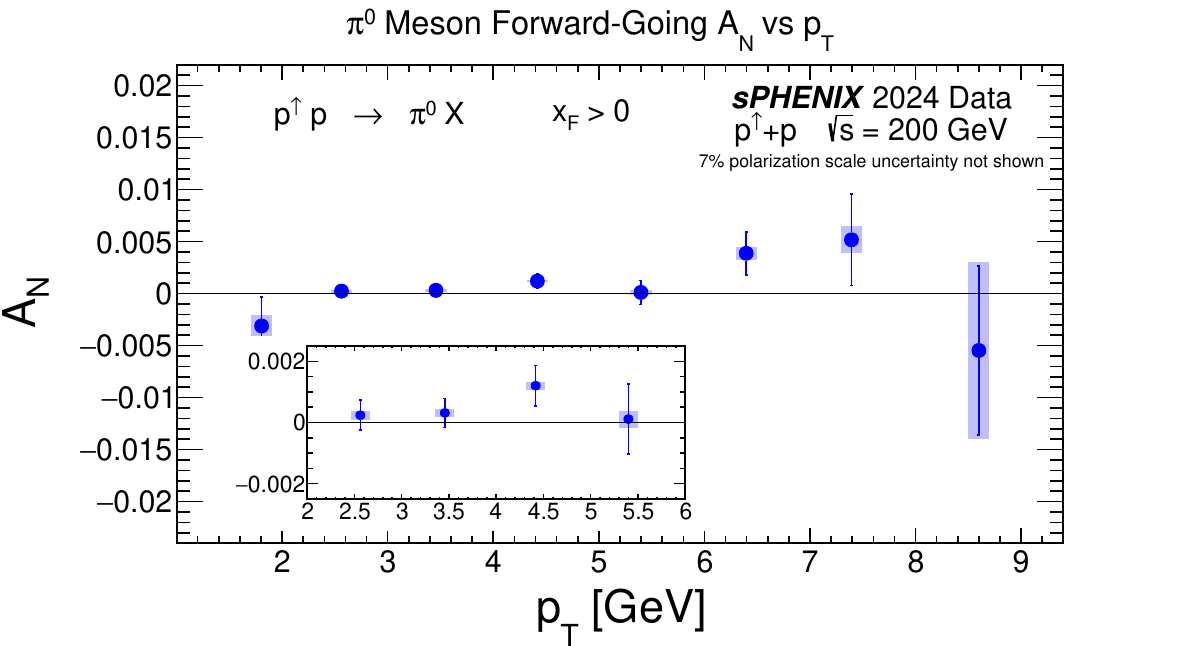}
    \caption{Forward-going $A_N$ for the $\pi^0$, plotted as a function of $p_T$. The error bars correspond to statistical uncertainties, while the light shaded bands show systematic uncertainties.}
    \label{fig:results_pi0_vs_pT}
\end{figure}

\begin{figure}
    \centering
    \includegraphics[page=3, width=0.8\textwidth]{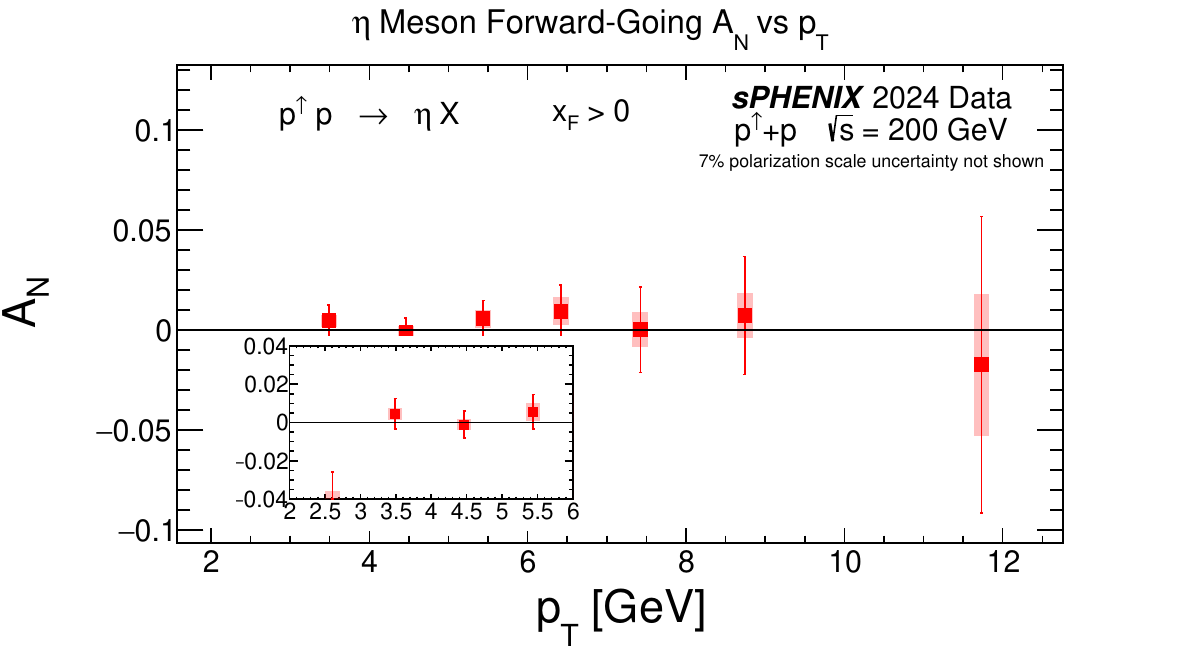}
    \caption{Forward-going $A_N$ for the $\eta$-meson, plotted as a function of $p_T$. The error bars correspond to statistical uncertainties, while the light shaded bands show systematic uncertainties.}
    \label{fig:results_eta_vs_pT}
\end{figure}

The main results of this analysis are the $p_T$-dependent corrected and beam-averaged TSSAs for $\pi^0$- and $\eta$-mesons produced in the forward direction relative to the polarized beam. These asymmetries, with both statistical (see Equation~\ref{eq:deltaAbg}) and systematic (see Chapter~\ref{ch:syst}) uncertainties, are shown in Figures~\ref{fig:results_pi0_vs_pT} and \ref{fig:results_eta_vs_pT} for the $\pi^0$- and $\eta$-mesons, respectively. The measured asymmetries are generally consistent with zero at or near the 1-$\sigma$ level. While some points differ from zero by more than 1-$\sigma$, this can be attributed to expected statistical fluctuation, rather than a non-zero asymmetry measurement.

\begin{figure}
    \centering
    \includegraphics[page=4, width=0.8\textwidth]{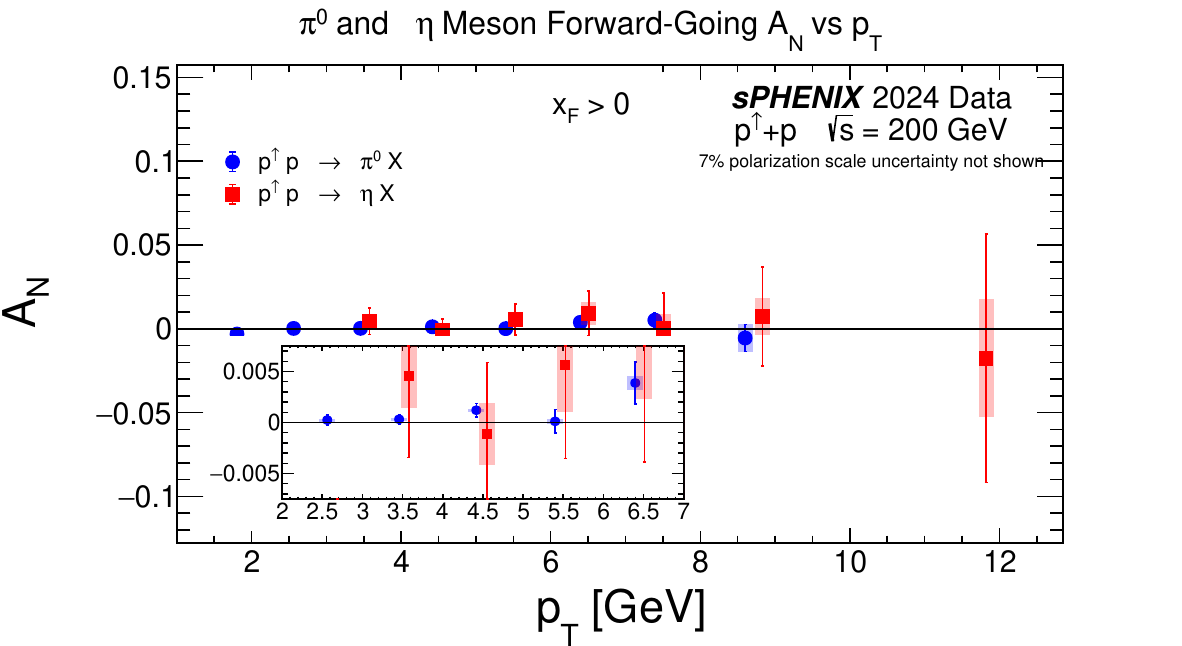}
    \caption{Forward-going $A_N$ for both the $\pi^0$- and $\eta$-mesons, plotted as a function of $p_T$.}
    \label{fig:results_pi0etacomparison_vs_pT}
\end{figure}

Figure~\ref{fig:results_pi0etacomparison_vs_pT} shows the $\pi^0$- and $\eta$-meson asymmetries together for direct comparison. As discussed in Chapter~\ref{ch:phys}, differences in TSSAs between the two mesons can shed light on the effects of meson mass and flavor content. However, as both sets of asymmetries are consistent with zero, this suggests that no such effects can be resolved in the kinematic region accessed by sPHENIX.


\begin{figure}[h]
    \centering
    \includegraphics[page=5, width=0.8\textwidth]{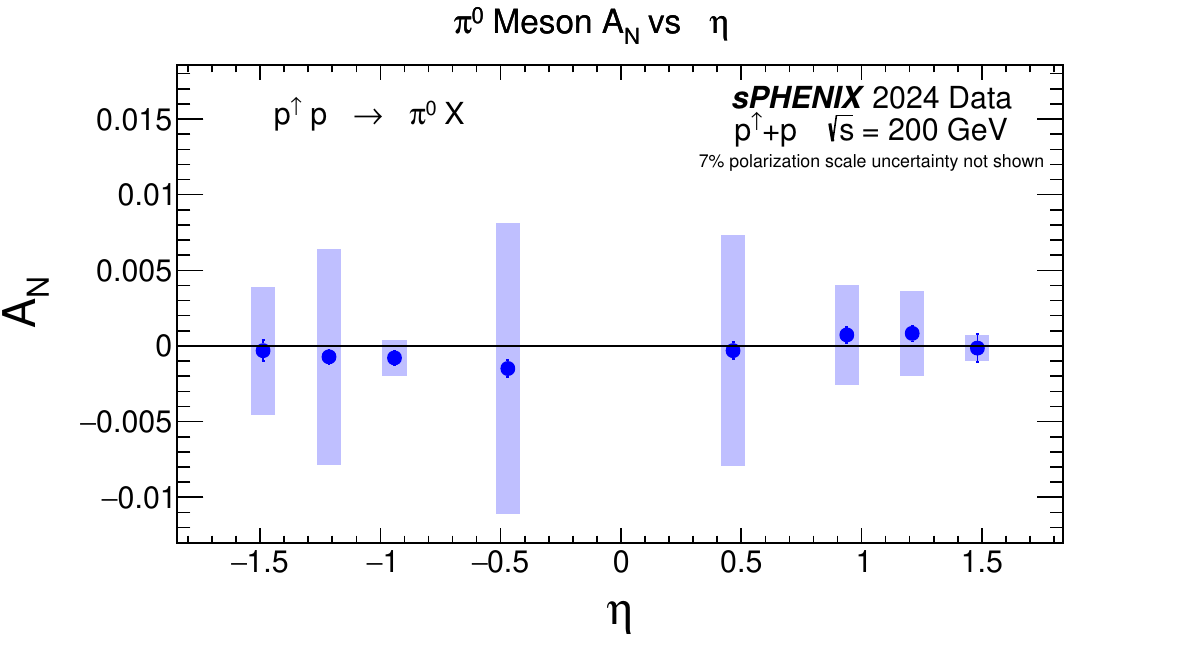}
    \caption{$A_N$ for the $\pi^0$, plotted as a function of pseudorapidity $\eta$. The error bars correspond to statistical uncertainties, while the light shaded bands show systematic uncertainties. Here $\eta$ is defined with respect to the polarized beam.}
    \label{fig:results_pi0_vs_eta}
\end{figure}

\begin{figure}
    \centering
    \includegraphics[page=6, width=0.8\textwidth]{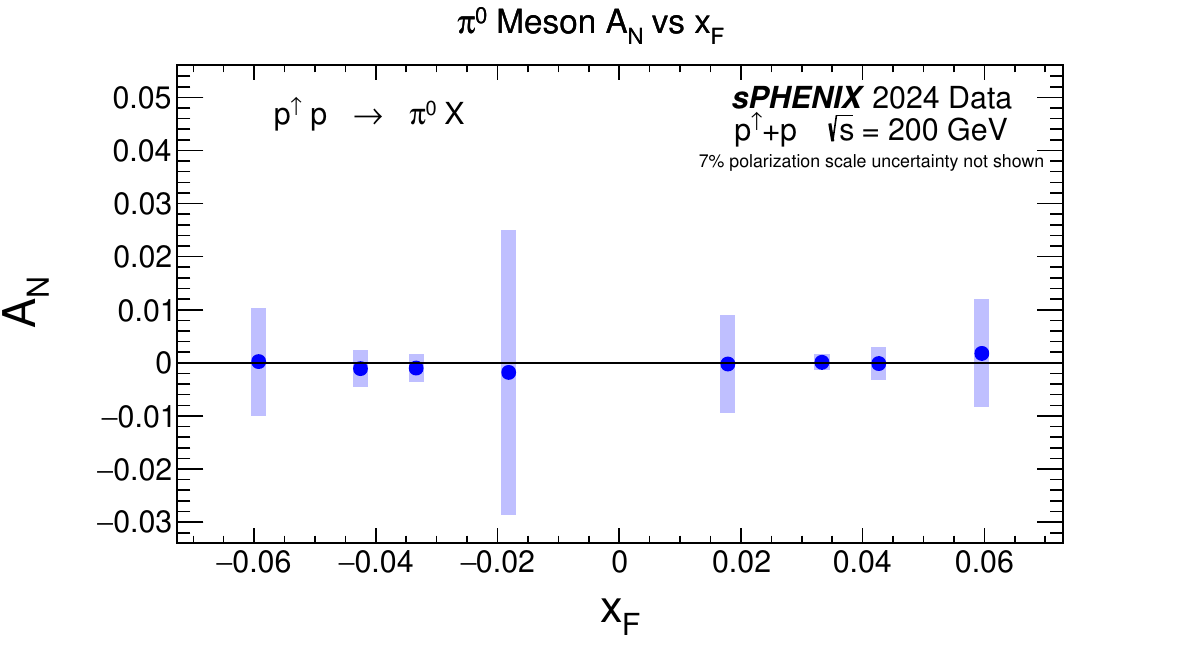}
    \caption{$A_N$ for the $\pi^0$, plotted as a function of $x_F$. The error bars correspond to statistical uncertainties, while the light shaded bands show systematic uncertainties.}
    \label{fig:results_pi0_vs_xF}
\end{figure}

\begin{figure}
    \centering
    \includegraphics[page=7, width=0.8\textwidth]{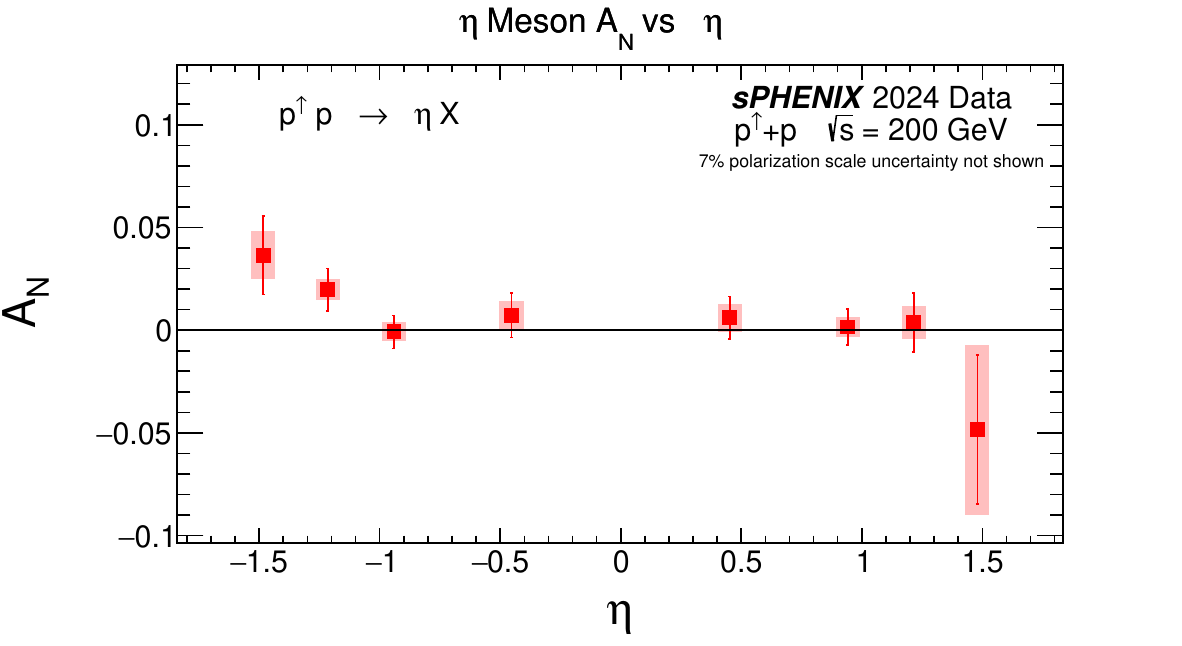}
    \caption{$A_N$ for the $\eta$-meson, plotted as a function of pseudorapidity $\eta$. The error bars correspond to statistical uncertainties, while the light shaded bands show systematic uncertainties. Here $\eta$ is defined with respect to the polarized beam.}
    \label{fig:results_eta_vs_eta}
\end{figure}

\begin{figure}
    \centering
    \includegraphics[page=8, width=0.8\textwidth]{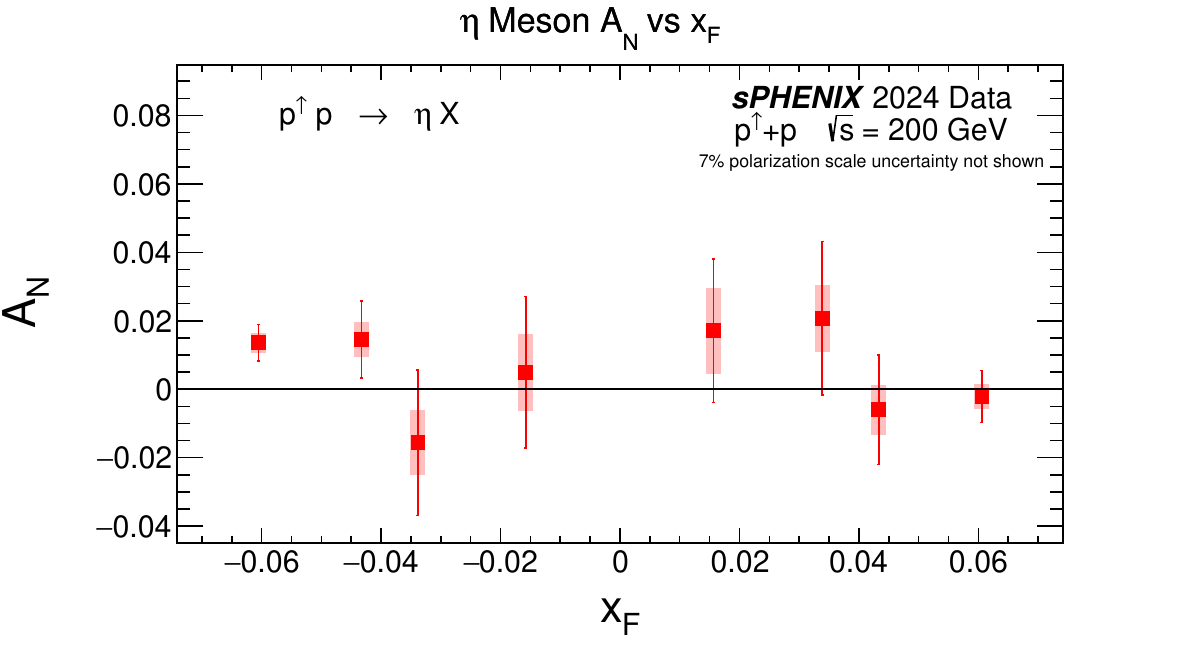}
    \caption{$A_N$ for the $\eta$-meson, plotted as a function of $x_F$. The error bars correspond to statistical uncertainties, while the light shaded bands show systematic uncertainties.}
    \label{fig:results_eta_vs_xF}
\end{figure}

In addition, the pseudorapidity- and $x$-Feynman-dependent TSSAs are shown in Figures~\ref{fig:results_pi0_vs_eta} and \ref{fig:results_pi0_vs_xF} for the $\pi^0$, and in Figures~\ref{fig:results_eta_vs_eta} and \ref{fig:results_eta_vs_xF} for the $\eta$-meson. Following the standard convention in the literature~\cite{RHICf_pi0_AN}, in these figures, positive (negative) values of $\eta$ and $x_F$ correspond to the forward (backward) direction relative to the polarized beam. These results show that both forward- and backward-going $\pi^0$ TSSAs are consistent with zero. Note the uncertainty for the $\pi^0$ TSSA is dominated by systematic uncertainty; this is due to the large $A_N^\mathrm{sys,bg}$ discussed in Section~\ref{sec:sys-eta-xF}. For the $\eta$-meson, the asymmetry at the largest negative $x_F$ deviates significantly from zero; when accounting for both statistical and systematic uncertainty, the deviation is roughly 2-$\sigma$. While such a deviation is unlikely to arise from purely statistical fluctuations, given that all other points are consistent with zero, we do not interpret this outlier as evidence of a non-zero TSSA in the backward direction. This finding is consistent with previous measurements of backward-going TSSAs for both charged and neutral hadrons~\cite{PHENIX_backward_AN}.


\begin{figure}[h]
    \centering
    \includegraphics[page=13, width=0.48\textwidth]{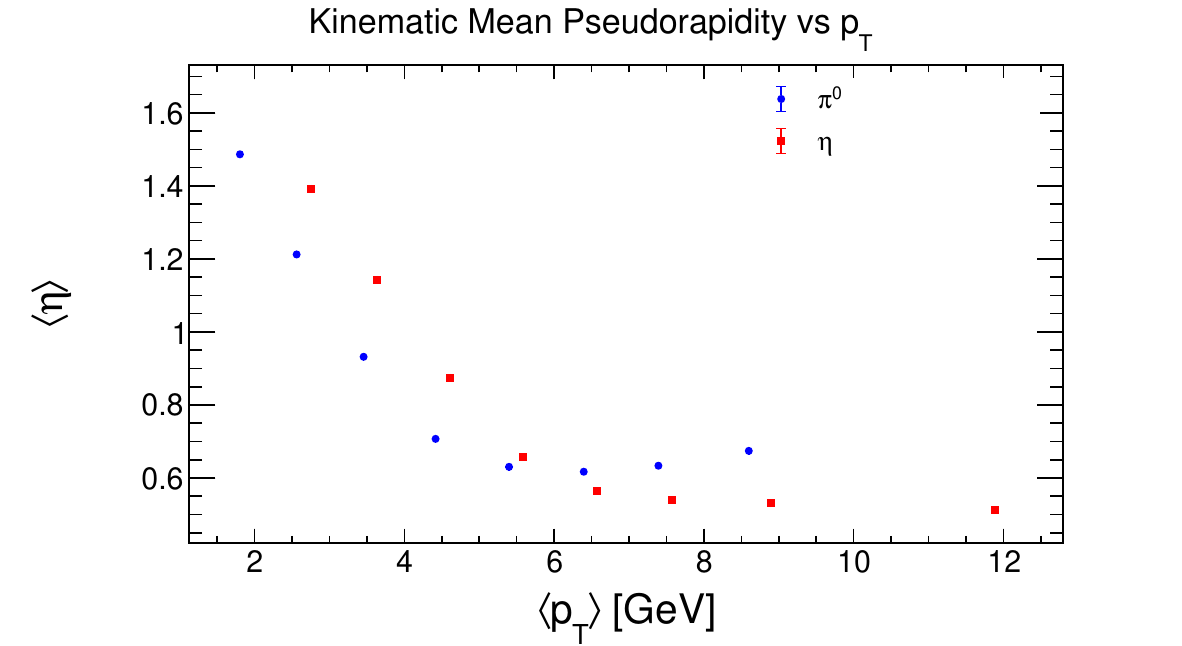}
    \includegraphics[page=14, width=0.48\textwidth]{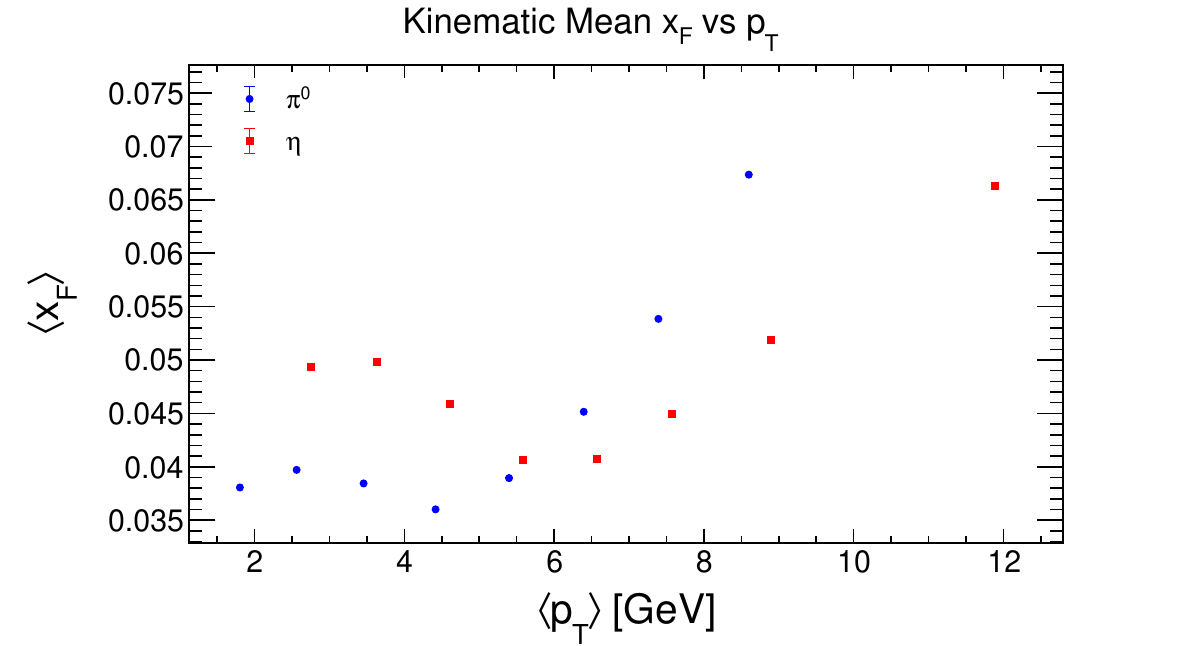}
    \caption{Kinematic mean values of pseudorapidity $\eta$ and $x_F$ in each $p_T$-bin for the $\pi^0$ and $\eta$-mesons.}
    \label{fig:results_kinematics_pT}
\end{figure}

\begin{figure}
    \centering
    \includegraphics[page=15, width=0.48\textwidth]{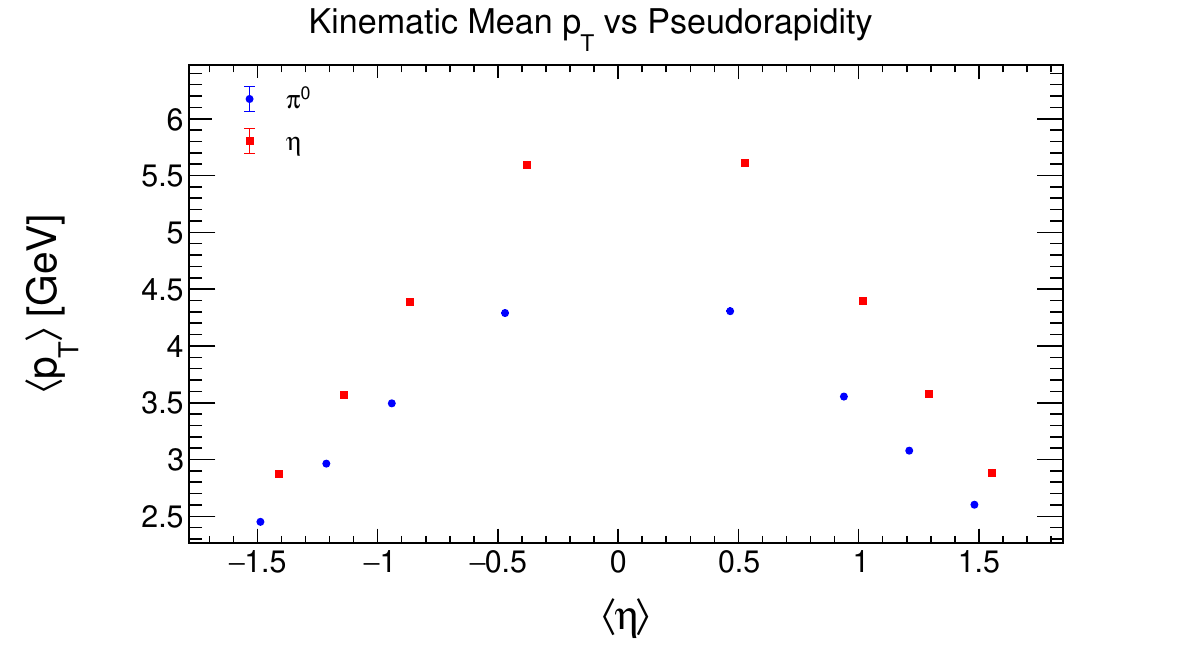}
    \includegraphics[page=16, width=0.48\textwidth]{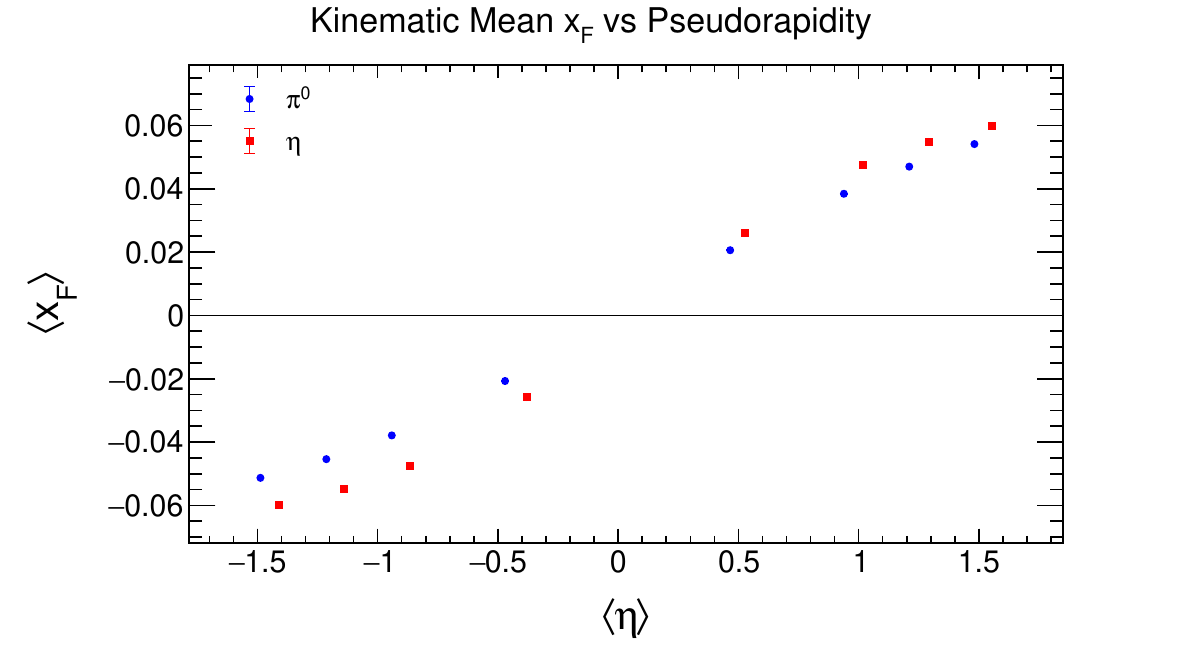}
    \caption{Kinematic mean values of $p_T$ and $x_F$ in each pseudorapidity-bin for the $\pi^0$ and $\eta$-mesons.}
    \label{fig:results_kinematics_eta}
\end{figure}

\begin{figure}
    \centering
    \includegraphics[page=17, width=0.48\textwidth]{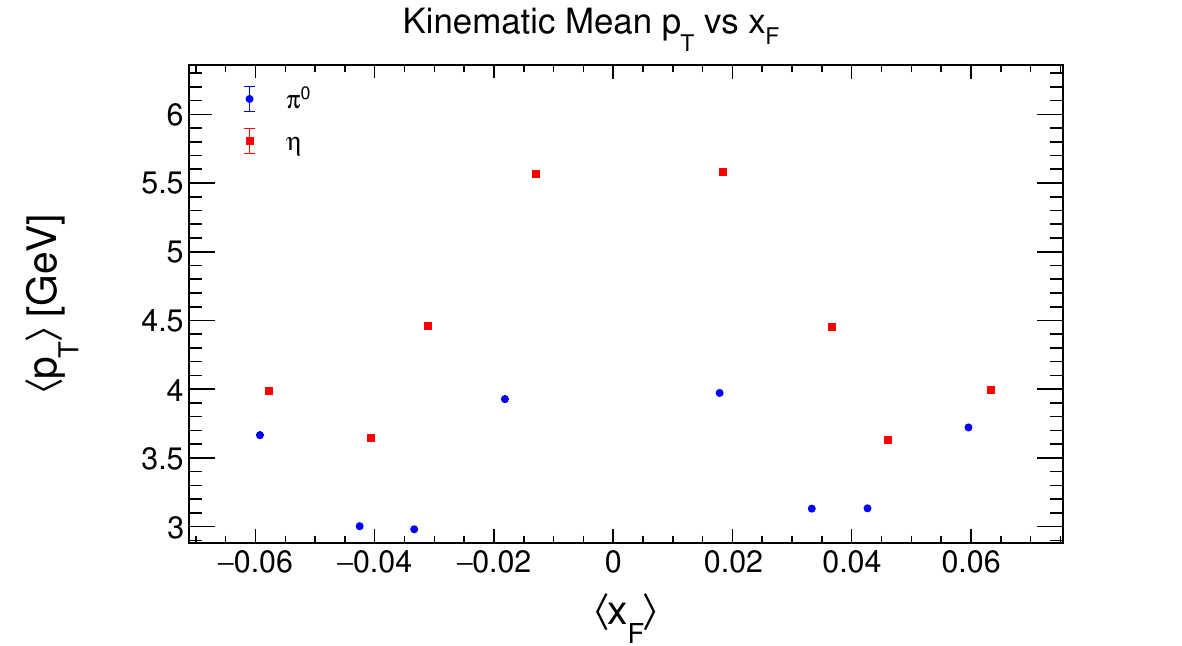}
    \includegraphics[page=18, width=0.48\textwidth]{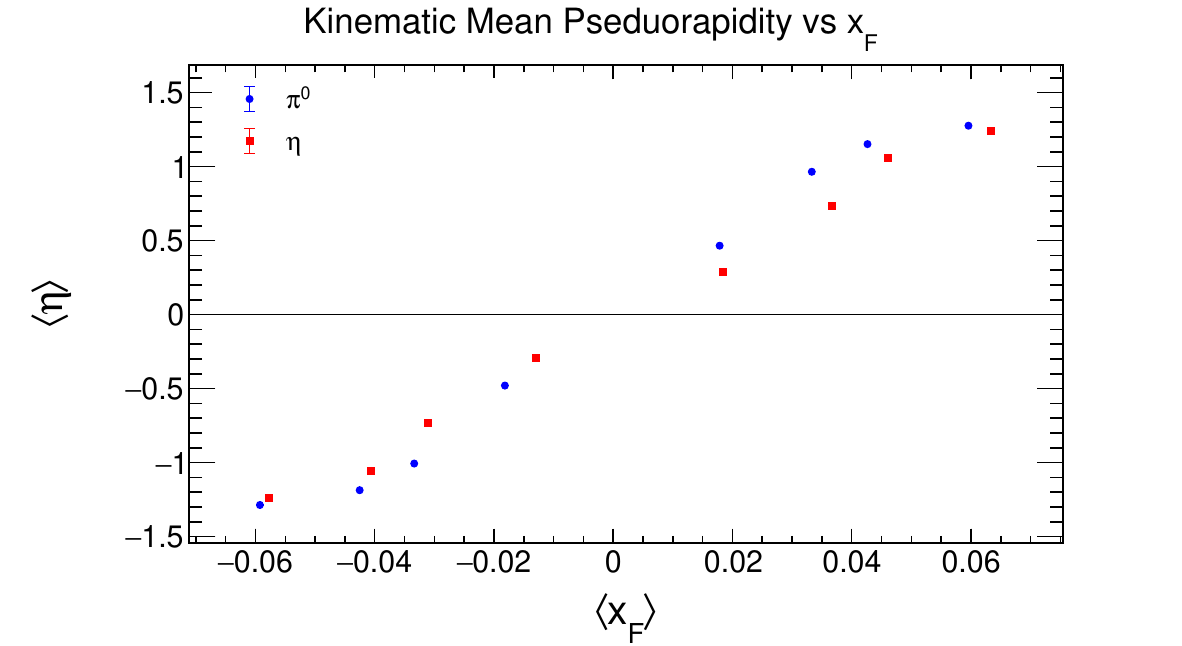}
    \caption{Kinematic mean values of $p_T$ and pseudorapidity $\eta$ in each $x_F$-bin for the $\pi^0$ and $\eta$-mesons.}
    \label{fig:results_kinematics_xF}
\end{figure}

Additional context for these results is given in Figures~\ref{fig:results_kinematics_pT}-\ref{fig:results_kinematics_xF}. These figures show the mean value of meson $p_T$, pseudorapidity, or $x_F$ in each kinematic bin. Figure~\ref{fig:results_kinematics_pT} shows the average pseudorapidity (left) and $x_F$ (right) for mesons in each $p_T$-bin; similarly Figure~\ref{fig:results_kinematics_eta} shows the average $p_T$ and $x_F$ in each pseudorapidity-bin, while Figure~\ref{fig:results_kinematics_xF} shows the average $p_T$ and $\eta$ in each $x_F$-bin. Note that these figures include both horizontal and vertical error bars. These represent the standard error on the mean, and are vanishingly small in all cases. Comparing these kinematic mean values to those shown in Figure~\ref{fig:global_pi0}, we see that the sPHENIX data covers a previously unmeasured region of the kinematic landscape. Therefore, this analysis provides beneficial new input for phenomenological models of $\pi^0$ and $\eta$-meson $A_N$ in the low-$x_F$ and moderate-$p_T$ region.


These results are consistent with previous measurements of neutral meson TSSAs in a similar kinematic region as that probed by sPHENIX. One previous measurement in particular, made by the PHENIX experiment during the 2015 RHIC run, bears several similarities with this sPHENIX measurement. As such, a more detailed comparison is given in the following section.

\clearpage

\section{Comparison with PHENIX}
\label{sec:phenix}

\begin{figure}[h]
    \centering
    \includegraphics[page=9,width=0.8\linewidth]{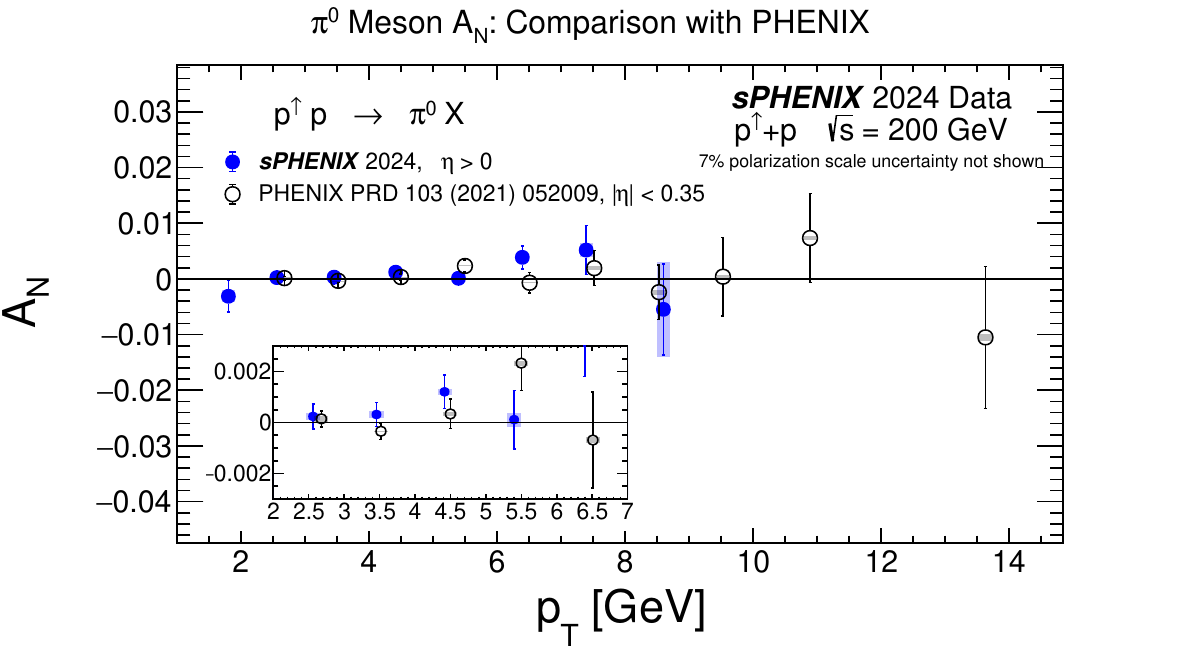}
    \caption{Beam-averaged $A_N$ for the $\pi^0$, plotted alongside the published PHENIX data \cite{PHENIX_A_N}.}
    \label{fig:pi0_pT_phenix}
\end{figure}

\begin{figure}
    \centering
    \includegraphics[page=10,width=0.8\linewidth]{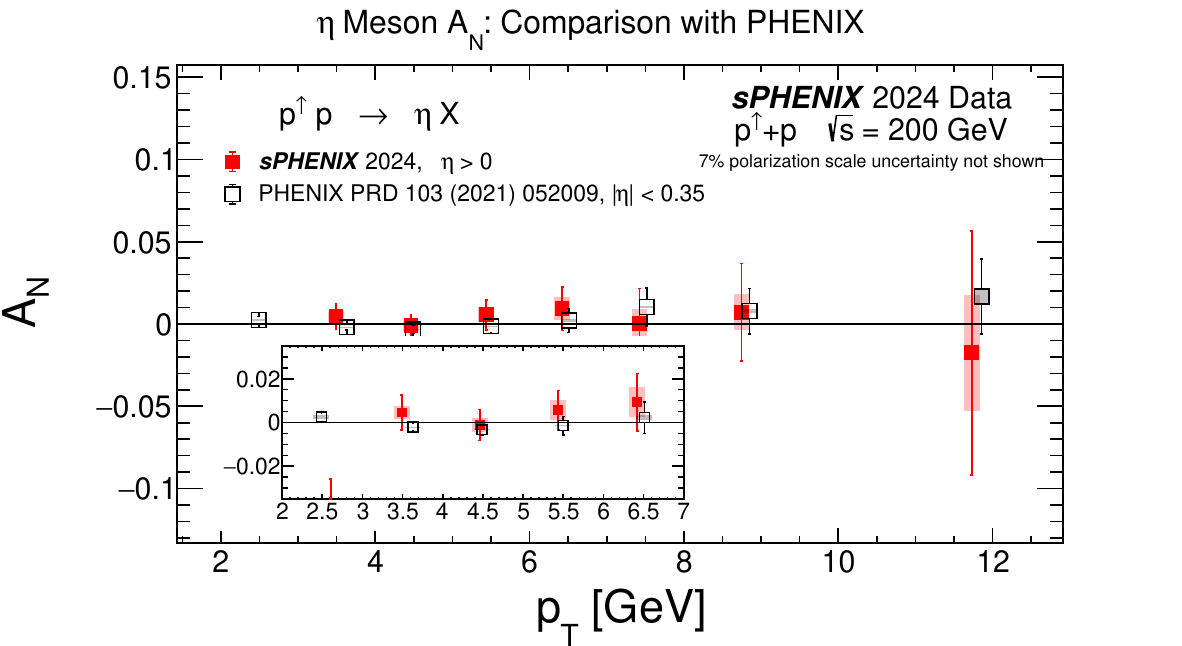}
    \caption{Beam-averaged $A_N$ for the $\eta$-meson, plotted alongside the published PHENIX data \cite{PHENIX_A_N}.}
    \label{fig:eta_pT_phenix}
\end{figure}

A comparison between the results of this analysis and the PHENIX results published in \cite{PHENIX_A_N} is shown in Figure~\ref{fig:pi0_pT_phenix} for the $\pi^0$ and Figure~\ref{fig:eta_pT_phenix} for the $\eta$-meson. Overall, both sets of results are consistent with a zero asymmetry measurement. Of particular note, this analysis is able to extend the $\pi^0$ $A_N$ measurement to lower $p_T$ than previously achieved by PHENIX\footnote{An earlier measurement by PHENIX did include $p_T$ as low as 1 GeV\cite{PhysRevLett.95.202001}}. However, the uncertainties in the PHENIX results are smaller than those in this analysis. This can be attributed to a combination of factors; key differences between the two measurements are detailed below.


\begin{itemize}
    \item \underline{Detector acceptance}: The mid-rapidity PHENIX electromagnetic calorimeter covered the pseudorapidity range $|\eta| < 0.35$. It was segmented into two arms, each covering an azimuthal angle of $\Delta \phi = \pi/2$. In contrast the sPHENIX EMCal has full azimuthal coverage and a nominal pseudorapidity range of $|\eta| < 1.1$. As noted in Section~\ref{sec:diphoton_cuts}, the effective range of reconstructed di-photons in this analysis is $|\eta| < 3.0$.
    \item \underline{Pseudorapidity selection}: As a consequence of the limited pseudorapidity acceptance of their electromagnetic calorimeter, the PHENIX analysis does not distinguish between mesons produced forward versus backward relative to the polarized beam; it includes all mesons in $|\eta| < 0.35$. In contrast, this analysis considers the $p_T$-dependence of only those mesons produced forward of the polarized beam, i.e. $0 < \eta < 3.0$.
    \item \underline{Trigger configuration}: Both analyses rely on their respective experiment's high-energy photon triggers to select events. However, while the lowest-energy photon trigger in sPHENIX in 2024 was at 3 GeV, the corresponding PHENIX trigger in 2015 was at only 1.5 GeV. Due to the trigger-cluster matching requirement (see Section~\ref{sec:cluster_cuts}), this implies that the available statistics at low-$p_T$ in the sPHENIX data set is significantly reduced compared to that of PHENIX.
    \item \underline{$\pi^0$ cluster merging}: The PHENIX electromagnetic calorimeter system was composed of two distinct types of calorimeter: six sectors of lead-scintillator with segmentation $\Delta\eta\times\Delta\phi = 0.011\times0.011$, and two sectors of lead-glass with segmentation $0.008\times0.008$. In both cases, the PHENIX segmentation is finer than that of the sPHENIX EMCal ($0.024\times0.024$). In addition, the PHENIX electromagnetic calorimeter was positioned roughly three times further from the beamline than that of sPHENIX. Accordingly, the physical distance between neighboring towers was significantly larger in PHENIX. Both of these factors imply that for two photons from a $\pi^0$ decay with fixed angular separation $\Delta R$, the number of towers separating the edges of their respective clusters is greater in PHENIX than in sPHENIX. Thus, PHENIX was less sensitive to the effect of cluster merging and was able to reliably reconstruct $\pi^0$s up to significantly higher transverse momentum than is possible in this analysis. Note that this effect is only relevant for the $\pi^0$: due to its larger mass, cluster merging is not a limiting factor for $\eta$-meson reconstruction.
    \item \underline{Background Correction}: While both analyses use the same method of correcting for background contamination (Equation~\ref{eq:bkgrcorr}), the background fractions measured by PHENIX are significantly lower than those observed in sPHENIX. The PHENIX background fractions for the $\pi^0$ ($\eta$-meson) ranged from 6-10\% (47-71\%), compared to 26-61\% (82-94\%) in sPHENIX. This is due in large part to the charged track rejection used by PHENIX. The PHENIX analysis used information from tracking detectors to veto clusters which could be geometrically matched to charged particle tracks, thereby reducing the background contribution due to clusters originating from electrons, positrons and hadronic interactions in the electromagnetic calorimeter. In contrast, the sPHENIX tracking system was still in the commissioning phase for the majority of the 2024 RHIC run. At present, high-quality track information is only available in a small subset of the sPHENIX data. Accordingly, this analysis is unable to reject charged tracks and therefore the meson signal-to-background ratio is much poorer than that of the PHENIX analysis. Accordingly, the background correction disproportionately enhances the statistical uncertainties in this analysis.
    \item \underline{Data volume}: The data set used by PHENIX comprises an integrated luminosity of approximately 60 $\mathrm{pb}^{-1}$. In contrast, the sPHENIX data used in this analysis is estimated to comprise roughly 35 $\mathrm{pb}^{-1}$ of the roughly 107 $\mathrm{pb}^{-1}$ total collected in 2024. While the sPHENIX integrated luminosity is smaller by nearly a factor of two, it should be noted that this is offset by the larger sPHENIX EMCal acceptance (roughly a factor of six). These factors act in opposite directions, so one may naively expect the total sPHENIX statistics to be larger by a factor of three. However, this does not translate directly to a reduction of statistical uncertainties due to the other considerations listed above.
\end{itemize}

Despite these differences, the PHENIX analysis is the existing published data most similar to this analysis. Both experiments operated using the same accelerator with the same polarized-proton beam configuration and the same center-of-mass energy; both measured particle production in the ``barrel'' or mid-rapidity region, in contrast with other experiments measuring TSSAs in the forward or far-forward regions~\cite{STAR_pi0_AN, STAR_pi0_eta_AN, RHICf_neutron_AN, STAR_jet_AN2}; and both analyses consider similar formulas for calculating $A_N$, and the same sources of systematic uncertainty. That the two analyses yield consistent results, despite the differences in kinematical coverage, will provide valuable information for theorists in understanding these asymmetries.

\chapter{Summary and Outlook}\label{ch:summary}
\section{Summary}
This thesis has presented a new measurement of transverse single-spin asymmetries in inclusive $\pi^0$- and $\eta$-meson production in $\sqrt{s} = 200$ GeV $p^\uparrow + p$ collisions. This is the first measurement of spin-dependent asymmetries at the sPHENIX experiment, currently the world's newest collider detector. As mentioned in Chapter~\ref{ch:intro}, this measurement also constitutes the first public sPHENIX result using the 2024 data~\cite{ConfNote}.

The current theoretical understanding of transverse-spin-dependent phenomena in nucleon scattering processes has been summarized. At present, the origins of TSSAs have not been conclusively identified, but are believed to be related to transverse parton momentum and spin correlations as well as multi-parton interference effects. This measurement, probing a previously unexplored region of the kinematic landscape, will add valuable information to the global TSSA data and aid phenomenologists in better understanding how these effects contribute to the observed asymmetries.

The analysis procedure has been described in detail, beginning with the reconstruction of $\pi^0$- and $\eta$-mesons from clusters in the sPHENIX electromagnetic calorimeter. The raw asymmetry is extracted for each meson using two distinct calculation methods. The raw asymmetry is then corrected for beam polarization and background contamination, yielding a physics asymmetry $A_N$. Systematic uncertainties related to the calculation method, background correction, and detector effects are assessed.

The final measured TSSAs have been presented in bins of transverse momentum, pseudorapidity and $x$-Feynman. These asymmetries are consistent with zero in the kinematic region accessible with sPHENIX. This result is consistent with previous measurements in the mid-rapidity region.

\section{Future Analysis Opportunities}
While this analysis is complete in its current form, there are several opportunities for potential improvements, listed below.

\begin{itemize}
    \item \underline{Unanalyzed data}: While sPHENIX collected an estimated 107 pb${}^{-1}$ of collision data in 2024, only about 50\% of this data is presently available due to issues with data production. As these issues are resolved, more data will become available to incorporate into the analysis. The final physics publication associated with this analysis will use the full 2024 data set. As the statistical uncertainty on $A_N$ scales as $1/\sqrt{N}$, the final uncertainties are expected to be scaled by roughly $1/\sqrt{2} \approx 71\%$.

    In addition, the current iteration of the analysis uses only photon-triggered events. One area for potential improvement is to also make use of minimum bias events. Doing so can help to improve meson statistics at low-$p_T$. However this must be handled carefully as to avoid introducing a physics bias by using multiple triggers to measure the same kinematic region. A careful study is needed to determine whether the minimum bias trigger would offer an increase in statistics compared to the photon triggers at low-$p_T$.
    \item \underline{Constant offset in relative luminosity asymmetry}: As discussed in Section~\ref{sec:rawresults}, a constant offset is observed in the $\phi$-dependent raw asymmetries, most prominently at low-$p_T$. Further study is needed to understand this behavior. Understanding, and, if possible, controlling this offset would give us greater confidence in the relative luminosity asymmetry calculation.
    \item \underline{Meson signal-to-background ratio}: The unfavorably high background fraction, particularly for the $\eta$-meson, is a dominating source of statistical uncertainty in this analysis. There are several ways to potentially improve on this:
    \begin{itemize}
        \item Using tracking detector information to reject clusters matching charged particle tracks, as done by PHENIX~\cite{PHENIX_A_N}, would significantly reduce background contributions. However, as much less data was collected in 2024 with the sPHENIX tracking system fully operational (only around 13 pb${}^{-1}$), a study is necessary to determine whether the reduced background fraction would compensate for the reduction in available statistics.
        \item As mentioned in Section~\ref{sec:cluster_cuts}, the cluster $\chi^2$ metric has been found to be an ineffective discriminator between single-photon showers and background. However, the authors of the sPHENIX $\pi^0$ and $\eta$-meson cross section and spectra analysis~\cite{ian-ppg11} have developed an alternative metric for this purpose, using a more modern method which is better tuned for sPHENIX. Incorporating this updated cluster probability metric into this analysis is an item of ongoing study.
        \item A final possibility for reducing the background fractions found in this analysis is to vary the size of the meson signal windows. The current version of the analysis uses a 3-$\sigma$ window around the respective meson peaks. Instead using a smaller window, e.g. 2- or 1.5-$\sigma$, would result in a higher signal-to-background ratio, at the cost of losing some true meson statistics. A detailed study is needed to determine what size signal window minimizes the statistical uncertainty on $A_N$.
    \end{itemize}
    \item \underline{Background systematic uncertainty}: The alternative di-photon mass background fitting procedure described in Section~\ref{sec:sys-bg} was tuned primarily to give reasonable background predictions in the $\eta$-meson region. Further tuning in the $\pi^0$ region, or complete separation of the two regions, may yield better agreement between the main and alternative fitting methods. This would in turn reduce the dominant systematic uncertainty in the $\eta$ and $x_F$ binnings.
    
    Another possibility for these binnings is to instead use an alternative fit based on a polynomial background term. The first iteration of this analysis used such a fit and found unacceptable disagreement between data and model mass distributions for the $p_T$ binning. However in the $\eta$ and $x_F$ binnings, the variation in the mass distributions is far less pronounced, and the polynomial fit may provide results more consistent with the main analysis threshold function fit.
    \item \underline{Comparison to theoretical predictions}: While phenomenological curves for meson $A_N$ in RHIC $p+p$ collisions exist~\cite{PHENIX_A_N}, these predictions pertain to the PHENIX detector acceptance ($|\eta| < 0.35$). The sPHENIX analyzers will contact theory groups to obtain updated curves for the sPHENIX kinematic region, so a theory-data comparison may be included in the final physics publication.
\end{itemize}

We conclude by noting some prominent examples of other analyses enabled by the 2024 sPHENIX data. In addition to the meson $A_N$ measured in this analysis, a parallel measurement for jet $A_N$ is currently in progress in sPHENIX. Unlike hadron asymmetry measurements, the jet asymmetry is not sensitive to the Collins effect. Therefore the jet $A_N$ measurement offers a unique probe of the Sivers effect and related CT3 interference effects. A similar $A_N$ measurement may also be performed for direct photon production, i.e. for photons produced directly in the hard scattering process. Related measurements of spin-dependent azimuthal correlations between di-hadrons and di-jets are also possible. Finally, information from the tracking system combined with calorimeter jet detection can be used to study identified hadron-in-jet azimuthal correlations, thereby probing the TMD Collins effect directly.


\backmatter

\printbibliography[heading=bibintoc,title={References}]

\clearpage
\setcounter{counterforappendices}{\value{page}}
\mainmatter
\setcounter{page}{\value{counterforappendices}}

\appendix

\chapter{Backward-Going Asymmetries}\label{app_backward}
\begin{figure}
    \centering
    \includegraphics[page=237, width=0.48\linewidth]{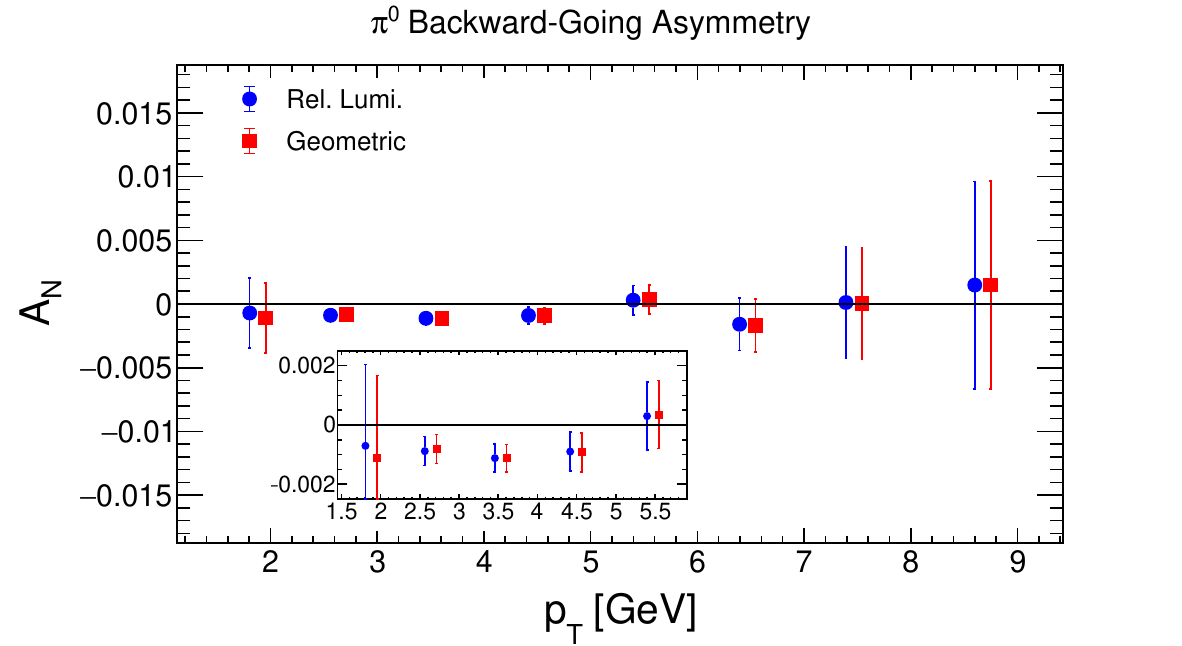}
    \includegraphics[page=303, width=0.48\linewidth]{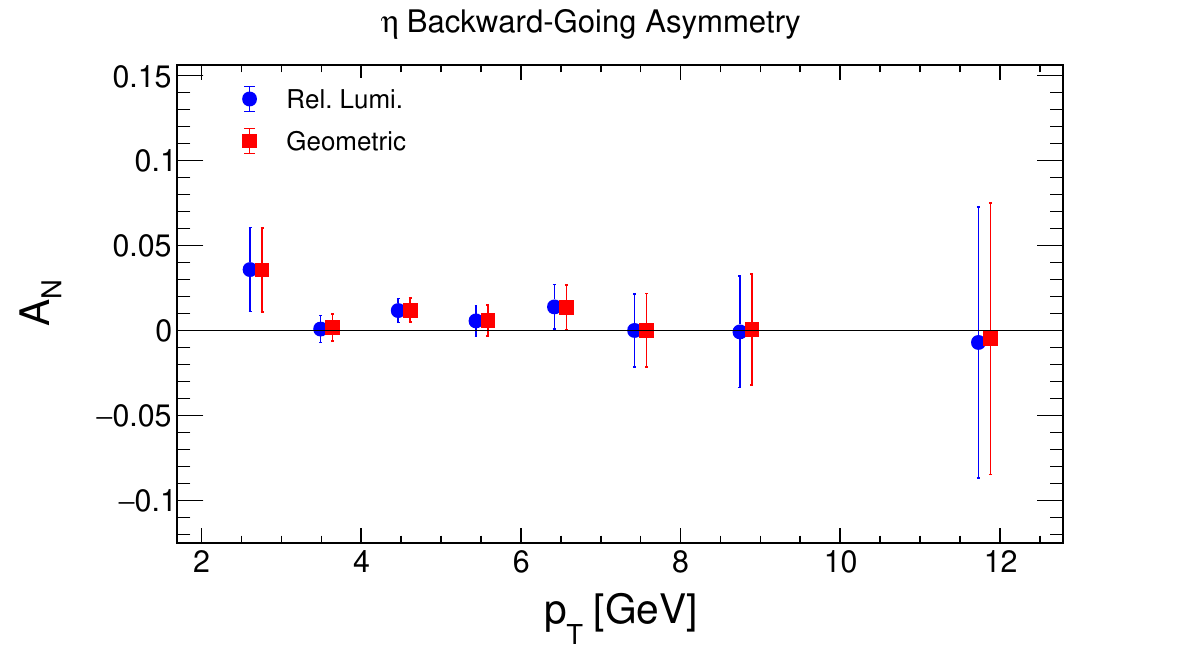}
    \caption{Beam-averaged $A_N$ for $\pi^0$ (left) and $\eta$ (right) mesons produced in the backward direction relative to the polarized beam.}
    \label{fig:backward}
\end{figure}

The corrected and beam-averaged asymmetries for mesons produced in the backward direction relative to the polarized beam are shown in Figure~\ref{fig:backward}. From prior measurements at negative $x_F$ \cite{PHENIX_backward_AN}, these backward asymmetries are expected to vanish. Most of the measured asymmetries are compatible with zero at or just above the 1-$\sigma$ level. The most significant deviation differs from zero at the 1.5-$\sigma$ level.

\begin{figure}
    \centering
    \includegraphics[page=381, width=0.48\linewidth]{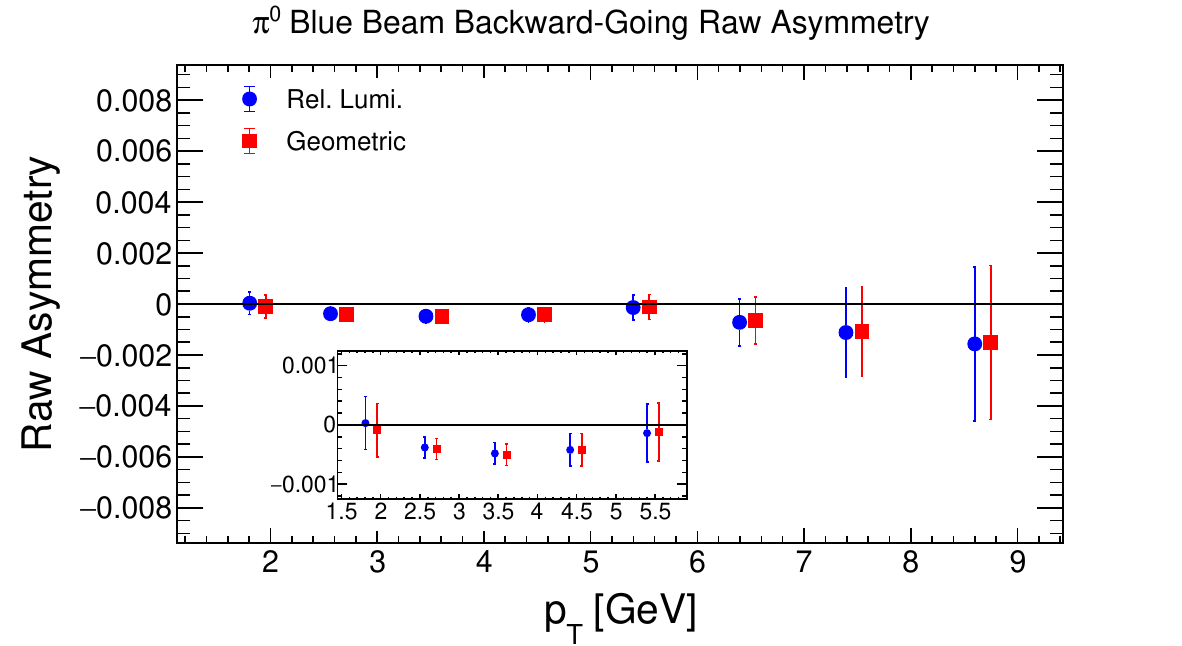}
    \includegraphics[page=382, width=0.48\linewidth]{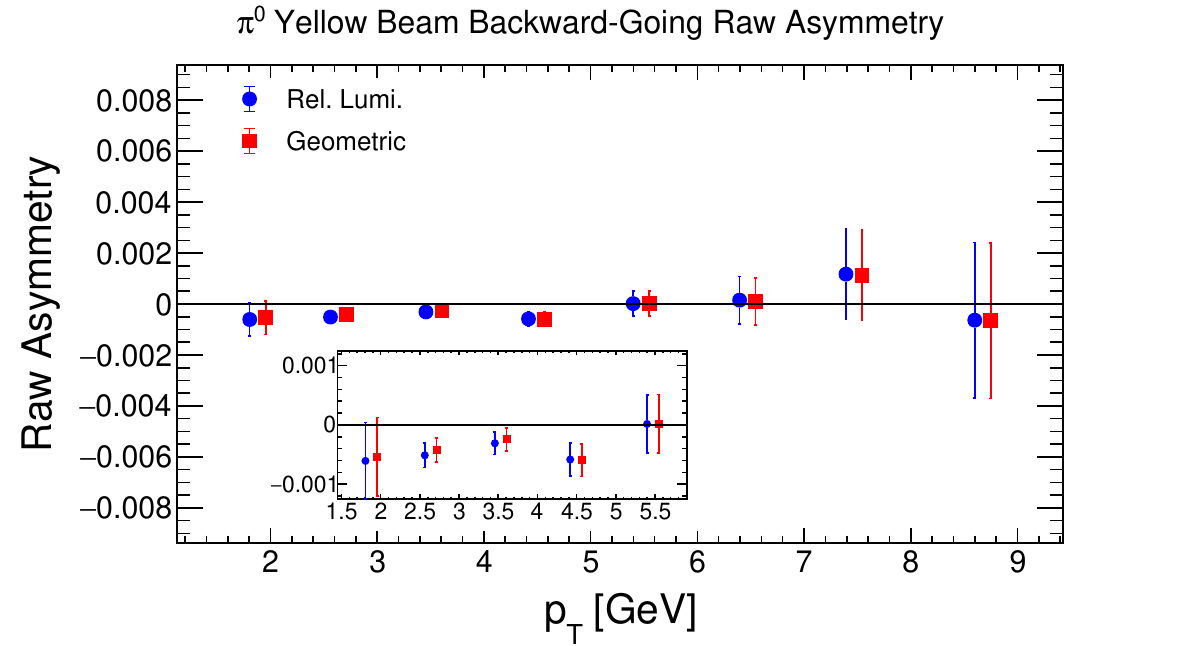}
    \caption{Raw asymmetries for di-photons falling within the $\pi^0$ signal region, produced in the backward direction relative to the Blue beam (left) and Yellow beam (right).}
    \label{fig:pi0_raw_backward}
\end{figure}

\begin{figure}
    \centering
    \includegraphics[page=383, width=0.48\linewidth]{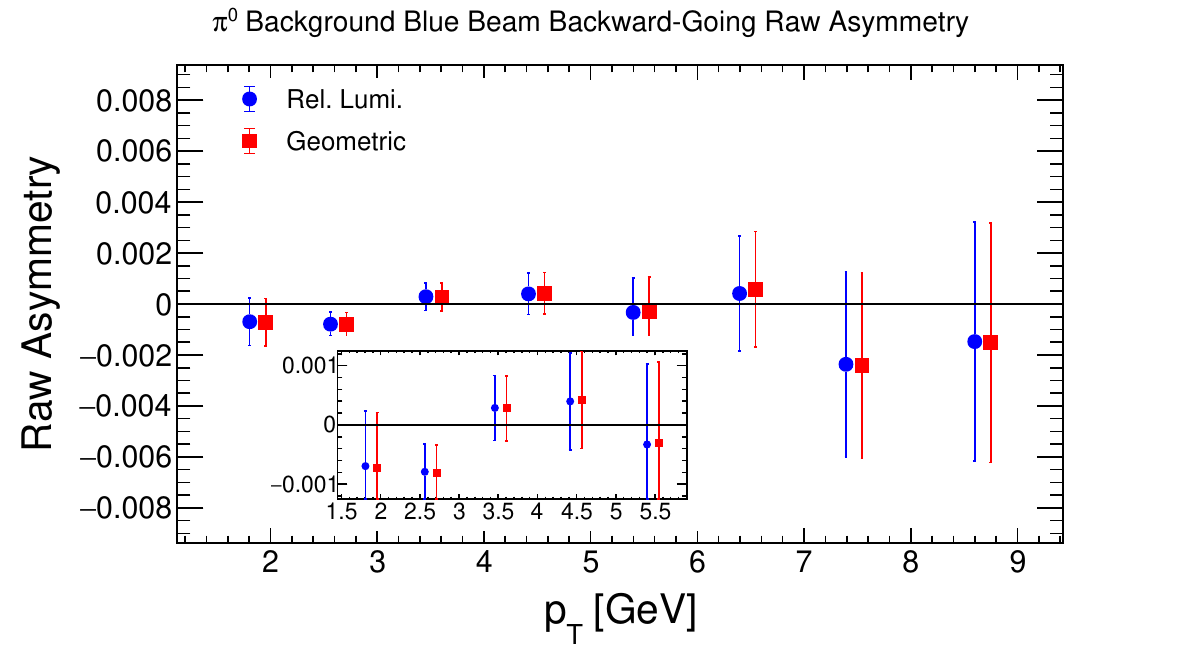}
    \includegraphics[page=384, width=0.48\linewidth]{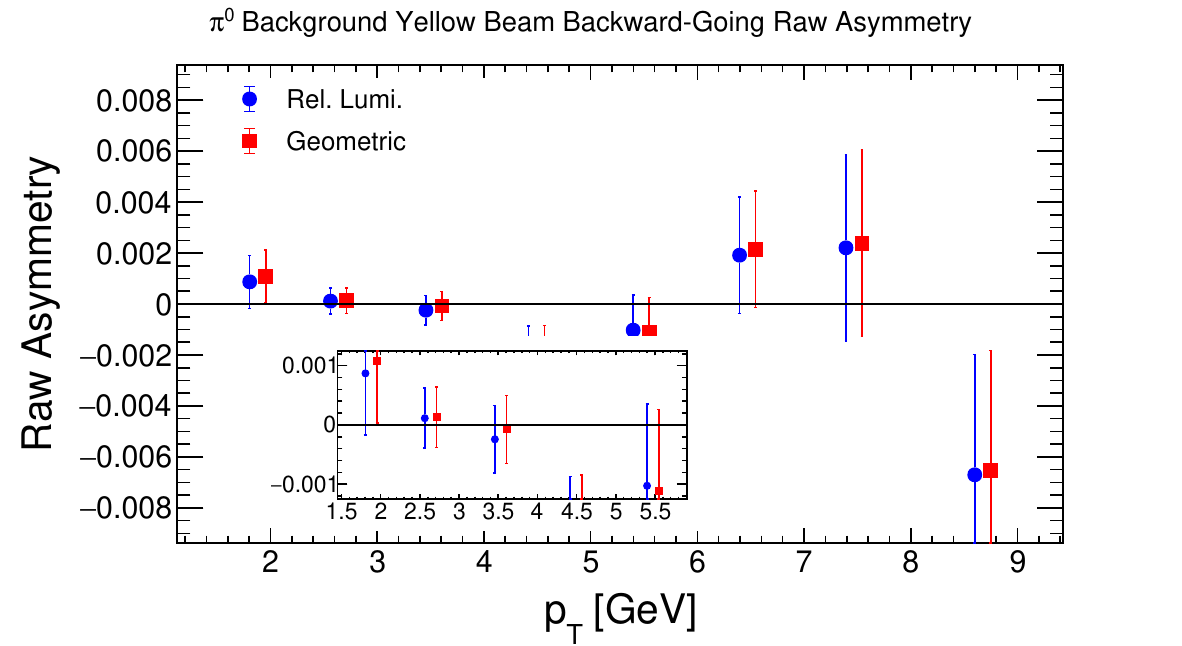}
    \caption{Raw asymmetries for di-photons falling within the $\pi^0$ background regions, produced in the backward direction relative to the Blue beam (left) and Yellow beam (right).}
    \label{fig:pi0bkgr_raw_backward}
\end{figure}

\begin{figure}
    \centering
    \includegraphics[page=389, width=0.48\linewidth]{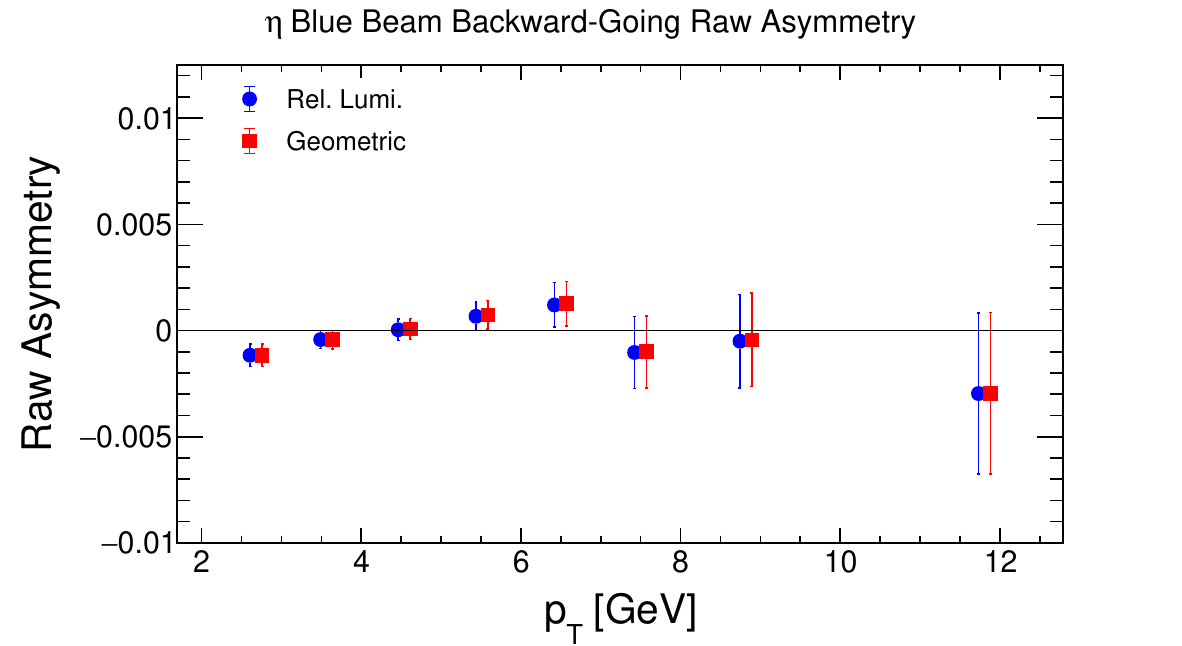}
    \includegraphics[page=390, width=0.48\linewidth]{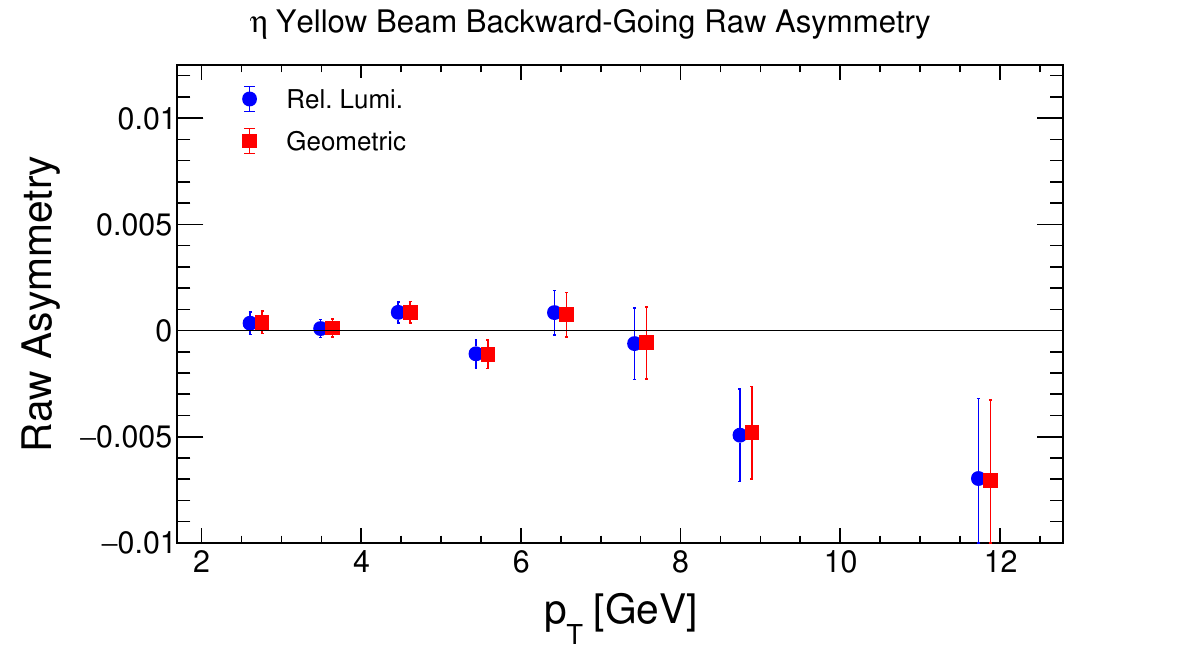}
    \caption{Raw asymmetries for di-photons falling within the $\eta$-meson signal region, produced in the backward direction relative to the Blue beam (left) and Yellow beam (right).}
    \label{fig:eta_raw_backward}
\end{figure}

\begin{figure}
    \centering
    \includegraphics[page=391, width=0.48\linewidth]{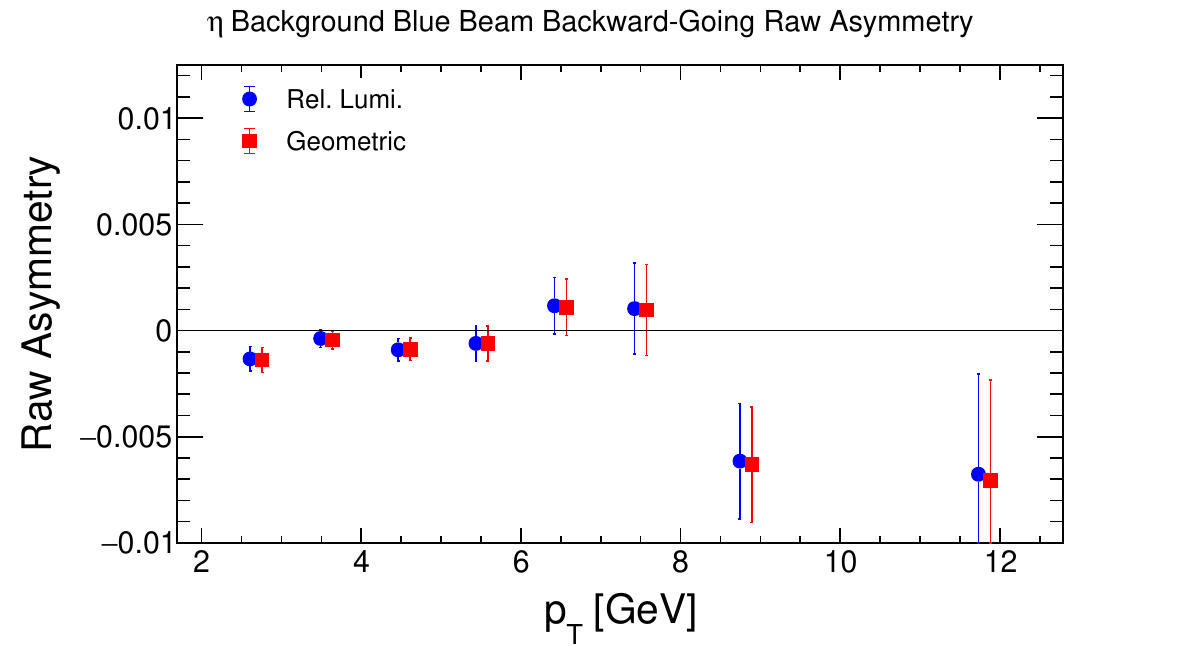}
    \includegraphics[page=392, width=0.48\linewidth]{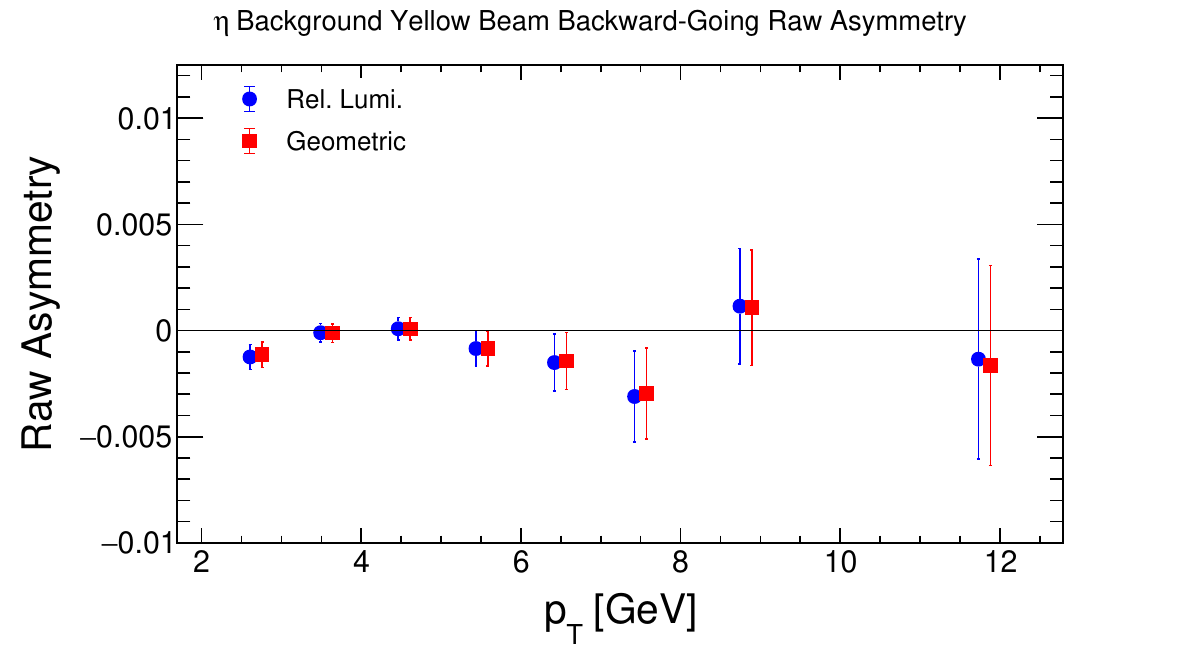}
    \caption{Raw asymmetries for di-photons falling within the $\eta$-meson background regions, produced in the backward direction relative to the Blue beam (left) and Yellow beam (right).}
    \label{fig:etabkgr_raw_backward}
\end{figure}

\begin{figure}
    \centering
    \includegraphics[page=385, width=0.48\linewidth]{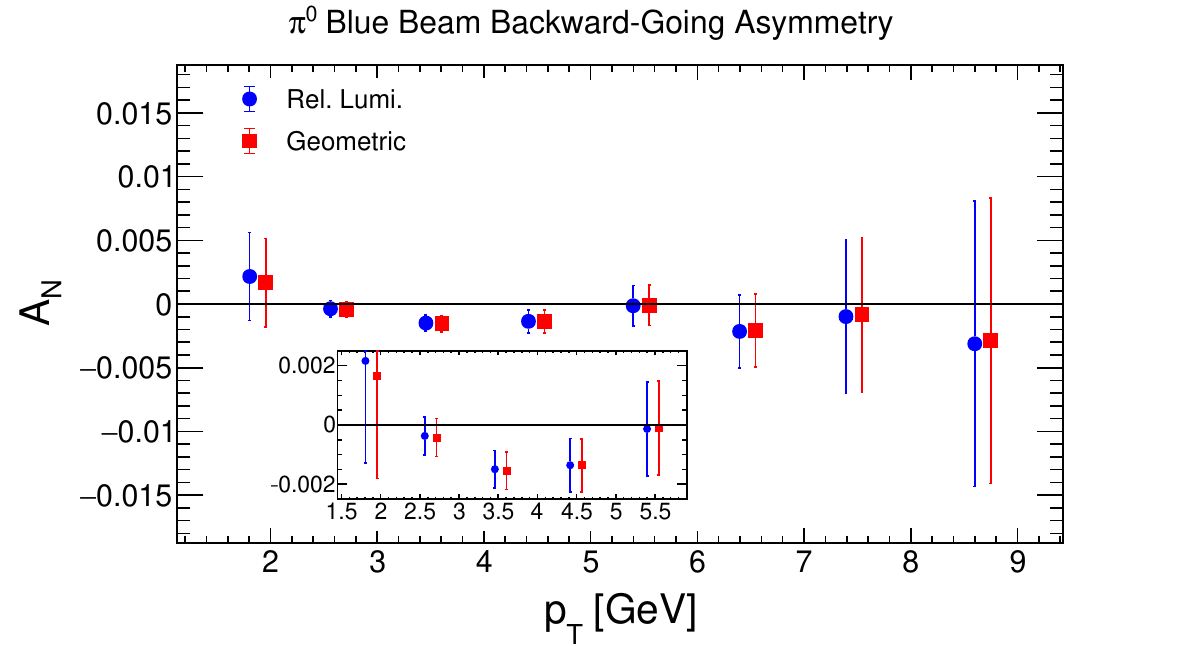}
    \includegraphics[page=386, width=0.48\linewidth]{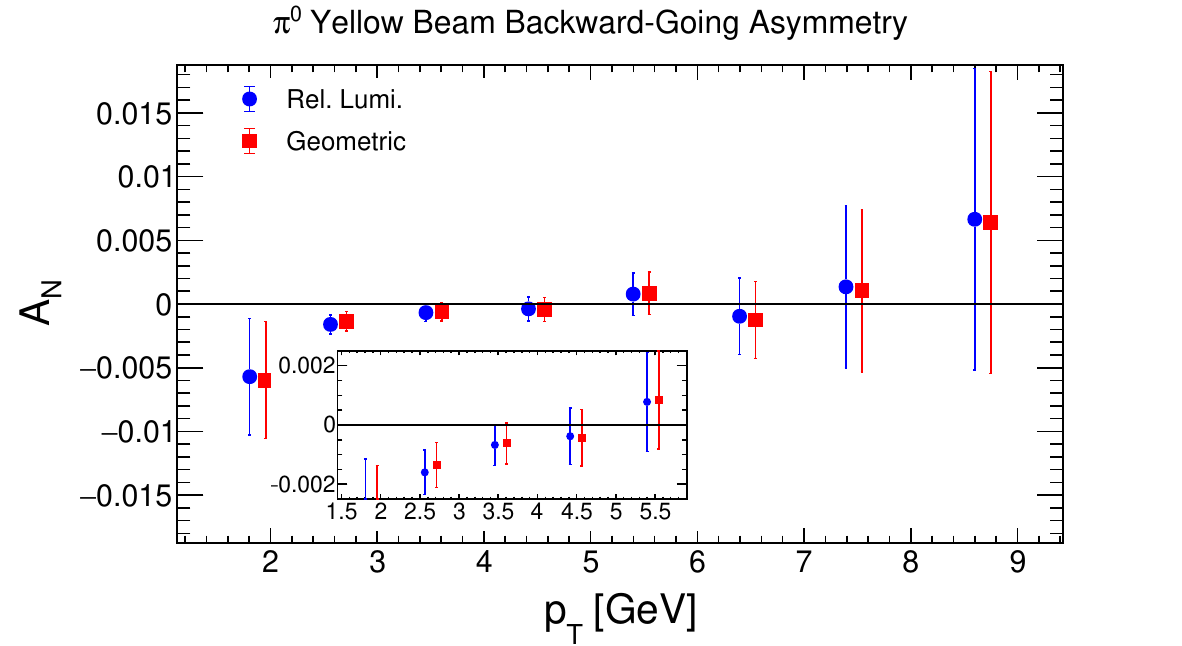}
    \caption{Corrected $A_N$ for $\pi^0$s produced in the backward direction relative to the Blue beam (left) and Yellow beam (right).}
    \label{fig:pi0_blueyellow_backward}
\end{figure}

\begin{figure}
    \centering
    \includegraphics[page=393, width=0.48\linewidth]{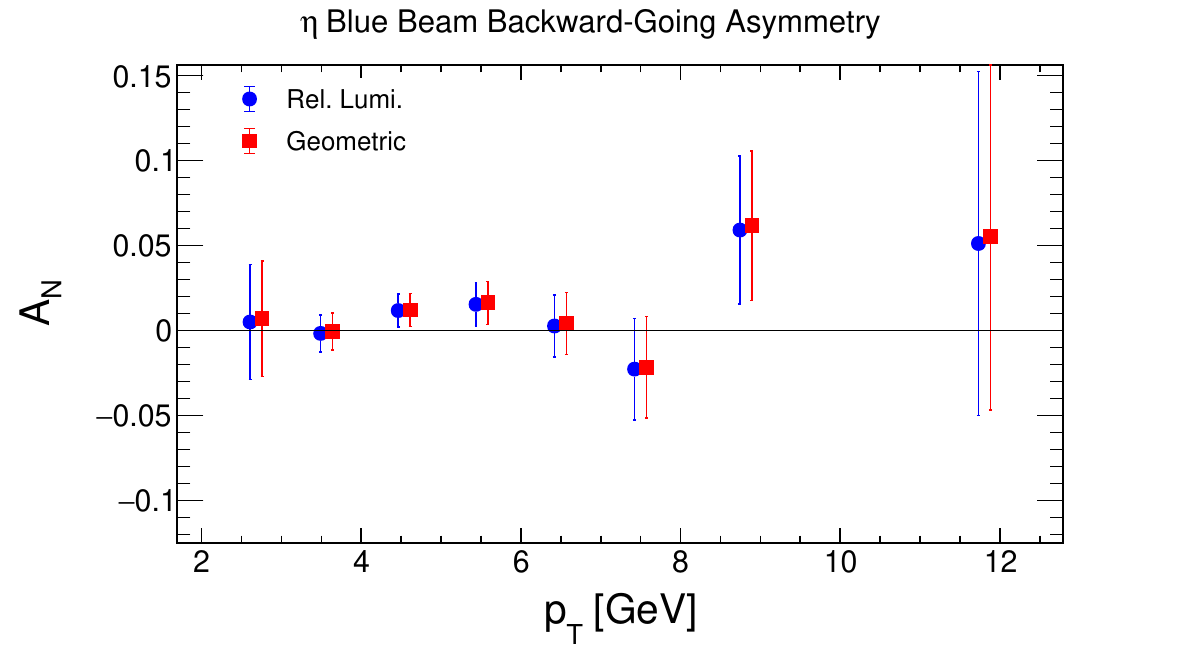}
    \includegraphics[page=394, width=0.48\linewidth]{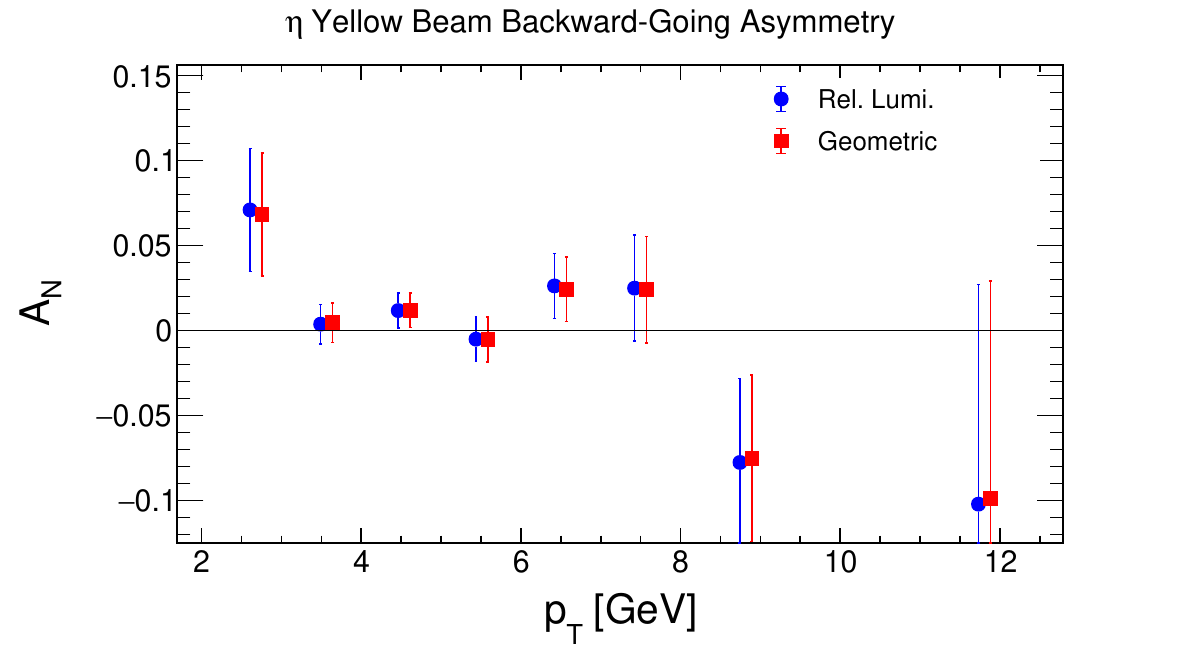}
    \caption{Corrected $A_N$ for $\eta$-mesons produced in the backward direction relative to the Blue beam (left) and Yellow beam (right).}
    \label{fig:eta_blueyellow_backward}
\end{figure}

The raw asymmetries entering into these corrected asymmetries are shown in Figures~\ref{fig:pi0_raw_backward}-\ref{fig:pi0bkgr_raw_backward} for the $\pi^0$ and Figures~\ref{fig:eta_raw_backward}-\ref{fig:etabkgr_raw_backward} for the $\eta$-meson. The corrected asymmetries are shown for Blue and Yellow beams individually in Figure~\ref{fig:pi0_blueyellow_backward} for the $\pi^0$ and Figure~\ref{fig:eta_blueyellow_backward} for the $\eta$-meson. As in the beam-averaged results, we see most of these asymmetries are compatible with zero at the 1-$\sigma$ level. Only 20 out of 96 data points deviate from zero by significantly more than 1-$\sigma$, and no points deviate by more than 2-$\sigma$. This deviation is within expectation from statistical fluctuation, and we conclude that no non-zero asymmetry is observed in the backward direction.

\clearpage

\end{document}